\documentclass[12pt,twoside]{report}
\usepackage[pdftex]{graphicx}     	
\usepackage{mystyles/mythesis}  	
\usepackage{graphicx}
\usepackage[pdftex]{geometry}
\usepackage{amssymb}             
\usepackage{mystyles/mydef}      
\usepackage{url}                 
\usepackage{afterpage}           
\usepackage{mystyles/deluxetable}   
\usepackage{mystyles/natbib}     
\usepackage{setspace}            
\usepackage{hyperref}            
\usepackage{lscape}              

\newcommand{\smallstretch}{1.0} 
\newcommand{\textstretch}{1.10}  	
\newcommand{\foretextstretch}{1.20}  
\newcommand{\gparindent}{3ex}    
\newcommand{\Msun}[0]{\ensuremath{M_{\odot}}}

\begin{document}
\renewcommand{\baselinestretch}{\textstretch}
\pagenumbering{roman}
\pagestyle{empty}
\renewcommand{\baselinestretch}{1.2}\small\normalsize
\vspace*{-0.2in}
\thispagestyle{empty}
\begin{center}
\rule[0pt]{5.6in}{1pt}
\vskip -12pt
\rule[0pt]{5.8in}{3pt}
\vskip 23pt
{\LARGE \sc Radio Pulsars in Binary Systems}
\vskip 13pt
\rule[0pt]{5.8in}{3pt}
\vskip -14pt
\rule[0pt]{5.6in}{1pt}
\end{center}
\vspace{20pt}
\begin{center}
{\Large Ren\'e Paul Breton}\\
\vspace{20pt}
Department of Physics \\
McGill University \\
Montr\'eal, Qu\'ebec \\
Canada \\
\vspace{20pt}
December 2008\\
\end{center}
\vfill
\vfill
\begin{center}
A Thesis submitted to McGill University\\
in partial fulfillment of the requirements of the degree of\\
Doctor of Philosophy\\
\end{center}
\vspace{0.0in}
\begin{center}
{\footnotesize \copyright\ Ren\'e P. Breton, 2008}
\end{center}

\cleardoublepage
\renewcommand{\baselinestretch}{\smallstretch}\small\normalsize
\pagestyle{empty}

\vspace*{3cm}

\begin{center}
Pour mon p\`ere, ma m\`ere et mon fr\`ere.\\
Votre support depuis toujours est un cadeau inestimable.
\end{center}

\cleardoublepage
\pagestyle{headings}
\tableofcontents
\newpage
\listoffigures
\newpage
\listoftables
\newpage
\renewcommand{\baselinestretch}{\smallstretch}\small\normalsize
\chapter*{Abstract}
\addcontentsline{toc}{chapter}{\protect\hspace{3.5ex}{Abstract}}
\renewcommand{\baselinestretch}{\foretextstretch}\small\normalsize

This thesis focuses on the study of binary radio pulsars, their evolution and some specific use of their properties to investigate fundamental physics such as general relativity and other gravitational theories. The work that we present here is organized in three main parts.

First, we report on the study of PSR~J1744$-$3922, a binary pulsar presenting a peculiar `flickering' flux behavior as well as spin and orbital properties that do not correspond to the expectations of standard evolution scenarios. We investigated the nature of this flux behavior. We also studied the pulsar's properties in relationship to the binary radio pulsar population and proposed the existence of an as yet unidentified class of binary pulsars.

Second, we conducted an in-depth analysis of the eclipses in the relativistic double pulsar system PSR~J0737$-$3039A/B. During these eclipses, the `A' pulsar partly disappears for $\sim 30$\,s behind its companion, `B'. The eclipse light curve displays a complex structure of flux modulation that is synchronized with the rotation of pulsar B. We worked on improving our understanding of the eclipse phenomenology and more particularly the modulation phenomenon. From our modeling of the eclipses, we precisely determined the geometry of pulsar B in space and used this information to study the temporal behavior of the eclipses, which revealed that pulsar B precesses around the angular momentum of the system in a way that is consistent with the prediction of general relativity.

Third, we searched for the signature of latitudinal aberration in the pulse profile of pulsar A in PSR~J0737$-$3039A/B. This relativistic effect should cause a periodic variation in the separation between the two pulse components of pulsar A on an orbital time scale. The non-detection of this effect allows us to put an upper limit on its amplitude, which constrains the geometry of pulsar A with respect to our line of sight as well as its emission geometry.

\newpage
\renewcommand{\baselinestretch}{\smallstretch}\small\normalsize
\chapter*{R\'esum\'e}
\addcontentsline{toc}{chapter}{\protect\hspace{3.5ex}{R\'esum\'e}}
\renewcommand{\baselinestretch}{\foretextstretch}\small\normalsize

Cette th\`ese se concentre sur l'\'etude des pulsars en syst\`emes binaires, leur \'evolution, ainsi que certains usages de leurs propri\'et\'es pour comprendre la physique fondamentale dont la relativit\'e g\'en\'erale et les th\'eories gravitationnelles alternatives. Le travail de cette th\`ese comprend trois parties principales.

En premier lieu, nous pr\'esentons l'\'etude de PSR~J1744$-$3922, un pulsar binaire d\'emontrant d'\'etranges fluctuations d'intensit\'e lumineuse ainsi que des propri\'et\'es orbitales et de rotation qui ne correspondent pas aux pr\'edictions des sc\'enarios \'evolutifs conventionnels. Nous analysons d'abord les fluctuations d'intensit\'e lumineuse. Nous \'etudions ensuite la nature de ce pulsar en relation avec la population de pulsars radio en syst\`emes binaires et proposons l'existence d'une classe de pulsars binaires qui n'avait pas encore \'et\'e mise \`a jour.

Deuxi\`emement, nous avons r\'ealis\'e une analyse en profondeur des \'eclipses du pulsar double relativiste PSR~J0737$-$3039A/B. Durant ces \'eclipses, le pulsar `A' dispara\^it partiellement pendant une trentaine de secondes derri\`ere son compagnon, `B'. La courbe de lumi\`ere des \'eclipses montre une complexe structure de modulation d'intensit\'e qui est synchronis\'ee avec la rotation du pulsar B. Les travaux pr\'esent\'es ici ont pour but de mieux comprendre la ph\'enom\'enologie des \'eclipses et visent plus particuli\`erement le ph\'enom\`ene de modulation. La mod\'elisation des \'eclipses nous a permis de pr\'ecis\'ement d\'eterminer la g\'eom\'etrie du pulsar B dans l'espace et d'en d\'eduire son \'evolution temporelle. Nous concluons que le pulsar B subit une pr\'ecession de son axe de rotation autour du moment angulaire du syst\`eme selon un taux et une direction en accord avec la pr\'ediction de la relativit\'e g\'en\'erale.

Pour conclure, nous avons recherch\'e la pr\'esence d'aberration latitudinale dans le profil du pulse du pulsar A, toujours dans le double pulsar PSR~J0737$-$3039A/B. Cet effet relativiste devrait causer une variation p\'eriodique de la s\'eparation entre les deux composantes du pulse \`a l'\'echelle de la p\'eriode orbitale. Malgr\'e une non-d\'etection, la limite sup\'erieure d\'eriv\'ee pour l'amplitude de cet effet permet de contraindre la g\'eom\'etrie du pulsar A par rapport \`a notre ligne de vis\'ee de m\^eme que la g\'eom\'etrie de son \'emission radio.
\newpage
\cleardoublepage
\renewcommand{\baselinestretch}{\smallstretch}\small\normalsize
\chapter*{Acknowledgments}
\markboth{{Acknowledgments}}{{Acknowledgments}}
\addcontentsline{toc}{chapter}{\protect\hspace{3.5ex}{Acknowledgments}}
\renewcommand{\baselinestretch}{\foretextstretch}\small\normalsize

I am truly indebted to my Ph.D. advisor Vicky Kaspi who, in the first place, took me in her research group five years ago despite that she already had more students to supervise than one would ``normally'' handle. She possesses many great qualities but I particularly admire her passion for astrophysics, and the rigor and determination that she puts into her work. I shall always remember the many iterations of corrections that she made on my papers, posters and proposals. Obtaining \emph{``Vicky's approval''} on a work means that it has necessarily reached a high quality standard. I came to realize that this is what made her an internationally respected scientist and I did my best to learn from her mentoring. I could never thank her enough for believing that I could carry on this double pulsar project.

Great people often attract great people and Vicky makes no exception. The McGill Pulsar Group have been populated by amazing scientists and I benefited from the invaluable support from them. Former postdoctoral fellows Scott Ransom --- thanks for sharing your amazing technical skills to me ---, Mallory Roberts --- many crazy, but yet so creative ideas ---, Maxim Lyutikov --- it all starts from a back of the envelope calculation --- and fellow graduate students Jason Hessels, Fotis Gavril, Cindy Tam, Maggie Livingstone and Rim Dib, and undergraduate student Claude-Andr\'e Faucher-Gigu\`ere kindly helped me adapt and become part of the group. As some were leaving the group --- sadly --- others great ones arrived: Marjorie Gonzalez, David Champion, Cees Bassa, Zhongxiang Wang, WeiWei Zhu, Anne Archibald and Patrick Lazarus. At another level, I have had the chance to collaborate with several scientists (Michael Kramer, Maura McLaughlin, Ingrid Stairs, Robert Ferdman, Fernando Camilo, Andrea Possenti, Martin Durant, Pierre Bergeron and Andrew Faulkner) from all around the world and this was a thrilling experience.

I spent a considerable fraction of the last five years to work, eat, live and sleep in my office. It certainly made things easier to be surrounded by nice office mates. Thanks to Fran\,cois Fillion-Gourdeau and the huge crowd from the ``349'' office and to the great folks from the ``meat locker'' in the 225: Kelsey Hoffman, Alex van Engelen, John Dudley, Sebastien Guillot, Dan O'Donnell, Christian Haakonsen and Joshi Wamamoto. Special mention to Marjorie, with whom I had the pleasure to share both offices.

The grad student life at McGill's Physics Department has been fantastic because of the lively people making it. Whether it was about having one (or many beers) at Thompson House, playing volleyball and ice hockey, or getting ``huuuuge'' at the gym, you were definitely essential: Medhi El Ouali, Till Hogardon, Rhiannon Gwyn, Philippe Roy, and everybody else... the list could go on forever, almost.

A special thank goes to Paul Mercure, our beloved HEP network administrator, who installed, fixed and rebooted computers for me so many times! It was also great to count on the support from staff in the front office and  from lab technicians Michel Beauchamp and Steve Godbout with whom I had to interact when I was the Physics 101/102 Head TA. I have had very nice scientific and informal conversations with many faculty members, in particular with professors from the astrophysics group. I should certainly thank Gil Holder for the many chats we had about Markov Chains and Bayesian statistics and Ken Ragan as well as Zaven Altounian who were in charge of the Physics 101/102 courses.

On a totally different ground, I would like to mention my gratitude to my closest friends Jonathan Duquette and Jens Kroeger with whom I chatted about physics, philosophy, meaning of life, etc., while enjoying a tea, as well as Mario Belzile whose talent for cooking chicken wings makes Canadiens hockey games way better.

When I stop and look back, I see many persons who, in one way or another, illuminated the way to where I stand now: my high scool teacher Marcel Gaudreault somehow managed to get me involved in science projects and participate to Expo-Sciences; Gabriel Forest, who left this world too early, was the most passionate science and astronomy communicator that I have ever met; Sophie Izmiroglu and her parents Rolande and Unal have been like a second family throughout most of my graduate life; Luc Tremblay, who tought me the key aspects of everyday' physics such as standing waves in beer bottles. I am also thankful to Laurent Drissen for giving me my first chance in astrophysics. His enthusiasm about astrophysics is contagious and not only is he very good at popularizing astronomy, he is also one of the rare astrophysicists that I know who can locate constellations in the sky. Also, thanks to Eddy Szczerbinski and Rhiannon for their comments on different parts of the thesis.

All my love and gratitude to Kate Clayson who performed the painful task of reading the first draft of my thesis. Her support during the thesis write-up certainly saved me from madness and the wonderful time we spent together made me quickly forget about the long hours of work.

\emph{Finalement, je suis infiniment reconnaissant aupr\`es de mes parents, de mon fr\`ere et du reste de ma famille. Vous avez toujours \'et\'e impliqu\'es et int\'eress\'es par mes nombreux projets. Il est toujours plus facile de croire en soi lorsque les personnes les plus importantes \`a nos yeux croient d'abord en nous.}

\newpage
\renewcommand{\baselinestretch}{\smallstretch}\small\normalsize
\chapter*{Contributions of Authors}
\markboth{{Contributions of Authors}}{{Contributions of Authors}}
\addcontentsline{toc}{chapter}{\protect\hspace{3.5ex}{Contributions of Authors}}
\renewcommand{\baselinestretch}{\foretextstretch}\small\normalsize

As one can expect, my Ph.D. thesis supervisor Vicky Kaspi was involved in all stages of my doctoral work. She originally suggested that I work on the projects that eventually became Chapters~\ref{c:1744} and \ref{c:0737_eclipse}. She provided me, among many other things, extremely valuable scientific advices and outstanding comments on early drafts of the work presented in this thesis.

{\bf Chapter~\ref{c:1744}}

This work was originally published as: R. P. Breton, M. S. E. Roberts, S. M. Ransom, V. M. Kaspi, M. Durant, P. Bergeron, and A. J. Faulkner. \emph{The Unusual Binary Pulsar PSR J1744$-$3922: Radio Flux Variability, Near-Infrared Observation, and Evolution}. ApJ, 661:1073–1083, June 2007

The contribution of the co-authors is as follows. Ransom provided the timing of the pulsar as well as several scripts useful to the analysis (most of them are related to the {\tt PRESTO} pulsar analysis package). Durant provided the CFHT near-IR observation and made an independent photometry analysis of the near-IR data as well as the write-up about the near-IR observation for the original paper. Bergeron provided white dwarf cooling curves. Faulkner provided part of the Parkes data set. Roberts and Kaspi were particularly involved in discussions related to the interpretation of the results. Roberts, Ransom and Kaspi provided most of the GBT and Parkes observations, which were conducted as part of a parallel project. Co-authors also provided very useful feedback on the write-up of the paper as well as valuable discussions over the course of the research.

{\bf Chapter~\ref{c:0737_eclipse}}

Part of this work --- the 820\,MHz eclipse modeling and the derived measurement of relativistic spin precession --- was originally published as: R. P. Breton, V. M. Kaspi, M. Kramer, M. A. McLaughlin, M. Lyutikov, S. M. Ransom, I. H. Stairs, R. D. Ferdman, F. Camilo, and A. Possenti. \emph{Relativistic Spin Precession in the Double Pulsar}. Science, 321, 104, July 2008. The remaining part of this chapter will be published as a second paper complementing the Science paper and will focus on the phenomenological aspects of the eclipse, the multi-frequency observations and the geometrical consequences of the eclipse modeling.

The contribution of the co-authors is as follow. Kaspi performed the original multi-frequency analysis of the double pulsar eclipses \citep{krb+04}. Kramer provided part of the write-up of the original Science paper, in particular the section about the theory-independent test of gravity involving the spin-orbit coupling constant $\left( \frac{c^2 \sigma_{\rm B}}{G} \right)$. McLaughlin performed the original modulation analysis of the double pulsar eclipses \citep{mll+04}. Lyutikov developed the original double pulsar eclipse model \citep{lt05} and provided an example code calculating an eclipse profile as well as useful discussions about the eclipse model. Kramer, McLaughlin, Stairs, Ferdman, Camilo and Possenti provided most of the double pulsar data as well as the timing solution \citep{ksm+06}. Ransom provided several analysis scripts (related to the {\tt PRESTO} pulsar analysis package). Co-authors also provided very useful feedback on the write-up of the published paper as well as valuable discussions over the course of the research.

{\bf Chapter~\ref{c:0737_aberration}}

This work is part of a larger investigation of pulsar A's pulse profile, looking for signs of latitudinal aberration on the orbital time scale (presented here) but also for long term changes potentially related to relativistic spin precession. This additional work is independently made by Robert D. Ferdman and other collaborators and should eventually be published in a peer reviewed journal.

\newpage
\pagestyle{empty}
\cleardoublepage
\renewcommand{\baselinestretch}{\foretextstretch}\small\normalsize
\renewcommand{\baselinestretch}{1.2}\small\normalsize
\vspace*{-0.2in}
\thispagestyle{empty}
\begin{center}
\rule[0pt]{5.6in}{1pt}
\vskip -12pt
\rule[0pt]{5.8in}{3pt}
\vskip 23pt
{\LARGE \sc Radio Pulsars in Binary Systems}
\vskip 13pt
\rule[0pt]{5.8in}{3pt}
\vskip -14pt
\rule[0pt]{5.6in}{1pt}
\end{center}
\vspace{-15pt}

\cleardoublepage
\pagenumbering{arabic}
\renewcommand{\baselinestretch}{\textstretch}\small\normalsize
\pagestyle{headings}


\chapter{Introduction}

\begin{flushright}
 \emph{``Nous sommes des poussi\`eres d'\'etoiles.''}\\
 Poussi\`ere d'\'etoiles, Hubert Reeves
 \vspace{0.5in}
\end{flushright}

\section{In a Nutshell...}
When asked about the topic of my doctoral thesis, I sometimes joke and answer that I am a stellar taxidermist or a space coroner. Then I mention more seriously that I study pulsars. As most people have no idea what they are, I explain that pulsars are (generally) formed when massive stars die in supernova explosions and, if not massive enough to leave behind a black hole, the remnant of the stellar core becomes a very dense object known as a \emph{neutron star}. What is a neutron star? Imagine an object that has 1.5 times the mass of the Sun, but which is compressed enough so that its size becomes comparable to that of Montreal Island, about $10 - 15$~km in radius. Under such extreme physical conditions, `normal' matter is crushed and turns into a bulk of neutrons. Such an object is essentially a neutron star.

Many neutron stars emit a (more or less) narrow beam of radio light along their magnetic axis (see Figure~\ref{f:lighthouse}). Since the spin axis about which the neutron star rotates is not necessarily aligned with the magnetic axis, its radio beam sweeps around the sky just like the beacon of a lighthouse. Distant observers will see the star blinking if they are lucky enough to be located on the path illuminated by the light beacon, hence the name of \emph{pulsars}, which is a contraction of `pulsating stars'.

\afterpage{
\clearpage

\begin{figure}
 \centering
 \includegraphics[width=5in]{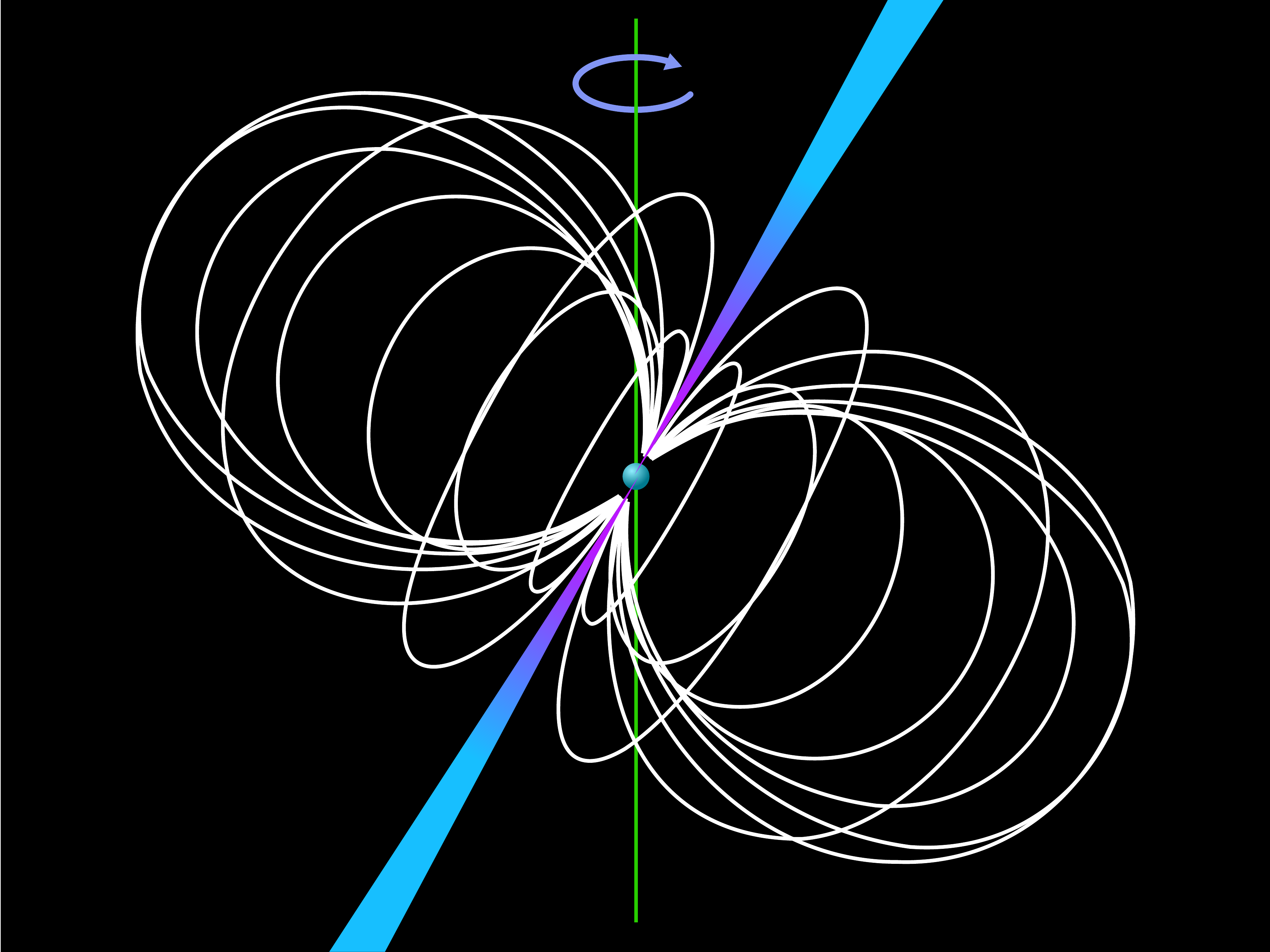}
 \caption[Artistic representation of a pulsar]{Artistic representation of a pulsar. The neutron star is the blue sphere at the center. The spin axis is represented by a vertical line while the magnetic axis is inclined about $30^\circ$ to the right. The white lines depict the magnetosphere of the pulsar, where each line represents a constant magnetic field. The two cones centered on the magnetic poles illustrate the emission beams from which the radio waves are believed to be emitted. Credit: Wikipedia, GNU Free Documentation License.}
 \label{f:lighthouse}
\end{figure}

\clearpage
}

In addition to being very compact and massive, pulsars also spin extremely fast. In fact, many pulsars spin faster than kitchen blenders --- several hundreds rotations per second!

Through their exotic nature, pulsars and neutron stars readily display their potential as promising tools for astrophysicists and physicists. They permit the investigation of physics that would otherwise be out of reach in laboratory experiments and hence can nicely complement our knowledge in the fields of condensed matter, nuclear physics, relativity and gravity, quantum mechanics, and electromagnetism. Not only can pulsars be extraordinary tools for physics, they are also interesting astrophysical objects. Studying their population and properties is directly related to understanding galactic dynamics and stellar evolution as they are forensic evidence of its past history \citep{hll+05,kh02}. Pulsars also have a significant impact on their immediate environment: magnetars --- a class of ultra-high-magnetic field pulsars visible in X-rays --- experience extreme outbursts that can affect the Earth's magnetosphere even if they are located several thousand light-years away \citep{iri05,hbs+05,pbg+05,gkg+05,ccr+05}. Young pulsars also affect their immediate environment and can power nebulae with their energetic relativistic wind \citep{sla08}. These pulsar wind nebulae appear to play an important role of `cosmic accelerator' for cosmic rays and some of them are among the brightest Galactic sources in high-energy $\gamma$-rays \citep{fls06,aab+07}.

The emission from radio pulsars can be compared to that of lasers since it results from a coherent process and it is generally highly linearly polarized. This, and the fact that they emit periodic, narrow pulses, prove to be powerful tools. From pulsar observations, one can calculate precisely their position and, sometimes, their parallax and proper motion; estimate their age, magnetic field strength, and distance; determine the average magnetic field along our line of sight and thus map out the global magnetic field structure of our Galaxy; try to detect very low-frequency gravitational waves; and more \citep{lk04}.

Many of these achievements come about from the precise timing of pulsars. From such monitoring, one can also detect the presence of bodies that are gravitationally bound to pulsars. Hence, according to the ATNF pulsar catalogue \citep{atnf}, about 8\% of pulsars are found in binary systems. Sometimes, they even host planetary companions \citep{wf92,bfs93}. PSR~B1257+12 has four planetary companions, two of which are Earth-mass bodies; these were the first extrasolar planets discovered \citep{wf92}.

The study of binary radio pulsars also has a broad extent of implications encompassing the scientific motivations enumerated above. Binary pulsars can help perfect theories of stellar evolution in binary systems because they provide the final link in the chain \citep{sta04b,vbj+05}. Also, using binary pulsar timing and observations of their companions, one can accurately measure pulsar masses, which are central to understand their internal structure \citep{nic06,lp07,tc99}. Finally, several binary pulsars, in particular those in relativistic systems, provide high-precision tests for general relativity and alternative theories of gravity \citep{ksm+06,ht75,dt92a}.

\section{Historical Background}
\subsection{The \emph{Great} Discovery}\label{s:pulsar_discovery}
It all began in 1967 in a field on the British countryside, a few miles southeast from Cambridge\footnote{Unless specific citations are provided, sources used to write this subsection are \citep{gho07,lg06,lk04,bel77}.}. It was just a simple scratch on a piece of paper made by a chart recorder. It was a scratch among many others, but this one had something different. When then-graduate student Jocelyn Bell saw it on the chart that was recording radio signals from outer space as well as picking up noise from the ground, unfortunately, she realized that this one was special. The scratch was clearly contrasting with known astronomical radio sources such as galaxies. On the other hand, it did not quite look like those produced by Earth-based radio interference either. Back in those days, a large fraction of the interference contaminating radio observations was originating from cars, or more exactly from the sparks of their engines.

What Jocelyn Bell noticed on that August 6 chart recording was the signal emitted by a radio pulsar. Without knowing, this finding would trigger what has now become four decades of pulsar astrophysics and from which blossomed great discoveries related to, among other things, the fate of massive stars and the validity of Einstein's general relativity in the extreme gravitational field regime.

To strangers wandering in the area of the Mullard Radio Astronomy Observatory (MRAO) (see Figure~\ref{f:mrao}), there was clearly some sort of experiment taking place. The whole field was covered with wood poles sticking out of the ground. Metal rods --- the dipole antennae --- were fixed to the poles and wires were connecting them. This strange rectangular forest covered no less than 4 acres of land and contained 2048 of these bizarre trees. There was also a shed in which the recording back ends, amplifiers and chart recorders were kept safe from weather hazards. This apparatus was in fact a radio telescope designed by Cambridge University Professor Anthony Hewish and built by himself and his team of students over the year of 1967. Sheep were also noticeable contributors, taking care of lawn mowing, since mechanical-powered lawnmowers would have produced undesirable radio interference.

\afterpage{
\clearpage

\begin{figure}
 \centering
 \includegraphics[width=6in]{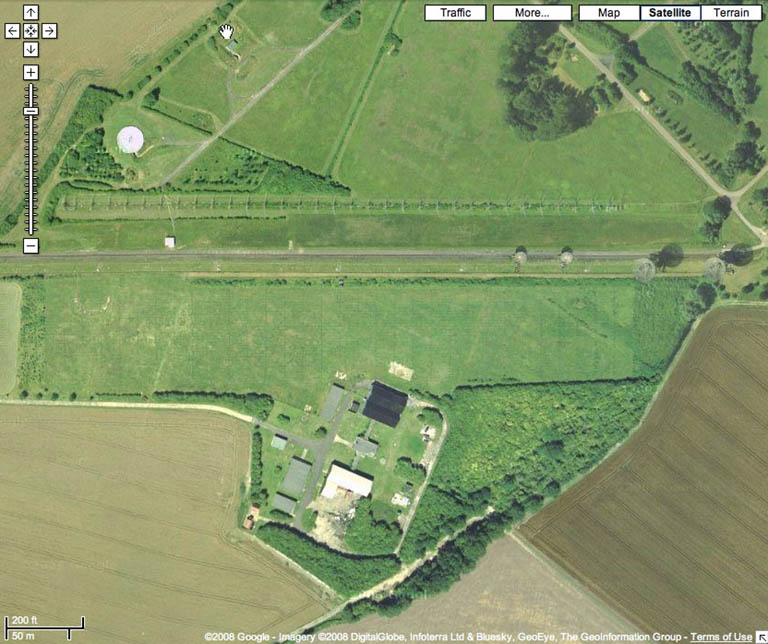}
 \caption[Satellite view of the Mullard Radio Astronomy Observatory]{Satellite view of the Mullard Radio Astronomy Observatory (MRAO) near Cambridge, UK. The Interplanetary Scintillation Array, used for the pulsar discovery lies in the lawn field at the center between the buildings and the horizontal track, which is the `One-Mile Telescope' used for aperture synthesis interferometry. Credit: Google Map.}
 \label{f:mrao}
\end{figure}

\clearpage
}

This new instrument called the \emph{Interplanetary Scintillation Array} aimed to study very different objects than pulsars. It was instead targeting galaxies, and more precisely quasars, which are the ultra-luminous cores of distant galaxies likely powered by a massive central black hole. A fraction of the quasar population is observable at radio wavelengths and because of their small size and large distance, they appear as point sources. Anthony Hewish was interested in investigating interplanetary scintillation using quasars \citep{hsw64}. This phenomenon is caused by the travel of light through ionized material. Density and ionization factor fluctuations introduce variations in the refraction index and hence result in variability of the light source intensity \citep{cr98}. A similar phenomenon can be observed when we look at an object located behind a BBQ grill; its shape is deformed because rays of light coming to us change direction as a consequence of refraction in the hot turbulent air. Scintillation of an astronomical radio source is produced by ionized gas that can be located in Earth's atmosphere, the Solar System, interstellar space or even intergalactic space if the source lies outside our Galaxy. It also requires a source having a small angular size because scintillation is caused by refraction which introduce a phase de-coherence of the light wave. The type of scintillation found in quasars originates from the interplanetary medium and Hewish realized he could use this phenomenon to constrain their angular dimension \citep{hsw64}. Interplanetary scintillation is effective only at low frequencies, $\sim$~327~MHz, and acts on short time scales, $\sim$~1~s. The Interplanetary Scintillation Array was therefore built specifically to detect short time scale intensity fluctuations and was sensitive to low flux density. Because of its design, the instrument was operating in transit mode; i.e. observing a fixed location in the sky and using the Earth's rotation to naturally cover a strip of the sky over the course of a day. So in short, this radio telescope was perfectly suited to find radio pulsars.

The August 6, 1967, detection of the first radio pulsar by Jocelyn Bell occurred only a month after she started acquiring data for her doctoral thesis on interplanetary scintillation of quasars. She identified the source among hundreds of meters of recorded paper data --- the chart recorder would spit out 30~m of paper every day --- and it subsequently came back at about the same time in the following days (see Figure~\ref{f:psr_discovery}). Even though Hewish initially dismissed the finding of his graduate student, thinking that it was terrestrial noise, he soon became convinced that the source was of astronomical origin. The transit time was occurring four minutes earlier every day and was therefore synchronized with the sidereal time, i.e. it was fixed with respect to the distant stars. They then decided to make future observations of this mysterious source at a higher time resolution --- Bell simply had to switch the chart recorder to a faster one just before the source became visible. More details about the source properties could therefore be investigated. Unfortunately, the source did not show up for several days, probably because of interstellar scintillation. When it finally showed up again at the end of November, the higher resolution recording revealed, to everyone's surprise, that it was emitting regular pulsations every 1.33 seconds. The discovery was shocking. How could a natural source be pulsating so periodically? The object was dubbed LGM-1, Little Green Man 1, since it was envisioned (in a rather speculative scenario) that pulsations could come from an intelligent form of extraterrestrial life.

\afterpage{
\clearpage

\begin{figure}
 \centering
 \includegraphics[width=4in]{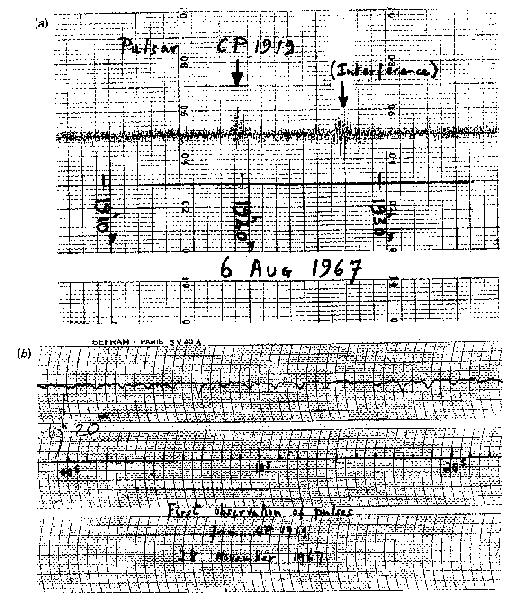}
 \caption[Discovery observation of the first pulsar]{Discovery observation of the first pulsar PSR~B1919+21 (formerly named CP~1919) by Jocelyn Bell in 1967. The upper image shows the original accidental detection, while the bottom one shows the first high time resolution detection indicating the presence of periodic pulsations. Credit: ATNF.}
 \label{f:psr_discovery}
\end{figure}

\clearpage
}

The mystery about the pulsating source took an interesting turn when, just before the Christmas break, Bell found another pulsating source in a different location of the sky. Further observations confirmed the detection and two more sources were also identified. All of them presented regular pulsations with periodicity of the order of one second. LGM-1 was not unique. It was then renamed CP~1919 after its right ascension (19h19m), while the other three were CP~1133, CP~0834 and CP~0950. After carefully gathering confirmation observations and making sure that the sources were real astronomical phenomena they announced their discovery to the scientific community. The discovery of CP~1919, and a mention to the three other sources, appeared on February 24, 1968, in Nature \citep{hbp+68}. It was followed, on April 13, by a second paper reporting details about the three other pulsating radio sources and providing further timing results from CP~1919 \citep{phb+68}.

The first pulsar paper is remarkably detailed \citep{hbp+68}. It contains a section describing the large pulsed flux variation of the signal, which is naturally expected from scintillation given the low frequency (80.5~MHz) and the small bandwidth (1~MHz) of the radio telescope. A very precise analysis of the arrival time of the pulses was made --- the timing of the periodicity of CP~1919 was better than one part in $10^{-5}$ during typical 6-h tracking observations while the relative error for their complete data set reached $20\times 10^{-7}$. \citet{hbp+68} were even able to observe the Doppler shift imprinted in the pulsation due to the orbital motion of the Earth around the Sun. They also ruled out with reasonable confidence the possibility that CP~1919 was orbiting another body because of the lack of an additional Doppler shift. Hence, the source was clearly located outside the Solar System and no parallax was detected to an upper limit of $2^{\prime}$, thus constraining its distance to be larger than $10^3$~AU. A shift in the arrival time of the pulse was also observed in the different frequency channels of the instrument. This was correctly interpreted as the delay caused by the frequency-dependent light-travel time in the ionized interstellar medium. The amount of delay is often referred to as the \emph{dispersion measure} and is proportional to the total column density of ionized material along the line of sight. Assuming a constant electron density in the interstellar medium, they found that the source was located about 65 parsecs away. According to the ATNF pulsar catalogue \citep{atnf}, the current distance estimate from the dispersion value is 2490 parsecs. Based on the duration of each pulse, they also concluded that the source could not exceed 4800~km in size. On the nature of the source, \citet{hbp+68} speculated that pulsations driven by the radial modes in a white dwarf or a neutron star was the most plausible explanation for the radio periodicity. While neutron stars had been theoretically predicted, no observational evidence had been made thus far. Following the radial pulsation hypothesis, they predicted that X-ray emission resulting from the hydrodynamic shock at the surface of the pulsating compact star might be observable.

The second paper highlighted that one of the other pulsars, CP~0950, has a `short' 0.25~s periodicity \citep{phb+68}. This was making the pulsating white dwarf scenario more unlikely as radial modes would have fundamental periods of about 1~s in white dwarfs. At higher densities, typical of that of neutron stars, shorter time scales were still possible.

The exact nature of pulsars remained unclear, however, until the end of 1968 when neutron star theory and pulsar observations finally converged. Pulsars are indeed neutron stars and some of them are even visible in X-rays, but the mechanism responsible for their radio and X-ray pulsations is not radial pulsation. The rotation of the neutron star itself combined with non-isotropic radiation is what produces the pulsation phenomenon. Among the contributing factors unveiling their true nature was the discovery, in the months following the first pulsar article, of two of the most famous pulsars: the pulsar associated with the Vela Nebula, discovered by \citet{lvm68}, and the Crab Nebula pulsar, by \citet{ric68}. It became apparent from the large range of pulse periods, in particular from the short periods of the Vela and the Crab pulsars --- 89 and 33 ms, respectively --- that the radial pulsation hypothesis presented major difficulties. In particular, it could not account for the whole population with a single type of object, either white dwarf or pulsar. Nevertheless, the original intuition of \citet{hbp+68} about the nature of pulsars was therefore remarkably close to reality and their work pioneered subsequent research in the field.

The name \emph{pulsar} was first coined in an article of the Daily Telegraph, 5 March 1968, in an interview with Anthony Hewish about the initial discovery \citep{hew68}.

\subsection{Theoretical Precursor Work}
Baade and Zwicky originally introduced the concept of neutron stars in 1934, less than three years after the discovery of the neutron by Sir James Chadwick. At the time, the duo was interested in observations of novae and they soon recognized that this type of event could be divided in two subclasses that they called \emph{novae} and \emph{super-novae} (now spelled supernovae) \citep{bz34a}. They correctly identified novae as the result of the sudden thermonuclear hydrogen burning at the surface of an accreting white dwarf in a binary system, while they proposed that supernovae followed from \emph{the transition from an ordinary star into a body of considerably smaller mass} \citep{bz34a}. Supernovae are much more energetic and brighter than novae --- the average peak absolute magnitudes are $M_B = -19.6 \pm 0.2$ for (type Ia) supernovae \citep{bra98} and $M_B \sim 8$ for novae \citep{dbk+06}. In a following publication, \citet{bz34b} went further and mentioned that the small bodies formed via supernovae would be neutron stars. It is now established that supernovae are themselves separable in two main sub-types that result from the core collapse of massive stars or from the collapse of heavy accreting white dwarfs reaching the Chandrasekhar mass limit \citep{co96}. The remnants of these supernovae are either neutron stars or black holes depending on the supernova type and the mass of the progenitor star.

Interestingly, this theoretical work did not seem to catch the attention of observers for more than 30 years until the first radio pulsar was discovered by a fortunate chance. Physicists, mainly those interested in nuclear physics and relativity took over the work of Baade and Zwicky. During this `transition' period, there have been a few milestone papers marking the development of neutron star theory. \citet{ov39} proposed an equation of state for neutron stars based on the work of \citet{tol39}. Their work on the internal structure of neutron stars led to what is now called the \emph{Tolman-Oppenheimer-Volkoff equation of state}. It would allow the prediction of masses, densities and radii of neutron stars well before they were even observed. Later, Hoyle, Narlikar and Wheeler suggested that large magnetic fields, of the order of $10^{10}$~G, could exist in neutron stars such as the one that was thought to exist at the center of the Crab Nebula \citep{hnw64}. This field would naturally emerge from the conservation of magnetic flux of a core-collapsing star having Sun-like magnetic dipolar field strength. They also addressed the issue related to the dissipation of the dynamical energy generated during the collapse by suggesting that it would be radiated away via electromagnetic waves. A simple estimate for the Crab Nebula indicated that it could be the source of energy driving the expansion of the nebula.

Ironically, the theoretical work that really bridged neutron stars and pulsars has been conducted and published at the same time that pulsars were discovered, and without even knowing about it. \citet{pac67} pointed out that highly magnetized neutron stars would generate electric currents and emit electromagnetic waves as they spin. He showed that the energy output could power the Crab Nebula. Shortly after, \citet{gol68} published his independent research restating the same general ideas as \citet{pac67}. He very clearly identified that rapidly rotating, highly magnetized neutron stars would emit a radio light beacon that would sweep across space because of the star's rotation about its axis. The canonical pulsar lighthouse model was born. He also predicted that pulsars should steadily slow down as a consequence of magnetic dipole radiation. Not long after, \citet{rc69} were confirming this idea with the announcement that the period of the Crab Pulsar was gradually increasing.

\subsection{Parallel Discoveries}
In parallel to the pulsar discovery by the Cambridge University group, there are many stories of `pre-discovery' detections of pulsars. These observations either went unnoticed then or were misinterpreted until after the `original' discovery paper came out. For example, an amateur astronomer might have seen pulsations from the Crab Pulsar while looking at the Crab Nebula with the optical telescope of the University of Chicago \citep{bru07}. The strange behavior that the woman noticed was dismissed immediately by Elliot Moore, the observatory astronomer, who thought it was normal atmospheric scintillation. It seems that some radio astronomers probably also observed giant pulses from the Crab Pulsar while studying the nebula. None of them, however, realized that it could have anything to do with neutron stars, which were predicted to exist at the center of such a supernova remnant.

US Air Force sergeant Charles Schisler made another early pulsar detection during the summer of 1967 \citep{bru07,sch08}; this was at the same time as Jocelyn Bell conducted her first `accidental' pulsar observations. Schisler was then a radar operator at the Clear Air Force Station in Alaska. This base was part of the Ballistic Missile Early Warning System that was aimed to detect missiles that would be launched from Siberia during the Cold War. The radar was specifically conceived to be sensitive to man-made pulses bouncing off these hypothetical missiles. As he was monitoring the radar screen one day, Schisler noticed a faint signal that was not related to anything he had seen before. He repeatedly saw the same source afterwards and found that it would appear 4 minutes earlier every day. His rudimentary astronomy knowledge helped him in concluding that the source was located in outer space since it was synchronized with the sidereal time. After carefully determining the source's celestial position to satisfy his own personal curiosity, Schisler queried the help of a University of Alaska astronomy professor. They came to the conclusion that the source was coincident with the location of the Crab Nebula. While continuing his work at the Alaskan radar base, Schisler observed about a dozen of these astronomical sources. Unfortunately, because of the nature of his military work, he waited until 2007 when the radar base was decommissioned and demolished to publicly talk about his observations.

\section{Thesis Outline}
In this thesis, we present the results of research about binary radio pulsars. Our studies involves a mixture of observational and theoretical work that aims to investigate peculiar properties of binary radio pulsars in order to obtain a better understanding of their nature and evolution but also to use them as tools to test general relativity. Most of the data used in this thesis were obtained at radio wavelengths with the Green Bank and the Parkes telescopes. Some of the radio observations have been conducted by the author of this thesis while a considerable part of the data set was shared by co-investigators in these projects. We also performed complementary follow up observations in other parts of the electromagnetic spectrum, namely at near-infrared and optical wavelengths, with various instruments such as the Gemini, Canada-France-Hawaii and SOAR observatories. In this thesis, we shall cover the following elements:

\subsubsection{Chapter~\ref{c:theory}}
In this chapter the basic elements of neutron star and pulsar theory are presented. First, we describe the nature of neutron stars and then introduce pulsars as a particular subset of this more general class of stellar object. We present the fundamental pulsar properties in the context of their supporting observational evidence and summarize the main tools and techniques used to observe and time radio pulsars. These are of great importance in order to structure a coherent theory of binary pulsar evolution and relate it to the observed population. We shall make use of this background information throughout this thesis and particularly in Chapter~\ref{c:1744}, which focuses on binary evolution. We outline how the peculiar properties of binary pulsars can be used to test general relativity and other theories of gravity. Finally, we review the main characteristics of the double pulsar PSR~J0737$-$3039A/B, which is the binary pulsar we study in Chapters~\ref{c:0737_eclipse} and \ref{c:0737_aberration}.

\subsubsection{Chapter~\ref{c:1744}}
In this chapter, we present the results of the study of the binary pulsar PSR~J1744$-$3922. The goal of this work was two-fold. First, we examine the peculiar flickering behavior of the radio emission from this pulsar and investigate the possible reasons that could cause it to fluctuate. Second, we analyze the properties of this pulsar in relationship with the existing population of binary pulsar and the evolution mechanisms forming them. It appears that PSR~J1744$-$3922 challenges some of the standard binary scenarios and we describe how this pulsar can fit in this picture.

\subsubsection{Chapter~\ref{c:0737_eclipse}}
The double pulsar PSR~J0737$-$3039A/B is a unique binary system in which two radio pulsars orbit each other. Once per orbit, one of the pulsars eclipses its companion for about 30~s. This chapter presents an extensive study of these eclipses. We perform a thorough analysis of the phenomenological properties of the eclipses and report on the results of a  quantitative modeling. This work yields the precise measurement of relativistic spin precession of the pulsar creating the eclipses. We also look at some of the interesting consequences emerging from the determination of the pulsar and the system geometry that arise from the eclipse modeling.

\subsubsection{Chapter~\ref{c:0737_aberration}}
Special and general relativity not only affect the dynamical motion of pulsars, they may also manifest as changes in the structure of their pulse profiles. Latitudinal aberration, for instance, causes the angle between our line of sight vector and the magnetic moment of a pulsar in a binary system to be different from what it would be if the pulsar was at rest. This kind of effect is generally too small to be visible. Fortunately, the double pulsar PSR~J0737$-$3039A/B displays characteristics such as a short orbital period and an almost perfectly edge-on orbit that, once they are combined to the properties of the `A' pulsar, could potentially be detected. This chapter reports on geometrical constraints obtained on pulsar A from the limit on the non-detectability of the latitudinal aberration effect.

\subsubsection{Chapter~\ref{c:conclusion}}
In this chapter, we summarize the essential contributions made by the research presented in this thesis and draw some general conclusions. We also briefly discuss the future work that shall follow this thesis work.


\chapter{Neutron Stars and Pulsars}\label{c:theory}

\begin{flushright}
 \begin{singlespace}
 \emph{``There are those burned out stars who implode into silence after parading in\\
 the sky after such choreography what would they wish to speak of anyway."}
 \end{singlespace}
 White Dwarfs, Michael Ondaatje
 \vspace{0.5in}
\end{flushright}

\section{Neutron Stars: An Overview}\label{s:ns_overview}
Neutron stars are among the most exotic and extreme objects populating our Universe. In the standard picture, neutron stars are the remnants of massive stars ending their life as supernovae.\footnote{Other formation mechanisms exist and will be described in \S\,\ref{s:binary_evolution}} They are small, about 10~km in radius, and their density is comparable to that of atomic nuclei --- typically $10^{14}$~g~cm$^{-3}$ \citep{lg06}. They generally also have large surface magnetic fields, ranging from $10^8$ to $10^{15}$~G \citep{gho07}. Although these properties appear unreal, they naturally emerge from simple physics conservation laws. In this section, we will review the basics of neutron star theory.

Normal stars are macroscopically stable because two main forces oppose each other. On one hand gravity pulls matter together while on the other hand thermal pressure (radiation and heat) produced by the nuclear reaction in the core of the star counteracts it. It is the fine equilibrium between both forces that holds stars together and prevents them from collapsing or flying apart. When the energy budget of a star changes, say because the rate or type of nuclear reaction varies, it will readjust its internal structure and size in order to reach the force balance again \citep{hkt04}.

Thermal energy is neither the only nor the most efficient type of negative pressure that can counteract gravity. In a totally different class of stars, composed of \emph{degenerate} matter, it is the quantum forces themselves that stand against gravity via Pauli's exclusion principle \citep{cha67}.

A first kind of degenerate stellar remnant is called a \emph{white dwarf}. The degeneracy pressure of electrons supports these stellar remnants \citep{cha67}. They are formed by the collapse of the core of a low mass, Sun-like star after the helium burning stage. These stars are not massive enough to ignite carbon burning that leads to the formation of an iron core. With no source of energy to support the gravitational force the core of the star shrinks until the electron degeneracy force becomes comparable. Electrons are fermions and hence are differentiable particles. No two fermions can be in the same quantum state. Therefore, the wavefunctions of electrons start overlapping in high-density conditions and any extra electron added to the system has to occupy a higher energy state in order to be differentiable \citep{gri95}.

The condition that electrons fulfill is analogous to a negative pressure as electrons are forced to occupy higher energy states and hence acquire larger momentum. This way, one can calculate the equilibrium force between the electron degeneracy pressure and gravity. To do so, one has to solve the classical hydrostatic equations \citep{dd92}:
\begin{equation}
 \frac{d{\rm P}}{dR} = -\frac{M(R)G}{R^2} \rho_0 \,,
\end{equation}
and
\begin{equation}
 \frac{dM(R)}{dR} = 4\pi R^2\rho_0 \,,
\end{equation}
for a given equation of state, which describes the pressure (P) versus density ($\rho$) behavior in the star.

In the 1930's, Subrahmanyan Chandrasekhar had conducted pioneer work related to white dwarfs that had important implications for neutron star theory \citep{cha31a,cha31b,cha31c,cha35}. He solved the hydrostatic equations of a Fermi degenerate gas having a polytropic equation of state, that is ${\rm P} \propto K\rho^\gamma$, for both the non-relativistic ($\gamma = 5/3$) and ultra-relativistic ($\gamma = 4/3$) cases.

In the non-relativistic case, $p = mv$, and the equation of state is given by
\begin{equation}
 {\rm P} = \frac{1}{20}\left(\frac{3}{\pi}\right)^{2/3} \frac{h^2}{m(\mu H)^{5/3}} \, \rho^{5/3} \,,
\end{equation}
where $h$ is the Planck constant, $m$ is the electron mass, $H$ is the mass of the proton, and $\mu$ is the molecular weight.

In the ultra-relativistic case, $p = E/c$, and
\begin{equation}
 {\rm P} = \left(\frac{3}{\pi}\right)^{1/3} \frac{hc}{8(\mu H)^{4/3}} \, \rho^{4/3} \,.
\end{equation}

Chandrasekhar found that $R \propto M^{-1/3}$ and $R = $\,\emph{constant} for the non-relativistic and ultra-relativistic cases, respectively. These solutions are quite remarkable given their counter-intuitive behavior. The more matter that is added to the degenerate star, the smaller it becomes. When entering the ultra-relativistic regime, the star ``saturates'' and its size does not depend on its mass anymore.

If the gravitational force increases further, hydrostatic equilibrium can no longer exist. At the point where the Fermi energy --- the energy of the highest occupied eigenstate --- reaches some critical value, inverse $\beta$-decay naturally occurs even in stable nuclei: $p + e + \Delta E = n + \nu_e$ \citep{gho07}. Thus ``normal'' matter transmutes into neutron degenerate matter. A new degeneracy pressure balance, similar to that of white dwarfs, can then occur but this time neutrons, which are also fermions, are responsible.

Neutron-degenerate matter has a larger fermi energy than electron-degenerate matter. Hence, \emph{neutron stars} can reach much higher densities than white dwarfs and this explains their smaller size: about 10~km vs. 10$^4$~km, respectively \citep{gho07}. Because of their very high densities, neutron stars are relativistic objects. We can easily demonstrate that about $10-20$\% of the mass of a neutron star lies in its gravitational binding energy \citep{wil01} simply by comparing the gravitational potential of a sphere having a uniform density,
\begin{equation}
 U = \frac{3}{5}\frac{GM^2}{R} \, ,
\end{equation}
with its rest mass energy
\begin{equation}
 E = Mc^2 \,.
\end{equation}

In fact, if the gravitational pressure becomes larger than the neutron degeneracy pressure, nothing would prevent the object from collapsing until it becomes a black hole. The size of a black hole is given by the Schwarzschild radius: $R_S = \frac{2GM}{c^2}$ \citep{sch16a,sch16b}. In comparison, the radius of a 1-2~\Msun neutron star is only a few times the size of a black hole.

\subsection{Internal Structure and Equation of State}\label{s:eos}
Although one may think \emph{a priori} that neutron stars are made of neutrons only, their structure is in fact more complex. In this subsection, we present the essential elements of neutron star structure. We shall not attempt, however, to extensively review this particular field of neutron star astrophysics since it is beyond the scope of this thesis. We refer readers interested in this topic to \citet{hei02,lp01,lp04,lp07}, for example.

The quest for understanding the internal structure of neutron stars started shortly after their existence was speculated by \citet{bz34a,bz34b}, well before any observational evidence was found. Besides the work of \citet{cha31a}, considerable work was originally done by \citet{ov39}. They investigated the equilibrium conditions of a spherically symmetric distribution of matter in the general relativistic framework using the method developed by \citet{tol39} and considered the case of a cold gas of degenerate neutrons.

Significant progress has been made on the internal structure since the early days of neutron star astrophysics. Quantitative models exist to describe their equation of state, that is, the relationship between the distribution of mass as a function of radius, $M(R)$, or, equivalently, the pressure as a function of density, ${\rm P}(\rho)$. Such modern equations of state were perfected using advances made in nuclear physics --- QCD in particular --- which allow quantitative treatment of nucleon-nucleon interactions, calculations of binding energies, and more \citep{hei02}. The heritage of this research yields a broad family of models describing the relationship between cooling and luminosity of neutron stars, as well as other properties linked to their structure. Despite all these efforts, the exact nature of neutron stars still remains an open question. Answering it will require further observational breakthroughs such as simultaneous measurements of masses and radii for example. This would allow one to narrow down the phase space predicted by different equations of state and eventually identify the correct theory.

In the canonical model, a neutron star is a body consisting of four main internal regions \citep{yp04,lp04} that are possibly covered by a thin atmosphere having a few centimeters thickness, whose composition may be hydrogen or a mixture of heavier nuclei such as oxygen or even iron \citep{zps96}. The internal structure is organized as follows \citep{yp04,lp04,wnr07} (see Figure~\ref{f:ns_structure}): 1. The outermost region, called the \emph{outer crust}, is made of ions and electrons. Ions tend to form a solid lattice under the strong Coulomb coupling and their neutron content increases inward as the larger pressure favors inverse $\beta$-decay. 2. The transition to the next zone, called the \emph{inner crust}, is marked by the neutron drip, which happens when the density reaches $4 \times 10^{11}$~g~cm$^{-3}$. At this point, neutrons start ``leaking'' out of nuclei. The inner crust possesses a large fraction of free neutrons, whose occurrence increases with density. They mix with free electrons and neutron-rich nuclei. 3. The \emph{outer core} starts where nuclei disappear. In this region, free neutrons are the main constituents of matter. A non-negligible fraction of electrons and protons are present, as well as traces of muons. Interactions in the outer core are dominated by nuclear forces. 4. Finally, the \emph{inner core} marks the innermost region of a neutron star. The extent of this region is not clearly defined, nor is the state of matter.

\afterpage{
\clearpage

\begin{figure}
 \centering
 \includegraphics[width=6in]{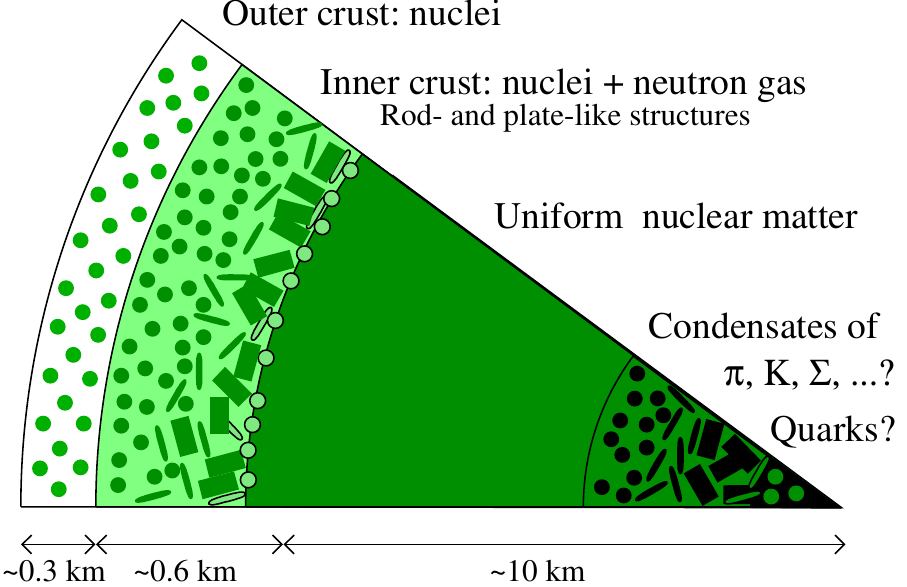}
 \caption[Cross-section view of a typical neutron star]{Cross-section view of a typical neutron star showing the four main internal regions: the outer crust, the inner crust, the outer core and the inner core. Credit: \citet{hei02}.}
 \label{f:ns_structure}
\end{figure}

\clearpage
}

The main differences between competing ``classical''\footnote{Other ``non-classical'' models have been proposed such as bare quark stars, also known as ``strange'' stars \citep{afo86,hzs86}.} equation of state models arise from the description of the inner core nature. These models vary from no significant changes with respect to the outer core structure, to a variety of more or less exotic states of matter: quark matter, hyperonic matter, pion condensate and kaon condensate, for example \citep{yp04}. These models yield distinct mass-radius relationships as well as other observable effects such as different cooling behaviors \citep{lp04}. Whether or not these high-density phase transitions really take place in nature can be probed by studying neutron stars. Ideally, one would work from joint measurements of mass, radius, age and luminosity made for several neutron stars. In a practical way, determining these properties is very challenging. The simultaneous measurement of several of these quantities, combining both high precision and reliability, in a single neutron star has not been done yet.

Many other fascinating phenomena related to the neutron star structure exist. For example, neutron star crusts and outer cores are known to contain superfluid vortices \citep{lp07,yp04}. It seems that the unpinning of these vortices could trigger ``starquakes'' and cause temporary changes in their rotational behavior \citep{lp04}. This scenario has been proposed to explain the sudden increases in the rotational frequency observed in some young pulsars. The mechanism through which these ``glitches'' happen is relatively poorly understood and still lacks self-consistent quantitative modeling.

\subsection{Magnetic Field}
Among the other important questions related to the interior physics of neutron stars is the origin of their large magnetic fields. Typical surface magnetic field strength in neutron stars, such as in radio pulsars, is $10^{12}$~G. Magnetic field values are inferred from the spin of pulsars, assuming they are slowing down because they emit dipolar electromagnetic radiation (see \S\,\ref{s:pulsar_overview}). If, \emph{a priori}, this seems extremely large, bear in mind the small size of neutron stars; the implied magnetic field moment is $\sim 10^{20}$~W~m, which is comparable to that of ordinary stars and planets --- the Sun has a magnetic moment $10^{22}$~W~m and the Earth $10^{16}$~W~m \citep{co96}. Therefore, a simple magnetic flux conservation argument is sufficient to explain the apparent large magnetic field strength values of neutron stars.

Even though we just argued that neutron stars' magnetic fields are not unreasonable, the intricate details of their origin are not perfectly understood. In particular, the large span of observed values, ranging from $10^8$ to $10^{15}$~G, is puzzling since it suggests that some intervening factors dramatically influence this property. What does exactly happen when the transition from a ``normal'' star to a neutron star occurs? Does the spread in observed magnetic fields simply reflect differences already existing in their progenitors?

At the lower end of the distribution of magnetic fields, in millisecond pulsars (MSPs), there is evidence for external factors intervening after the neutron star birth. Although the physical mechanism responsible for the lower observed magnetic field, $10^8$ vs. $10^{12}$~G, is unknown, the connection with binary systems and, more precisely, a mass accretion episode, appears to be well established \citep[see][and \S\,\ref{s:gc_pulsars}]{bv91}. This phenomenon will be investigated in more detail in Chapter~\ref{c:1744}.

Conversely, the origin of the high magnetic fields in neutron stars, those of magnetars more specifically (see \S\,\ref{s:magnetars}), is yet to be demonstrated. The solution to this question would help shedding light on the link relating ``normal'' radio pulsars and ``magnetars''.

Most theories about the origin of neutron star magnetic fields fall in two categories: either the large magnetic field results from the conservation of magnetic flux from a progenitor star that already has a large magnetic field \citep{fw95}, or there is some magnetic dynamo mechanism arising from instabilities inside the progenitor during its collapse that permits the generation of a large magnetic field \citep{td93}.

The magnetic fields of neutron stars are at the centre stage of neutron star astrophysics. It is a key element closely related to several of their observable properties such as the radio and X-ray pulsar emission, the spin-down of rotation-powered pulsars and magnetar bursts. In these processes, not only is the strength of the magnetic field important, but also its structural properties. In radio pulsars, for instance, the large-scale dipolar magnetic field is directly connected to the lighthouse emission model.

\section{Pulsars: An Overview}\label{s:pulsar_overview}
Pulsars represent a particular class of rapidly spinning neutron stars that we observe as periodic emitters of electromagnetic radiation. Radio pulsars, i.e. those visible at radio wavelengths, are the canonical type and by far the most common pulsars. Pulsars also have now been observed over the entire spectrum, from radio to high-energy gamma rays, including X-rays, optical and infrared. Although the physical process responsible for the pulsed emission varies across the spectrum, they all share the common behavior of pulsating, which originates from the rotation of the neutron star about its spin axis. Timing the pulsations of a pulsar therefore allows the determination of its spin frequency.

Even though pulsars come in different varieties, visible in various parts of the electromagnetic spectrum and possessing a rich set of attributes, we can distinguish two main groups: rotation-powered and non-rotation-powered pulsars. In this section, we shall present a brief overview of the different kinds of pulsars with a particular emphasis on radio pulsars, which are the type of neutron stars studied in this thesis.

\subsection{Rotation-Powered Pulsars}\label{s:rotation_powered}
As their name suggests, rotation-powered pulsars owe their energy output to their rotation. That is \citep{gho07},
\begin{equation}
 E_{\rm rot} = \frac{1}{2} I \Omega^2 \approx 2 \times 10^{46} \frac{I_{45}}{P_s^2} \quad {\rm erg} \,,
\end{equation}
where $I_{45}$ is the moment of inertia in units of $10^{45}$~g~cm$^2$ and $P_s = \frac{2\pi}{\Omega}$ is the spin period in seconds\footnote{To avoid confusion, we shall refer to $P$ as the spin period in a generic way whereas $P_s$ refers to the spin period in seconds when equations are expressing calculated quantities.}.

The classical rotation-powered pulsar model assumes that the magnetic field structure outside the pulsar is dipolar, at least at large scales (several hundred neutron star radii)\footnote{See \citet{hew72} for an early review of the `classical' pulsar model and properties.} \citep{go69}. Under this assumption, such a rotating bar-magnet should experience a regular spin-down due to electromagnetic radiation at a frequency corresponding to its rotational frequency \citep{jac75}. The loss in spin-down energy can be written \citep{gho07,lk04}
\begin{eqnarray}
 \dot{E}_{\rm rot} = \frac{\Omega^4}{6c^3} B^2 R^6 \sin^2 \alpha
  &\approx & 10^{31} \frac{B_{12}^2 R_6^6}{P_s^4} \sin^2 \alpha \quad {\rm erg}\,\, {\rm s}^{-1} \,,\\
  &\approx & 3.95 \times 10^{31} I_{45} \frac{\dot{P}_{-15}}{P_s^3} \quad {\rm erg}\,\, {\rm s}^{-1} \,,
\end{eqnarray}
with $B_{12}$ being the dipolar magnetic field strength at the neutron star surface in units of $10^{12}$~G, $R_6$ the neutron star radius in units of $10^6$~cm, $\alpha$ the angle between the dipole moment and the spin axis, and $\dot{P}_{-15}$ the period derivative in units of $10^{-15}$~s/s.

In this framework, one can also relate the period derivative to the period as \citep{gho07}
\begin{equation}
 \dot{\Omega} = -K \Omega^3 \,,
\end{equation}
with
\begin{equation}
 K = \frac{B^2 R^6 \sin^2 \alpha}{3Ic^3} \,.
\end{equation}
Note that this relationship holds true for a spinning dipole in vacuum only.

More generally, one can write \citep{gho07}
\begin{equation}
 \dot{\Omega} = -K \Omega^n \,,
\end{equation}
where
\begin{equation}\label{eq:braking_index}
 n = \frac{\Omega \ddot \Omega}{\dot \Omega}
\end{equation}
is called the braking index of the pulsar. The braking index can be measured for young pulsars, as they spin-down faster than old pulsars, if a sufficiently long timing baseline allows the determination of the second period derivative. So far, all measurements led to values less than 3, the value predicted for a perfect dipole in vacuum \citep{lkg+07}. A magnetic field structure different from dipolar, time evolution of the magnetic field or a significantly ionized medium around the pulsar instead of vacuum may explain these lower values \citep{lkg+07}.

By integrating the above equation one can determine the spin-down time \citep{gho07}
\begin{equation}
 t = - \frac{\Omega}{(n-1)\dot \Omega} \left[ 1 - \left( \frac{\Omega}{\Omega_0} \right) ^{n-1} \right].
\end{equation}
In the above equation, $\Omega$ and $\Omega_0$ are the current and initial spin frequencies, respectively. Under the assumption that $\Omega \ll \Omega_0$, the spin-down time of a perfect dipole ($n = 3$) becomes \citep{lk04}
\begin{equation}
 \tau_c = - \frac{\Omega}{2 \dot \Omega} = \frac{P}{2 \dot P} \approx 15.8 \frac{P_s}{\dot P_{-15}} \quad {\rm Myr} \,.
\end{equation}
The last equation is called the \emph{characteristic age} of the pulsar and is easily inferred from the most fundamental timing parameters: $P$ and $\dot P$.

Finally, one can derive a relationship between the dipolar magnetic field strength at the surface of the pulsar and its period and period derivative \citep{lk04}
\begin{equation}\label{e:magnetic_field}
 B = \sqrt{\frac{3c^3}{8\pi^2}\frac{I}{R^6 \sin^2 \alpha} P \dot P} \approx 3.2 \times 10^{12} \sqrt{P_s \dot P_{-15}} \quad {\rm G} \,.
\end{equation}

Another important concept related to rotation-powered pulsars, as well as to all kinds of pulsars in general, is what is referred to as the \emph{light cylinder}. If one assumes that the magnetic field lines of a pulsar co-rotate rigidly with its surface, there is a distance beyond which the co-rotation velocity exceeds the speed of light. From this condition, one can define an imaginary cylinder of radius \citep{lk04}
\begin{equation}
 R_{\rm LC} = \frac{c}{\Omega} = \frac{cP}{2\pi} \approx 4.77 \times 10^4 P_s \quad {\rm km}
\end{equation}
centered about the spin axis of the pulsar. Any field line that does not close within this region remains open.

It is also practical to represent the pulsar population on a diagram showing their spin period derivatives as a function of their spin periods since they are the two key observable parameters (see Figure~\ref{f:p_pdot}). Such a diagram is commonly referred to as the $P-\dot P$ diagram and readily shows the fundamental pulsar properties such as their ages, magnetic field and spin-down luminosity. The $P-\dot P$ diagram is to pulsars what the Hertzsprung-Russell is to stars. In the $P-\dot P$ diagram of Figure~\ref{f:p_pdot}, `normal' radio pulsars are displayed as yellow dots (see \S\,\ref{s:radio_pulsars}). Young pulsars such as the Crab and Vela form in the upper left side of the diagram. They eventually spin down and move toward the `pulsar island'. We observe that several of these young pulsars are associated with supernova remnants (SNR), which are marked with green stars in Figure~\ref{f:p_pdot}. Several of them are also energetic enough to be visible in X-rays (see \S\,\ref{s:xray_pulsars}). In the upper right part of the diagram are the soft $\gamma$-ray repeaters (SGRs) and the anomalous X-ray pulsars (AXPs), which are marked by open red triangles (see \S\,\ref{s:magnetars}). These young pulsars are powered by their extreme magnetic fields. Finally, binary pulsars are indicated by red dotted circles (see \S\,\ref{s:binary}). Most of them lie at the lower left part of the diagram, which implies that they are old, rapidly spinning and have low magnetic fields. In the standard picture, these pulsars have been `recycled' by mass transfer from their companion \citep{bv91}. The `death line' represents the line below which the electric potential drop across the magnetic poles would not be large enough to enable the pair production responsible for the radio emission \citep{rs75}. The one pulsar below the `death line' in Figure~\ref{f:p_pdot} is PSR~J2144$-$3933. This 8.51-s pulsar was initially thought to have a spin period of 2.84\,s, namely because pulsation was detected only every third pulse period \citep{ymj99}. It has the longest spin period and the lowest spin-down luminosity of all known radio pulsars. PSR~J2144$-$3933 challenges pulsar models since typical neutron stars are not expected to possess a large enough electric potential drop to generate pair creation necessary to power radio emission \citep{ymj99}.

\afterpage{
\clearpage

\begin{figure}
 \centering
 \includegraphics[width=6in]{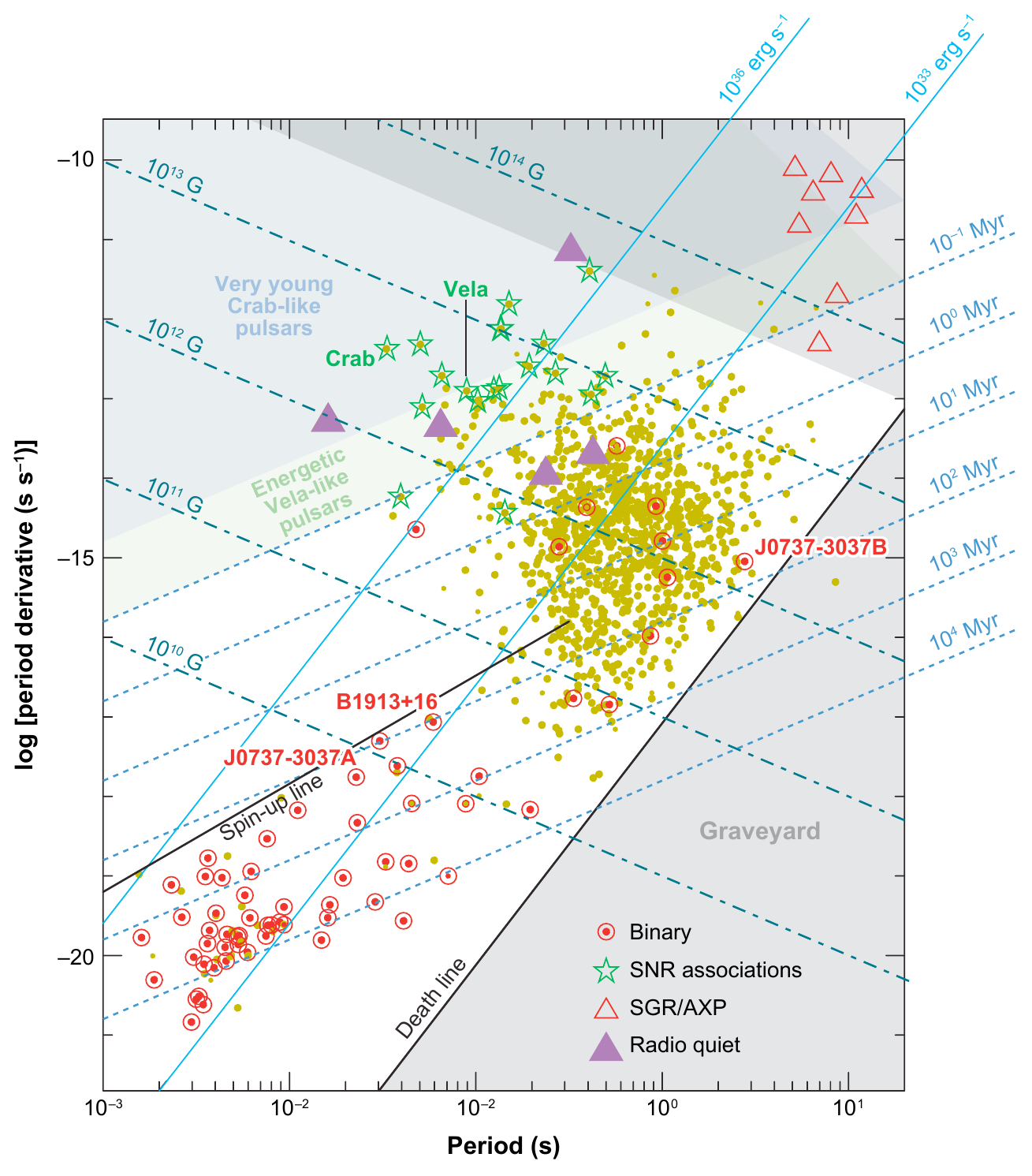}
 \caption[The $P-\dot{P}$ diagram]{The $P-\dot{P}$ diagram is that standard tool to present the entire pulsar population (including pulsars that are not visible at radio wavelengths). From their spin period ($P$) and spin period derivative ($\dot{P}$), and assuming they are perfect rotating magnetic dipoles, one can infer their magnetic field strength, their age and their spin-down energy. Credit: \citet{ks08}.}
 \label{f:p_pdot}
\end{figure}

\clearpage
}

\subsubsection{Radio Pulsars}\label{s:radio_pulsars}
Radio pulsars are the most common type of pulsars. They are found with spin periods ranging from 1.399~ms for the fastest currently known\footnote{Incidentally, the fastest pulsar was found at McGill University by then fellow Pulsar Group Ph.D. candidate Jason Hessels.} \citep{hrs+06} to 8.51~s for the slowest \citep{ymj99}. Their inferred magnetic fields vary between $10^8$ and $10^{14}$~G and their inferred ages between a few hundred years and a few billion years\footnote{The oldest pulsars are also the fastest millisecond pulsars. Because they are thought to experience a spin-up phase from ``recycling'' in a binary system, their estimated age from the timing might differ from their true age.}. Their population is very diverse and comprises young energetic pulsars such as the Crab and Vela pulsars as well as very old, rapidly rotating pulsars in binary systems.

The pulsed radio emission that one observes does not result from the time-varying magnetic field (the spinning bar magnet). If it was the case, the observed radio light would be monochromatic and have a frequency corresponding to the spin frequency of the pulsar. Instead, radio emission is created by relativistic electrons accelerated along the open magnetic field lines of the pulsar. They emit synchrotron radiation as they gyrate around the open magnetic field lines under the Lorentz force \citep{gol68,pac68} and this produces an emission cone along the magnetic pole. A pulsation is visible each time the emission cone sweeps across our line of sight just as the fan-beam of a light house illuminates the seashore. Important pieces of the puzzle are still missing, however, in order to complete the radio emission model \citep{lyu08}.

\subsubsection{X-Ray Pulsars}\label{s:xray_pulsars}
X-ray (rotation-powered) pulsars are typically young, energetic pulsars that emit pulsations in X-rays but, sometimes, also in other high energy bands such as $\gamma$-rays. They owe the production of high-energy photons to their high spin-down rates, which deposit a large amount of energy in the surrounding magnetosphere. The combination of rapid spin period and large magnetic field is ideal to accelerate charged particles efficiently. The accelerating electric potential drop across their open field lines is estimated to be \citep{rs75}:
\begin{equation}
\Phi \simeq \frac{\Omega^2 B R^3}{2c^2} \approx 6 \times 10^{12} \frac{B_{12}}{P^2} \quad {\rm eV} \,.
\end{equation}
Particles accelerated at large Lorentz factors can produce energetic $\gamma$-ray photons. In turn, these photons can efficiently generate electron-positron pair cascades, that is, the interaction of two $\gamma$-ray photons produces an electron-positron pair ($\gamma + \gamma \rightarrow e^+ + e^-$), which can then be accelerated in the pulsar's magnetic field and also pair-produce \citep{hl06}. Depending on models, two such photons can generate several hundreds of these pairs --- this quantity is referred to as the particle multiplicity.

Various physical processes (inverse Compton scattering, synchrotron and cyclotron emission, curvature radiation, etc.) may be responsible for the high-energy emission of these pulsars and produce characteristic spectral components observable in the X-ray but at lower and higher energies as well \citep{har07}. The exact physical location and the intricate details of particle creation and acceleration are still not perfectly understood (see Figure~\ref{f:emission_regions}). One family of models considers that acceleration occurs close to the surface near the \emph{polar caps}, a region where the open field lines emerge from the surface \citep{har07}. In the main competing class of models, acceleration occurs further away in the magnetosphere in the \emph{outer gap} or the \emph{slot gap} \citep{har07}. The former region lies along the transition region between the open and closed field lines where the induced electric field and the local magnetic field cancel out ($\vec E \cdot \vec B = 0$). The latter region is located on the border of the last open field lines.

\afterpage{
\clearpage

\begin{figure}
 \centering
 \includegraphics[width=6in]{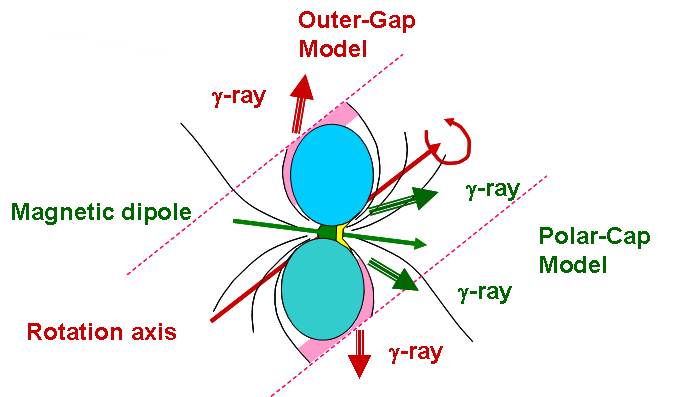}
 \caption[Schematic view of a pulsar's emission regions]{Schematic view of a pulsar displaying the different high-energy emission regions. The magnetic field lines wrapping at the light cylinder, marked by dashed red lines, define the closed field lines region (colored in blue). The outer gaps, in pink, are located around the first open field lines while the polar caps lie straight above the magnetic pole. Credit: PoGOLite Collaboration Webpage.}
 \label{f:emission_regions}
\end{figure}

\clearpage
}

Whereas X-ray pulsars tend to be young, rapidly spinning neutron stars with ``typical'' magnetic fields, some of them were recently found with high magnetic fields ($\sim 10^{13}$ -- $10^{14}$~G) and long spin periods ($\sim 1$~s) \citep{saf08,gkc+07,km05}. It is not perfectly clear how these pulsars are related to other X-ray and radio pulsars yet, but it is reasonable to believe that they slowed down rapidly because of their high magnetic fields. Even though their large energy output can be accounted for by their loss in rotational energy, they lie in the same neighborhood of the $P-\dot P$ diagram as magnetars (see Figure~\ref{f:p_pdot} and, below, \S\,\ref{s:magnetars}). Are these high-magnetic field pulsars some transitional state between normal radio pulsars and magnetars? Are they the natural extension of a diverse population having a wide range of initial properties? The recent observation of a magnetar-like burst from one of these X-ray pulsars by \citet{ggg+08} definitely brought solid evidence bridging the gap between the two families. Additional discoveries will be needed, however, to fully understand their interrelationship.

Several X-ray pulsars also display pulsed emission at other wavelengths such as radio, infrared and optical. This is the case for the Crab Pulsar \citep{wf83}. In these pulsars, the X-ray and radio pulse profiles sometimes look similar, with aligned phases, but this is not always the case \citep{wf83,ocg+08}. Many X-ray pulsars have very sinusoidal pulse profiles. Their spectra generally contain a thermal component as well as a non-thermal component, and are often fitted by a black body plus a power-law component. Although the high-energy photons certainly interact in the magnetosphere, it appears that the pulsation might be associated with a hot spot or some thermal inhomogeneity at the surface of the neutron star \citep{gkp06}.

\subsubsection{Rotating Radio Transients (RRATs)}
Rotating radio transients (RRATs) were identified for the first time by \citet{mll+06} in the reprocessing of the Parkes Multibeam Pulsar Survey using a new search technique called a \emph{single pulse search}. The technique consists of searching raw, dedispersed time series by convolving them with a kernel function --- typically a top-hat function, but other choices are possible --- in order to identify data sequences having values significantly above the noise level. In the original discovery paper, 11 radio transients were found using this technique. For all of them but one, it was possible to assess that they had a regular spin period. In one case, a significant period derivative was even found, thus allowing one to infer its magnetic field strength: $5 \times 10^{13}$~G.

This discovery considerably changed the approach to searching pulsars as it suggests the existence of a large population of neutron stars emitting radio pulses in an intermittent way \citep{mll+06}. In this case, conventional Fourier techniques would generally fail to detect these RRATs because many of them typically have very low burst duty cycles.

The nature of these RRATs is unclear and several ongoing research studies try to address this problem. They may simply represent the more extreme part of the distribution of pulsars showing nulls and/or drifting and/or mode switching behaviors. This hypothesis may certainly account for some of the RRATs since more sensitive observations revealed persistent radio pulsations in a handful of them \citep{mcl06}. It was also shown that some pulsars displaying important subpulse variations or mode switching would appear as RRATs if they were located further away \citep{wsr+06}. More research in this field will be needed to shed light on the RRAT phenomenon and its connection to conventional radio pulsars.

\subsection{Non-Rotation-Powered Pulsars}
Among pulsars that are not powered by their rotational energy are the accretion-powered pulsars and the magnetars. Below is a succinct overview of these two sub-classes of pulsars.

\subsubsection{Accretion-Powered Pulsars}
Accretion-powered pulsars are found in binary systems. They also often appear as low-mass X-ray binaries (LMXBs) and high-mass X-ray binaries (HMXBs). As their names indicate, these pulsars accrete mass from a binary companion, typically via Roche-lobe overflow or wind accretion \citep{gho07}. Accretion-powered pulsars can show spin-down or spin-up behaviors depending or whether the accretion mechanism transfers angular momentum to or away from the pulsar. They present spin periods ranging from a few milliseconds to several hundred seconds \citep{lcc+05}. Accretion millisecond pulsars in LMXBs are close siblings to MSPs which are believed to be their descendants (see more details in \S\,\ref{s:binary}) \citep{acr+82,rs82}. The pulsations coming from these pulsars, observable in X-rays, are produced by the reprocessed thermal emission from their surface, which is heated when the accreted mass, streamed by the magnetic field, lands at the polar caps \citep{shk07}.

\subsubsection{Magnetars}\label{s:magnetars}
Magnetar is the generic name for the family of pulsars comprising Anomalous X-ray Pulsars (AXPs) and Soft Gamma-Ray Repeaters (SGRs). These pulsars experience episodic burst and outburst events during which their flux can increase by as much as three orders of magnitude above their pre-burst level \citep{wt06}. After these high-level activities, magnetars relax back to a quiescent level. Because of the large dynamic range in flux, a large fraction of the magnetar population might be very difficult to detect in their quiescent state. For this reason, blind searches for new magnetars are generally ineffective and instead they are found by serendipitous observations or in all-sky monitors while they burst \citep{ims+04}. The relatively low, though somewhat unknown, duty cycle of magnetars lets us suppose that they may represent an important fraction of the Galactic neutron star population --- maybe as large as the radio pulsar population, according to optimistic estimates \citep{wt06}.

Magnetars are characterized by large period derivatives and long spin periods (between 2 and 12~s according to the online magnetar catalogue\footnote{\url{http://www.physics.mcgill.ca/~pulsar/magnetar/main.html}}). This implies young characteristic ages and enormous dipolar magnetic fields ($10^{14}$ to $10^{15}$~G). Their observed luminosity in outburst, and even in quiescence for some of them, is much larger than would be obtained from 100\% efficiency conversion of their spin-down energy to X-ray emission and hence they cannot be powered by their rotation \citep{kas07}. Two main models originally attempted to explain their nature \citep{wt06}. First, it was proposed that magnetars could be powered by accretion from a mass-transferring companion or by a fallback disc left over after the supernova explosion. Very stringent upper limits have been set using timing on the mass of possible companions and show that all known magnetars are isolated. Also, no trace of disc that would be massive enough to trigger outbursts has been found. In contrast, the \emph{magnetar} scenario \citep{dt92b}, in which the enormous magnetic field of these neutron stars powers their emission, gained observational evidence over time and is now well established.

AXPs and SGRs were historically identified as different kinds of objects since they displayed slightly different properties. However, with the increasing number of sources --- as of October 2008 there are now 6 known SGRs and 9 known AXPs --- and their long-term monitoring, it seems that AXPs and SGRs overlap very much. The discovery of SGR-like bursts from the AXP 1E1048.1$-$5937 by \citet{gkw02} really closed the gap between the two classes and showed that they both are manifestations of magnetars, perhaps linked by an evolutionary path.

A tremendous amount of observational and theoretical progress has been made over recent years, yet no fully self-consistent physical model has been able to explain their properties as a whole, such as their spectrum and timing. Magnetars display a wide variety of properties and more observations will be needed to sort out the exact picture. Qualitatively, however, it is generally believed that magnetic stress accumulated in the crust causes a slow twisting of the magnetic field outside the neutron star \citep{tlk02,td95}. When the twisting reaches a critical level, an event similar to a starquake happens and is accompanied by a magnetic reconnection (see Figure~\ref{f:magnetar_sketch}). This deposits an extremely large amount of energy in the magnetar's magnetosphere, maybe as a fireball. Such an event would be followed by a reorganization of the external magnetic field structure.

\afterpage{
\clearpage

\begin{figure}
 \centering
 \includegraphics[width=6in]{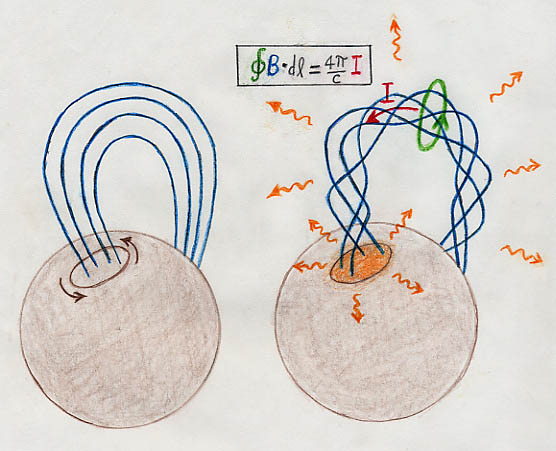}
 \caption[Magnetar's magnetic field lines twisting]{Illustration of the twisting of the magnetic field lines of a magnetar that leads to their outburst. Credit: Sketch from Robert C. Duncan (http://solomon.as.utexas.edu/~duncan/magnetar.html).}
 \label{f:magnetar_sketch}
\end{figure}

\clearpage
}

\section{Measurements of Pulsar Properties}
\subsection{Mass}
Determining pulsar masses is essential to address a number of questions related to pulsars, but also to other fields of physics and astrophysics. One of the most sought answers related to neutron stars concerns their internal structure and, more specifically, their equation of state (see \S\,\ref{s:eos}). Tackling the equation of state of neutron stars would in principle require simultaneously measuring radii and masses for several neutron stars, and quantifying their age and luminosity would also be desirable \citep{lp01,lp04,lp07}. In fact, being able to do so for just one neutron star would itself be a great help! Although this might become possible some day, significant progress can already be made from mass measurements only. Equations of state generally predict minimum and/or maximum masses and therefore bracketing the range of possible masses can help to exclude several models \citep{lp07}.

Pulsar mass measurements are also useful to investigate their formation and evolution. Accurately measuring masses for isolated pulsars is very challenging. It would certainly be interesting to relate their mass to their progenitor and determine whether or not pulsars are all born with a ``universal'' mass. Masses of pulsars in binary systems are more easily obtained and can provide information about differences arising from evolutionary channels \citep{vbj+05,vdh07}. As we shall see in \S\,\ref{s:binary}, pulsars that have shorter periods and lower magnetic fields are believed to have been recycled more, or more efficiently, than those having intermediate spin periods and magnetic fields \citep{brr+07}. A very important question that arises is whether the amount of mass accreted plays a crucial role in the spin up process and how much is needed in order to recycle a pulsar to fast spin periods\citep{wij97}. Answering this question would not only help understanding binary evolution better, it would also partly answer questions related to accretion efficiency as well as the process through which the binary pulsars' magnetic fields dramatically decrease when they get recycled \citep{bis06}.

The sample of currently available pulsar masses is rather small and besides a handful of binary pulsars it has a limited precision. Nevertheless, this sparse data set has already proven to be extremely helpful. One of the most striking features of the early mass measurements is that the mass distribution was comprised within a relatively narrow range of masses centered around $1.35 \pm 0.04$~\Msun \citep{tc99}. More recent measurements, however, appears to yield a larger range (see Figure~\ref{f:psr_mass}). Also, the fact that some pulsars could be as massive as $2.08 \pm 0.19$~\Msun \citep{fwv+08} appears to favor the `stiff' equations of state at high densities and would probably rule out exotic models such as quark stars.

\afterpage{
\clearpage

\begin{figure}
 \centering
 \includegraphics[width=5in]{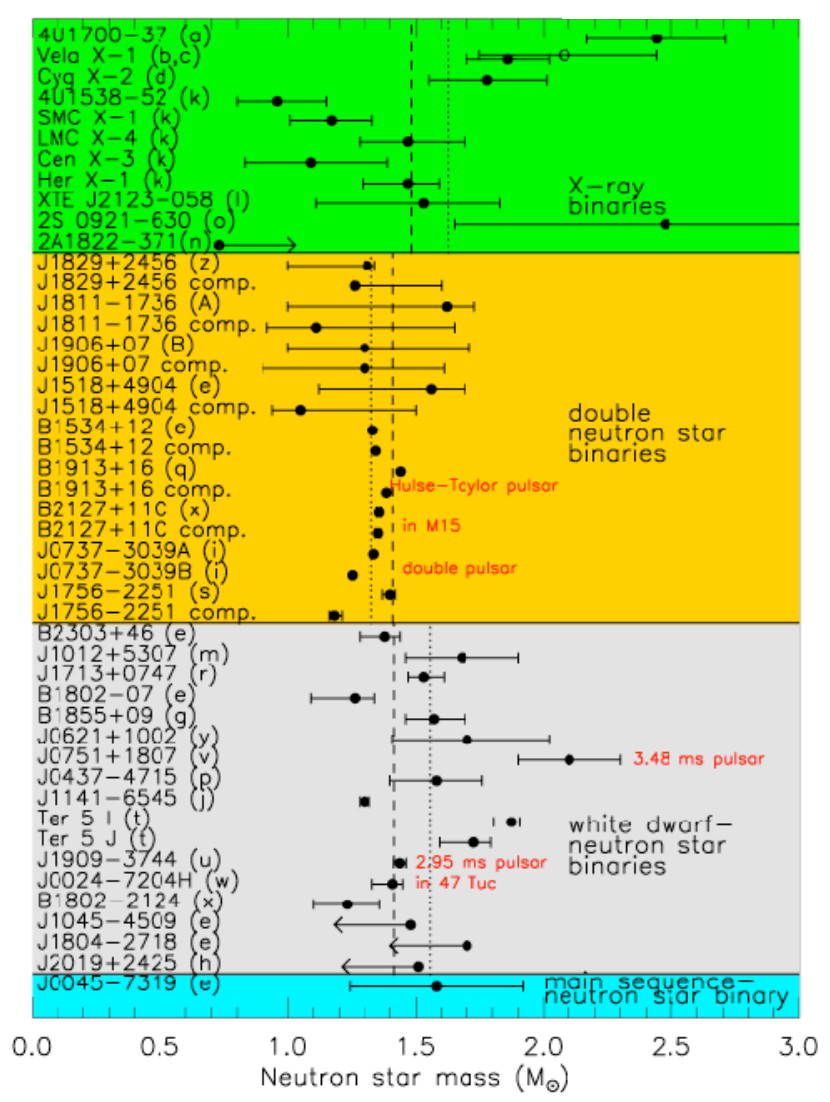}
 \caption[Neutron star mass measurements]{Neutron star mass measurements for X-ray binaries (green), double neutron star binaries (yellow), white-dwarf $-$ neutron star binaries (gray) and main sequence $-$ neutron star binaries (blue). Credit: \citet{lat07}.}
 \label{f:psr_mass}
\end{figure}

\clearpage
}

Pulsar masses are determined in several ways. Pulsar timing offers the most precise and reliable way of measuring masses for pulsars in binary systems \citep{nsk08}. As we shall explain in \S\,\ref{s:relativistic_timing}, under the assumption that general relativity is valid, the measurement of two `post-Keplerian' parameters allows one to solve for both masses in the system. Despite the great quality of these measurements, the number of binary systems that present measurable post-Keplerian parameters is small because they require a relativistic orbit and/or a favorable mass/geometry configuration. So far, post-Keplerian parameters have been observed in 21 binary systems of which 7 yielded measurements of pulsars' masses\footnote{Note that two or more post-Keplerian parameters must be measured in order to determine the mass of a pulsar in a binary system.} \citep[as of 2006, c.f.][]{nic06}. The most precise estimate obtained so far is for the double pulsar PSR~J0737$-$3039A/B, which contains two radio pulsars. The mass of both of them has been measured with an unparalleled precision: 1.3381(7) and 1.2489(7)~\Msun, respectively \citep{ksm+06}.

Another great tool for measuring the mass of a pulsar relies on the observation of its companion at optical and/or near-infrared wavelengths \citep{vbj+05}. The Doppler shift of the spectral lines induced by the radial component of the orbital motion allows one to solve for the total mass of the system. One can then combine it with a post-Keplerian parameter, when available, or with the mass of the companion inferred from our knowledge of stellar structures and atmospheres in order to obtain the mass of the pulsar (see Figure~\ref{f:wd_fit} and \citet{bvk+06}). In fact, the majority of binary pulsars are in orbit with white dwarf companions (see \S\,\ref{s:binary_population}). These degenerate stars are ideal since their internal structure and atmosphere are ``simple'' compared to that of neutron stars and ``normal'' stars and hence existing models are very accurate. Fitting their spectrum, particularly their prominent Balmer absorption lines, allows one to infer their surface gravity and temperature \citep{ber01,blr01,bsw95}. Once a cooling model is applied, this yields a precise mass determination of the white dwarf. Similar measurements can be obtained from other kinds of stellar-type companions such as OB(e) stars, though the mass determination from their spectra is much less accurate. When the system's parameters are over-constrained, one can alternatively use this information to validate white dwarf and stellar models.

\afterpage{
\clearpage

\begin{figure}
 \centering
 \includegraphics[width=6in]{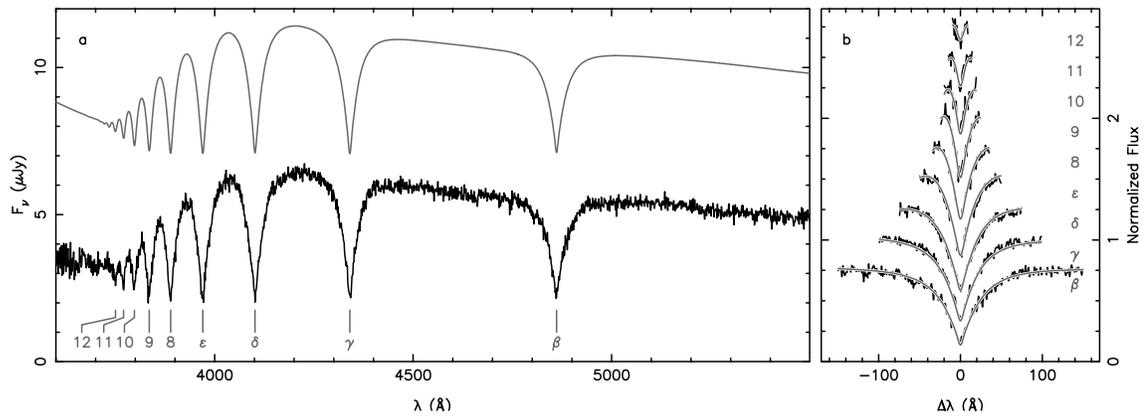}
 \caption[Spectral fit of a white dwarf's atmosphere]{Spectra of the white dwarf companion to PSR~J1911$-$5958A (lower left panel) compared to the best-fit white dwarf atmosphere model (upper right panel). The Balmer lines series is showed in the right panel with the best-fit curves superimposed to the data. One can infer the effective temperature and surface gravity, from which the mass and radius can be determined. Credit: \citet{bvk+06}.}
 \label{f:wd_fit}
\end{figure}

\clearpage
}

Other mass estimates can be obtained from observations of accreting X-ray binaries displaying quasi-periodic oscillations (QPOs) \citep{zyk+07,mlp98,vdk06} or derived from redshift measurements of spectral line features in type I X-ray bursts \citep{wb07}. In either case, these alternative ways of measuring masses are model-dependent and do not rival the precision attained by dynamical techniques.

\subsection{Radius}
So far, radius measurements for neutron stars have been very difficult. In contrast with mass measurement, timing cannot be used as a proxy to directly infer the radius of a neutron star. The lack of radius measurements for pulsars having well determined masses has been a limiting factor in the quest for determining their equation of state.

Other types of neutron stars, however, offer interesting means of determining their size. In low-mass X-ray binaries for example, a neutron star accretes mass from a companion. This process is not continuous and the neutron star will undergo active accretion phases interrupted by periods of quiescence. When such a neutron star is quiescent, there is either radiation of the energy accumulated in its crust or generated by the residual low-rate, radial accretion at its surface \citep{wb07}. In either case, the X-ray thermal emission originates from the neutron star's atmosphere. The spectrum will therefore depend on the atmosphere composition as well as the structure and strength of the neutron star's magnetic field \citep{lat07}. Typical modeling assumes a hydrogen-rich atmosphere and neglects the effect of the magnetic field \citep{wb07}. The latter assumption is justified since these measurements are carried out on old neutron stars that present low magnetic fields ($\sim 10^8$~G), which are negligible in this context. In general, one can determine the redshifted radiation temperature, $T_{\infty}$, from the spectral analysis (see Figure~\ref{f:quiescent}). If the distance is reliably known --- nearby neutron stars with parallactic distance or globular cluster neutron stars are used most of the time --- one can deduce the redshifted radius of the neutron star, $R_{\infty}$, using $F_{\infty} = (R_{\infty}/d)^2 \sigma T_{\infty}^4$ \citep{wb07}. Under an assumption or independent measurement of the surface gravity or neutron star mass, one obtains the proper radius of the neutron star $R_{\infty} = R/\sqrt{1-2GM/Rc^2}$ \citep{wb07}.

\afterpage{
\clearpage

\begin{figure}
 \centering
 \includegraphics[width=6in]{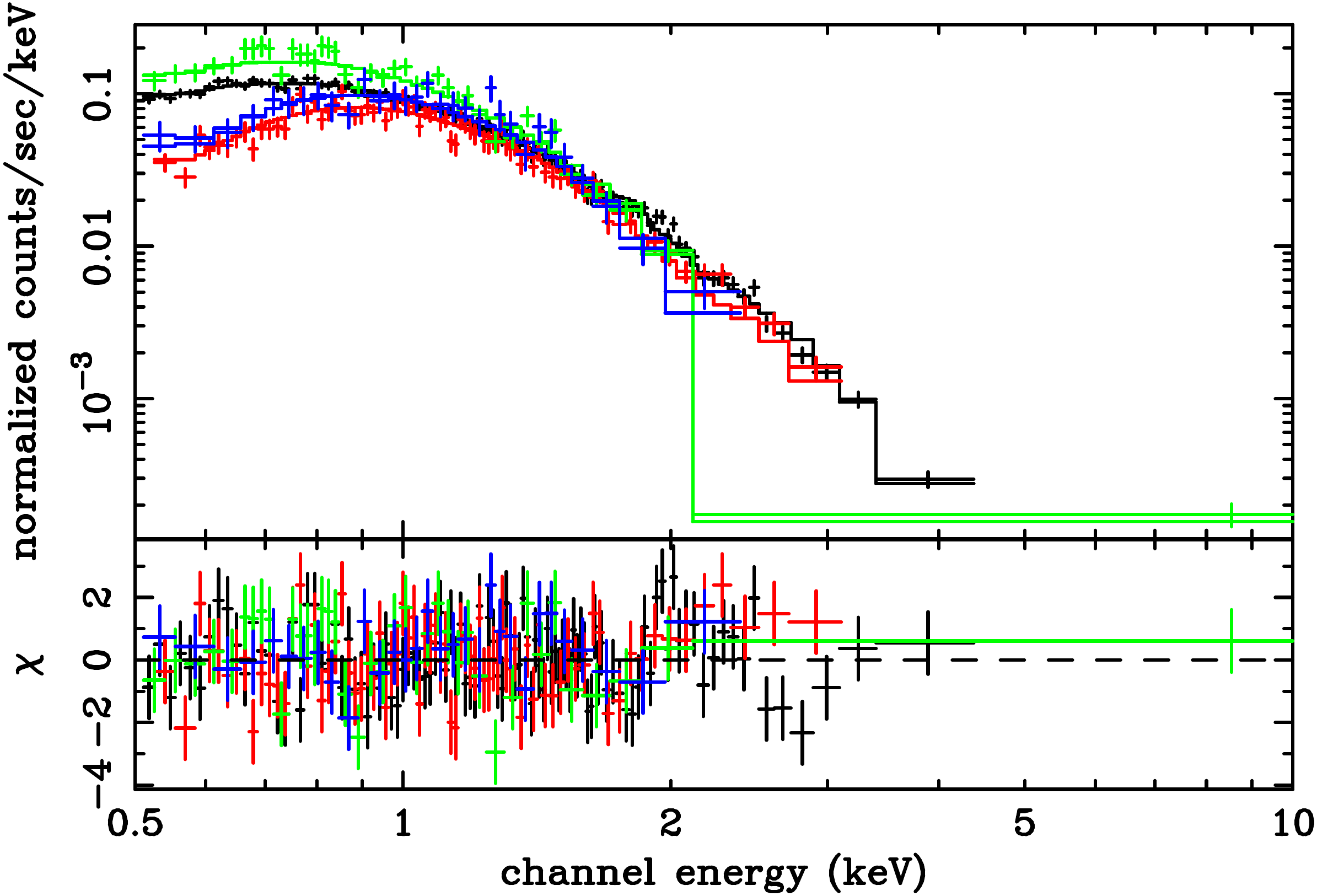}
 \caption[X-ray spectrum of a quiescent LMXB]{X-ray spectrum of a quiescent low-mass x-ray binary in the globular cluster 47~Tuc observed with the Chandra X-ray Telescope. The detector subarrays are indicated with different colors as they do not have the same response. The data were fitted by the hydrogen atmosphere NSATMOS model and photoelectric absorption, which allows to constrain the mass-radius ratio. Credit: \citet{hrn+06}.}
 \label{f:quiescent}
\end{figure}

\clearpage
}

As we can see, observations of quiescent, old neutron stars provide excellent tools for studying their equation of state. The caveat here, as for the non-dynamical mass determination techniques discussed previously, is that they rely on several assumptions such as atmosphere models and distance estimates.

\subsection{Magnetic Field}
The magnetic field of pulsars is a fundamental ingredient responsible for their pulsed emission in the various energy bands that are observed. It controls the way that the rotation-powered pulsars spin down \citep{gol68}, and it appears to play a pivotal role in binary evolution \citep{bis06}. In magnetars, the magnetic field even provides the main source of energy \citep{td95}. Measuring the magnetic field strength of pulsars is therefore interesting but understanding its structure is also relevant for explaining how they are generated, how they are coupled to the internal structure of the neutron star, how emission mechanisms work and how accretion and recycling proceeds.

In rotation-powered pulsars, magnetic field strength values are inferred from the pulsar spin-down (see Equation~\ref{e:magnetic_field}). These measurements are based on the assumption that pulsars are spinning magnetic dipoles and that they spin in a vacuum \citep{go69}. Although this model is thought to provide reliable order-of-magnitude results, it is desirable to obtain independent measurements. The timing-based magnetic field method is also extended to magnetars despite the fact that they are clearly not powered by dipolar radiation.

At the present time, there is no direct means of determining the magnetic field strength of radio pulsars. In X-ray pulsars and X-ray binaries, however, absorption lines are sometimes detectable and proper identification of these lines can yield an independent estimate of the magnetic field. For instance, the electron-positron cyclotron resonance frequency is proportional to the magnetic field strength and is equal to the electron rest-mass energy (511 keV) when the field reaches $4.414 \times 10^{13}$~G \citep{gho07}. Depending on the exact magnetic field strength, one could in principle see absorption features at the fundamental energy or at higher-order harmonics.

Spectral features that appear to be electron cyclotron lines have been detected in about a dozen accreting X-ray binaries \citep{sta03b,hl06}. They yield an estimated $\sim 10^{12}$~G magnetic fields, which is consistent with spin-down measurements in radio pulsars. In magnetars, detections of spectral features \citep{gdk08,ris+03,isp03,gkw02} indicate that this class of neutron stars would have $\sim 10^{15}$~G magnetic fields if the features are interpreted as proton cyclotron resonance. Further indirect evidence of pulsar magnetic field values have been obtained in other ways such as spectral modeling of magnetar X-ray emission \citep{gog08}.

Investigating the magnetic field structure of pulsars is probably even more difficult than inferring their strength. The relative success of the pulsar spin-down model at predicting the right magnetic fields and braking indices are certainly excellent accomplishments of the dipole model, as is the pulsar lighthouse model at explaining general radio pulsar shapes and polarization properties \citep{rc69b,hm01}. Nevertheless, they do not constitute unique proof of perfect dipolar structures and hence obtaining more ``direct'' evidence is desirable.

Some of the most significant progress has been made not by studying pulsars themselves but, instead, by making use of the nebulae that surround the young, energetic ones such as the Crab Pulsar. These \emph{pulsar wind nebulae} (PWNs) are created by the large amount of radiative energy and energetic particles deposited by pulsars in the surrounding interstellar medium (see Figure~\ref{f:pwn}). Several of these PWNs, which are particularly bright in X-rays, display toroidal and jet-like structures \citep{buc08}. Detailed magnetohydrodynamics simulations suggest that their morphology closely follows the structure of the pulsar's magnetic field \citep{kl04}.

\afterpage{
\clearpage

\begin{figure}
 \centering
 \includegraphics[width=6in]{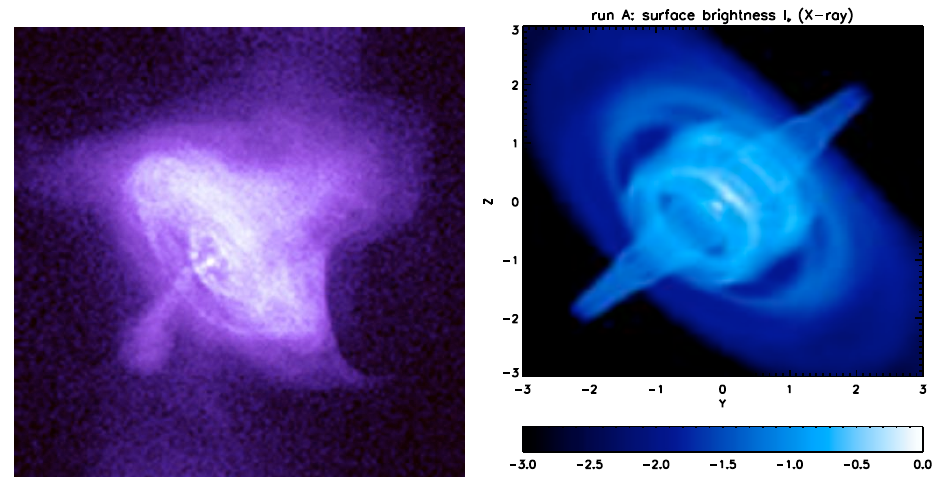}
 \caption[The Crab's pulsar wind nebula]{Comparison of the Crab pulsar wind nebula observed in X-ray by Chandra (left) with a simulated X-ray brightness map (right). Models are able to reproduce general features of pulsar wind nebulae such are toroidal and ring-like structures as well as dipolar jets. Credit: \citet{vda+07} and NASA/CXC/SAO.}
 \label{f:pwn}
\end{figure}

\clearpage
}

More recently, observations and modeling of magnetospheric eclipses in the double pulsar PSR~J0737$-$3039A/B opened up a new, direct way of probing the magnetic field structure of pulsars \citep{lt05,bkk+08}. This constitutes, in fact, part of the material that we shall present in Chapter~\ref{c:0737_eclipse}.

\section{Radio Telescopes}\label{s:radio_telescopes}
This thesis essentially concentrates on radio pulsars in binary systems. Most of the data used in the original research presented here were therefore collected with radio telescopes. More specifically, two radio telescopes have been used: the 100-meter Robert C. Byrd radio telescope (also known at the GBT) in Green Bank, West Virginia, United States (see Figure~\ref{f:gbt}); and the 64-meter Parkes radio telescope in New South Wales, Australia (see Figure~\ref{f:parkes}). In this section, we shall briefly describe the general principle behind radio observation of pulsars\footnote{This section is based on the information found in \citet{lg06,cr08,whi96}}.

\afterpage{
\clearpage

\begin{figure}
 \centering
 \includegraphics[width=6in]{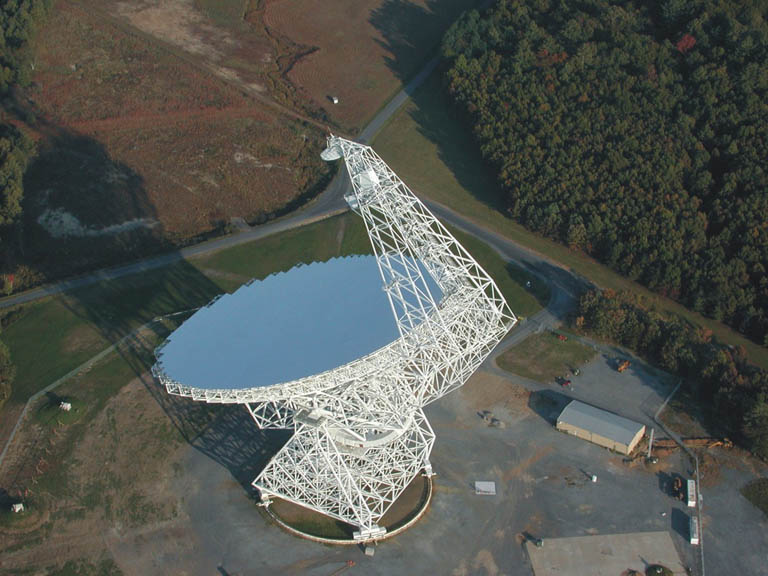}
 \caption[The Green Bank Telescope]{The 100-meter Robert C. Byrd radio telescope in Green Bank, West Virginia. Credit: NRAO / AUI / NSF.}
 \label{f:gbt}
\end{figure}

\clearpage

\begin{figure}
 \centering
 \includegraphics[width=6in]{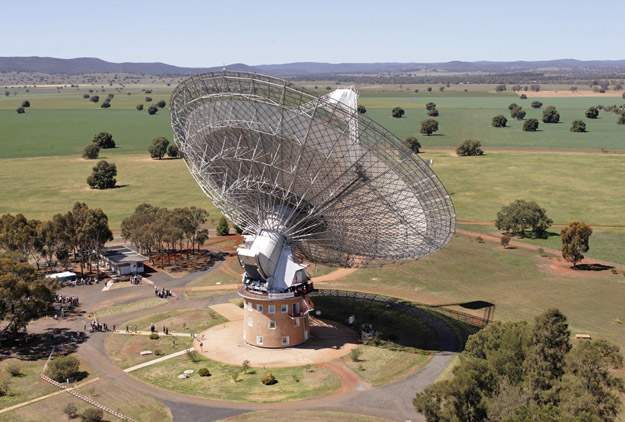}
 \caption[The Parkes Telescopse]{The 64-meter Parkes radio telescope in Parkes, New South Wales, Australia. Credit: Shaun Amy, CSIRO.}
 \label{f:parkes}
\end{figure}

\clearpage
}

Pulsar observations are generally conducted on single ``dish'' radio telescopes. Just like more conventional optical telescopes, the aim of the primary reflector is to collect radio waves over a large effective area and redirect them toward a device converting the radio signal into an electrical signal. The larger the effective collecting area, the more sensitive the radio telescope is at detecting weak sources. Since pulsars are relatively weak point sources, no imaging capability is required to observe them. Therefore, single dish antennae are perfect instruments since they offer great sensitivity. All the photons are generally collected by one receiver; in other words, these single dish antennae are one-pixel cameras.

The radio signal is generally sampled by a wave-guide feed sensitive to the two orthogonal polarizations of light. Each polarization then follows a separate channel. A block diagram showing a typical (simplified) back-end system of a radio telescope is showed in Figure~\ref{f:heterodyne}. The detected radio frequency (RF) is first amplified and then sent into the receiver. Most modern radio telescopes use superheterodyne receivers that allow the whole system, except the front-end, to work in a fixed, narrow range of frequencies. The RF amplified signal is first mixed with the signal generated by a local oscillator (LO). The results of this process are four signals coming out of the mixer and having frequencies corresponding to the original RF signal ($f_{RF}$), the original LO signal ($f_{LO}$), and two signals consisting of the sum and the difference of the RF and LO signals: $f_{RF} + f_{LO}$ and $f_{RF} - f_{LO}$. This technique can be used to convert the high frequency RF signal into an intermediate frequency (IF). Lower frequency signals are more efficiently transported and easier to analyze by electrical devices than higher frequency ones. By filtering the output one can therefore isolate the IF, which is usually chosen to be the difference signal: $f_{IF} = f_{RF}-f_{LO}$. By tuning the frequency of the local oscillator, one can always manage to obtain the same $f_{IF}$ regardless of the RF signal and thus the receiver can be optimized to work around this frequency. After the step of filtering, the IF is amplified. Additional steps of mixing with another local oscillator, filtering and amplifying are sometimes done in order to improve the quality of the received signal. Finally, the signal can be demodulated in order to recover the baseband signal. Sampling devices can then record the signal. Alternatively, it can also go to filter banks or correlators directly without being demodulated.

\afterpage{
\clearpage

\begin{figure}
 \centering
 \includegraphics[width=6in]{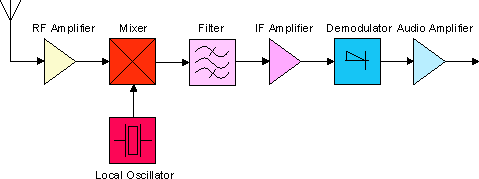}
 \caption[A superheterodyne receiver]{Block diagram showing the main components of a superheterodyne receiver. Credit: Wikipedia, GNU Free Documentation License.}
 \label{f:heterodyne}
\end{figure}

\clearpage
}

\section{Pulsar Timing}\label{s:timing}
One of the characteristics that make pulsars unique astrophysical objects is the periodic emission of electromagnetic waves. The pulsar science framework is based on this property and timing pulsars with an astonishing precision permits inference of several physical quantities. Pulsar timing is itself a complex task and we shall describe in this section the general principles of this fine art\footnote{Unless otherwise stated, references used to write this section are \citet{lk04,lg06}}.

When doing pulsar timing, one essentially attempts to associate the observed pulses from a pulsar, or more strictly speaking their times of arrival (TOAs), with a model describing their past behavior, from which future TOAs can be predicted. In order to obtain a TOA, one has to fold the pulsar time series using a preliminary timing solution. This process allows one to stack individual subpulses and construct a profile having a signal significantly above the noise level. Bear in mind that the pulsar signal is periodic whereas the noise should in principle be random and uncorrelated over an extended amount of time. This explains how data folding permits the detection of pulsars with individual subpulses much lower than the noise level. Depending on the characteristics of the pulsar being timed (isolated or in binary system, flux density, etc.), the folded profile accumulates between a few pulses to several minutes of observation. Then, a template representing the pulse profile --- often a high signal-to-noise profile --- is cross-correlated with the folded profile in order to find the precise time corresponding to the location of a fiducial reference feature. Once TOAs are generated, one fits them to the timing model. Timing is generally an iterative process in which new TOAs are added to the old ones and the timing solution is improved from the existing one. TOAs can then be regenerated if the new timing solution significantly changes the folded profile used to generate them. Several ingredients are involved in pulsar timing models but we can generally separate them into two fundamental categories: extrinsic and intrinsic components.

Two of the most widely used pulsar analysis packages are {\tt PRESTO}, developed by Scott Ransom \citep{rem02}, and {\tt SIGPROC}, developed by Duncan Lorimer \citep{sig}. We made use of both pulsar packages, and more particularly {\tt PRESTO}, for the processing of the data presented in this thesis. Pulsar timing is almost exclusively done using the {\tt TEMPO} software, which was written by Taylor, Manchester, Nice, Weisberg, Irwin, Wex
and others\footnote{\url{http://www.atnf.csiro.au/research/pulsar/tempo/}}. More recently, a revamped version of this software called {\tt TEMPO~2} has mainly been written by \citet{hem06}\footnote{\url{http://www.atnf.csiro.au/research/pulsar/tempo2/}} in order to allow for extended timing capabilities such as the \emph{pulsar timing array}, which is used in order to detect extremely low-frequency gravity waves \citep{jhv+06}.

\afterpage{
\clearpage

\begin{figure}
 \centering
 \includegraphics[width=5.5in]{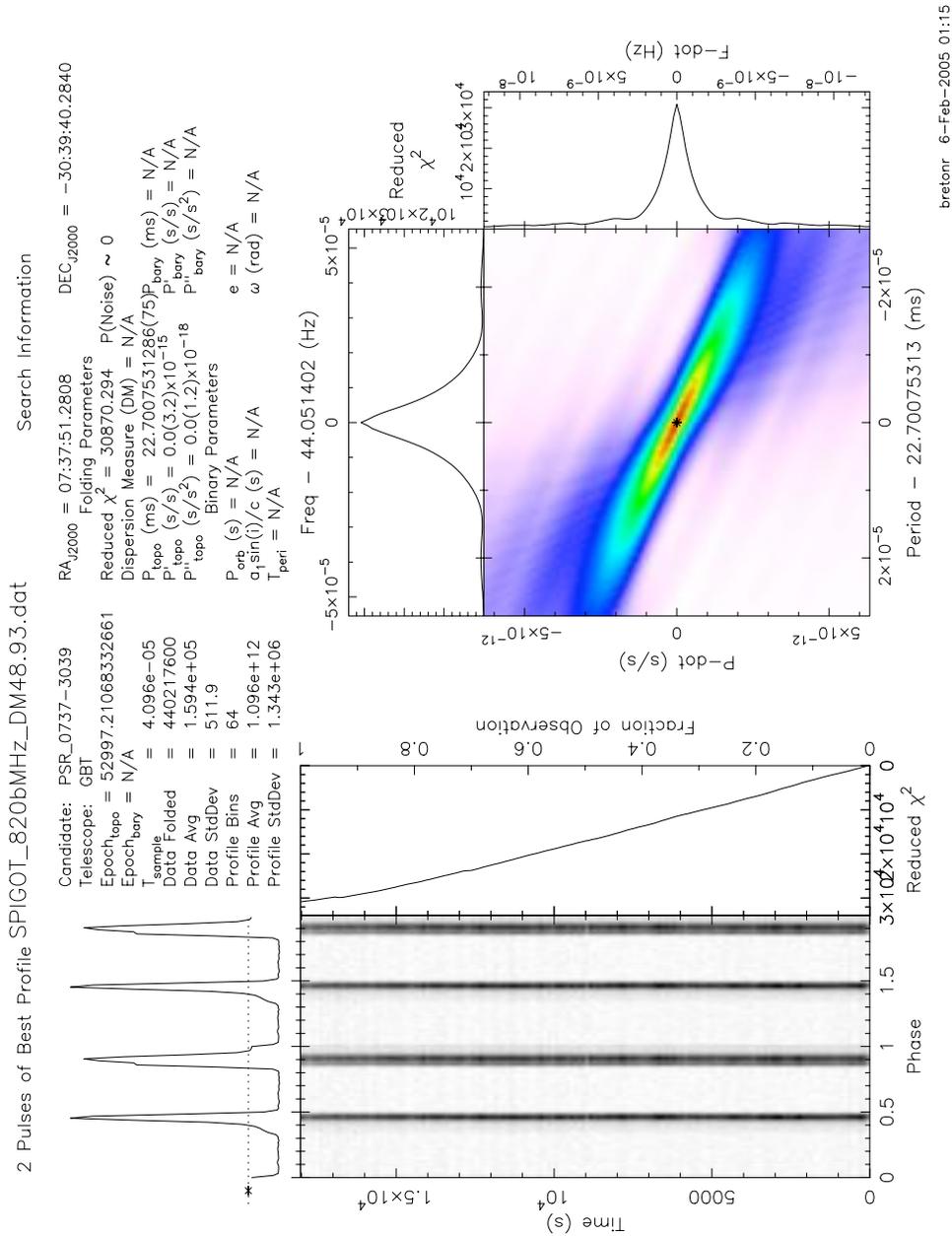}
 \caption[Typical output of pulsar data]{Typical output of pulsar data, here PSR~J0737-3039A, made by the pulsar analysis package {\tt PRESTO} \citep{rem02}. General information about the pulsar is presented at the top of the page and a figure displaying the chi-square of a search around the fold period and period derivative is showed in the lower right corner. The data are folded and the summed profile is displayed on the top left corner, while the sub-fold intervals are shown below.}
 \label{f:presto}
\end{figure}

\clearpage
}

\subsection{Extrinsic Timing Components}
To first order, pulsar timing is relatively simple: pulsars spin and emit beams of light that intersect our line of sight once per rotation. Although it is in fact more complicated than that, let us assume for now that some pulsars intrinsically emits perfectly periodic signals. As seen from a ground-based observatory, the time interval between pulses would vary over the course of a year because of the motion of the Earth around the Sun. The Earth is therefore not a suitable choice for pulsar timing and choosing the Solar System barycenter (SSB) is certainly more appropriate.

One can relate the TOA of a pulse received by an observatory on Earth, $t_{\rm topo}$ (topocentric time), to one measured at the SSB, $t_{\rm SSB}$ (barycentric time), as follows \citep{lk04}:
\begin{equation}
 t_{\rm SSB} = t_{\rm topo} + t_{\rm corr} - \frac{\Delta D}{\nu^2} + \Delta_{R\odot} + \Delta_{S\odot} + \Delta_{E\odot} \,.
\end{equation}

The first correction, $t_{\rm corr}$, is a time correction that synchronizes the observatory clock to the Terrestrial Time international time standard, TT(BIPM), which is maintained by the Bureau International de Poids et Mesures (BIPM) based on the average time of a large number of atomic clocks.

The second correction, $\frac{\Delta D}{\nu^2}$, is the dispersion measure correction (see Figure~\ref{f:dm}). As we discussed previously, light travels at a different speed in an ionized medium because of the change in refraction index. This traveling speed is frequency-dependent and varies as the square of the observed frequency, $\nu$. When barycentering a pulsar's TOA, one corrects the observed time of a signal arriving at an infinitely high frequency or, in other words, as if space were empty. In this correction, $\Delta D  = {\cal D} \, {\rm DM}$, where ${\cal D} = \frac{e^2}{2\pi m_e c} = 4.148808 \times 10^3$~MHz$^2$~pc$^{-1}$~cm$^3$~s is a constant and DM is a measurable quantity called the \emph{dispersion measure}. The dispersion measure corresponds to the integrated column density of free electrons along the line of sight, ${\rm DM} = \int_0^d n_e {\rm d}l$\quad pc\,cm$^{-3}$, and can be determined experimentally using the frequency-dependent behavior of $\Delta D$. If the observed bandwidth is sufficiently large compared with the dispersive delay, the signal will suffer a smearing across the band. This can be corrected ``incoherently'' by breaking the bandwidth into small frequency channels and applying a delay to each of them. Alternatively, one can use a ``coherent'' dedispersion system combined to the recording back-end that will apply the delay continuously across the bandwidth and avoid the frequency channel discretization approximation.

\afterpage{
\clearpage

\begin{figure}
 \centering
 \includegraphics[width=5.5in]{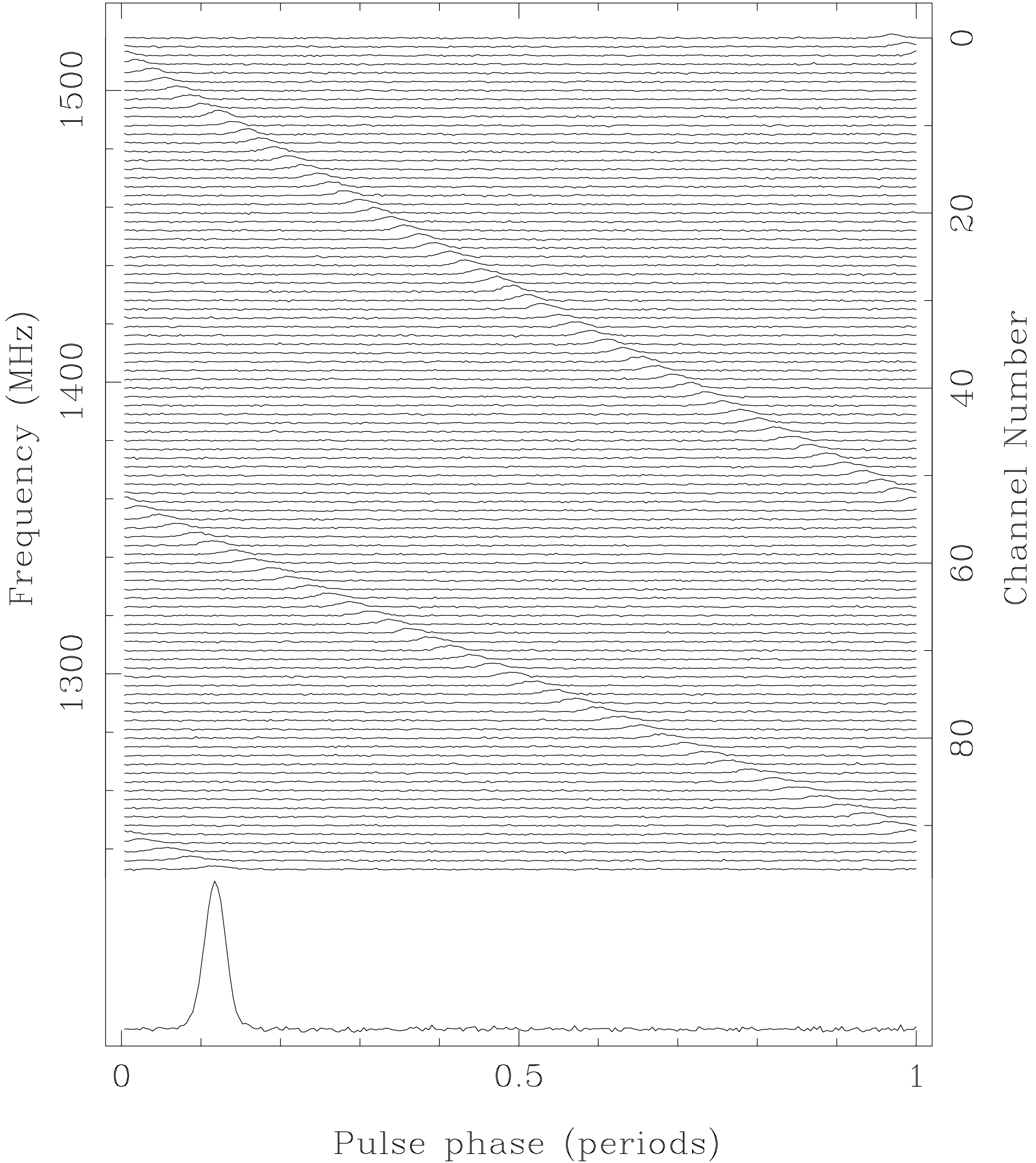}
 \caption[Pulsar signal dedisperion]{A typical pulsar signal dispersed by the ionized interstellar medium. The data are folded at the spin period in each frequency channel individually and presented with the observed frequency on the vertical axis and the pulse phase on the horizontal axis. We observe that the signal arrives with a delay (i.e. drift toward the right) as the frequency decreases. Credit: Handbook of Pulsar Astronomy, \citet{lk04}.}
 \label{f:dm}
\end{figure}

\clearpage
}

The third correction, $\Delta_{R\odot}$, is called the R\"omer delay. It is a classical correction to account for the finite speed-of-light travel time between the observatory and the SSB. This correction considers the precise motion of the Earth and includes perturbations from all major bodies in the Solar System. Another component of the R\"omer delay arises from the fact that observatories are not at the center of the Earth, hence their geographical location with respect to the center of mass of the Earth must also be included.

The fourth correction, $\Delta_{S\odot}$, is called the Shapiro delay. It is known from Einstein's theory of general relativity that matter curves space-time. Since light travels on straight lines (geodesics), the curvature generated by the Sun and other bodies of the Solar System increases the effective distance traveled by photons to us. The signature of this delay depends on the geometry; it varies according to the location of the body responsible for space-time curvature with respect to our line of sight to the source. As one may expect, the Shapiro delay becomes larger when the projected distance is smaller. The Shapiro delay can be decomposed into two independent parameters: the `shape' parameter describes the angular extent of the delay; and the `range' parameter scales its amplitude. In the Solar System, the Sun induces a delay that can reach $\sim 120$~$\mu$s whereas for Jupiter it goes up to $\sim 200$~ns. As for the R\"omer delay, contributions from all significant sources have to be summed up.

Finally, the last correction, $\Delta_{E\odot}$, is the Einstein delay. This delay results from the combined effect of time dilation, mainly due to the motion of the Earth in its orbit, and the gravitational redshift due to the gravitational potential of the Sun, Earth and other main bodies of the Solar System. A particular contribution to the Einstein delay arises from the ellipticity of the Earth's orbit. Since our distance to the Sun varies, the gravitational force at the Earth changes and induces a variation in the flow of time with respect to distant observers.

Since the arrival times of pulses undergo different kinds of delays that are position-dependent, it is possible to exploit this particular property in order to derive the pulsar position from timing even though the radio telescopes that are being used to observe them do not have imaging capabilities. For nearby pulsars, parallaxes can be obtained using the very long baseline of the Earth's orbit. In certain cases, pulsar timing parallactic distances have been measured up to several kiloparsecs \citep{hll+05}. Additionally, proper motion is sometimes observed for pulsars having apparent large transverse velocities. Timing parallaxes and proper motion measurements are more accurately determined for pulsars located away from the ecliptic plane as the R\"omer delay, which is the leading correction, scales as $\cos \beta$, where $\beta$ is the ecliptic latitude.

\subsection{Intrinsic Timing Components}
As we can see, even for a simple periodic signal, pulsar timing requires one to deal with several high-precision details. Intrinsically, pulsars' timing behavior should be relatively simple; they spin at a very stable rate and gradually slow down due to the conversion of their rotational energy into electromagnetic radiation. As we have described in \S\,\ref{s:rotation_powered}, dipole radiation, as well as other forms of rotational slow down, have well defined behaviors relating the period, the first-order, and the higher-order period derivatives. In a practical way, however, barycentered TOAs are empirically fitted to a period signal that can have multiple order derivatives. The goal of the process is to ``phase connect'' the TOAs; i.e. to fit the observed TAOs such that each rotation of the pulsar is accounted for without losing phase coherence even if there are observational gaps.

In young pulsars, for which the spin-down rate is large, the second period derivative has physical significance and allows one to determine the braking index (see Equation~\ref{eq:braking_index}). For most pulsars, however, the timing behavior beyond the first period derivative is dominated by noise. This noise is generally very `red', i.e. it has a power density proportional to $f^{-\beta}$ with $\beta > 0$. In old MSPs, part of the noise may be attributed to external causes such as gravitational waves, dispersion measure variations, clock errors and inaccuracies in the solar system ephemeris. Most of these external sources should present well-defined correlations with the timing noise of other old MSPs \citep{hlk06}. \citet{hlk06} put stringent limits on the importance of these contributions and showed that some the noise in old MSPs is intrinsic to them \citep{hlk06}. In young pulsars, which have spin-down rates $\sim 10^{5-10}$ times larger than that of MSPs, external sources are negligible and hence their timing noise is dominated by intrinsic factors.

Pulsars, and more specifically young ones, sometimes experience `extreme' forms of timing anomalies known as \emph{glitches} \citep{rm69}. These events are characterized by a sudden, quasi-instantaneous\footnote{A `live' observation of a glitch in the Vela Pulsar revealed that the transition time from the original to the new period was less than 40~s \citep{dml02}} increase of the pulsar spin frequency occasionally accompanied by a change in the spin-down rate, which sometimes relaxes back to its pre-glitch rate \citep{lss00} (see Figure \ref{f:glitch}). The physical reason for glitches is not known precisely but it is generally agreed that since no significant pulse profile changes are observed, something must be occurring inside the neutron star such as the unpinning of superfluid vortices that would transfer angular momentum outward to the crust and cause the spin-up \citep{lss00}.

\afterpage{
\clearpage

\begin{figure}
 \centering
 \includegraphics[width=6in]{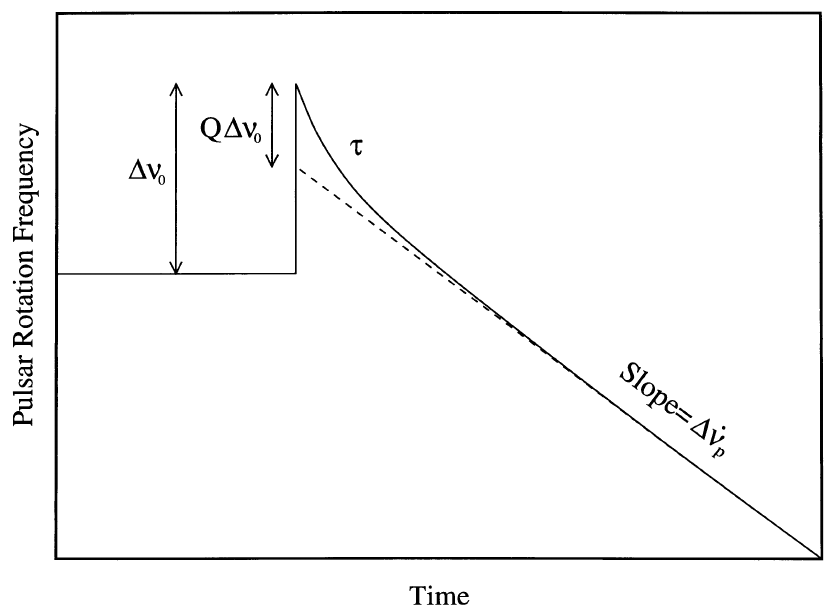}
 \caption[Representation of an idealized glitch]{Representation of an idealized glitch. The initial rotational frequency is constant because the constant spin-down (first period derivative) has been subtracted. When the glitch occurs, it causes a sudden $\Delta \nu_0$ increase in rotational frequency, which is often followed by a exponential recovery $Q \Delta \nu_0$. In the post-glitch epoch, the spin down may also be faster larger than initially by a fraction $\Delta \dot \nu_0$. Credit: \citet{lss00}.}
 \label{f:glitch}
\end{figure}

\clearpage
}

\subsection{Binary Pulsar Timing}\label{s:binary_timing}

When pulsars are in binary systems, additional delays affect their TOAs, notably because of their motion around the center of mass. In addition to the SSB corrections presented above, one has to convert the observed binary pulsar TOAs to its center-of-mass frame of reference using another set of corrections similar to those introduced above.

\subsubsection{Non-relativistic Systems}
When the orbital velocity is much smaller than the speed of light, Newtonian dynamics accurately account for the behavior of binary pulsars. In this situation, the R\"omer delay essentially reduces to the Doppler shift induced by the orbital motion around the center of mass and is described by Kepler's laws. Five parameters called the \emph{Keplerian parameters} are used to characterize  these orbits (see Figure~\ref{f:keplerian_orbit}): 1) the orbital period, $P_b$; 2) the eccentricity, $e$; 3) the projected semi-major axis, $a_p \sin i$; 4) the longitude of periastron (also known as periapsis), $\omega$; 5) the epoch of passage at periastron, $T_0$. Note that the orbital inclination, $i$, can generally not be determined from the timing of a non-relativistic binary pulsar as the orbital Doppler shift only affects the line-of-sight velocity component. Consequently, the measurable Keplerian parameters can be combined using Kepler's third law in order to obtain a quantity called the mass function of the system:
\begin{equation}
 f = \frac{(m_c \sin i)^3}{(m_p + m_c)^2} = \left( a \sin i \right)^3 \left(\frac{2 \pi}{P_b}\right)^2 \frac{1}{T_{\odot}} \quad \Msun \,,
\end{equation}
where $m_p$ and $m_c$ are the mass of the pulsar and its companion, respectively, in solar mass, $P_b$ is in days, $a \sin i$ is in light-second, and $T_{\odot} = 4.925490947$~$\mu$s. The mass function can be useful at determining the minimum mass of a companion (for $i = 90^{\circ}$) when a pulsar mass is assumed.

\afterpage{
\clearpage

\begin{figure}
 \centering
 \includegraphics[width=6in]{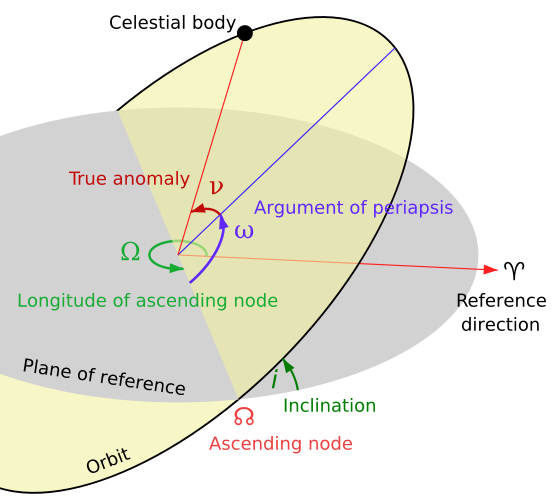}
 \caption[Schematic view of a Keplerian orbit]{Schematic view of a Keplerian orbit showing the Keplerian parameters that are used to describe the orbit. Credit: Wikipedia, GNU Free Documentation License.}
 \label{f:keplerian_orbit}
\end{figure}

\clearpage
}

In principle a sixth Keplerian parameter called the position angle (also known as argument) of the ascending node, $\Omega_{\rm asc}$ ($\Omega$ in Figure~\ref{f:keplerian_orbit}) describes the orientation of the orbit in the sky plane. As it requires supplemental knowledge about the geometry of the system, which is rarely measurable, and it does not provide relevant additional dynamical information, it is generally neglected.

\subsubsection{Relativistic Systems}\label{s:relativistic_timing}
In systems with larger orbital velocities, Newton's equations of motion fail to perfectly explain the orbital dynamics and treating these relativistic systems requires the incorporation of higher-order corrections to the classical motion. The most common way to proceed consists of using the \emph{post-Keplerian (PK) parameters}, which are purely phenomenological corrections to the orbit see \citep[see][for a complete description of relativistic binary pulsar timing]{dt92a}\footnote{The post-Keplerian (PK) formalism must not be confused with the parametrized post-Newtonian (PPN) formalism. The former is based on a phenomenological description of the orbital motion. It aims to extend the concept of Keplerian orbits to relativistic systems by accounting for secular variations of the Keplerian parameters and various other effects such as gravitational redshift, time dilation and aberration \citep{dt92a}. On the other hand, the PPN formalism is based on the metric representation of gravity \citep{wil01}. It accounts for the non-linearity of relativistic theories of gravity by expanding the metric in the regime of weak gravitational fields and slow velocities (typically ${\cal O}(v/c)$, ${\cal O}(v^2/c^2)$) around a flat space. Note that one can find an equivalence between the PK parameters and the PPN parameters \citep[see][]{dt92a}.}. Since this formalism does not rely on any particular theory of gravity, it gives one the freedom to establish the correspondence of these PK parameters to a given theory and test it. Within the framework of a theory of gravity, the PK parameters can be written as specific functions of the 5 Keplerian parameters as well as the two masses in the system. Of course, general relativity is a natural choice and we shall present a brief overview of the tests of gravity involving binary pulsars in \S\,\ref{s:gravity}.

In theory, there are several first-order and higher-order PK parameters that can be measured. So far, a total of five have been determined \citep{nic06,lbk+04}. They are: 1) the periastron advance, $\dot \omega$; 2) the gravitational redshift and time dilation, $\gamma$; 3) the orbital period decay, $\dot{P}_b$; 4-5) and the Shapiro delay `shape', $s$, and `range', $r$, parameters (see Figure~\ref{f:shapiro}). These PK parameters take the following form in general relativity \citep{lor05}:
\begin{equation}
 \begin{array}{rcl}
  \dot \omega &=& 3 \left( \frac{P_b}{2\pi} \right)^{-5/3} \left(T_{\odot} M\right)^{2/3} \left(1-e^2\right)^{-1} \,, \\
  \gamma &=& e \left( \frac{P_b}{2\pi} \right)^{1/3} T_{\odot}^{2/3} M^{-4/3} m_c \left(m_p + 2m_c\right) \,, \\
  \dot P_b &=& -\frac{192\pi}{5} \left( \frac{P_b}{2\pi} \right)^{-5/3} \left(1 + \frac{73}{24}e^2 + \frac{37}{96}e^4 \right) \left(1-e^2\right)^{-7/2} T_{\odot}^{5/3} m_p m_c M^{-1/3} \,, \\
  r &=& T_{\odot} m_c \,, \\
  s &=& x \left( \frac{P_b}{2\pi} \right)^{-2/3} T_{\odot}^{-1/3} M^{2/3} m_c^{-1} \,,  
 \end{array}
\end{equation}
where $M = m_c + m_p$, the total mass of the system.

The periastron advance is caused by the fact that relativistic orbits are not closed. The gravitational redshift arises from the `downhill'' and/or ``uphill'' travel of the photons in the potential well of the system in order to reach us, whereas the time dilation is due to the varying rate of time of accelerated frames of reference; both effects couple and are undistinguishable. The orbital period decay is a consequence of gravitational wave radiation that takes energy away from the system. Finally, the Shapiro delay is the same phenomenon as the SSB correction introduced above.

\afterpage{
\clearpage

\begin{figure}
 \centering
 \includegraphics[width=5in,angle=-90]{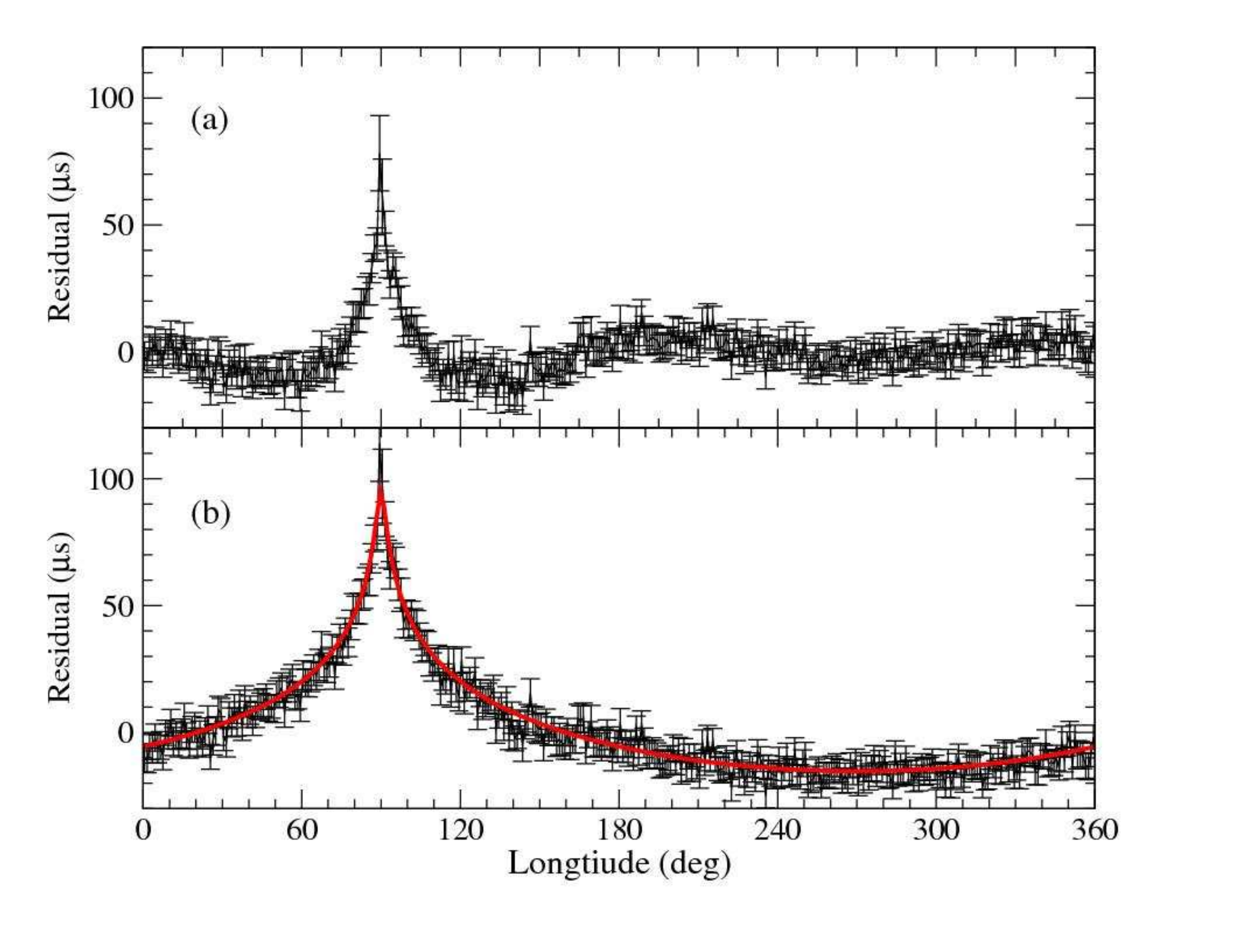}
 \caption[Effect of the Shapiro delay on the double pulsar]{Effect of the Shapiro delay on the double pulsar PSR~J0737$-$3039A/B timing. The plot shows the timing residuals as a function of orbital phase. In this figure, $90^\circ$ corresponds to the superior conjunction of pulsar A (i.e. pulsar A behind pulsar B). The upper panel displays the residuals for the timing model not including the effect of the Shapiro delay. The bottom panel show the residuals with the best-fit Shapiro delay curve in red. Credit: \citet{ksm+06}.}
 \label{f:shapiro}
\end{figure}

\clearpage
}

From the above PK parameters, $\dot \omega$ and $\gamma$ require eccentric orbits to be measurable, the Shapiro delay is visible for highly inclined (i.e. edge-on) orbits only, whereas $\dot{P}_b$ demands very short orbital periods. Varying the mass of the companion also affects the visibility of the PK parameters, and hence the favorable geometrical configuration of some systems made PK parameter measurements possible even in systems that do not appear to be very ``relativistic''.

\section{Binary Pulsars}\label{s:binary}
Binary pulsars constitute one of the predominant subsets of known pulsars and they are particularly relevant to radio pulsars and, of course, to accretion-powered pulsars. This thesis specifically focuses on binary radio pulsars and hence we shall provide a short review of binary pulsar evolution and classes of binary radio pulsars in this section.

Evolution with a binary companion, as opposed to being isolated, dramatically changes the observed properties of a pulsar. A quick look at the $P-\dot P$ diagram (see Figure~\ref{f:p_pdot}) is sufficient to realize that binary radio pulsars preferentially occupy the lower left corner of the diagram. That is, they generally have short spin periods, slow spin-downs and, correspondingly, low magnetic fields and large characteristic ages. A close look at the distribution of spin periods of binary and non-binary pulsars (see Figure~\ref{f:distribution_Binary}) does not leave any doubt: out of the 30 fastest spinning pulsars, 22 are found in binary systems\footnote{Based on the ATNF pulsar catalogue \citep{atnf}, as of September 2008.}. As we shall explain below, finding a high correlation between rapidly spinning pulsars and binarity is not a selection effect but a natural consequence of binary evolution.

\afterpage{
\clearpage

\begin{figure}
 \centering
 \includegraphics[width=6.5in]{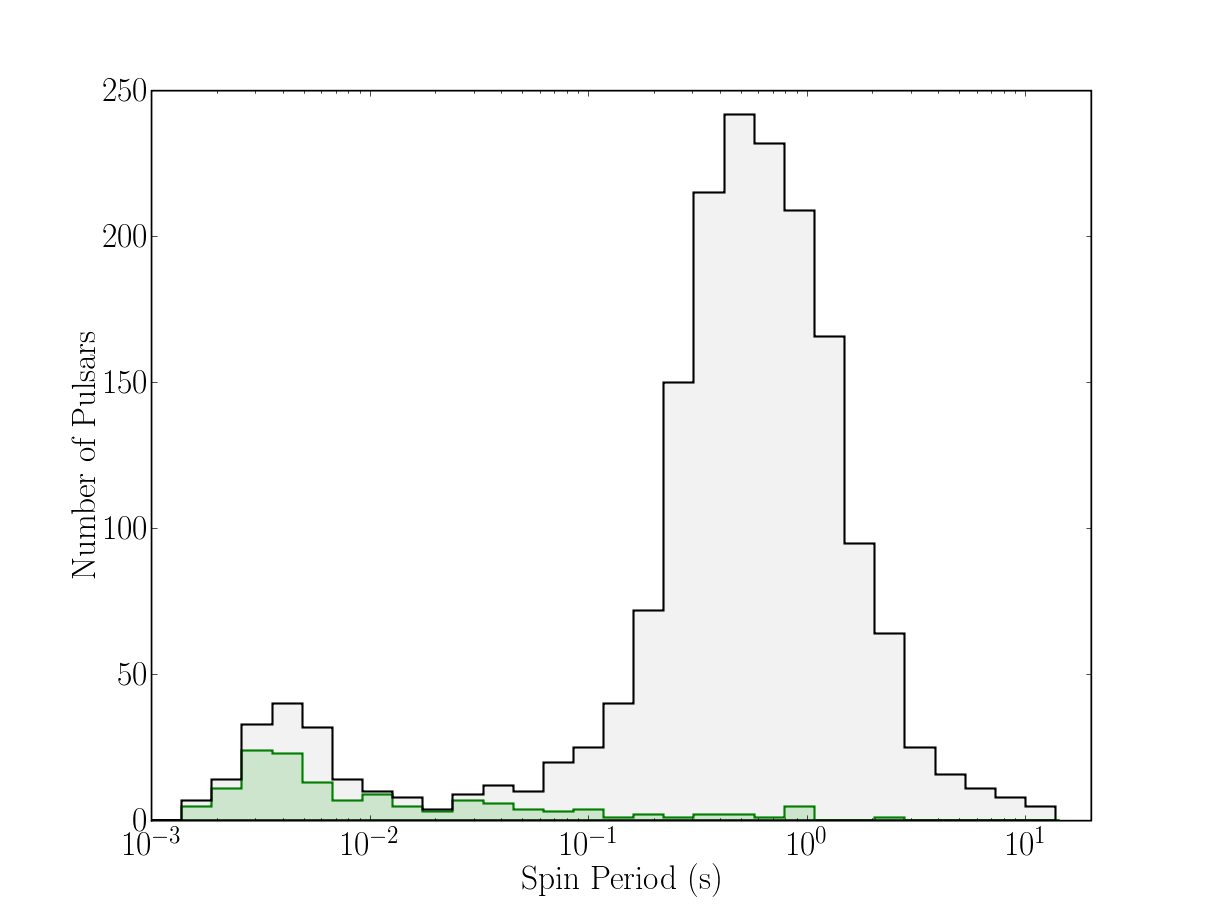}
 \caption[Distribution of spin periods]{Distribution of spin periods of the entire observed population of pulsars (gray) compared to that of binary pulsars only (green). Binary pulsars are preferentially found at short spin periods and constitute the majority of MSPs. Some of the isolated short period pulsars are young pulsars that have not spun down to long periods yet whereas the others are old pulsars, which might be descendants of binary systems that lost their companions somehow (see, for example, the black widow example in \S\,\ref{s:other_binaries}).}
 \label{f:distribution_Binary}
\end{figure}

\clearpage
}

The presence of a companion opens up a plethora of windows for studying pulsar physics and, even using them as astrophysical tools to test gravity and general relativity \citep{sta03a,sta04b}. This latter aspect will be covered in detail in \S\,\ref{s:gravity}. In this section, we will direct our focus on the basics of binary pulsar evolution and discuss some research aspects involving these systems.

\subsection{Binary Evolution}\label{s:binary_evolution}
As we mentioned in \S\,\ref{s:ns_overview}, in the classical picture, neutron stars are born from massive stars ending their life in supernovae. This scenario is appropriate when the progenitor star is isolated but stellar evolution becomes much more complex when two or more stars are gravitationally bound. In this situation, the evolution of each star is no longer independent but is, instead, coupled to that of its companion(s). Current surveys are inconclusive regarding the precise stellar multiplicity rate in the Milky Way and despite the fact that it appears to vary anywhere from $\sim$ 15 to 80\% and may depend on the spectral type \citep{lad06}, it nevertheless constitutes a non-negligible fraction of the Galactic population. It is therefore natural to expect that the pulsar population displays some observable properties inherited from binary evolution.

The interaction of stars in binary systems can manifest itself in different ways but the most crucial certainly is mass transfer. Mass transfer occurs in compact binaries in which the orbital separation is such that matter at the surface of one of the two stars becomes loosely bound and equally attracted by the companion (see Figure~\ref{f:roche}). We refer to this equipotential surface that delimitates the zone of influence of each body in a binary system as the Roche lobe, after the French astronomer who first studied this concept. If a star reaches the size of the Roche-lobe, matter will start being transfered to its companion via the inner Lagrangian point $L1$.

\afterpage{
\clearpage

\begin{figure}
 \centering
 \includegraphics[width=4.75in,angle=90]{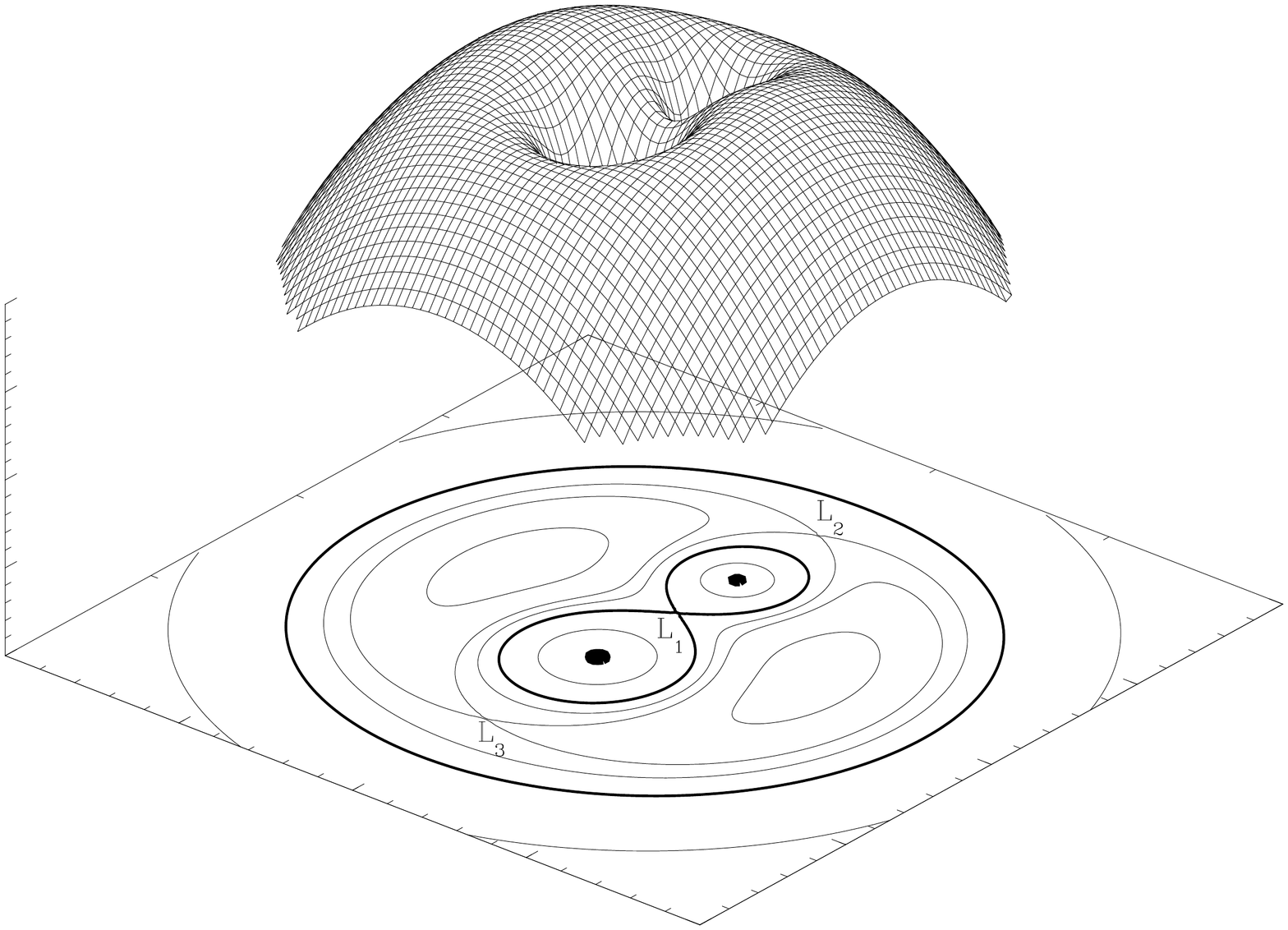}
 \caption[The Roche lobe equipotential surface]{Three-dimensional view of the equipotential gravitational surfaces between two orbiting bodies with a cross-section projection taken in the orbital plane of the system. The Roche-lobe surface, presenting the shape of an ``8'', marks the region of gravitational influence of each body in the system. Credit: \citep{vds06}.}
 \label{f:roche}
\end{figure}

\clearpage
}

The onset of mass transfer marks an important turning point in the binary system evolution since mass transfer plays a significant role on the evolution of the stars themselves and it is also accompanied by angular momentum transfer and non-conservative effects such as mass loss that change the orbital and rotational properties of the stars. According to \citet{gho07} and \citet{bv91}, the behavior of the interaction between the two stars mainly depends on: \emph{1) the evolutionary state of the core of the donor star at the onset of the mass transfer, 2) the donor star's envelope structure, and 3) the mass ratio}. \citet{kw67} introduced a classification scheme to identify the main evolutionary outcomes of a close binary system based on the above factors (see Figure~\ref{f:radius_evolution}): case A systems start mass transfer before the donor star ends hydrogen-core burning; case B systems start mass transfer between the end of hydrogen-core burning and before the ignition of helium-core burning; case C systems start mass transfer after the ignition of helium-core burning and before the ignition of carbon-core burning \citep{kw67,pac71,gho07}.

\afterpage{
\clearpage

\begin{figure}
 \centering
 \includegraphics[width=6in]{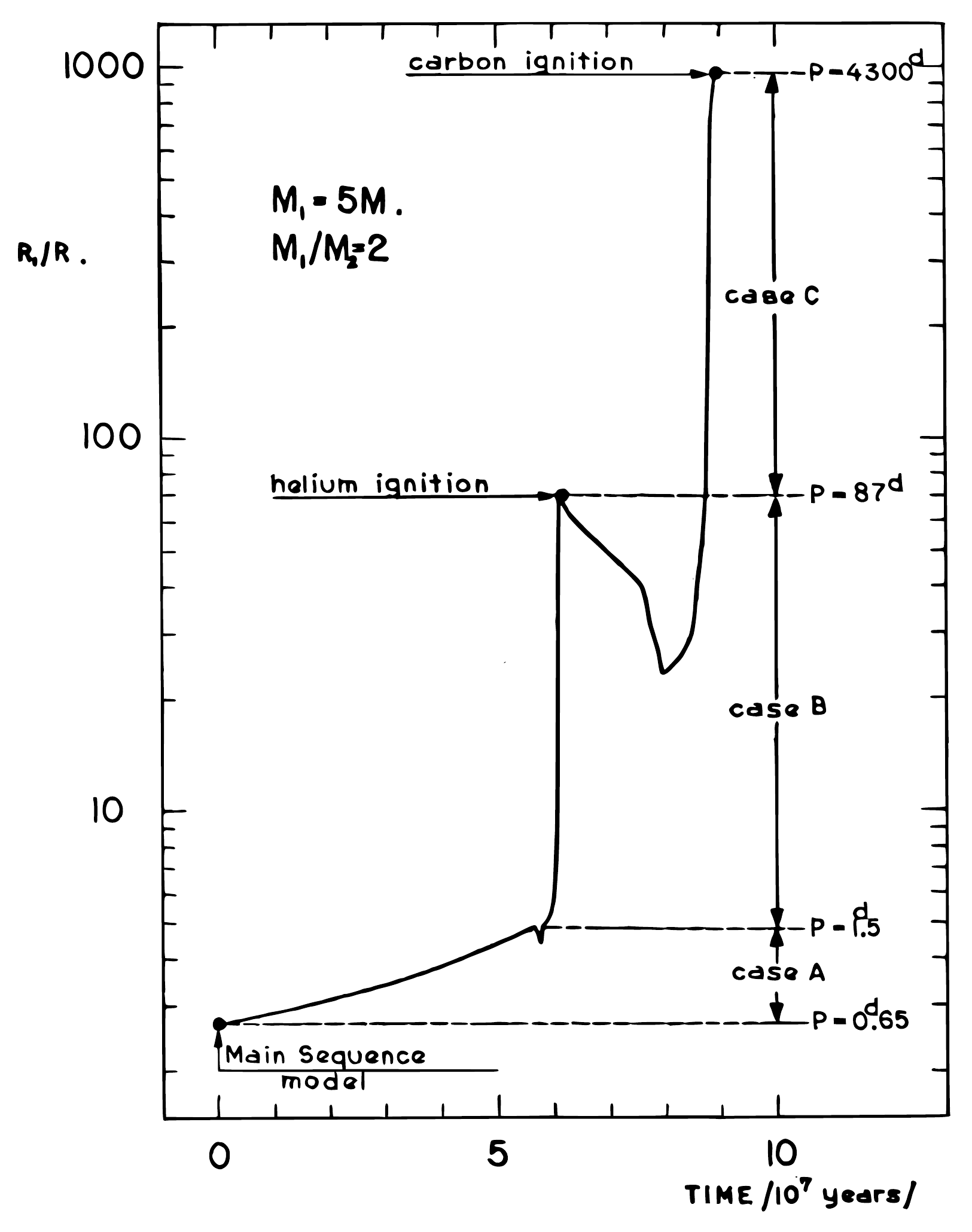}
 \caption[Radius evolution of a 5\,\Msun star]{Time evolution of the radius of a $M_1=5$\,\Msun star. The transitions between the three main evolutionary cases (A, B and C) are indicated along with the orbital period corresponding to a Roche-lobe having the size of the star, assuming an initial mass ratio $q=M_1/M_2=0.5$. Credit: \citet{pac71}.}
 \label{f:radius_evolution}
\end{figure}

\clearpage
}

The evolution of binary systems leading to the formation of binary pulsars usually involves a phase of mass transfer from the pulsar progenitor to its companion. Unless the pulsar's companion is also a neutron star or a black hole, the pulsar's progenitor is generally the star that was initially more massive since it evolves first. As we shall see in \S\,\ref{s:binary_population}, most binary pulsars comprise evolved companions and rare are the systems with non-evolved stars. In fact, all binary pulsars having short spin periods and low magnetic fields --- that is, the majority of the binary pulsar population --- present a common aspect: they all have evolved companions. This cocktail of properties leads to the conclusion that these pulsars have been ``recycled'' by their companions. In this process, a ``normal'' or ``old'' pulsar gets rejuvenated to a short spin period following the transfer of angular momentum while it is accreting from its companion. As we mentioned in \S\,\ref{s:binary}, it appears that the magnetic field of the pulsar also decays as a result of this process.

Despite that the intricate details of binary evolution depend on the evolutionary class to which the system belongs, the general behavior of the mass transfer is regulated by whether it is conservative or non-conservative (see Figure~\ref{f:evolution_cases}). Conservative mass transfer is associated with stars that have not developed a deep convective envelope. In such a case, the star restores its hydrostatic equilibrium on a thermal timescale, i.e. almost instantaneously \citep{bv91}. The star then has a smaller size than before and should therefore stop filling its Roche lobe. Since the star becomes out of thermal equilibrium, it will slowly expand to reach a new equilibrium on a thermal timescale and this should allow it to fill its Roche lobe again \citep{gho07}. In a practical way, this process acts in a feedback loop and this allows the star to transfer mass to its companion in a stable way.

\afterpage{
\clearpage

\begin{figure}
 \centering
 \includegraphics[width=6in]{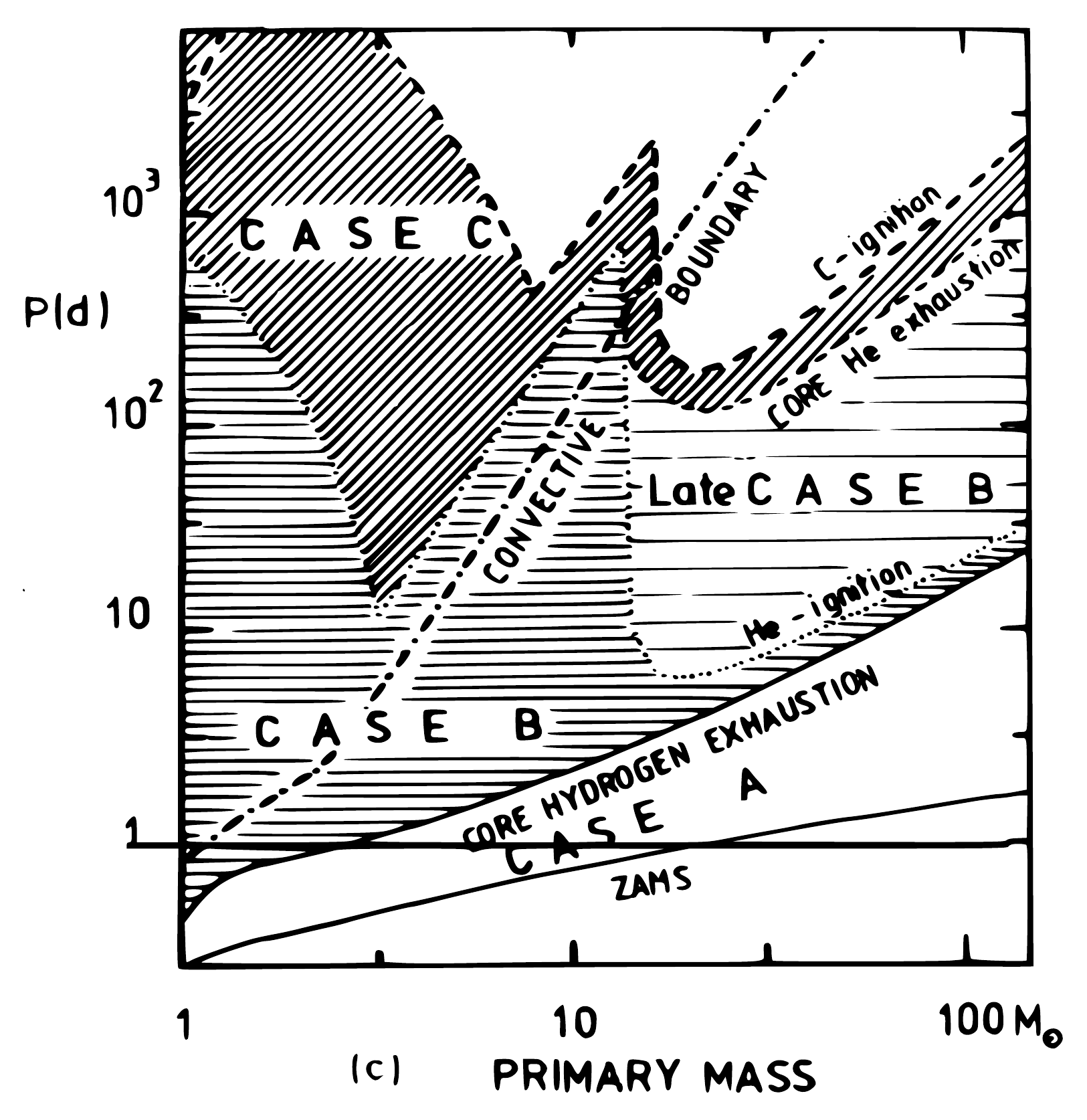}
 \caption[Evolution-type phase space]{The type of binary evolution depends on the donor star, $M_1$, and the orbital period at the onset of the mass transfer, assuming an initial companion mass ratio $q=M_1/M_2=0.5$. The convective boundary delimits systems that can evolve through stable Roche-lobe overflow from those entering unstable mass transfer in common envelope evolution. Credit: \citet{bv91}.}
 \label{f:evolution_cases}
\end{figure}

\clearpage
}

The equation of state of stars having a deep convective envelope is different than those having a radiative structure. The volume of these stars is nearly inversely proportional to their mass and hence as they transfer mass to their companion they tend to become larger on a dynamical timescale \citep{bv91}. In general, the change in orbital distance induced by the mass transfer is not fast enough in order to let the Roche-lobe radius accommodate the size of the star. Consequently, the star will enter a cataclysmic, highly non-conservative mass transfer phase called common envelope in which it becomes so large that the companion lies inside its envelope \citep{bv91}.

\subsection{Binary Radio Pulsar Population}\label{s:binary_population}
As we have argued before, different evolutionary paths lead to pulsars with different properties and companions \citep{bv91,sta04b,vbj+05}. One may choose to base the binary pulsar phylogeny on either aspect --- the pulsar properties or the companion type --- but it appears that both are complementary and constitute more or less two ways of viewing the same thing (see Figure~\ref{f:population}). In this subsection, we shall present an overview of the main classes of observed binary radio pulsars following that presented in \citet{vbj+05}, with some additions from the review by \citet{sta04b}.

\afterpage{
\clearpage

\begin{figure}
 \centering
 \includegraphics[width=6in]{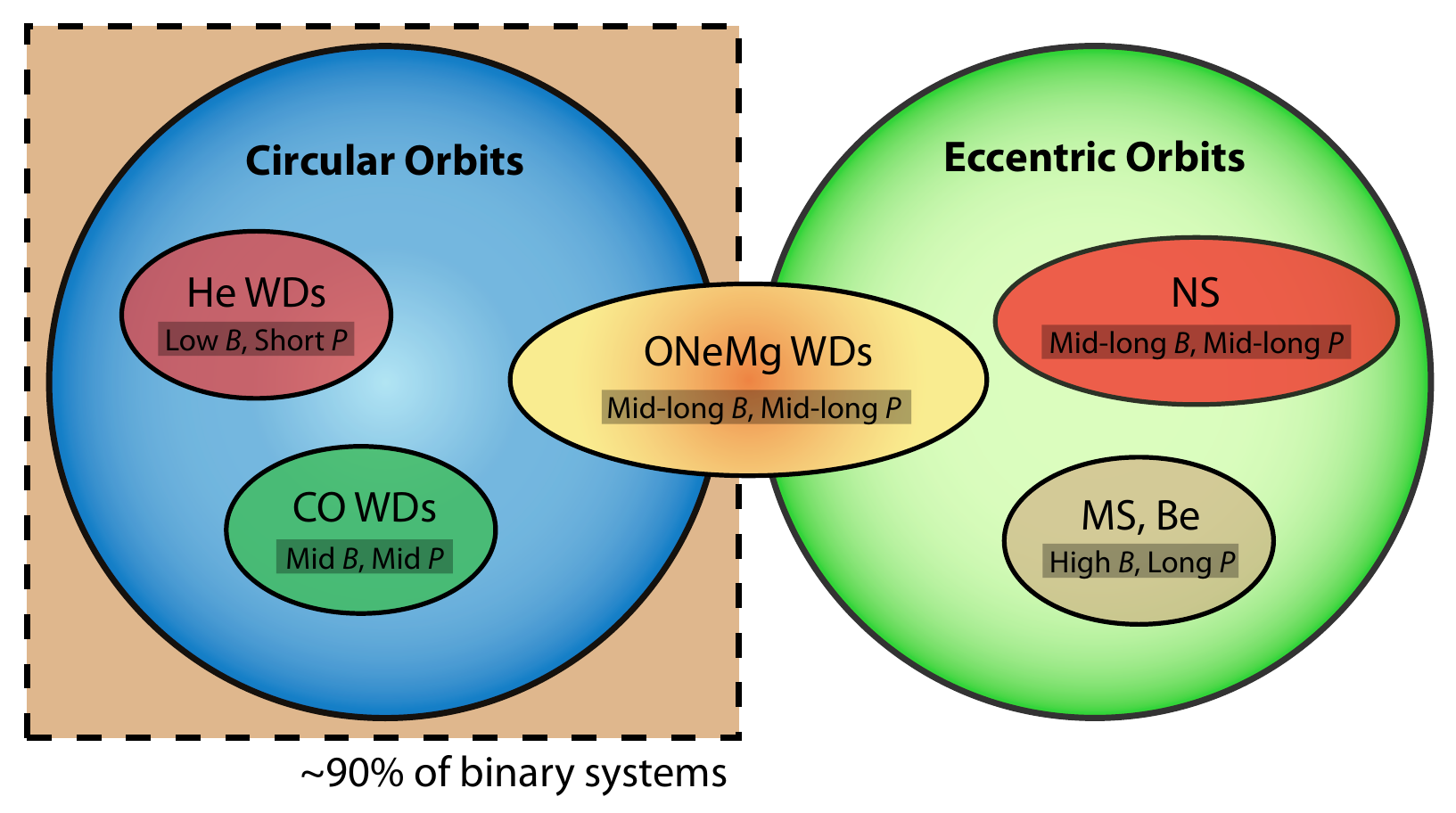}
 \caption[The binary pulsar population]{Venn diagram representing the binary pulsar population according to the nature of their companions and summarizing their main properties. The spin periods and magnetic fields are general qualitative remarks; short, mid-long and long spin periods being a few milliseconds, a few tens of milliseconds and a few hundreds of milliseconds, respectively, while low, mid and high magnetic fields are in the range of $10^{8-9}$, $10^{9-10}$ and $10^{11-12}$\,G, respectively. Therefore, `standard', isolated neutron stars would have `long' spin periods and `high' magnetic fields.}
 \label{f:population}
\end{figure}

\clearpage
}

\subsubsection{Pulsar + Main Sequence Star}
In binary systems consisting of pulsars with main sequence companions, the more massive star evolved faster and became the observed pulsar after exploding in a supernova. This type of binary is rare and they are usually found with massive, early spectral-type companions such as B stars. There is a strong observational bias against finding these systems as main sequence companions have limited lifetimes and will eventually start transferring mass to the pulsar. In several systems, the companion is also likely going to hide the pulsar or quench its radio emission if it has wind.

\begin{itemize}
\item {\bf Pulse period:} Typical of that of isolated pulsars.
\item {\bf Magnetic field:} Typical of that of isolated pulsars.
\end{itemize}

\subsubsection{Recycled Pulsar + He WD}\label{s:psr_hewd}
Following the nomenclature adopted by \citet{sta04b}, we shall also refer to this class as \emph{case ${\cal A}$}\footnote{In order to avoid confusion with the standard binary evolution nomenclature, we differentiate \emph{case A} introduced by \citet{kw67,pac71} from \emph{case ${\cal A}$} used here.}. Pulsars with helium white dwarf (He WD) companions typically evolve from \emph{low-mass X-ray binaries (LMXBs)} in which a neutron star accretes from a low-mass main-sequence star filling its Roche lobe. Initially, the system probably had a large mass ratio and the more massive star became a pulsar after the supernova. The less massive secondary then kept evolving until it filled its Roche lobe, which marks the onset of mass transfer of the LMXB phase. This generally occurs when the secondary ascends the giant branch and then becomes large. External factors may also contribute, though in a much lesser extent, to accelerating the evolution process. For example, heating caused by the newly born neutron star's radiation may bloat the secondary's envelope \citep{vkk+00} and magnetic braking may reduce the orbital separation \citep{es00,bd05}.

He WD progenitors are low-mass stars ($\lesssim 2.8$~\Msun), which, after the hydrogen-shell ignition, have helium cores that become degenerate and slowly climb the giant branch until they experience the helium flash \citep{bv91}. Theoretically, the evolutionary timescale of such stars forming $\lesssim 0.45$~\Msun He WDs is comparable to or longer than the Hubble time. Binary evolution, however, significantly accelerates the process; as the secondary loses mass to the benefit of the neutron star, its internal structure and the rate of nuclear reactions re-adjust to compensate for the mass transfer, which induces a variation of the orbital separation and, consequently, of the Roche-lobe radius \citep{ts99,taa04}. It is therefore possible to form $\lesssim 0.45$~\Msun He WDs in binary systems.

\begin{itemize}
\item {\bf Pulse period:} Steady and stable accretion occurring over a long timescale allows very fast, fully recycled millisecond pulsars ($P \lesssim 10$~ms).
\item {\bf Magnetic field:} Very low magnetic field ($\sim 10^8$~G).
\item {\bf Orbital period:} A correlation is predicted between the mass of the He WD and the orbital period of the system \citep{rpj+95}. The core mass of the progenitor is related to the size of its envelope and since mass transfer happens via stable Roche-lobe overflow, which is directly related to the orbital separation, the companion's radius is required to match that of the Roche lobe.
\item {\bf Orbital eccentricity:} A correlation is predicted between the eccentricity and the orbital period as a result of the damping of convective eddies in the white dwarf progenitor's envelope. This damping excites the epicyclic motion in the orbit and stochastically maintains the eccentricity to a larger value than one would expect from the tidal circularization \citep{phi92}.
\end{itemize}

\subsubsection{Recycled Pulsar + CO/ONeMg WD}
Following the nomenclature adopted by \citet{sta04b}, we shall also refer to this class as \emph{case ${\cal B}$}\footnote{In order to avoid confusion with the standard binary evolution nomenclature, we differentiate \emph{case B} introduced by \citet{kw67,pac71} from \emph{case ${\cal B}$} used here.}. Pulsars with carbon-oxygen white dwarf (CO WD) or oxygen-neon-magnesium white dwarf (ONeMg WD) companions experience more cataclysmic evolutions than their relatives from \emph{case ${\cal A}$}. Because the secondary star of these systems is initially more massive, helium-burning ignition will occur shortly after the start of hydrogen-shell burning and hence the structure of the star is considerably changed. Stable mass transfer via the inner Lagrangian point will generally not occur for an extended period because the star cannot dynamically readjust fast enough to match the Roche lobe. Instead, the secondary expands beyond the Roche lobe and the system enters an unstable \emph{common envelope}. The neutron star and the core of its companion will then in-spiral because of the viscous forces created by the surrounding envelope. The orbital binding energy is transferred into the envelope until enough is deposited to expel it, leaving the neutron star and a bare CO or ONeMg-rich core, which ultimately becomes a CO/ONeMg WD surrounded by a helium or hydrogen layer, depending on the evolutionary stage of the secondary at the onset of the common envelope. These systems are typically descendants of HMXBs or \emph{intermediate-mass X-ray binaries (IMXBs)} \citep{clm+01}.

\begin{itemize}
\item {\bf Pulse period:} Unstable mass transfer is believed to be less efficient and to span a shorter amount of time than stable Roche-lobe overflow. Consequently, pulsars are mildly recycled to intermediate spin periods ($10 \lesssim P \lesssim 100$~ms).
\item {\bf Magnetic field:} Intermediate between isolated pulsars and MSPs ($\sim 10^9$~G).
\item {\bf Orbital period:} No particular correlation is expected but the orbital separation should be relatively small because of the in-spiral experienced during the common envelope phase.
\item {\bf Orbital eccentricity:} No particular correlation is expected but the orbital eccentricity should be ``larger'' than for the \emph{recycled pulsar + He WD} class because the circularization should be less efficient for common envelope than stable mass transfer.
\end{itemize}

\subsubsection{Recycled Pulsar + Neutron Star}
The natural extension of the previous type of binary pulsars from the \emph{case ${\cal B}$} class is one with a companion massive enough to ignite triple alpha nuclear reactions and build up iron in its core. In such a case, the secondary would eventually explode as a supernova and become a neutron star just like the primary. During the mass transfer phase that leads to the formation of the second neutron star, these systems are generally observed as \emph{high-mass X-ray binaries (HMXBs)}. Like the above class, which eventually forms CO/ONeMg WDs, this type of binary experiences unstable mass transfer and common envelope evolution. This binary class is fundamentally different from those having lighter companions because a second supernova occurs in the system. Even though the orbit may have circularized during the last mass transfer stage, the kick imparted by the second supernova explosion will generally result in a large orbital eccentricity.

\begin{itemize}
\item {\bf Pulse period:} Similar as for the \emph{case ${\cal B}$} class --- mildly recycled ($10 \lesssim P \lesssim 100$~ms).
\item {\bf Magnetic field:} Similar as for the \emph{case ${\cal B}$} class --- intermediate magnetic field ($\sim 10^9$~G).
\item {\bf Orbital eccentricity:} The second supernova explosion produces a kick that, for most configurations, is likely going to impart a large orbital eccentricity. Some systems, however, appear to have a relatively ``small'' eccentricity \citep{vdh07}.
\end{itemize}

In these systems, the observed pulsar is the first-born neutron star and because it accreted mass from its companion, it experienced partial recycling. Note that the second-born neutron star may also emit radio pulsations. Its visibility depends on the viewing geometry of the system since its radio beam would also have to sweep across our line of sight. If it were the case, this second-born neutron star would appear with characteristics similar to those of an isolated pulsar because it did not experience recycling.

The double pulsar PSR~J0737$-$3039A/B is the only known case of a binary system in which both neutron stars are observable radio pulsars \citep{lbk+04}. Recent studies by \citet{dv04} and \citet{std+06} suggest that the double pulsar might have experienced a slightly modified version of the above formation mechanism. Rather than the conventional supernova explosion of a massive star having an iron core, the second-born pulsar would result from an electron-capture induced collapse \citep{std+06,vdh07}. This hypothesis is supported by the low orbital eccentricity and the probable small kick imparted to the system after the formation of the second pulsar. The evolution prior to the formation of the second pulsar, however, remains sensibly unchanged.

\subsubsection{Pulsar + Neutron Star}
This class of binary pulsar is very similar to the \emph{recycled pulsar + neutron stars} class. Here the observed pulsar is the second-born neutron star and it has not experienced recycling.

\begin{itemize}
\item {\bf Pulse period:} Similar to that of isolated pulsars.
\item {\bf Magnetic field:} Similar to that of isolated pulsars.
\item {\bf Orbital eccentricity:} The second supernova explosion produces a kick that, for most configurations, is likely going to impart a large orbital eccentricity.
\end{itemize}

Note that a double neutron star system may appear as a \emph{recycled pulsar + neutron star}, a \emph{pulsar + neutron star} or a \emph{recycled pulsar + pulsar} system, depending on whether they intrinsically emit radio pulsations and also on the viewing geometry.

\subsubsection{Pulsar + CO/ONeMg WD}
As we explained before, in the early evolution of a binary the more massive star evolves faster than its companion. If the primary star is massive enough, it will explode as a supernova and leave behind a neutron star remnant. It is important to bear in mind, however, that during the giant phase of the primary, mass transfer from the primary to the secondary will likely occur if the orbital separation permits. Such an event can dramatically change the course of the evolution of the system. For instance, if both stars have intermediate masses below the threshold for forming neutron stars, mass transfer during this stage can significantly increase the mass of the secondary. In this case, the primary would eventually exhaust its envelope and become a massive CO/ONeMg WD. Then, the newly massive secondary would experience a supernova explosion and become a neutron star. In this scenario, the pulsar is born from the secondary star, not the primary \citep{sta04b}.

\begin{itemize}
\item {\bf Pulse period:} Since the pulsar is formed after the white dwarf and does not accrete mass, it should resemble ``normal'' isolated pulsars.
\item {\bf Magnetic field:} Since the pulsar is formed after the white dwarf and does not accrete mass, it should resemble ``normal'' isolated pulsars.
\item {\bf Orbital eccentricity:} The eccentricity is likely to be large given that no mass transfer helped circularize the orbit after the supernova.
\end{itemize}

\subsection{Other Pulsar Binaries}\label{s:other_binaries}
There are pulsar binaries that fail to fit into one of the above categories. Obviously, binary evolution is a complicated process and even if the above classification succeeds in grouping the majority of binary radio pulsars, there is a continuum of properties rather than discrete characteristics. External factors and peculiar properties of the binary members can play an important role. As we shall discuss in Chapter~\ref{c:1744}, there may also exist alternative evolution channels that are more rare but yet common to several binary pulsars.

\subsubsection{Black Widow Pulsars}
PSR~B1957+20 is the classical example of peculiar binary systems \citep{fst88}. The radio pulsar in this system is an energetic MSP that has been fully recycled by its companion. It has a 1.6~ms period and a low $1.67 \times 10^8$~G magnetic field, typical of \emph{recycled pulsar + He WD} systems. Its companion, however, has an extremely low mass, $0.02$~\Msun, and optical observations \citep{kdf88,fgl+88,vac+88} reveal that the nearly Roche-lobe filling companion has been ablated by the strong wind of the pulsar, which lies in a close 9.2-hr orbit \citep{peb+88}. In $H_\alpha$, a bow shock is clearly visible and confirms the above scenario \citep{kh88}. After the discovery of PSR~B1957+20, which has been coined the \emph{black widow pulsar}, people thought they had solved the problem of isolated MSPs: recycled pulsars that are isolated simply blasted away their companion \citep{vv88}. Although this is a fair proposal, only a handful of such black widow pulsars have been found \citep{sbb96,kbr+05} and so the question is still open as to whether this mechanism can account for all the isolated MSPs. If not, it is not clear how these pulsars ended up single.

\subsubsection{Globular Cluster Pulsars}\label{s:gc_pulsars}
The situation of pulsars in globular clusters clearly contrasts with that of pulsars in the Galactic field. As we show in Figure~\ref{f:distribution_all}, the two populations significantly diverge: globular clusters, as opposed to the Galactic field, preferentially form rapidly spinning pulsars. The dichotomy also appears in the rate of binary systems, which is higher in globular clusters as well --- $\sim$52\% are in binaries, as opposed to only $\sim$7\% for the entire pulsar population\footnote{According to the ATNF catalogue \citep{atnf} at the time when this thesis was written.} (see Figures~\ref{f:distribution_GC_Binary} and \ref{f:distribution_Binary_GCBinary}). Interactions between stars over the course of the evolution of these globular cluster pulsars can explain the difference. Not only do numerical simulations indicate that globular clusters are rich in primordial binary stars \citep{ibf+05}, that is, stars that are born in binary systems, they also have large dynamical interaction rates \citep{ihr+08}. In fact, it appears that a large fraction of the globular cluster neutron star binaries are not with their original companion \citep{ihr+08}; stars and pulsars that are isolated may end up in a binary system after a capture, while binary systems may be disrupted, altered or ejected from the cluster. The dynamics particular to globular clusters manifests itself in the large number of MSPs per unit mass compared to the Galactic field but also in the large fraction of highly eccentric binary MSPs --- 10\% of the 130 known globular cluster pulsars are in binary systems with eccentricities $e > 0.2$ --- that is 1000 times more efficient per unit mass than the Galactic disk \citep{crl+08}.

\afterpage{
\clearpage

\begin{figure}
 \centering
 \includegraphics[width=6.5in]{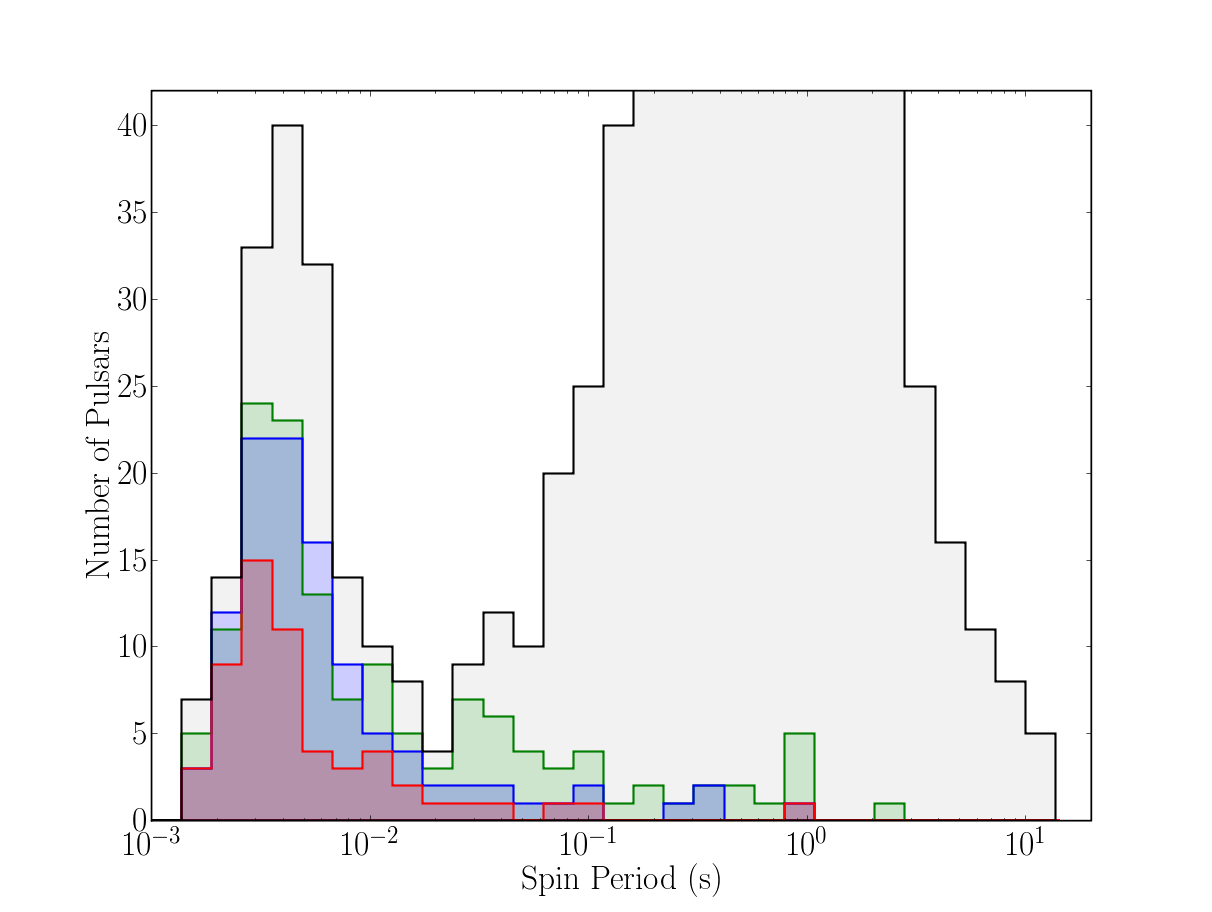}
 \caption[Distribution of spin periods: binaries, GC and GC binaries]{Distribution of spin periods of the entire observed population of pulsars (gray) compared to that of all binary pulsars (green), of pulsars in globular clusters (blue) and binary pulsars in globular clusters (red). Globular clusters preferentially form binary pulsars and pulsars having short spin periods.}
 \label{f:distribution_all}
\end{figure}

\clearpage

\begin{figure}
 \centering
 \includegraphics[width=6.5in]{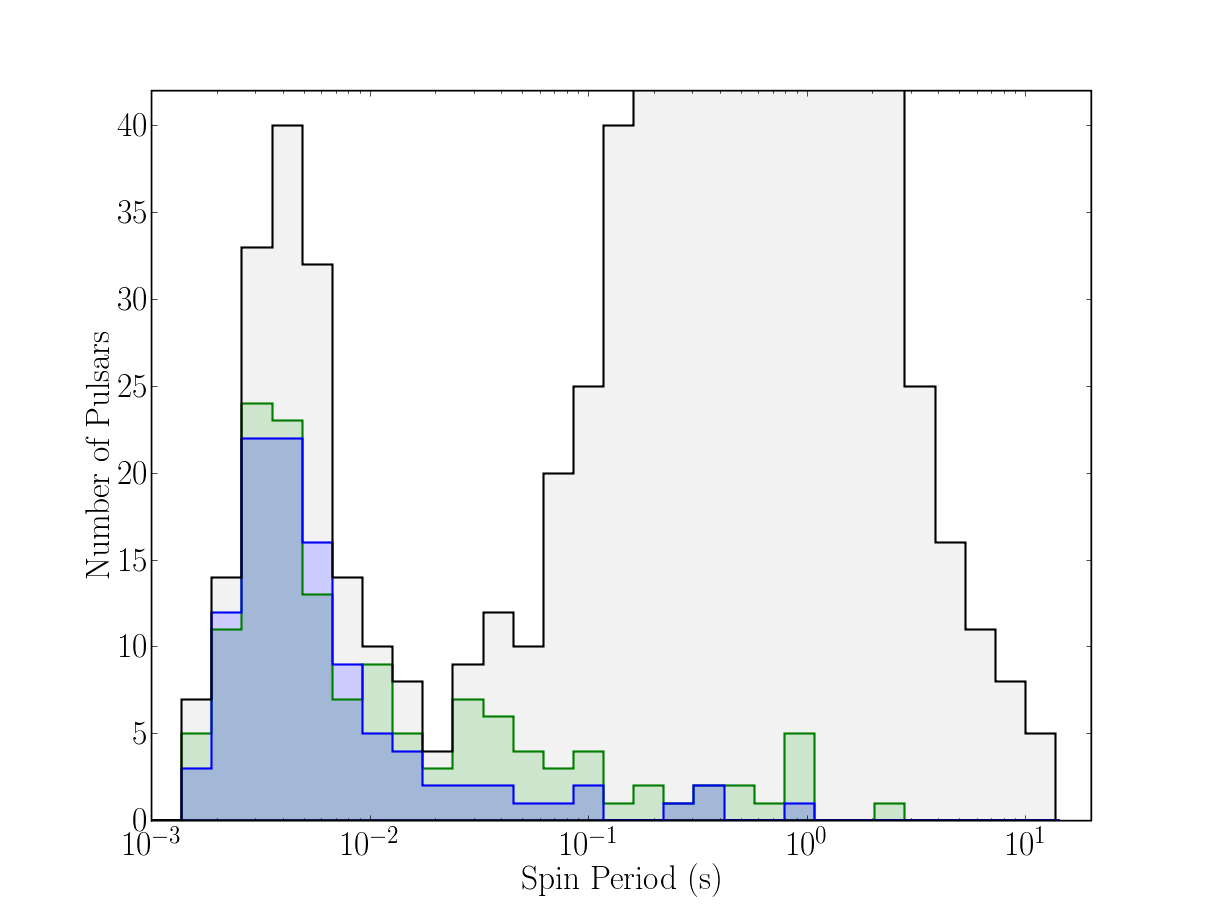}
 \caption[Distribution of spin periods: binaries and GC]{Distribution of spin periods of the entire observed population of pulsars (gray) compared to that of all binary pulsars (green) and pulsars in globular clusters (blue). Globular clusters preferentially form binary pulsars. Note, however, that even though the two distributions are very similar, all pulsars in globular clusters are not necessarily in binary systems and vice versa (see Figures~\ref{f:distribution_all} and \ref{f:distribution_Binary_GCBinary}).}
 \label{f:distribution_GC_Binary}
\end{figure}

\clearpage

\begin{figure}
 \centering
 \includegraphics[width=6.5in]{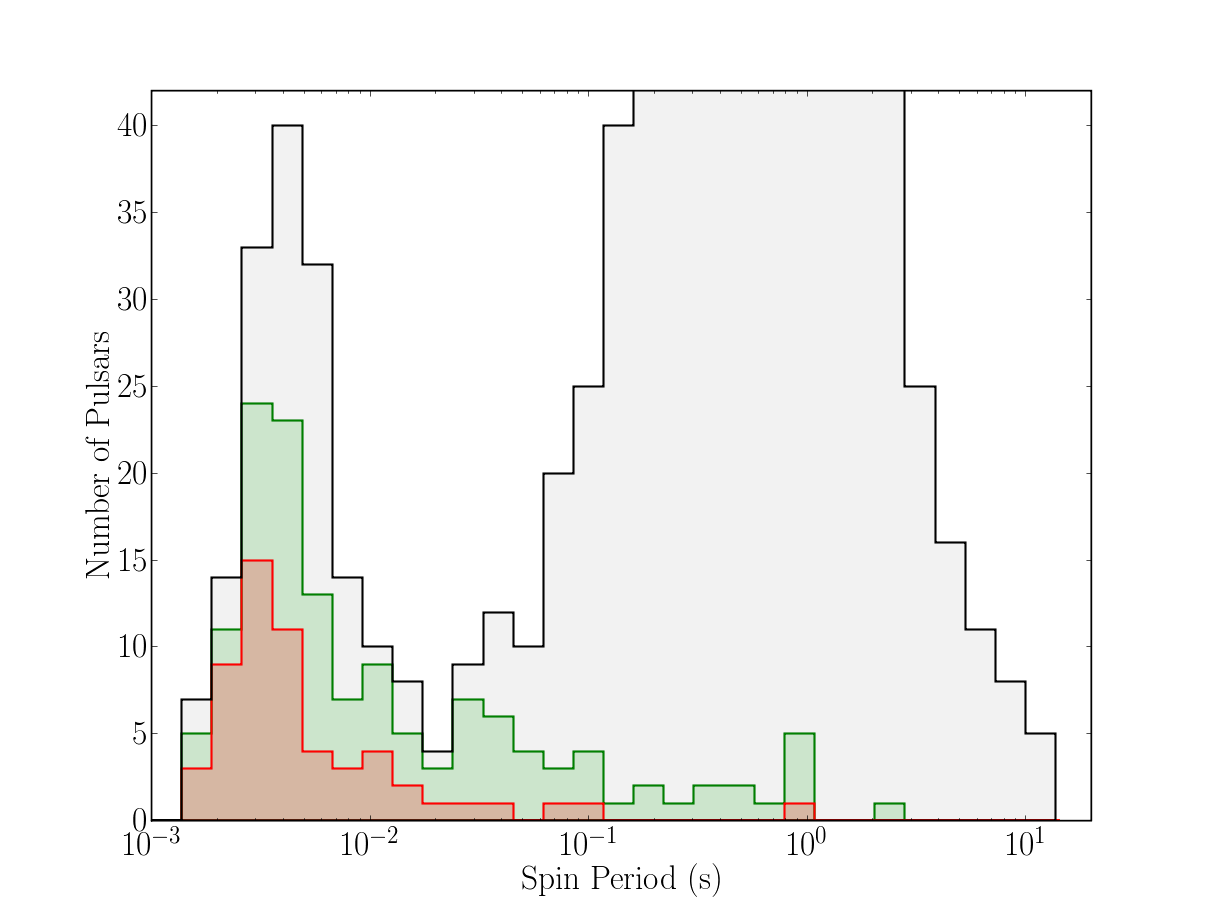}
 \caption[Distribution of spin periods: binaries and GC binaries]{Distribution of spin periods of the entire observed population of pulsars (gray) compared to that of all binary pulsars (green) and binary pulsars in globular clusters (red). An important fraction of the binary pulsar population is found in globular clusters; globular clusters are more efficient at forming binary pulsars than the Galactic field.}
 \label{f:distribution_Binary_GCBinary}
\end{figure}

\clearpage

}

One particular consequence of the virial theorem applied to binary interactions states that the most energetically favorable configuration tends to make hard binaries harder and soft binaries softer \citep{ben06}. This is also known as Heggie's Law \citep{heg75}. In short, if the binding energy of a system is larger than that of an incoming field star (hard binary case), then the energy after the encounter would increase whereas it would decrease if the energy of the system is lower (soft binary case). The latter scenario can be visualized as follows: imagine a loosely bound system being hit by a high velocity projectile; chances are that the system will be less bounded, and maybe even disrupted, after the collision. The general rule of thumb is that when a binary system encounters a field star, the lighter component of the binary is often ejected while the more massive incoming star or stellar remnant binds to the heavier component of the system \citep{ben06}. In a binary-binary interaction, the more massive components will likely end up together while the lighter components often get ejected.

As we can see, the dynamics of globular cluster evolution are quite complex and companion exchange might happen at different stages of the pulsar's life or prior to its formation. Because pulsars are compact and relatively massive objects, they preferentially bind with other stars and therefore have better chances of getting recycled.

\subsubsection{Peculiar Binaries}
More recently, the discovery of PSR~J1903+0327 also showed an example of non-conventional evolution \citep{crl+08}. This MSP shows evidence of full recycling since it has a short 2.15~ms spin period and a low $2.0 \times 10^{8}$~G magnetic field. However, it has a relatively massive companion ($M = 1.05$~\Msun) if one assumes that the observed periastron advance is entirely due to general relativity, and its orbital eccentricity is large ($e = 0.44$). If this binary pulsar had been in a globular cluster, its properties would just appear as a normal consequence of dynamical interaction. It turns out, however, that PSR~J1903+0327 is in the Galactic field. Its evolution must therefore have been quite peculiar. Three possible scenarios are discussed by \citet{crl+08}: 1) the pulsar did not get recycled but was actually born with a fast spin period and a low magnetic field; 2) the pulsar was ejected from a globular cluster; 3) the pulsar is in a triple system and the timing only reveals the signature of the inner companion, which would likely be a white dwarf, whereas the optical counterpart is associated with the third companion in a much wider, highly inclined orbit. Such an inclined orbit is required in order to explain the large eccentricity of the inner binary via the Kozai mechanism.

\section{Binary Pulsars as Benchmarks for Gravity Theories}\label{s:gravity}
Binary radio pulsars are extraordinary tools for studying general relativity and alternative theories of gravity. They owe their success to a combination of physical and observational properties that make them nearly ideal laboratories for this type of study. First, pulsars are neutron stars and because they are extremely compact, they can be treated as point-mass particles. When two neutron stars orbit each other, the system is fully described by dynamical equations of motion and other effects like tides are negligible. Second, radio pulsars can be timed with very high precision that sometimes rivals with the stability of atomic clocks. Hence, their orbital motion is easily monitored via the Doppler shift imprinted in the TOAs of their pulses.

Despite the fact that the tests of gravitational theories involving binary radio pulsars are currently less accurate than solar-system experiments, they yield qualitatively different and complementary information since they are conducted in the environment of strong-field gravity \citep{wil93} (see also \S\,\ref{a:strong_field}). General relativity predicts that the behavior of gravity is independent of the nature of the bodies involved. However, this is not generally true when gravity is treated in a generic way that encompasses more general classes of theories \citep{de92a,de92b,de96a,de96b}. In this context, pulsars are privileged laboratories since their large binding energy, which represents between 10 and 20\% of their rest-mass energy, allows putting limits on theories themselves but also on how they can differentiate from possible common behavior in the low-field gravity regime.

\subsection{``Classical'' Tests of Gravity}\label{s:tests_gr}
Most binary pulsars have WD companions in very circular, non-relativistic orbits \citep[see][ and \S\,\ref{s:binary_population}, for reviews about the different types of binary pulsars]{sta04b,vbj+05}. Ironically, the first discovered binary pulsar, PSR~B1913+16, found by \citet{ht75}, is one of the rare pulsar binaries containing two neutron stars in an eccentric, relativistic orbit\footnote{They constitute about $5-10$\% of the observed binary population \citep{atnf,sta04b}.}.

As we pointed out in \S\,\ref{s:relativistic_timing}, an accurate description of the orbital motion of relativistic binaries require the addition of PK parameters \citep{dt92a} to the standard five Keplerian parameters used for classical orbits \citep{tfm79}. Within the framework of a given theory of gravity these PK parameters are specific functions of the five measurable Keplerian parameters and the two unknown masses in the system \citep{dt92a}. Measuring two PK parameters allows one to solve for the two masses and any extra PK parameter yields a test for theories of gravity \citep{sta03a}.

For relativistic binary pulsars, tests of gravity are usually illustrated using \emph{mass-mass diagrams}, which display the mass of the two bodies against each other (see Figure~\ref{f:mass_mass}). When a particular theory of gravity is considered, PK parameters are represented by lines on a mass-mass diagram and they must intersect, within uncertainties, at a common point if the theory is valid.

In the Hulse-Taylor pulsar, three PK parameters have been successfully measured \citep{wt05}. One of these parameters, the rate of change of the orbital period is caused by the orbital damping due to energy loss through gravitational wave emission \citep{tfm79,tw82} (see Figure~\ref{f:decay}). This was the first indirect evidence of gravitational waves emission and Hulse and Taylor received the 1993 Physics Nobel Prize in recognition of their discovery.

\afterpage{
\clearpage

\begin{figure}
 \centering
 \includegraphics[width=5in]{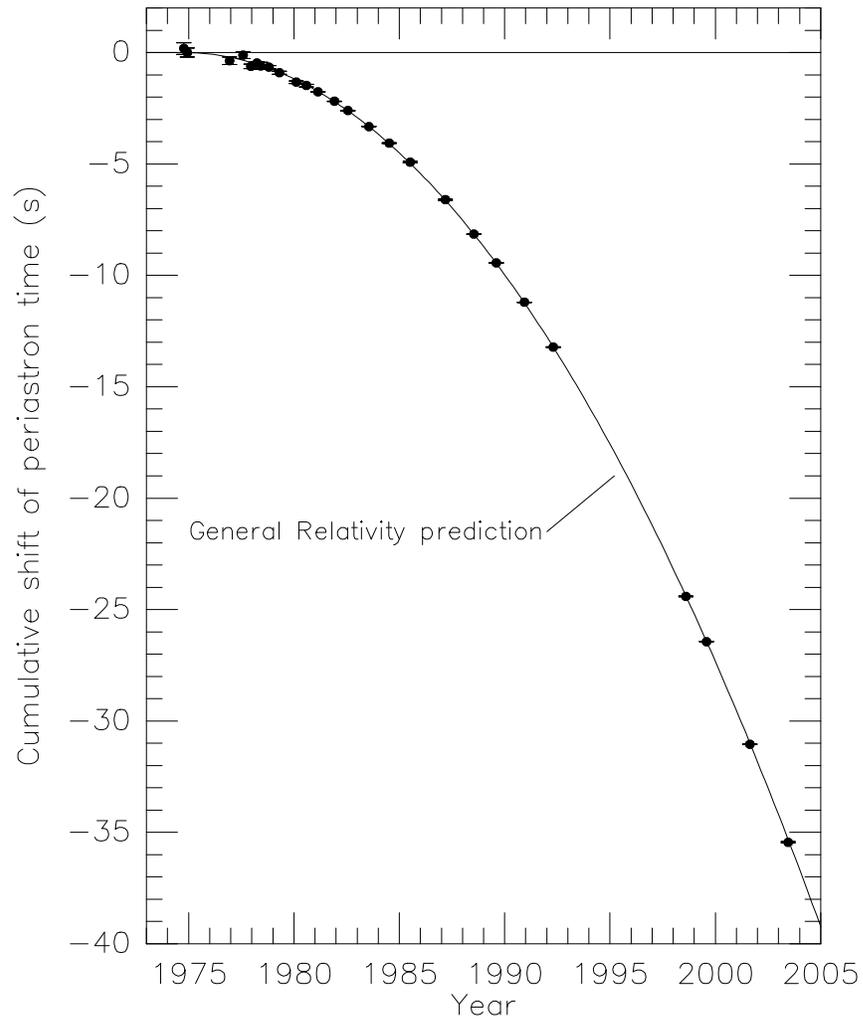}
 \caption[Orbital period decay of the Hulse-Taylor pulsar]{Orbital period decay of the Hulse-Taylor pulsar PSR~B1913+16. Each orbit, the periastron passage occurs earlier as a result of the shrinking of the orbit. The plot shows the cumulative shift as a function of time. The data points indicate the measurement whereas the solid line is the prediction of general relativity. Credit: \citet{wt05}.}
 \label{f:decay}
\end{figure}

\clearpage
}

There are about two dozens pulsars with at least one measured PK parameter and more than half a dozen of them permit one or more tests of gravity because they have more than three measured PK parameters \citep{nic06}. The current best test of general relativity has been made in the double pulsar PSR~J0737$-$3039A/B by \citet{ksm+06} using the PK Shapiro `shape' parameter. The observation agrees with the predicted value to within an uncertainty of 0.05\%.

\subsection{Other Tests of Gravity}\label{s:other_tests}
Other tests of general relativity and gravity have been performed using binary pulsars and are described in the review by \citet{sta03a}. Many of them are related to different aspects of the Strong Equivalence Principle. They include tests of preferred-frame effects, preferred-location effect, non-conservation of momentum and gravitational Stark effect, which are all derived from the high-timing precision of pulsars. Changes in the Chandrasekhar mass, which could imply a time variation of the fine-structure constant, has also been tested by comparing masses of pulsars having different characteristic ages. Finally, relativistic theories of gravity generally predicts spin-orbit and spin-spin coupling. These two effects should introduce high-order secular orbital variations that might be observed someday, but they also cause the pulsar's spin-axis to precess. Evidence of relativistic precession due to spin-orbit coupling was qualitatively observed as morphologic changes in the pulse profile of some relativistic pulsars \citep{wrt89,kws03} and was qualitatively measured to low precision using a combination of pulse morphology and polarization measurement by \citet{sta04a}. In Chapter~\ref{c:0737_eclipse} we shall present a high-precision quantitative measurement using a novel approach.

\section{The Double Pulsar}
More than 30 years after the discovery of the first relativistic binary pulsar by Hulse and Taylor, about ten of these systems have been found. As we explained in the previous section, these pulsars have provided excellent tests of general relativity in the strong-field regime \citep{sta03a,sta04b}. The most celebrated relativistic binary pulsar, PSR~J0737$-$3039A/B, is commonly known as the ``double pulsar'' because both neutron stars in this system are observable radio pulsars. The first member of this system, hereafter denoted pulsar A, is a 23-ms pulsar found in 2003 by \citet{bdp+03} in Parkes Multibeam Pulsar Survey data. Pulsar B, the 2.8-s companion, went unnoticed for several weeks until \citet{lbk+04} found a second pulsar signal in the data having exactly the same orbital period but a half-period shift in its radial velocity signature. Pulsar B is an extremely dim radio pulsar except in two particular regions of the orbit where it becomes much brighter, hence why it did not clearly appear in the original analysis.

The potential of the double pulsar for testing gravity was immediately recognized. Its 2.4-hr orbital period is the shortest among all known relativistic binary pulsars and consequently it is the most relativistic double neutron star system. Indeed, it took only three years of timing observations to achieve, and supersede, the precision reached at testing general relativity that is obtained in other relativistic binary pulsars that have been observed over a much longer period of time \citep{ksm+06}. A record five post-Keplerian parameters have been precisely measured from the radio timing of pulsar A. Furthermore, as the companion is also a pulsar, it is possible to independently measure the projected semi-major axis for each member of the system. The ratio of their projected semi-major axis provides a theory-independent measurement of their mass ratio, thus enabling an additional fourth test of general relativity in this system only \citep{ksm+06}. The future of double pulsar timing is promising as it could permit to measure second-order PK parameters. Also, as the double pulsar lies considerably closer to Earth than other known double neutron star systems, the contribution from the Galactic potential acceleration to its apparent orbital period decay amounts to only 0.02\% of the total value and therefore presents less limitation that the Hulse-Taylor pulsar.

The double pulsar not only amalgamates great properties that make it a wonderful duet of relativistic pulsars to time, it also exhibits several unique phenomena -- eclipses of pulsar A \citep[see, e.g.][]{mll+04,krb+04}, orbital modulation of the pulsed flux from pulsar B \citep[see, e.g.][]{mkl+04,lyu05}, subpulse drifting from pulsar B synchronized with pulsar A's rotational frequency \citep{mkl+04}, pulsed X-ray emission from pulsar A and pulsar B \citep[see, e.g.][]{cpb04,pdm+04,mcb+04,gm04,cgm+07,prm+08,ptd+08} -- that are clear indication of the interaction between the two pulsars. Some of these phenomena, the eclipses and a relativistic aberration phenomenon, will be investigated in more details in Chapter~\ref{c:0737_eclipse} and \ref{c:0737_aberration}. An excellent review of the recent studies related to the double pulsar has been made by \citet{ks08}.


\chapter{The Unusual Binary Pulsar PSR~J1744$-$3922}\label{c:1744}

\begin{flushright}
 \emph{``Genius is one percent inspiration and ninety-nine percent perspiration.''}\\
 Thomas Alva Edison
 \vspace{0.5in}
\end{flushright}

This chapter presents the study of the binary pulsar PSR~J1744$-$3922. This pulsar exhibits a highly variable pulsed radio emission as well as an unusual combination of spin and orbital characteristics compared to typical recycled pulsars. We report on a statistical multi-frequency study of the pulsed radio flux variability which suggests that this phenomenon is extrinsic to the pulsar and possibly tied to the companion, although not strongly correlated with orbital phase. We also investigate the nature of this pulsar, which presents unexplained properties in the context of binary evolution, and suggest that it belongs to a previously misidentified class of binary pulsars. Near-infrared observations allowed us to detect a possible companion's counterpart and appears to support alternative evolutionary scenarios.

This work was originally published as: R. P. Breton, M. S. E. Roberts, S. M. Ransom, V. M. Kaspi, M. Durant, P. Bergeron, and A. J. Faulkner. \emph{The Unusual Binary Pulsar PSR J1744$-$3922: Radio Flux Variability, Near-Infrared Observation, and Evolution}. ApJ, 661:1073–1083, June 2007

\section{Introduction} \label{s:intro}
A mid-Galactic latitude pulsar survey with the Parkes Radio Telescope \citep{crh+06} detected three new pulsars in binary systems, none of which easily fits within the standard evolutionary scenarios proposed for the majority of recycled pulsars (see a review of binary pulsar evolution in \S\,\ref{s:binary_evolution}). One of them, PSR~J1744$-$3922, was independently discovered during the reprocessing of the Parkes Multibeam Pulsar Survey data \citep{fsk+04}. This 172-ms binary pulsar has a relatively high surface dipole magnetic field strength ($B = 1.7 \times 10^{10} \, {\rm G}$, see Equation~\ref{e:magnetic_field}) suggesting it is mildly recycled. However, it appears to have a very light companion (minimum mass $0.085 \,\Msun$) in a tight and nearly circular 4.6-hr orbit (see Table~\ref{t:orbparams}). This type of orbit and companion are typical of those of fully recycled pulsars (which we define as pulsars with $P\lesssim 8$\,ms and $B\lesssim 10^9$\,G) with He white dwarf (WD) companions (see \S\,\ref{s:psr_hewd}). Thus, why PSR~J1744$-$3922 escaped being fully recycled is a puzzle.

\afterpage{
\clearpage

\begin{table}
 \centering
 \begin{tabular}{lc}
 \hline
 \hline \\ [-1.8ex]
 \multicolumn{1}{c}{Parameter} & Value \\ [0.3ex]
 \hline \\ [-1.3ex]
 Orbital Period, $P_b$ (days)            \dotfill & 0.19140635(1) \\
 Projected Semi-Major Axis, $x$ (lt-s) \dotfill & 0.21228(5) \\
 Orbital Period Derivative, $|\dot P_b|$ (s\,s$^{-1}$) \dotfill
 & $<2 \times 10^{-10}$ \\
 Projected Semi-Major Axis Derivative, $|\dot x|$ (lt-s\,s$^{-1}$) \dotfill
 & $<7 \times 10^{-12}$ \\ [0.5ex]
 \hline \\ [-1.8ex]
 \multicolumn{2}{c}{Derived Parameters} \\ [0.3ex]
 \hline \\ [-1.3ex]
 Eccentricity, $e$                           \dotfill & $<$0.001 \\
 Mass Function, $f_1$ (\Msun)                  \dotfill & 0.0002804(2) \\
 Minimum Companion Mass, $M_c$ (\Msun)         \dotfill & $\geq$\,0.085 \\
 Surface Dipole Magnetic Field, $B$ (G) \dotfill & $1.7 \times 10^{10}$ \\
 Spin Down Energy Loss Rate (erg\,s$^{-1}$) \dotfill & $1.2 \times 10^{31}$ \\
 Characteristic Age, (yr) \dotfill & $1.7 \times 10^9$ \\ [0.5ex]
 \hline
 \end{tabular}
 \caption[Orbital and timing parameters for PSR~J1744$-$3922]{Orbital and timing parameters for PSR~J1744$-$3922. Please refer to \S\,\ref{s:binary_timing} for a description of orbital and timing parameters. Numbers in parentheses represent twice the formal errors in the least significant digits as determined by {\tt TEMPO} after scaling the TOA errors such that the reduced-$\chi^2$ of the fit was unity. The pulsar is assumed to have mass 1.4\,\Msun. Values reported from \citet{rrh+08}.}
 \label{t:orbparams}
\end{table}

\clearpage
}

In addition to this atypical combination of spin and orbital properties, PSR~J1744$-$3922 exhibits strong pulsed radio flux modulations, making the pulsar undetectable at 1400\,MHz for lengths of time ranging from a few tens of seconds to tens of minutes. It has been suggested by \citet{fsk+04} that this behavior might be the nulling phenomenon seen in a handful of slow, isolated pulsars. Nulling is a broad-band, if not total, interruption of the radio emission for a temporary period of time. On the other hand, although nulling could affect pulsars in binary systems as well, many binary pulsars vary due to external effects such as eclipses. Such external effects might explain PSR~J1744$-$3922's variability as well.

In this chapter, we first report on multi-frequency observations of PSR~J1744$-$3922 that suggest the radio variability is not intrinsic to the pulsar. However, our analysis does not show strong evidence of a correlation between radio flux and orbital phase, as one might expect from traditional eclipse-like variability. We then report on our infrared search for a counterpart of the companion using the Canada-France-Hawaii Telescope (CFHT). This observation has identified a $\textrm{K}^{\prime}=19.30(15)$~\footnote{Throughout this thesis, numbers in parentheses denote $1\sigma$ errors on the last significant digits.} star at the position determined by the radio timing observations. Finally, we examine why other properties of this pulsar make it incompatible with standard evolutionary scenarios, and identify a few other systems which have similar characteristics. The addition of PSR~J1744$-$3922 as an extreme case among this group motivates us to identify a possible new class of binary pulsars. We propose several possible evolutionary channels which might produce members of this class and explain how the nature of the companion to PSR~J1744$-$3922 could be used to constrain the origin of these systems.

\section{Pulsed Radio Flux Variability} \label{s:analysis}
The observed average pulsed radio emission from a pulsar can fluctuate for several different reasons. These include effects from the pulsar itself, as in nulling \citep[e.g.][]{bac70}, its environment, as in eclipsing binary pulsars \citep[e.g.][]{fgl+88}, or the interstellar medium, as in scintillation \citep[e.g.][]{ric70}\footnote{The scintillation of pulsars is related to the same physical mechanism as the scintillation of quasars, which accidentally led to the discovery of pulsars (see \S\,\ref{s:pulsar_discovery})}. In the case of PSR~J1744$-$3922, during a typical 1400\,MHz observation, the radio emission seems to turn on and off randomly on timescales varying from tens of seconds to tens of minutes (see Figure~\ref{f:profile} for sample folded profiles). In many instances, the pulsar is undetectable through entire observations ($\sim 15$\,min). In a previous analysis of these radio flux modulations, \citet{fsk+04} concluded on the basis of observations at 1400\,MHz that PSR~J1744$-$3922 is probably a pulsar experiencing pulse nulls. As nulling is a broad-band phenomenon \citep[e.g.][]{bsw+81}, we decided to investigate the frequency dependence of the fluctuations in order to test the nulling hypothesis.

\afterpage{
\clearpage

\begin{figure}
 \centering \leavevmode
 \includegraphics[width=6in]{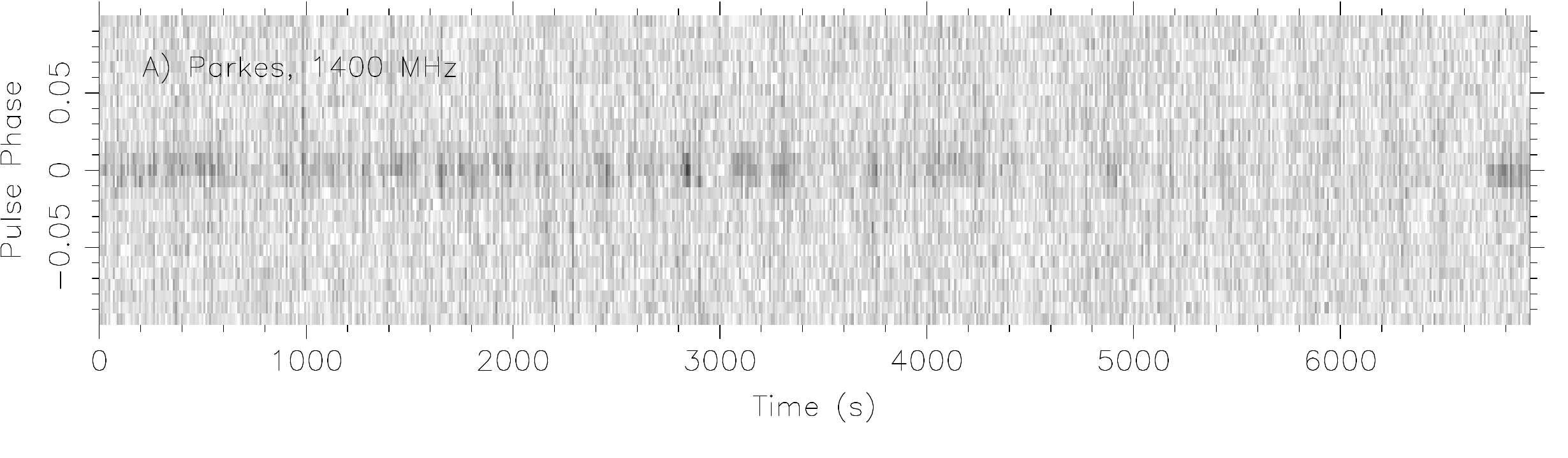}
 \includegraphics[width=6in]{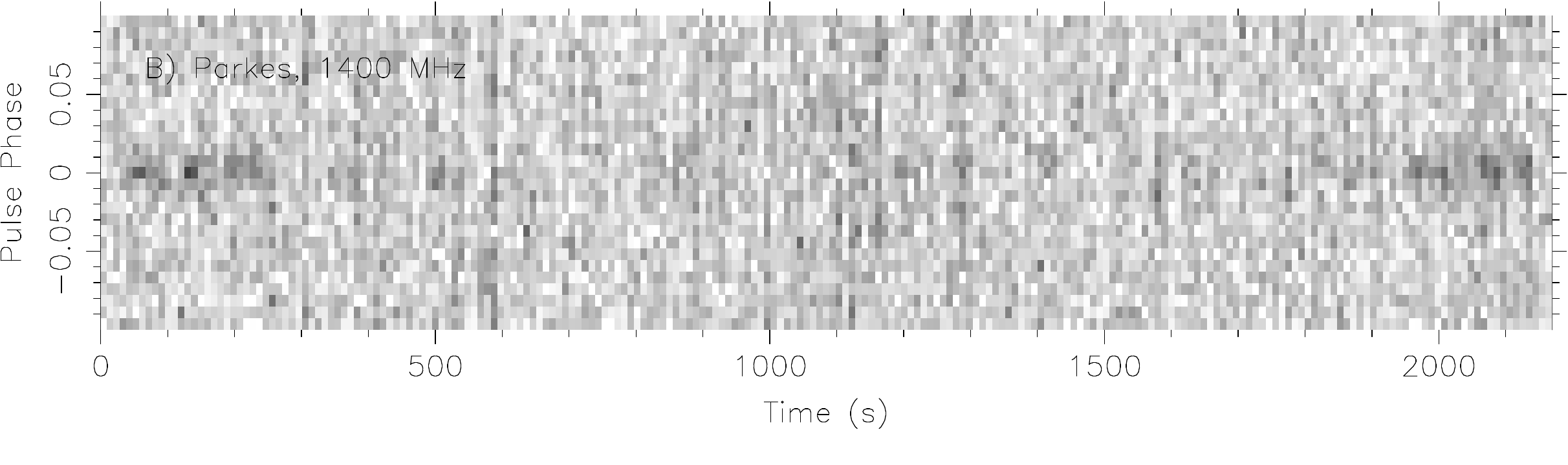}
 \includegraphics[width=6in]{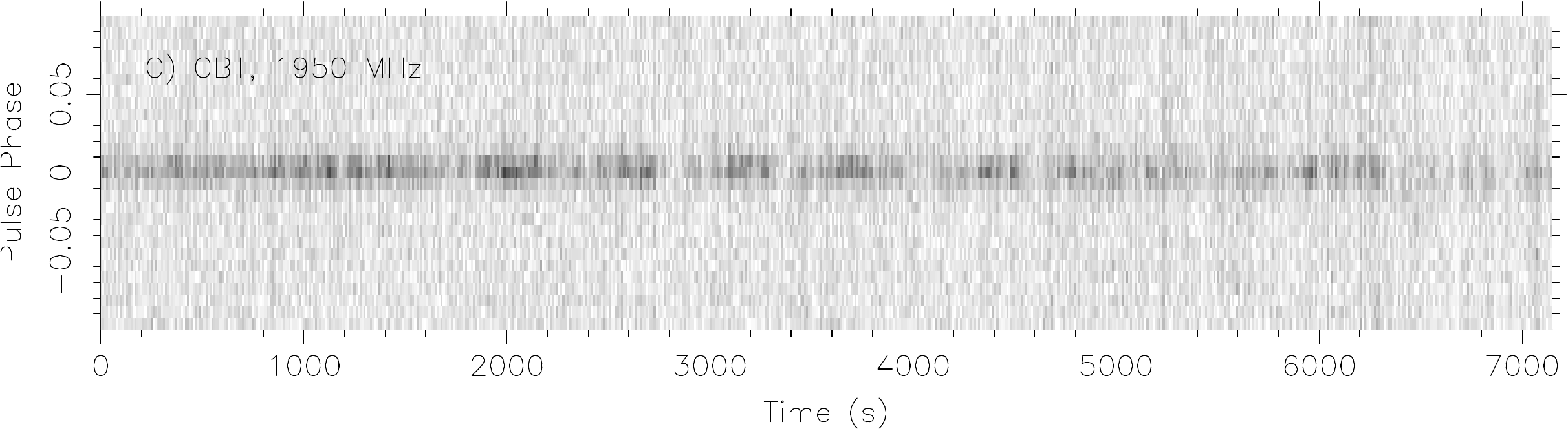}
 \caption[Sample folded intensity profiles of PSR~J1744$-$3922]{Sample folded intensity profiles of PSR~J1744$-$3922 as a function of time for two 1400\,MHz observations at Parkes with 576\,MHz bandwidth (panel A \& B) and a 1950\,MHz observation at GBT with 600\,MHz bandwidth (panel C). The grayscale represents the intensity of the signal, with darker regions being brighter. The center of the gaussian-like pulse profile should appear at the pulse phase 0.}
 \label{f:profile}
\end{figure}

\clearpage
}

\subsection{Data and Procedure} \label{ss:procedure}
The work we report is based on an extended dataset combining both the data reported independently by \citet{fsk+04} and by \citet{rrh+08}. A total of 112 radio timing observations of PSR~J1744$-$3922 were made at the Parkes and the Green Bank telescopes between 2003 June and 2006 January (see \S\,\ref{s:radio_telescopes} for more details about radio telescopes and pulsar observing systems). Relevant details for the current study are summarized in Table~\ref{t:setups} and we refer to the above two papers for more details about the observational setups and timing results.

\afterpage{
\clearpage

\begin{table}
 \begin{center}
 \begin{tabular}{cccccccc}
 \hline
 \hline \\ [-1.8ex]
 $\nu_{center}$\tablenotemark{a} &
 BW\tablenotemark{b} &
 Num. chan.\tablenotemark{c} &
 Sampling &
 $T_{sys}$\tablenotemark{d} &
 Gain &
 T\tablenotemark{e} &
 Num. obs.\tablenotemark{f} \\ 
 MHz            & MHz               &
                & $\mu$s            &
 K              & K/Jy              &
 min            &                   \\ [0.5ex]
 \hline \\ [-1.8ex]
 \multicolumn{8}{c}{Parkes Telescope} \\ [0.3ex]
 \hline \\ [-1.3ex]
 680/2900\tablenotemark{g} & 56/576 & 256/192 & 500/500 & 68/31 & 0.625/0.59 & 20/20 & 5 \\
 1375 & 288 & 96   & 500   & 28 & 0.71 & 17 & 69 \\
 1400 & 576 & 192  & 250   & 32   & 0.71 & 15 & 13 \\ [0.5ex]
 \hline \\ [-1.8ex]
 \multicolumn{8}{c}{Green Bank Telescope} \\ [0.3ex]
 \hline \\ [-1.3ex]
 820  & 48  & 96   & 72    & 37   & 2.0  & 18  & 4  \\
 1400 & 96  & 96   & 72    & 24   & 2.0  & 25  & 19 \\
 1850 & 96  & 96   & 72    & 22   & 1.9  & 60  & 3  \\
 1950 & 600 & 768  & 81.92 & 22   & 1.9  & 210 & 4  \\
 4600 & 800 & 1024 & 81.92 & 20   & 1.85 & 257 & 1  \\ [0.5ex]
 \hline
 \end{tabular}
 \end{center}
 \caption[Receiver temperatures and gains for the Parkes and Green Bank telescopes]{Receiver temperatures and gains are estimated operating values. Parkes values are provided by J. Reynolds (2006, private comm.). The system temperature corresponds to the sum of the receiver temperature and the sky temperature, which is determined from the 408\,MHz all-sky survey and converted to other frequencies by assuming a power-law spectrum having a spectral index $-2.6$ \citep{hss+82,hss+95}.}
 \tablenotetext{a}{Central frequency of the receiver.}
 \tablenotetext{b}{Observing bandwidth.}
 \tablenotetext{c}{Number of frequency channels.}
 \tablenotetext{d}{System temperature.}
 \tablenotetext{e}{Average total integration time per observation.}
 \tablenotetext{f}{Number of observations.}
 \tablenotetext{g}{Observations were made simultaneously at these two frequencies.}
 \label{t:setups}
\end{table}

\clearpage
}

For the purpose of studying the radio emission variability, we made time series of the pulsed flux intensity. We dedispersed the data at the pulsar's dispersion measure (DM) of 148.5\,pc\,cm$^{-3}$ and then folded the resulting time series in 10-s intervals using the timing ephemerides from \citet{rrh+08} and the pulsar analysis package {\tt PRESTO} \citep[][please also refer to \S\,\ref{s:timing} for more details about pulsar timing]{rem02}. For each observation, the pulse phase was determined from the profile averaged over the entire observation. We fit each 10-s interval of the folded pulse profile with a constant baseline plus a Gaussian of variable amplitude having a fixed width at the predetermined phase. A Gaussian FWHM=0.01964$P$, where $P$ is the pulse period, nicely fits the profile averaged over many observations in the frequency range 680-4600\,MHz. Errors on the best-fit amplitudes returned by our least-square minimization procedure were scaled under the assumption that the off-pulse region RMS represents the total system noise.

Although no flux standard has been observed, we obtained pulsed flux density estimates by using the radiometer equation and scale the observed off-pulse RMS levels by the predicted noise levels. The system temperature was assumed to be the sum of the receiver temperature (provided in Table \ref{t:setups}) and the sky temperature, which we determined from the 408\,MHz all-sky survey and converted to other frequencies by assuming a power-law spectrum having a spectral index $-2.6$ \citep{hss+82,hss+95}. Hence the predicted off-pulse RMS level is:
\begin{equation}
 {\rm RMS}_{\rm pred} = \frac{T_{\rm sys}}{G \sqrt{2B T_{\rm int}}} \,,
\end{equation}
where $G$ is the gain, $B$ the observed bandwidth (see Table \ref{t:setups}), $T_{\rm int}$ the integration time per pulse bin and $T_{\rm sys}$ is the system temperature.

We also accounted for the offset between the telescope pointing and the real position of the source (since in the early observations the best-fit timing position had not yet been determined) by approximating the telescope sensitivity to be an azimuthally symmetric Gaussian having a FWHM corresponding to the radio telescope beam size, which is 8.8$^{\prime}$ and 13.8$^{\prime}$ at 1400\,MHz for GBT and Parkes, respectively.

In this way, we generated flux time series for all 112 observations of PSR~J1744$-$3922 (see Figure~\ref{f:lightcurve} for examples). As we describe next, these results show that scintillation and nulling are highly unlikely to be the origin of the observed variability.

\afterpage{
\clearpage

\begin{figure}
 \centering
 \leavevmode
 \includegraphics[width=6in]{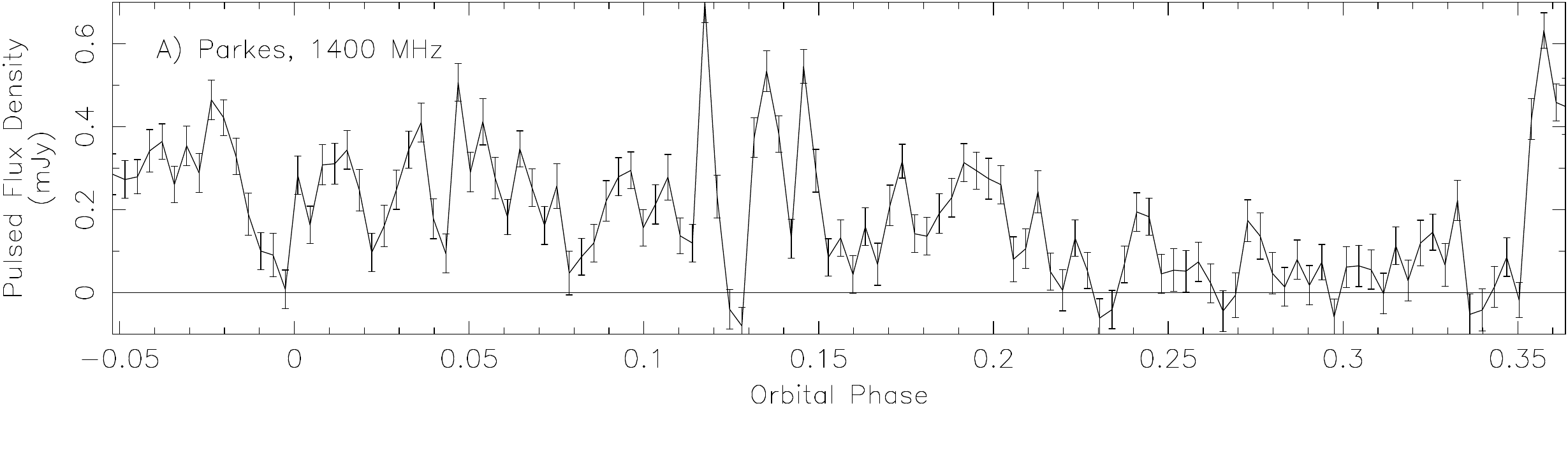}
 \includegraphics[width=6in]{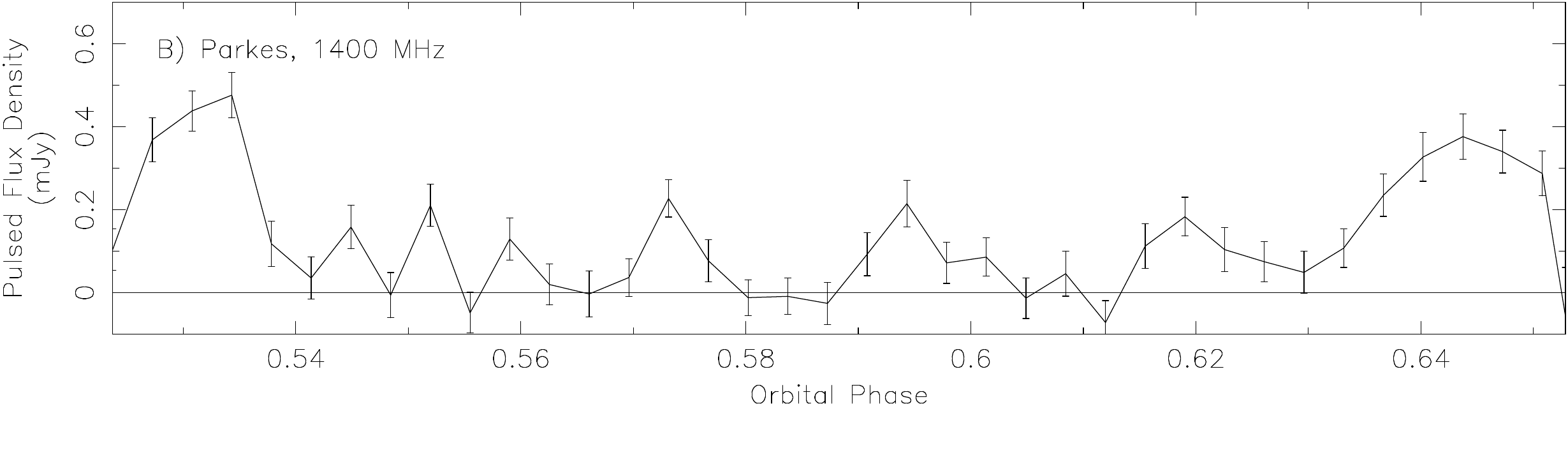}
 \includegraphics[width=6in]{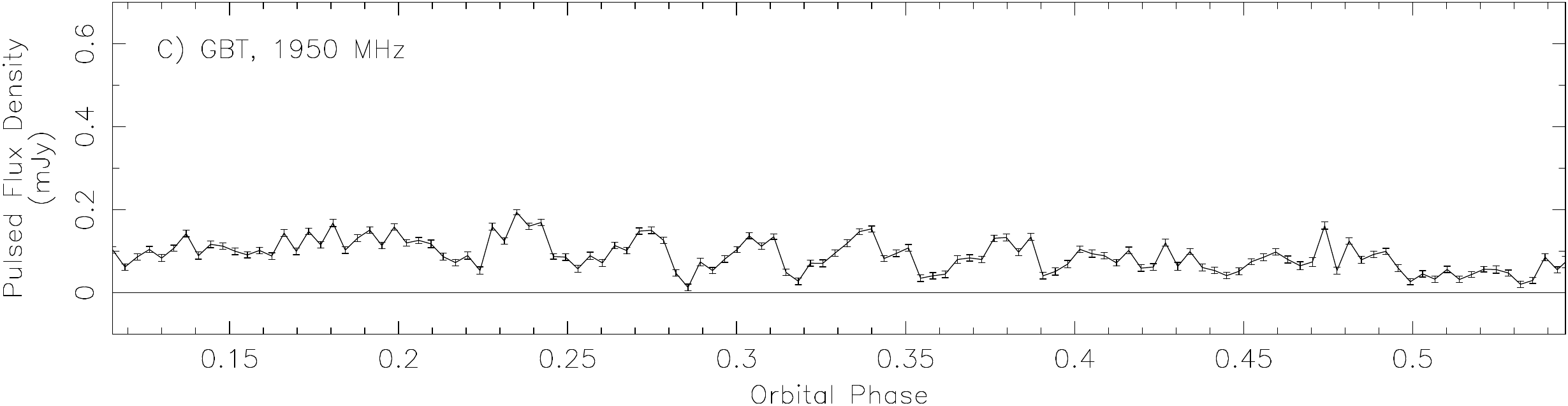}
 \caption[Sample radio pulsed flux density time series for PSR~J1744$-$3922]{Sample light curves of the pulsed radio flux density as a function of orbital phase. Each data point represents 60\,sec of data and orbital phases are defined so that 0.25 is when the companion is in front of the pulsar. Panel A, B and C are the two 1400\,MHz observations and the 1950\,MHz observation, respectively, displayed in Figure~\ref{f:profile}. We note that the flux drops below the detection limit at 1400\,MHz several times whereas it appears to be always above this threshold at 1950\,MHz. Also, as illustrated in Figure~\ref{f:histogram}, the variability is much stronger at lower frenquency.}
 \label{f:lightcurve}
\end{figure}

\clearpage
}

\subsection{Radio-frequency-dependent Variability} \label{ss:nuller}
As \citet{fsk+04} discussed previously, interstellar scintillation (ISS) is unlikely to be the source of fluctuations in PSR~J1744$-$3922. As we mentioned in \S\,\ref{s:pulsar_discovery}, scintillation is produced by a diffractive scattering medium along our line of sight and the typical diffractive scintillation timescale can be expressed as \citep[see][]{cr98}:
\begin{equation}
   \Delta t_d = 2.53 \times 10^4 \frac{D \Delta \nu _d}{\nu V_{\rm ISS}}
                \,\, \textrm{s},
\end{equation}
with $D$ the distance to the source in kpc, $\Delta \nu _d$ the decorrelation bandwidth in MHz, $\nu$ the observed frequency in GHz and $V_{ISS}$ the velocity of ISS diffraction pattern in km\,s$^{-1}$. For the 1400\,MHz observation shown in Figure~\ref{f:profile}, for example, the NE2001 model for the Galactic distribution of free electrons \citep{cl02} predicts $\Delta \nu_d = 0.01$\,MHz in the line of sight of PSR~J1744$-$3922 at a distance of 3.0\,kpc estimated from the DM. $V_{ISS}$ is typically dominated by the pulsar velocities, which are in the range 10-100\,km\,s$^{-1}$ for most binaries. Therefore, we estimate the scintillation timescale to be of the order of a few seconds to a minute at most. This could be compatible with the fast flux variations but can certainly not explain the extended periods where the pulsar goes undetected. Perhaps most importantly, strong scintillations will be averaged away since typical observing bandwidths are much larger than the decorrelation bandwidth, and therefore contain many ``scintles'' (i.e. intensity modulations). Such averaging effectively rules out the ISS hypothesis.

Another possibility is that the flux modulation is related to intrinsic nulling of the pulsar. Based on observations at 1400\,MHz only, \citet{fsk+04} identified it as the most likely explanation. Only observed in old, isolated pulsars so far, (though nothing prevents a binary pulsar from being a nuller) nulling is a broad-band, if not total, interruption of the radio emission \citep[e.g.][]{bsw+81}.

Considering the fraction of observations with no detection of radio emission at various frequencies (see Table~\ref{t:nodetection}), we note qualitatively that PSR~J1744$-$3922 is regularly undetectable at low frequencies but easily detectable at high. For instance, we obtained four Parkes observations at 680 and 2900\,MHz simultaneously in which the pulsar is detected twice at 2900\,MHz while remaining undetected at 680\,MHz. In addition, seven long GBT observations centered at 1850 and 1950\,MHz show highly variable emission but little evidence that the pulsar ever disappears completely (see Figure~\ref{f:lightcurve}). Clearly, however, observations could be biased by the relative instrumental sensitivity in each band and by the intrinsic spectrum of the pulsar.

\afterpage{
\clearpage

\begin{table}
 \begin{center}
 \begin{tabular}{ccccccc}
 \hline
 \hline \\ [-1.8ex]
 Frequency & \multicolumn{2}{l}{Parkes} &
 \multicolumn{2}{l}{GBT} &
 \multicolumn{2}{l}{Both$^{\rm c}$} \\
 MHz &
 \% & & \% & &
 \% & \\ [0.3ex]
 \hline \\ [-1.3ex]
 680\tablenotemark{a}   & 100  & (5)  & --   & (--) & 100  & (5)   \\
 820   & --   & (--) & 75 & (4)  & 75 & (4)   \\
 1400\tablenotemark{b}  & 33 & (72) & 32 & (19) & 33 & (91)  \\
 1850  & --   & (--) & 0  & (3)  & 0  & (3)   \\
 1950  & --   & (--) & 0  & (4)  & 0  & (4)   \\
 2900\tablenotemark{a}  & 50 & (4)  & --   & (--) & 50 & (4)   \\
 4600  & --   & (--) & 0  & (1)  & 0  & (1)   \\
 Total & 38 & (81) & 29 & (31) & 35 & (112) \\ [0.5ex]
 \hline
 \end{tabular}
 \end{center}
 \caption[Occurrence of non-detection of PSR~J1744$-$3922]{Percent of observations with no detection of PSR~J1744$-$3922. Numbers in parentheses represent the total number of observations for each band.}
 \tablenotetext{a}{Parkes observations at 680\,MHz and 2900\,MHz were simultaneous. Excessive RFI contamination prevents us from using one of the Parkes 2900\,MHz observations.}
 \tablenotetext{b}{Parkes observations at 1375\,MHz and 1400\,MHz were combined by discarding the non-overlapping part of the observed frequency band.}
 \tablenotetext{c}{Denotes the average of Parkes and GBT, weighted by the number of observations.}
 \label{t:nodetection}
\end{table}

\clearpage
}

To investigate the effect of instrumental sensitivity and spectral energy distribution, we analysed the pulsed flux densities at different frequencies. Measured values were estimated using the radiometer equation as explained in \S\,\ref{ss:procedure} and are displayed in Table~\ref{t:flux}. We note that \citet{fsk+04} reported a different flux density than ours at 1400\,MHz (0.20(3) vs. 0.11(3)\,mJy, respectively). The discrepancy could be because the average flux density changes depending on the amount of time PSR~J1744$-$3922 spends in its ``bright state'' during an observation. The large standard deviation (0.16\,mJy, see Table~\ref{t:flux}) at this frequency suggests that by restricting the calculation to observations for which PSR~J1744$-$3922 is nicely detected, a higher flux value can be obtained.

\afterpage{
\clearpage

\begin{table}
 \begin{center}
 \begin{tabular}{ccccc}
 \hline
 \hline \\ [-1.8ex]
 Frequency & Num. data\tablenotemark{a} & Average & Standard Deviation & Maximum \\
 MHz & & mJy & mJy & mJy \\ [0.5ex]
 \hline \\ [-1.3ex]
 680   &  40   & $<$\,0.07\tablenotemark{b} &  --  & --   \\
 820   &  20   & 0.10  &  0.06  & 0.22 \\
 1400  &  2244 & 0.11  &  0.16  & 0.45 \\
 1850  &  142  & 0.11  &  0.04  & 0.20 \\
 1950  &  852  & 0.08  &  0.04  & 0.19 \\
 2900  &  40   & 0.09  &  0.08  & 0.16 \\
 4600  &  258  & 0.006 &  0.012 & 0.03 \\ [0.5ex]
 \hline
 \end{tabular}
 \end{center}
 \caption[Estimated Pulsed Flux Density of PSR~J1744$-$3922]{Estimated Pulsed Flux Density of PSR~J1744$-$3922. Values were derived using the radiometer equation, implicitly assuming that the off-pulse RMS is a good estimate of the system temperature and that the sky emits according to the 408\,MHz all-sky survey \citep{hss+82,hss+95}. Relative errors are estimated to be $\sim 30$\,\%.}
 \tablenotetext{a}{Number of data points used at each frequency. The time resolution is 1 minute per data point.}
 \tablenotetext{b}{$3\sigma$ upper limit.}
 \label{t:flux}
\end{table}

\clearpage
}

For the simultaneous observations at 680 and 2900\,MHz in which the pulsar was not detected at 680\,MHz (see Table~\ref{t:setups}), we can put an interesting approximate lower limit on the spectral index, $\alpha$, of the pulsar if we assume that the minimum detectable pulsed flux at 680\,MHz is an upper limit to the pulsed flux at this frequency:
\begin{equation} \label{e:sensitivity}
   \alpha \gtrsim \frac{\log (S_{2900} / S_{680})}{\log (\nu _{2900} / \nu
   _{680})}
      \simeq 0.17 \,\, .
      \end{equation}

Such a value is extremely flat compared to that of the average population of pulsars, which has a spectral index of $-1.8 \pm 0.2$ \citep{mkk+00}. Thus either PSR~J1744$-$3922 has a spectrum very different from those of most pulsars and/or the flux variability is intrinsically frequency dependent.

The distribution of pulsed flux density values for all 1400 and 1950\,MHz observations is shown in Figure~\ref{f:histogram}. It is clear from the distribution at 1400\,MHz that the numerous non-detections are responsible for the peak below the sensitivity threshold. We also observe that the 1400\,MHz flux density has a higher average value and is much more variable than at 1950\,MHz. This frequency dependence suggests some sort of scattering mechanism with the unscattered flux level much higher than the observed average flux at 1400\,MHz, and argues against it being classical nulling. Therefore, we conclude that this unknown mechanism affecting the lower frequency flux might be responsible for the apparent flat spectrum derived from the simultaneous 680-2900\,MHz observation.

\afterpage{
\clearpage

\begin{figure}
 \centering \leavevmode
 \includegraphics[width=6in]{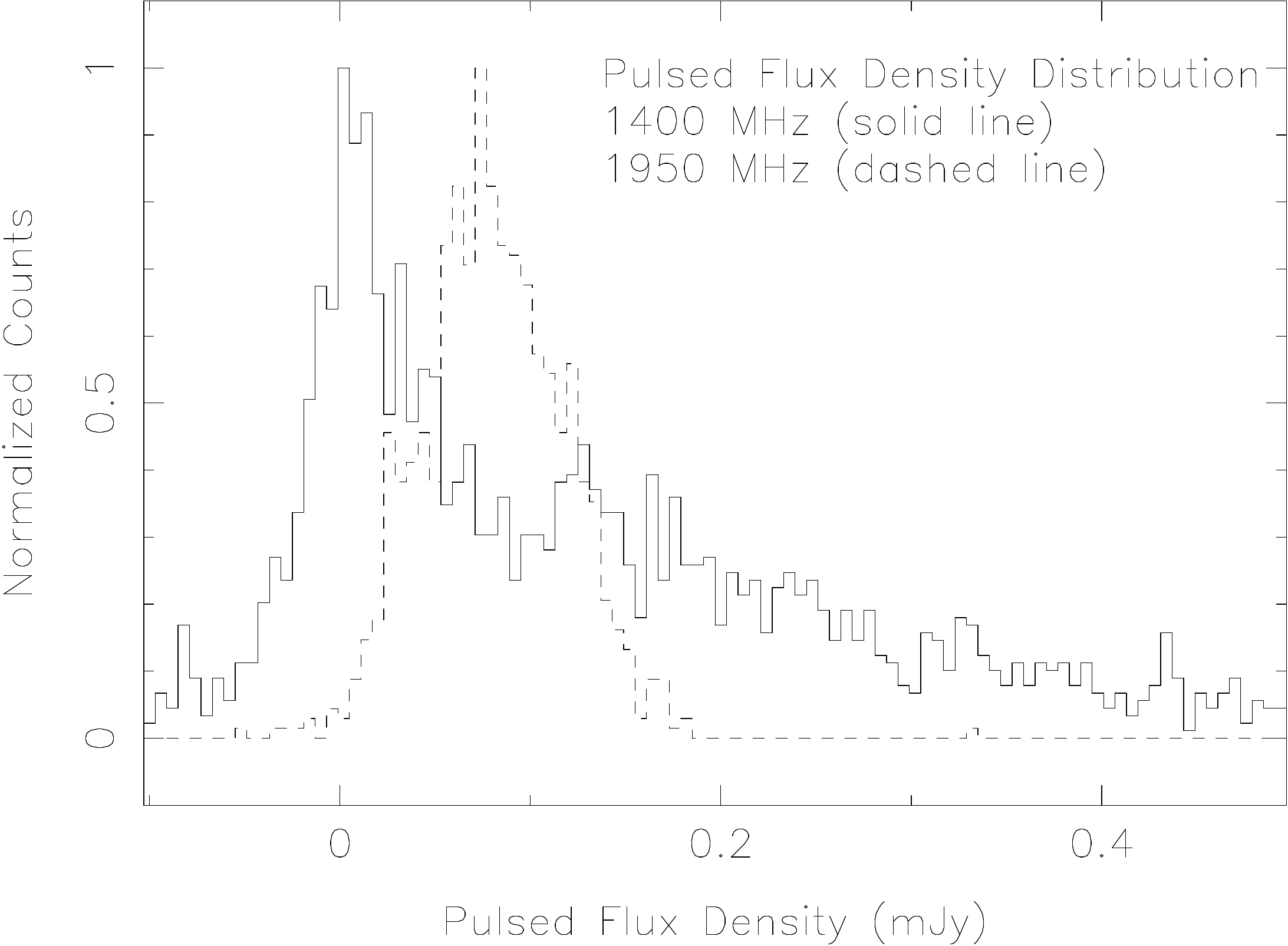}
 \caption[Distribution of the measured pulsed flux density of PSR~J1744$-$3922]{Distribution of the measured pulsed flux density values at 1400\,MHz (solid line) and at 1950\,MHz (dashed line). The $3\sigma$ sensitivity threshold is $\sim$ 0.02 and 0.01\,mJy at 1400 and 1950\,MHz, respectively. We observe that PSR~J1744$-$3922 rarely disappears at 1950\,MHz whereas there is a significant number of non-detections, centered about zero, at 1400\,MHz (negative values are reported when the pulsed flux density is below the telescope sensitivity, meaning the flux determination algorithm has fit noise). Also, we note that the pulsed flux density is more variable and spans over higher values at 1400\,MHz than at 1950\,MHz.}
 \label{f:histogram}
\end{figure}

\clearpage
}

Recent simultaneous multi-frequency observations of PSR~B1133+16, a well-established nuller, show that single pulse nulls do not always happen simultaneously at all frequencies \citep{bgk+07}. They also observe, however, that the overall null fraction does not present any evidence of frequency dependence, which might mean that sometimes nulls are simply delayed at some frequencies. In the case of PSR~J1744$-$3922, the S/N limits us to consider the pulsed flux over times corresponding to many pulse periods only. Therefore, the kind of non-simultaneous, frequency-dependent effect seen by \citet{bgk+07} is not relevant to our analysis and thus we expect the variability to be independent of frequency if really caused by nulling.

Although our flux measurements at other frequencies are not simultaneous, we can assume they are good statistical estimates of the normal flux of the pulsar and use them to characterize its spectrum. In an attempt to estimate an unbiased spectral index, we can use the approximate maximum flux value at each frequency. The 1400, 1850, 1950, 2900 and 4600\,MHz data give a spectral index lying between $-1.5$ and $-3.0$, which is similar to many known pulsars. However, it seems that the flux at 820\,MHz is much smaller than expected from a single power-law spectrum. Since flux variations are very important at low frequency and we only have a single detection at 820\,MHz, the reported maximum value is probably not representative of the real flux of the pulsar at this frequency.

In summary, the facts that the pulsar radio emission rarely drops below our detection threshold at 1950\,MHz and that the radio variability is frequency dependent, demonstrate that the flux modulation is probably not classical nulling. A non-nulling origin for the fluctuations at 1400\,MHz also explains why PSR~J1744$-$3922 does not fit in with expectations based on the correlations observed between null fraction and spin period \citep{big92}, and between null fraction and equivalent pulse width \citep{lw95}. In comparison with nullers, it has one of the smallest spin periods and a small pulse width ($\sim 3.4$\,ms), but one of the largest ``null'' fractions ($\sim 60$\% at 1400\,MHz). Since this pulsar is in a tight binary system, the possibility of influence by its companion is therefore an important alternative to consider.

\subsection{Orbital Correlation Analysis} \label{ss:orbital}
Even though a quick examination of the time series confirms that the flux decreases observed for PSR~J1744$-$3922 are not due to systematic eclipses of the pulsar by its companion, a more subtle orbital correlation could exist. To search for such an effect, we ask whether or not the pulsar is more likely to be detected at a particular orbital phase. For this analysis, we folded the time series in 1-min intervals, and defined the pulse as detected if the best-fit Gaussian amplitude was greater than its 1$\sigma$ uncertainty. In order to limit spectral effects, we restrict the analysis to observations covering the range 1237.5--1516.5\,MHz at Parkes\footnote{This is the common range of the observing modes centered at 1375 and 1400\,MHz.} and 1404.5--1497.5\,MHz at GBT.

The results of this analysis are shown in Figure~\ref{f:analysis}a. The histogram represents the fraction of detected pulses with respect to the total number observed, as a function of orbital phase. Errors were determined using Poisson statistics, implicitly assuming our assigning of each interval as a detection or non-detection is accurate.  There is a suggestion that PSR J1744$-$3922 is more often undetectable between phases $\sim 0.3-0.7$. The best-fit constant model gives a $\chi^2 /9 =5.85$ (the histogram has 10 orbital phase bins), which, if correct, would be highly significant. To test the accuracy of our errors, we performed a Monte Carlo simulation consisting of 10000 trials where we assigned each measurement a random orbital phase. The mean $\chi^2 /9 =0.37$ with a standard deviation of 0.18, suggesting our error estimates are significantly overestimated.

\afterpage{
\clearpage

\begin{figure}
 \centering
 \leavevmode
 \includegraphics[width=4.25in]{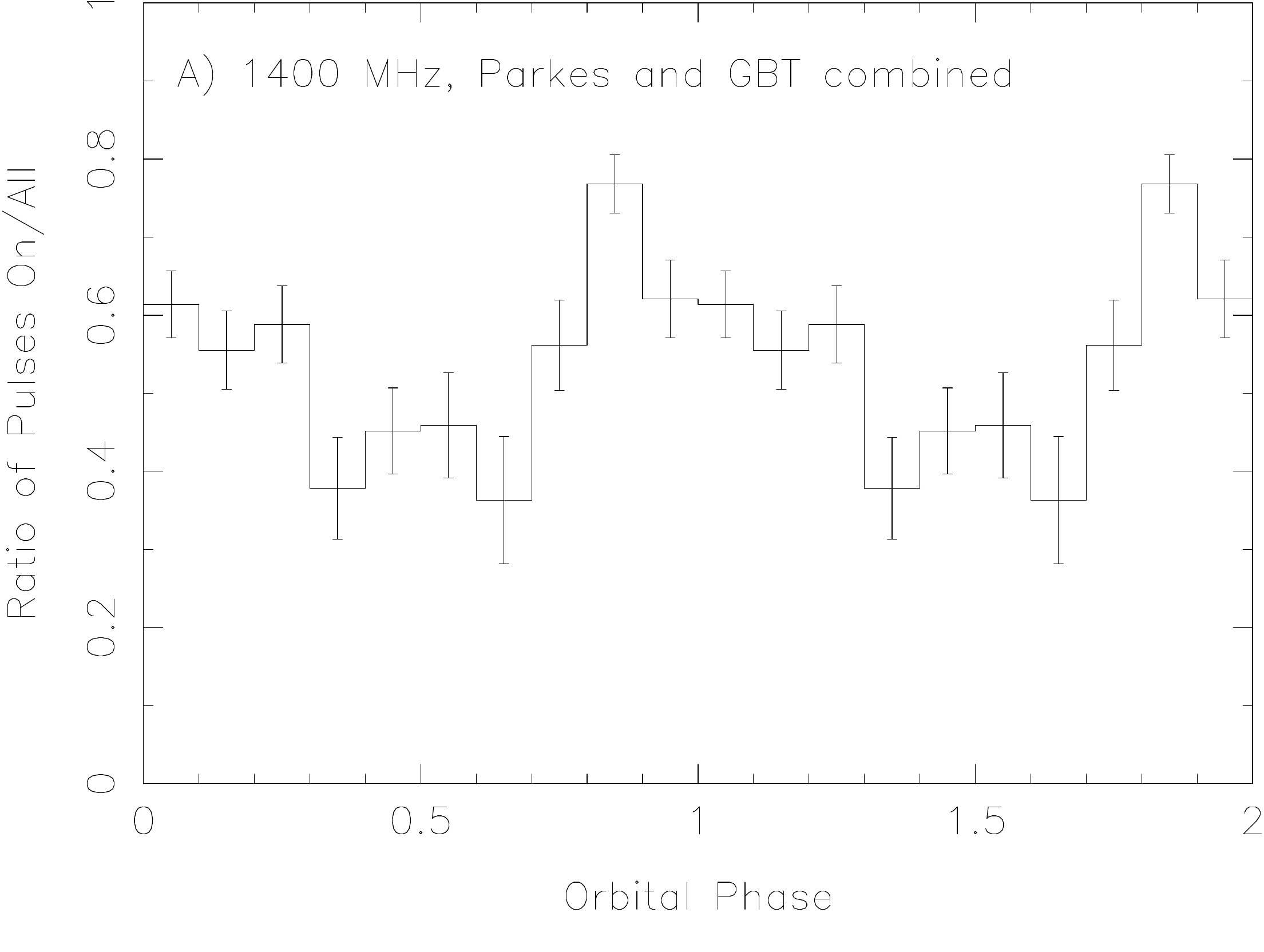}
 \includegraphics[width=4.25in]{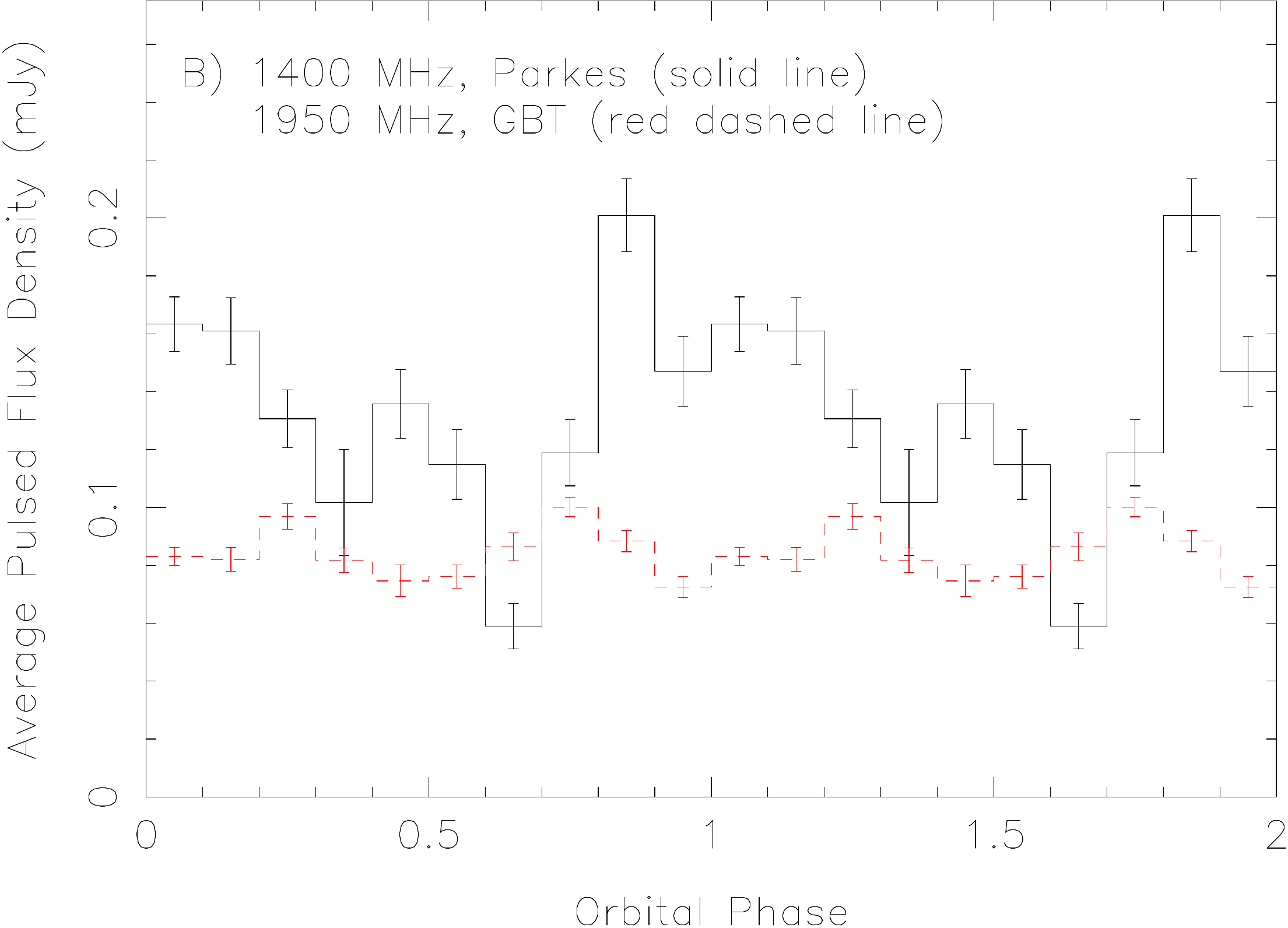}
 \caption[Orbital variability of the pulsed flux of PSR~J1744$-$3922]{Ratio of pulses detected with respect to the total number observed at different orbital phases for the 1400\,MHz Parkes and GBT combined (panel A) and a similar plot showing the average pulsed flux intensity (panel B) for the Parkes 1400\,MHz data (solid line) and the GBT 1950\,MHz data (red dashed line). Orbital phases are defined so that 0.25 is the pulsar's superior conjunction (i.e. the companion passes in front of the pulsar). The scale of the error bars represents the estimated errors without scaling from the results of the Monte Carlo simulations, hence they are likely underestimated.}
 \label{f:analysis}
\end{figure}

\clearpage
}

To investigate further, we performed an analysis similar to the previous one but for the pulsed flux density measured at each orbital phase averaged over all observations. We selected two subsets of data: the Parkes and GBT observations made at 1400 MHz and the GBT 1950 MHz observations. The first one includes 101 observations which are on average $\sim 15$ minutes long, but there are a few observations which are significantly longer. The latter subset includes four observations, two of which have full orbital coverage, one covering $\sim 75\%$ of the orbit and the last one $\sim 40\%$. Observations with no detection of PSR~J1744$-$3922 are assigned upper limit values of three times the background noise level (which is very small compared with the average pulse of the pulsar when it is on). Errors in each bin of the histogram are estimated from the RMS of the individual values in each orbital bin. Results are plotted in Figure~\ref{f:analysis}b. For the 1400 MHz data, a fit to a constant line has a $\chi^2 /9 =15.58$, and for the GBT 1950\,MHz data, we find a $\chi^2 /9 =5.59$. Randomizing the individual data points and folding them resulted in a $\chi^2 /9 = 1.02$ at both frequencies, suggesting our error estimates are reasonable. Although this analysis strongly rules out the constant model for our folded light curves, the shapes of these curves at 1400 and 1950\,MHz are not consistent with each other. This leads us to wonder how would an orbital correlation be possible and, in addition, be showing a different behavior at these two frequencies?

This could be a result of the paucity of observations (typically 3--5) at any given orbital phase. Therefore, random fluctuations in the flux on timescales of tens of minutes (which we see in the time series) would likely not be averaged out. To test whether or not the significant deviation from a constant model depends on the particular phasing of our observations, we again performed Monte Carlo simulations in which 10000 trial histograms similar to those shown in Figure~\ref{f:analysis} were generated from the real data by adding a random orbital phase shift at the beginning of each time series and then determining the $\chi^2 /9$ value for a constant model. From the resulting distribution of $\chi^2 /9$ values, we estimate the chance probability of obtaining the particular $\chi^2 /9$ values obtained, or higher, using the real orbital phases. For the on-off analysis, the probability is 0.102, or a formal $\chi^2 /9 =1.63$. For the 1400 and 1950\,MHz flux density light curves, the probabilities are 0.015 ($\chi^2 /9 =2.28$) and 0.212 ($\chi^2 /9 =1.34$). This suggests there may be some correlation with the orbit, but the shapes of our folded light curves are still dominated by more stochastic flux variations given our limited data set. We would expect a standard eclipse mechanism to make the pulsar dimmer when the companion is in front of it at phase 0.25, which does not seem to be the case. There may be large orbit to orbit variations in the eclipse depths, durations, and phases. This kind of behavior has been observed in other binary pulsar systems such as Ter5A, Ter5P, Ter5ad and NGC6397A \citep[see][for examples]{rhs+05,hrs+06,dpm+01}, but a much larger data set would be required to obtain a reliable average light curve to show if this is the case for PSR~J1744$-$3922. It is still possible that the short time scale pulsed flux variations tend to group in ``events" during which the pulse gets dimmer. These ``events" may last for a significant fraction of the orbital phase causing the apparent marginal orbital correlation given our limited statistics.  

\subsection{Accretion and mass loss limits} \label{ss:eclipse}
Many other systems are known to exhibit strong flux radio variations for which the orbital dependence is well established. One of them is PSR~B1718$-$19 \citep{lbh+93}. Interestingly, this pulsar has a low-mass companion and orbital properties similar to those of PSR~J1744$-$3922 and is also harder to detect at low frequency. At 408 and 606\,MHz, PSR~B1718$-$19 gets so dim that it is barely detectable ($\lesssim 0.1$\,mJy) during a large part of the orbit in spite of good instrument sensitivity and the large observed peak flux density (0.7 and 1.3 mJy, respectively). On the other hand, at 1404 and 1660\,MHz the orbital modulation of the average flux is much less important. This flickering is probably made by material left over by the wind of the companion, a bloated main sequence (MS) star \citep{jv05}. Although the companion of PSR~B1718$-$19 is not large enough to come near to filling its Roche-lobe, this could happen in a tighter binary system like that of PSR~J1744$-$3922. In this case, some kind of tidal stripping could be occuring, leaving material around the system. This could explain why the pulsar does not disappear at conjunction like PSR~B1718$-$19 does, but in a more stochastic way.

If the companion is losing mass, one might expect to observe small changes in the orbital parameters (\S\,\ref{s:binary_evolution}). From radio timing \citep[see][]{rrh+08}, we can set an upper limit of $|\dot P_b| < 2 \times 10^{-10}$\,s\,s$^{-1}$ and $|\dot x| < 7 \times 10^{-12}$\,lt-s\,s$^{-1}$ on the rate of change of the orbital period and the rate of change of the projected semi-major axis, respectively. We can use the latter quantity to evaluate the implied mass loss limit. For a circular orbit, which is a good approximation here, we can express the rate of change of the semi-major axis as \citep{ver93}:
\begin{equation}
   \frac{\dot a}{a} = 2 \frac{\dot J}{J} - 2 \frac{\dot M_{c}}{M_{c}}
                      \left(
		      1 - \frac{\beta M_{c}}{M_{p}} - 
		      \frac{(1-\beta) M_{c}}{2 (M_{p} + M_{c})} -
		      \alpha (1-\beta) \frac{M_{p}}{M_{p}+M_{c}} 
		      \right) \,\, ,
\end{equation}
where $x = a \sin i$, $\dot{x} = \dot{a} \sin i$, $M_{c}$ and $M_{p}$ are the mass of the companion and the pulsar, respectively, $J$ is the total angular momentum of the system, $\beta$ is the fraction of mass accreted by the pulsar and $\alpha$ is the specific angular momentum of the mass lost in units of the companion star's specific angular momentum.

For the case in which the total orbital angular momentum of the system is preserved ($\dot J = 0$) we can see that mass loss from the companion\footnote{Here we implicitly assume that $M_{c}<M_{p}$.} ($\dot M_{c} < 0$) would necessarily lead to a widening ($\dot a/a > 0$) of the orbit if: 1) the mass transfer is conservative ($\beta = 1$), or 2) the mass transfer is non-conservative ($\beta < 1$) and $\alpha < 1+M_{c}/(2M_{p})$ \citep[see][for more details]{ver93}.

By considering the conservative case, in which $|\dot M_{c}|=|\dot M_{p}|$, we obtain an upper limit on a possible mass accretion rate by the pulsar $|\dot M_{p}| \lesssim 3 \times 10^{-12} \,\Msun\,{\rm yr}^{-1}$. This would be even lower for the non-conservative case. Accretion onto the pulsar is possible if the corotation radius:
\begin{equation}
   r_{co} \simeq 500 P_{{\rm psr}}^{2/3} M_{1.4}^{1/3} \quad \textrm{km},
\end{equation}
is larger than the magnetospheric radius of the pulsar:
\begin{equation}
   r_{mag} \simeq 800 \left( \frac{B_{{\rm psr}}^{4} R^{12}_{10}}{M_{1.4}
             \dot M^2_{-12}} \right)^{1/7} \quad \textrm{km},
\end{equation}
where $P_{{\rm psr}}$ and $B_{{\rm psr}}$ are, respectively, the spin period and the surface dipole magnetic field in units of PSR~J1744$-$3922's $P$ and $B$ (172\,ms and $1.68 \times 10^{10}$\,G); $M_{1.4}$ is the mass of the pulsar in units of $1.4\,\Msun$; $R_{10}$ is the radius of the pulsar in units of 10\,km; and $\dot M_{-12}$ is the accretion rate in units of $10^{-12}$\,\Msun\,yr$^{-1}$. Given the upper limit on the mass accretion rate, and the likely conservative $R_{10}=1$ value for the neutron star radius, we find that $r_{mag}$ ($\gtrsim 600$\,km) is likely larger than $r_{co}$ ($\sim 500$\,km). This argues against any significant accretion occurring in the system.

That no significant accretion is occuring is also supported by an XMM-Newton observation that allows \citet{rrh+08} to put a conservative upper limit $\sim 2 \times 10^{31}$\,erg\,s$^{-1}$ on the unabsorbed X-ray flux from accretion in the 0.1-10\,keV range. Assuming
\begin{equation}
   L_{X} = \frac{\eta G M \dot{M}}{R} \, ,
\end{equation}
with a conversion efficiency of accretion energy into observed X-rays $\eta=0.1$, we get $\dot M_{p} \lesssim 2 \times 10^{-14} \, \Msun \, \textrm{yr}^{-1}$ for accretion at the surface of the pulsar and $\dot M_{p} \lesssim 2.4 \times 10^{-12} \, \Msun \, \textrm{yr}^{-1}$ if accretion is limited to the boundary of the magnetospheric radius. Therefore, it appears unlikely that PSR~J1744$-$3922 is accreting and, if the companion is losing mass, it is probably expelled away from the system.

\section{Infrared Observations} \label{s:infrared}
The radio flux variability, which might be due to material leaving the surface of the companion, and the atypical evolutionary characteristics of the pulsar (see \S\,\ref{s:evolution}) can be investigated further by observing its companion at optical or near-infrared wavelengths. We imaged the field of PSR~J1744$-$3922 at $\textrm{K}^{\prime}$-band on the night of 2005 April 19 using the Canada-France-Hawaii 3.6-m Telescope (CFHT) at Mauna Kea. The telescope is equipped with the Adaptive Optic Bonette (AOB) \citep{rsa+98}, which provides good correction for atmospheric seeing, and KIR, the 1024$\times$1024 pixel HAWAII infrared detector with 0.0348$''$ pixel scale. The total integration time was 30\,s$\times$59 integrations = 1770\,s.

We substracted a dark frame from each image, and then constructed a flat-field image from the median of the science frames. The images were then flat-fielded, registered and stacked to make the final image. The final stellar profile has a FWHM of 0.17$''$, degraded somewhat from the optimal correction provided by the AOB system due to a high airmass ($\sim 2.0$) and poor natural seeing. Figure~\ref{f:IRimage} shows the final image we obtained.

\afterpage{
\clearpage

\begin{figure}
 \centering
 \includegraphics[width=5in]{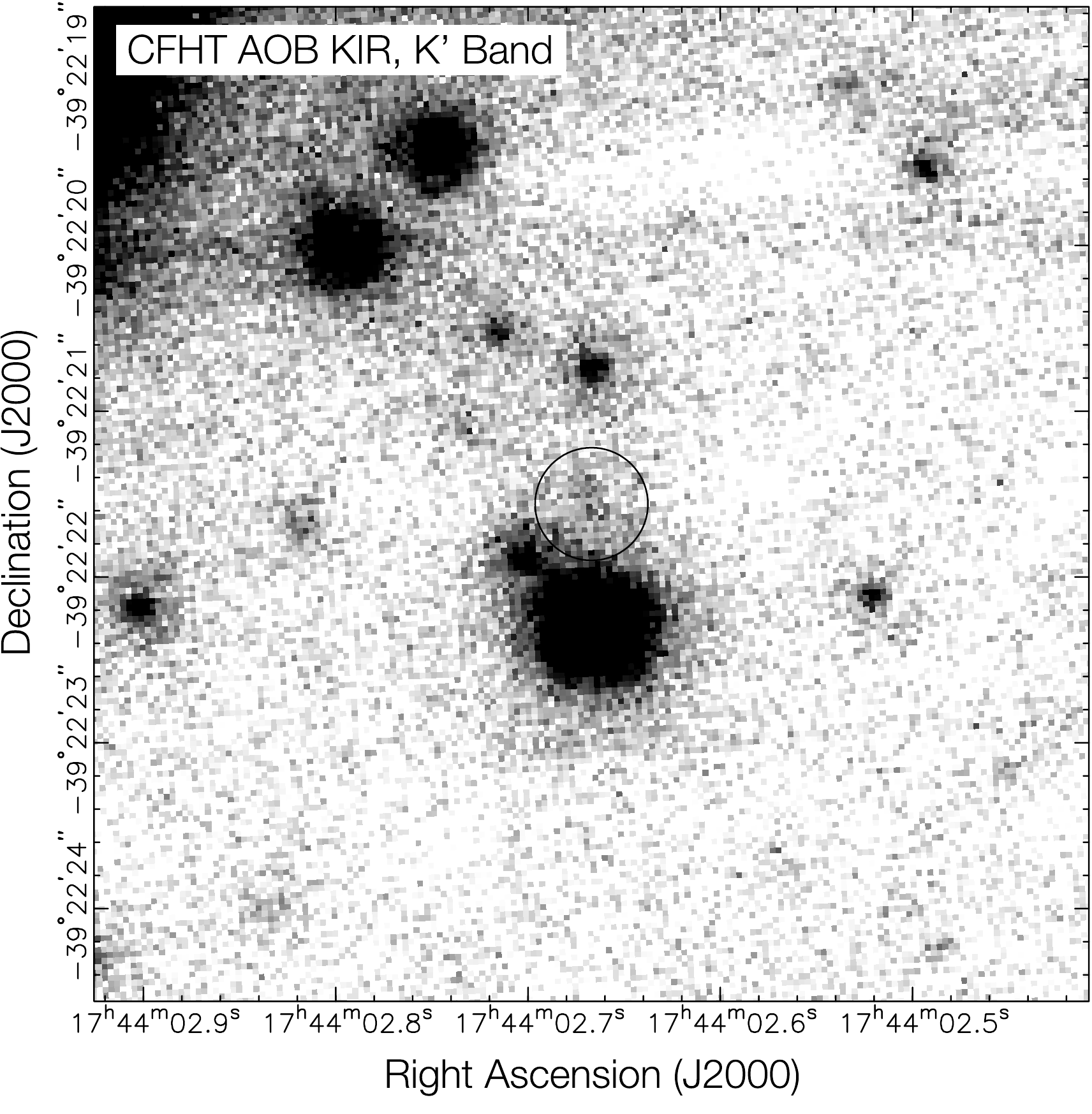}
 \caption[Near-infrared counterpart of PSR~J1744$-$3922's companion]{Near-infrared image of the field of PSR~J1744$-$3922 in the $\textrm{K}^{\prime}$ band, obtained with AOB KIR at CFHT. The $0.34''$ positional error circle ($3\sigma$ confidence) is shown, with the proposed counterpart at the centre.}
 \label{f:IRimage}
\end{figure}

\clearpage
}

We analyzed the final CFHT image using the standard routine {\tt daophot} \citep{ste87} for PSF fitting photometry, and calibrated the image using the standard star FS34 \citep{ch92}. To find the photometric zero point, we performed photometry on the standard star with a large aperture containing most of the flux, and applied an aperture correction for the PSF stars in the science image. Careful calibration of measurement errors has been done by adding artificial stars of known magnitude to blank parts of the image and then measuring their magnitude through the PSF fitting process along with the real stars. Thus, errors on the magnitude returned by {\tt daophot} can be rescaled by calculating the standard deviation for the added stars.

We found the astrometric solution for the image by cross-identifying five stars with the 2MASS catalogue \citep{scs+06}, fitting for scale, rotation and displacement. The final astrometric uncertainty is 0.34$''$ at the $3\sigma$ confidence level. This value depends on the matching to the reference stars because the error on the radio timing position of PSR~J1744$-$3922 is negligible, ($\sim 0.03''$). The final image is displayed in Figure~\ref{f:IRimage}, with the positional error circle centered at the radio position of PSR~J1744$-$3922: $\alpha=17^{\rm h} 44^{\rm m} 02^{\rm s} 667(1)$ and $\delta=-39^{\circ}22'21.52''(5)$. Only one star falls inside this circle, and for this we measure\footnote{Our calibration was made against the Vega photometric system.} $\textrm{K}^{\prime}= 19.30(15)$. We observe that, above the $3\sigma$ detection limit, the average stellar density is 0.079\,arcsec$^{-2}$ and hence the probability of a star falling in the error circle is only 2.9\,\%. Due to its positional coincidence and the low probability of chance superposition, we henceforth refer to this object as the possible counterpart to PSR~J1744$-$3922.

Unfortunately our near-infrared observation does not tightly constrain the nature of the companion to PSR~J1744$-$3922, mainly because of the uncertainties in the distance to the system and in the companion mass, as well as the fact that its temperature is unknown. Assuming the NE2001 electron density model \citep{cl02} is correct, we infer a DM distance of $3.0 \pm 0.6$\,kpc and hence a distance modulus ranging from 11.9 to 12.8. Using the three-dimensional Galactic extinction model of \citet{dcl03} we get a value of $A_{V} \simeq 1.9$ for a distance of 3.0\,kpc. Converting\footnote{For simplicity, and because the error on the magnitude is dominated by the distance estimate, we hereafter assume that K and $\textrm{K}^{\prime}$ are similar.} the inferred extinction to $\textrm{K}^{\prime}$ band gives $A_{K'} \simeq 0.21$ \citep[see][for conversion factors]{rl85}. Therefore the estimated absolute magnitude of the counterpart lies in the range $M_{K^{\prime}} \simeq [6.1,7.4]$.

We can evaluate how probable it is that the companion is a He WD, a typical low-mass companion in binary pulsar systems, since WD cooling models can put restrictions on the stellar mass and cooling age, given an observed flux. Figure~\ref{f:cooling} shows the absolute $\textrm{K}^{\prime}$ magnitude as a function of cooling age for He WDs of different masses. These cooling tracks were made by using WD atmosphere models based on the calculations of \citet{bsw95}, and thereafter improved by \citet{blr01} and \citet{ber01}, in combination with evolution sequences calculated by \citet{dsb+98}. Although the mass range of models available to us does not go below 0.179\,\Msun, we can deduce that to be so luminous, a He WD would need to have a very low mass and a cooling age much lower than the characteristic age of the pulsar (1.7\,Gyr). Even if the mass were equal to the lower limit of 0.08\,\Msun\ derived from radio timing, it seems unlikely that such a companion could be as old as the pulsar characteristic age. Therefore, if the companion is indeed a He WD, the pulsar's spin period must be close to the equilibrium spin frequency it reached at the end of the recycling process. In this case, its characteristic age is an overestimate of its true age.

\afterpage{
\clearpage

\begin{figure}
 \centering
 \leavevmode
 \includegraphics[width=6in]{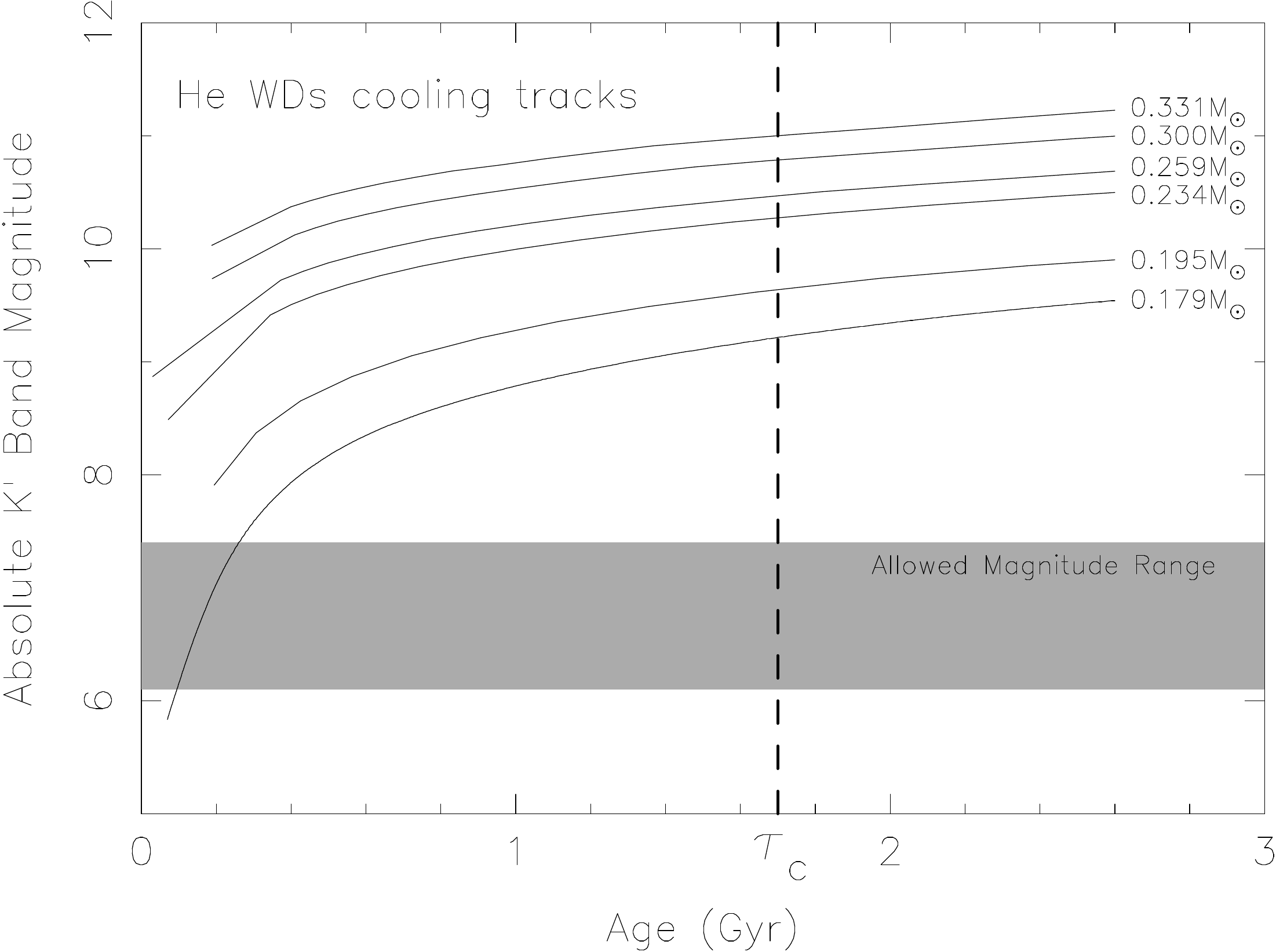}
 \caption[White dwarf cooling tracks]{Cooling tracks for He WDs made using the evolution sequences of \citet{dsb+98} and our atmosphere models \citep[see][for more details]{bsw95,blr01,ber01}. Lines show the cooling for constant masses of 0.179, 0.195, 0.234, 0.259, 0.300 and 0.331\,\Msun, from bottom to top, respectively. The shaded region is the restricted range of absolute $\textrm{K}^{\prime}$ band magnitude inferred from the CFHT data, and the dotted vertical line indicates the characteristic age of PSR~J1744$-$3922.}
 \label{f:cooling}
\end{figure}

\clearpage
}

Another possibility is that the companion is not a He WD. If the companion is a low-mass main sequence (MS) star, then the minimum mass required to match the lower limit on the absolute $\textrm{K}^{\prime}$ flux is $\sim$ 0.25\,\Msun, regardless of the pulsar age. Such a companion mass requires a favorable face-on orbit but this cannot be ruled out from the near-infrared and radio data. On the other hand, a lower-mass bloated MS star could be equally as bright, so this is also a possibility.

We cannot constrain the nature of the counterpart very well from a measurement in a single near-infrared filter. Ideally, obtaining a spectrum could yield: 1) a precise determination of the nature of the counterpart, 2) orbital Doppler shift measurements of the spectral lines which can prove the association as well as determine the mass ratio, 3) in the case of a white dwarf, an estimate of its cooling age from spectral line fitting.

\section{Binary Pulsar Evolution} \label{s:evolution}
The sporadic radio emission from PSR~J1744$-$3922 is not the only indication that there is something unusual about the pulsar's interaction with its companion. The pulsar is in a tight and low eccentricity ($e<0.001$) orbit with an apparently very light companion having a minimum mass of 0.08\,\Msun. On the other hand it has a relatively large surface magnetic field ($1.7\times 10^{10}$\,G) and an extremely long spin period (172\,ms) compared with other binary pulsars having low-mass companions (see \S\,\ref{s:binary_evolution} for a short review on binary pulsar evolution and \citet{sta04b,vbj+05} for other reviews from the literature). These properties, along with the relatively bright near-infrared counterpart of the companion, make it unusual and suggest it evolved differently than most binary pulsars.

As we argued in \S\,\ref{s:binary_population}, the nature of the companion plays a key role in determining the final spin and orbital characteristics of a pulsar binary system. Most of the binary pulsar population consists of pulsars with low-mass companions in low-eccentricity orbits --- the \emph{case ${\cal A}$} class (see \S\,\ref{s:binary_evolution}). They are neutron stars (NS) which were spun up to very short periods ($P \lesssim 8$\,ms) after accreting matter from a low-mass star during a long and steady Roche-lobe overflow (RLO) phase (\S\,\ref{s:binary_evolution} and \citealt{ts99}). As we argued in \S\,\ref{s:binary_evolution}, the recycling process is responsible for the partial suppression of the surface magnetic field to values of the order of $10^{8-9}$\,G. The most robust predictions of this model are the correlations linking the orbital period to the mass of the He WD \citep{rpj+95} (see Figure~\ref{f:mass_comp}) and the eccentricity to the orbital period \citep{phi92}.

\afterpage{
\clearpage

\begin{figure}
 \centering
 \leavevmode
 \includegraphics[width=6in]{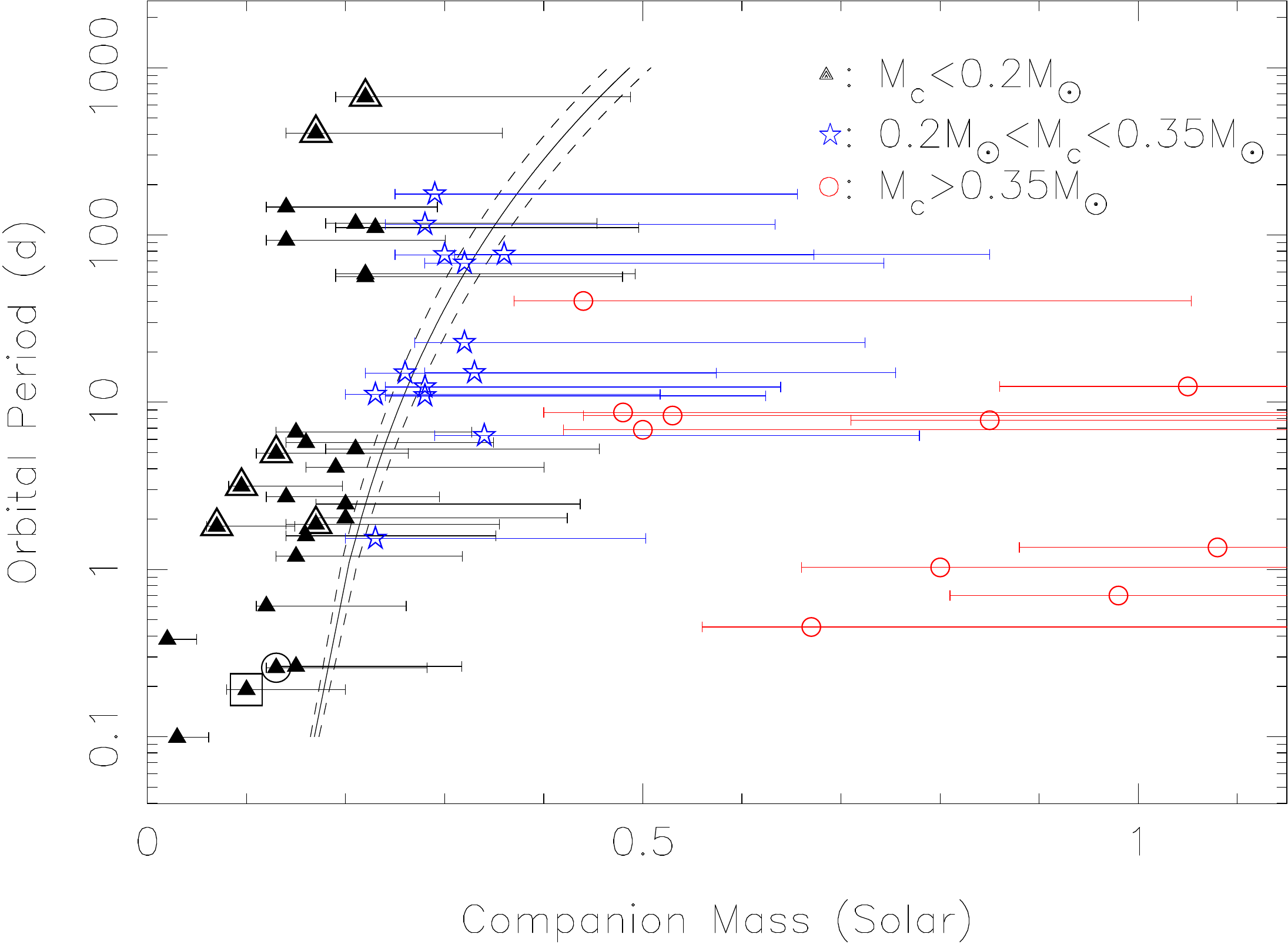}
 \caption[Orbital period versus companion mass for binary pulsars]{Orbital period versus companion mass for binary pulsars in the Galactic field in circular orbits ($e<0.01$). Symbols indicate the companion median mass (corresponding to $i=60^{\circ}$) and are coded according to the minimum mass: $M_{c}\leq 0.2 \,\Msun$, $0.2 \,\Msun < M_{c} \leq 0.35 \,\Msun$ and $M_{c} > 0.35 \,\Msun$ are black triangles, blue stars and red circles, respectively. Error bars cover the 90\,\%-probability mass range for randomly oriented orbits having $i=90^{\circ}$ to $i=26^{\circ}$. The curves are the predicted $P_b-M_{c}$ relationships for different metallicity progenitors \citep[from][]{ts99}. PSR~J1744$-$3922 is identified by a square outline and pulsars listed in Table~\ref{t:comparison} with triangle outlines. The plot also includes the globular cluster pulsar PSR~B1718$-$19 (marked by a circle outline) because it resembles PSR~J1744$-$3922 (see \S\,\ref{ss:eclipse}). Data from the ATNF Pulsar Catalogue \citep{atnf}.}
 \label{f:mass_comp}
\end{figure}

\clearpage
}

On the other hand, the \emph{case ${\cal B}$} channel is made of pulsars having more massive CO WD or ONeMg WD companions ($M_{c} \gtrsim 0.45 \Msun$) (\S\,\ref{s:binary_evolution}). Their intermediate mass progenitors did not sustain a stable RLO phase, instead evolving in a short-duration, non-conservative, common envelope (CE) phase during which the pulsar spiraled into its companion's envelope. This process only partly recycled the pulsar, leading to intermediate spin periods ($P \gtrsim 8$\,ms) and leaving a higher magnetic field ($\sim 10^{9-10}$\,G).

In Table~\ref{t:comparison}, the expected properties of systems resulting from these two evolutionary channels are compared with those of PSR~J1744$-$3922. Both scenarios fail to account for all the observed characteristics; this suggests a special evolution for PSR~J1744$-$3922. This can also be seen from a $P-B$ diagram (Figure~\ref{f:recycling_a}) made for binary pulsars in the Galactic field having circular orbits. As opposed to isolated and other kinds of non-recycled binary pulsars, there exists a very strong relationship linking $P$ and $B$ which is presumably related to the recycling process. To our knowledge, such a correlation has not been reported in the recent literature although it was indirectly found by \citet{vdh95} who reported a possible correlation between $P-P_b$ and $B-P_b$ for binary pulsars in circular orbits. The many binary pulsars discovered in recent years may be making it easier to appreciate. In Figure~\ref{f:recycling_a} we see that pulsars having light companions (\emph{case ${\cal A}$}) generally gather in the region of low magnetic field and short spin period whereas the \emph{case ${\cal B}$} pulsars lie in higher-valued regions. Surprisingly, of the six highest magnetic field pulsars, five of them, including PSR~J1744$-$3922, appear to have light companions. The remaining one, PSR~B0655+64, is an extreme system associated with the \emph{case ${\cal B}$} subclass since it has a massive WD companion \citep{vbj+05}. However, the \emph{case ${\cal B}$} subclass cannot accommodate the other five pulsars because, assuming random orbital inclinations, a simple statistical estimate gives less than a 0.1\,\% probability for all of them to be more massive than the required $0.45 \, \Msun$. For PSR~J1744$-$3922 in particular, the orbital inclination would need to be less than 12.5$^{\circ}$, which represents a 2.5\% probability.

\afterpage{
\clearpage

\begin{table}
 \begin{center}
 \begin{tabular}{lrrrrr}
 \hline
 \hline \\ [-1.8ex]
 Name & $P$ & $\log{B}$ & $P_b$ & $M_{c,min}$\tablenotemark{a} & Type \\
  & ms & G & days & \Msun & \\ [0.5ex]
 \hline \\ [-1.3ex]
 \emph{Case ${\cal A}$} & $\lesssim 8$ & 8-9 & -- & $\lesssim 0.45$ & He WD \\
 \emph{Case ${\cal B}$} & $\gtrsim 8$ & 9-10 & -- & $\gtrsim 0.45$ & CO WD \\
 \noalign{\vskip .8ex} \hline \noalign{\vskip .8ex}
 PSR~J1744$-$3922 & 172.44 & 10.22 & 0.19 & 0.08 & ? \\
 PSR~B1718$-$19\tablenotemark{b} & 1004.03 & 12.11 & 0.25 & 0.12 & Bloated MS\tablenotemark{c} \\
 PSR~B1831$-$00 & 520.95 & 10.87 & 1.81 & 0.06 & ? \\
 PSR~J1232$-$6501 & 88.28 & 9.93 & 1.86 & 0.14 & ? \\
 PSR~J1614$-$2318 & 33.50 & 9.14 & 3.15 & 0.08 & ? \\
 PSR~J1745$-$0952 & 19.37 & 9.51 & 4.94 & 0.11 & ? \\
 PSR~B1800$-$27 & 334.41 & 10.88 & 406.78 & 0.14 & ? \\
 PSR~J0407+1607 & 25.70 & 9.16 & 669.07 & 0.19 & ? \\ [0.5ex]
 \hline
 \end{tabular}
 \end{center}
 \caption[Characteristics of partly recycled binary pulsars]{Characteristics of partly recycled binary pulsars ($P_{s}>8$\,ms) in the Galactic field in low-eccentricity orbits ($e<0.01$) and having low-mass companions ($M_{c,min}<0.2 \, \Msun$)}
 \tablenotetext{a}{$M_{c,min}$ refers to the minimum mass of the companion corresponding to an orbital inclination angle of $90^\circ$ and assuming a mass for the pulsar of $1.35 \, \Msun$. }
 \tablenotetext{b}{In globular cluster NGC~6342.}
 \tablenotetext{c}{\citet{jv05}}
 \label{t:comparison}
\end{table}

\clearpage
}

\afterpage{
\clearpage

\begin{figure}
 \centering \leavevmode
 \includegraphics[width=6in]{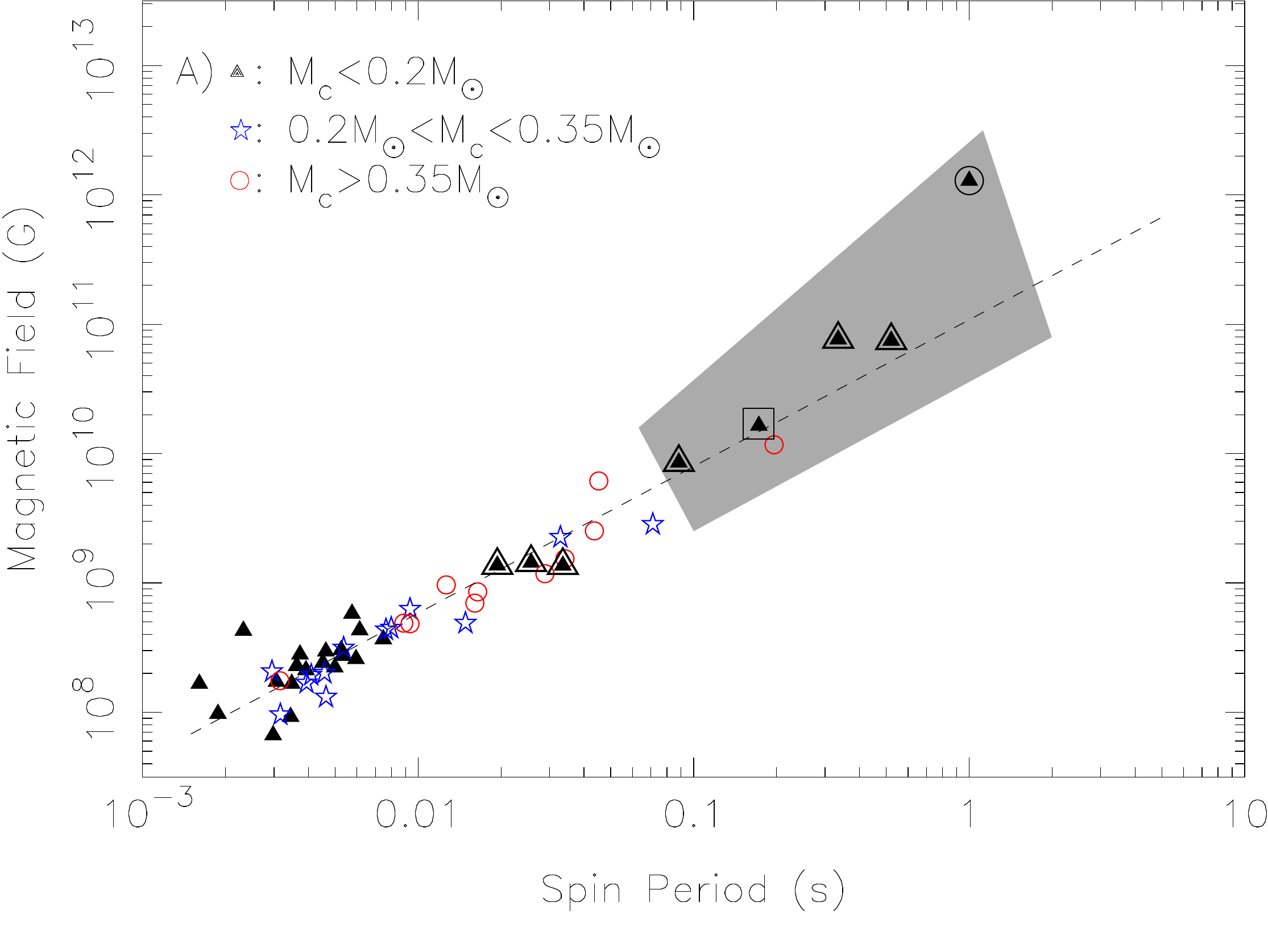}
 \caption[Magnetic field strength versus spin period for binary pulsars]{Inferred surface dipolar magnetic field strength versus spin period for binary pulsars in the Galactic field in circular orbits ($e<0.01$). Symbols are coded according to the companion minimum mass: $M_{c}\leq 0.2 \,\Msun$, $0.2 \,\Msun < M_{c} \leq 0.35 \,\Msun$ and $M_{c} > 0.35 \,\Msun$ are black triangles, blue stars and red circles, respectively. PSR~J1744$-$3922 is identified by a square outline and pulsars listed in Table~\ref{t:comparison} with triangle outlines. The shaded area is the approximate region of the proposed class of binary pulsars similar to PSR~J1744$-$3922. The plot also includes the globular cluster pulsar PSR~B1718$-$19 (marked by a circle outline) because it resembles PSR~J1744$-$3922 (see \S\,\ref{ss:eclipse}). The dashed line is the best-fit for a power-law, $B \propto P^{\alpha}$, with $\alpha = 1.13$. We excluded PSR~B1718$-$19 from the fit as it is in a globular cluster. Data from the ATNF Pulsar Catalogue \citep{atnf}.}
 \label{f:recycling_a}
\end{figure}

\clearpage
}

We \emph{also} report in Table~\ref{t:comparison} the principal characteristics of binary pulsars that appear to be partly recycled (e.g. $P>8$\,ms and $e<0.01$) but have companions likely not massive enough to be explained by the standard \emph{case ${\cal B}$} scenario. These pulsars have related properties and could have experienced similar evolutionary histories. Apart from their strange position in the $P-B$ diagram, they also stand out when we look at the $P_b-P$ relationship (Figure~\ref{f:recycling_b}). In this plot, we see that pulsars having low-mass companions (\emph{case ${\cal A}$}) occupy the bottom region, below $P \lesssim 8$\,ms, and their spin periods are more or less independent of the orbital period. This arises from the fact that recycling probably saturates for a given accretion rate and/or accretion mass \citep{ks04}. On the other hand, fewer constraints exist in this parameter space for pulsars with massive WD companions (\emph{case ${\cal B}$}). Their short CE evolution would limit the recycling efficiency and thus the final parameters are more sensitive to the initial conditions. Finally, we observe a third category made of relatively slow pulsars with low-mass companions in compact orbits. Neither the \emph{case ${\cal A}$} nor the \emph{case ${\cal B}$} scenarios is able to explain such properties, especially for the most extreme systems like PSR~J1744$-$3922 and PSR~B1831$-$00. Therefore, we suggest a new ``class'' of binary pulsars having the following properties: 1) long spin periods (in comparison to millisecond pulsars), 2) large surface magnetic fields ($\sim 10^{10-11}$\,G), 3) low-mass companions, likely $0.08-0.3 \, \Msun$, having nature yet to be determined, 4) low eccentricities, and possibly 5) short orbital periods ($\lesssim 5$\,d). On this last point, very wide orbit systems like PSR~B1800$-$27 and PSR~J0407+1607 might be explained by the standard \emph{case ${\cal A}$} scenario in which it is difficult to achieve an extended period of mass transfer from the companion to the pulsar.

\afterpage{
\clearpage

\begin{figure}
 \centering \leavevmode
 \includegraphics[width=6in]{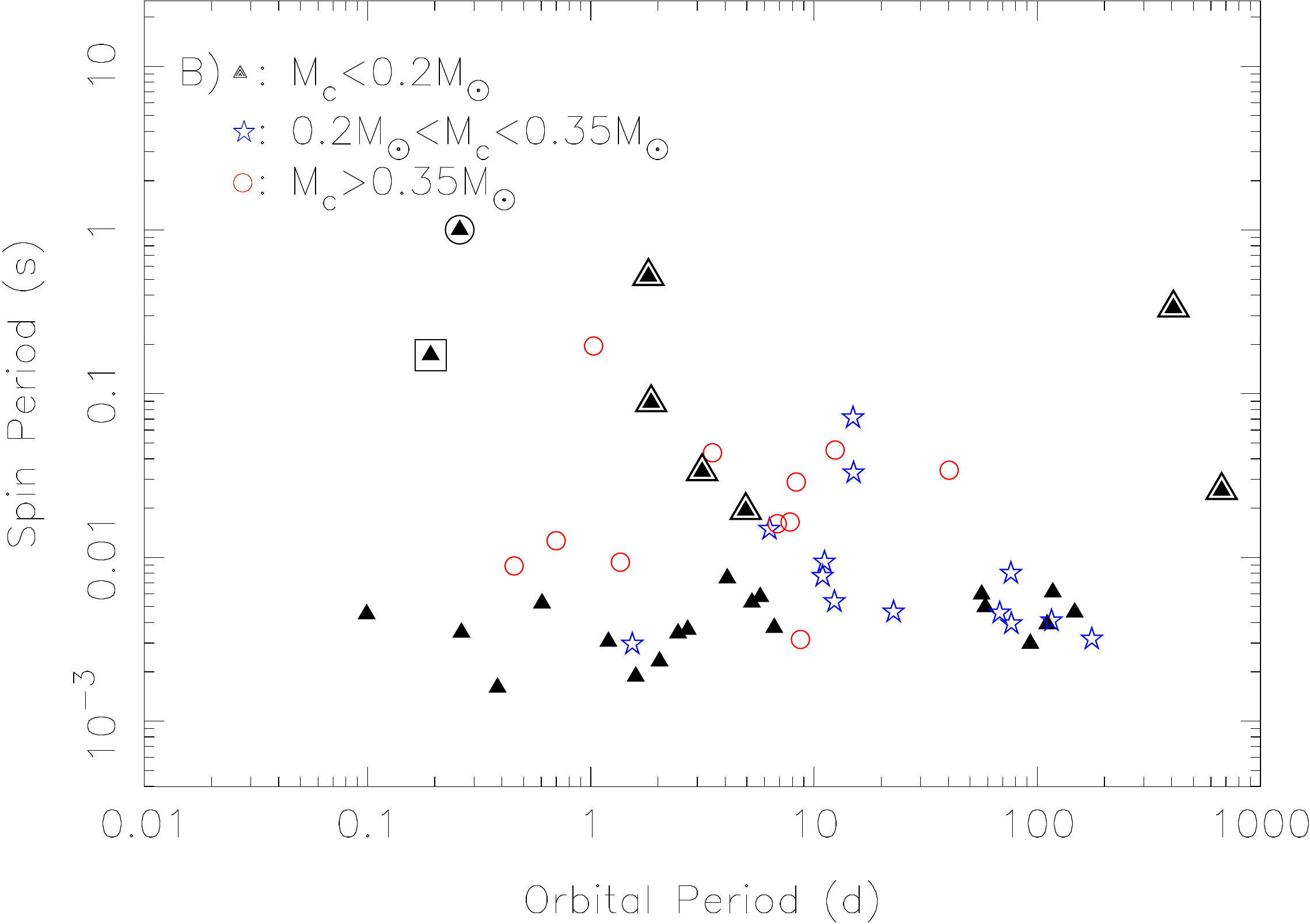}
 \caption[Spin period versus orbital period for binary pulsars]{Spin period versus orbital period for binary pulsars in the Galactic field in circular orbits ($e<0.01$). Symbols are coded according to the companion minimum mass: $M_{c}\leq 0.2 \,\Msun$, $0.2 \,\Msun < M_{c} \leq 0.35 \,\Msun$ and $M_{c} > 0.35 \,\Msun$ are black triangles, blue stars and red circles, respectively. PSR~J1744$-$3922 is identified by a square outline and pulsars listed in Table~\ref{t:comparison} with triangle outlines. The plot also include the globular cluster pulsar PSR~B1718$-$19 (marked by a circle outline) because it resembles PSR~J1744$-$3922 (see \S\,\ref{ss:eclipse}). Data from the ATNF Pulsar Catalogue \citep{atnf}.}
 \label{f:recycling_b}
\end{figure}

\clearpage
}

\section{Discussion} \label{s:discussion}
The existence of another ``class'' of binary pulsars is supported by the fact that several pulsars now occupy a region of the parameter space delimited by the spin period, orbital period, magnetic field and companion mass that seems inaccessible to the standard evolutionary channels. PSRs~J1744$-$3922, J1232$-$6501, B1718$-$19 and B1831$-$00 are certainly the most noticeable candidates. Although other studies also identified that some of these pulsars have strange characteristics \citep[see, e.g.][]{vdh95,sl00,eb01}, there is no consensus on their evolutionary histories. For instance the case of PSR~B1718$-$19 was considered somewhat unique because it is in a globular cluster and hence has possibly been partly perturbed, or even greatly changed, by stellar interactions \citep{esg96} (see \S\,\ref{s:gc_pulsars} for a discussion about binary pulsars in globular clusters). The discovery of PSR~J1744$-$3922, in the Galactic field, is important because it strengthens the possible connection between PSR~B1718$-$19 and other similar binary pulsars in the field. Unless they do not have WD companions, these pulsars were at least partially recycled because tidal circularization is needed to explain eccentricities of 0.01 and smaller. However some of them have eccentricities $10^{-3}-10^{-2}$, relatively large given their very short orbital periods, which is in contrast to tight \emph{case ${\cal A}$} systems having smaller eccentricity \citep{phi92}. In this section we speculate and put constraints on some scenarios that may explain this possible new class of pulsars.

\subsection{Recycled High Magnetic Field Pulsar Channel}\label{s:j1744_magnetar}
One possibility is that pulsars like PSR~J1744$-$3922 may initially have had a magnetar-strength magnetic field ($B\sim 10^{14}-10^{15}$\,G) (\S\,\ref{s:magnetars}). Such a pulsar could experience standard \emph{case ${\cal A}$} evolution involving conservative RLO but since the initial magnetic field is higher than for ordinary pulsars by 1--2 orders of magnitude, at the end of the recycling process the field might be $10^{10-11}$\,G instead, consistent with observations. Since the recycling mechanism appears to strongly correlate the final magnetic field with the final spin period, the pulsar would also have an unusually long spin period as well.

Whether this extrapolation to magnetars and high magnetic field pulsars is valid depends on accretion-induced decay models of the magnetic field strength. Among the proposed mechanisms is magnetic field burial, in which material accretes through the polar cap and, while piling up at the poles, exerts a latitudinal pressure gradient by trying to spread toward the equator. This effect tends to drag the field lines away from the poles, increasing the polar cap radius and decreasing the magnetic moment of the pulsar. \citet{pm04} show that the magnetic field would naturally ``freeze'' to a minimum stable strength once the amount of accreted mass exceeds some critical value, if the magnetospheric radius is comparable to the size of the neutron star. According to \citet{pay05}, this critical mass could reach up to 1\,\Msun\ for a $10^{15}$\,G NS and hence magnetic field suppression would be difficult to achieve due to the large amount of accretion mass required. However, accounting for other effects like the natural decay of magnetic field due to X-ray emission and high-energy bursts \citep[see][for a recent review]{wt06}, might make a partially suppressed final magnetic field of $~10^{10}$\,G plausible.

A recycled high magnetic field pulsar is expected to leave a He WD companion and follow the companion mass-orbital period relationship as for the normal \emph{case ${\cal A}$} systems. PSR~B1718$-$19 is excluded from this kind of evolution because its companion is a bloated MS star. All the other pulsars listed in Table~\ref{t:comparison} are potential members, albeit PSRs B1800$-$27 and J0407+1607 would need relatively face-on orbits ($i < 30$ and $24^\circ$, implying 13 and 9\% probability for randomly oriented orbits, respectively) to match the $P_b-M_{c}$ relationship (Figure~\ref{f:mass_comp}). Another interesting prediction of this scenario is that, because the recycling process can leave pulsars with relatively long spin periods, they might not be very different from the spin periods when accretion ceased. Thus, the real age could be much lower than the timing-based characteristic age, and the WD companions would have younger cooling ages as well.

The fraction of Galactic field binary pulsars in this class ($\sim$7/60) might naively be thought to be similar to the fraction of observed magnetars with respect to the ordinary pulsars ($\sim$10/1500). However, the observed population of magnetars suffers from severe selection effects because many of them appear to lie dormant, becoming observable for only brief intervals, like XTE~J1810$-$197 \citep{ims+04}. Therefore, we can relate these parameters as follows:
\begin{equation}
   \frac{N_{\mbox{\footnotesize class}}}{N_{\mbox{\footnotesize
   binary}}} = 
   \frac{N_{\mbox{\footnotesize obs. mag.}}}{N_{\mbox{\footnotesize radio}}}
   \frac{1}{f_{\mbox{\footnotesize quies}}} \,\, ,
\end{equation}
where $N_{\mbox{\footnotesize class}}$ is the number of pulsars in the new class, $N_{\mbox{\footnotesize binary}}$ is the number of binary pulsar systems, $N_{\mbox{\footnotesize obs. mag.}}$ is the number of observed magnetars, $N_{\mbox{\footnotesize radio}}$ is the number of radio pulsars and $f_{\mbox{\footnotesize quies}}$ the fraction of magnetars in quiescence.

Hence we estimate that, due to quiescence, the fraction of observed magnetars with respect to the total population is about $\frac{10}{1500}/\frac{7}{60} \sim 0.06$. This would make a total population of $\sim 175$ magnetars which is consistent with the possible $\sim 100$ in our Galaxy estimated by \citet{wt06}, who used $f_{\mbox{\footnotesize quies}} = 0.1$. Our crude calculation has many caveats: Is the binary magnetar population similar to the binary radio pulsar population? Can the short lifetime of magnetars and high magnetic field neutron stars limit the number of such recycled systems? Clearly, this latter point depends strongly on the time evolution of the self-induced decay of the magnetic field, which seems to operate on a time scale of a few tens of thousands of years \citep{kas04}.

\subsection{UCXB Evolutionary Track Channel} \label{s:ucxb}
There exists a class of neutron stars and CO/ONeMg WDs that accrete from light He WD donors in ultra tight (few tens of minutes) orbits. These X-ray emitters, known as ultracompact X-ray binaries (UCXBs), may be the result of wider $\sim 0.5$-day systems that have decayed due to gravitational wave radiation. A CE scenario has been proposed as a viable channel for forming CO/ONeMg WD -- He WD and NS -- He WD systems \citep{bt04}. In fact such a channel would be very similar to the \emph{case ${\cal B}$} scenario but for an initially much lighter companion. For companions not massive enough to experience the He flash, a CE phase is possible if the onset of mass transfer occurs late enough in the evolution so that the companion has reached the asymptotic giant branch \citep{bt04}. In this case, the RLO becomes unstable and it bifurcates from the standard \emph{case ${\cal A}$} track to the CE phase (see Figure~\ref{f:evolution_cases} for a diagram showing the different evolution types as a function of orbital period and donor's mass at the onset of the mass transfer).

Pulsars experiencing this evolution would be partially recycled, like \emph{case ${\cal B}$} systems, but they would have He WD companions. After this stage, only sufficiently tight systems with orbital periods of about one hour or less can evolve to become UCXBs within a Hubble time because gravitational decay is negligible for wider orbits. If PSR~J1744$-$3922-like pulsars belong to the long orbital separation high-end of the UCXB formation channel, we might expect to see more such pulsars at similar and smaller orbital separations than PSR~J1744$-$3922. It is possible, however, that the observed sample is biased: as the orbital separation decreases, wind and mass loss by the companion become more important and would make them more difficult to detect in classical pulsar surveys conducted at low frequency (i.e. $\sim$400\,MHz) where eclipses are more frequent and radio emission might simply turn off. The ongoing ALFA survey at Arecibo, at 1400\,MHz \citep{crh+06}, could therefore find several new pulsars like PSR~J1744$-$3922. Additionally, larger orbital accelerations make them more difficult to find and, since gravitational wave radiation varies as the fourth power of the orbital separation, their lifetimes are dramatically shorter.

In such a scenario, these pulsars might be born at a spin period that is comparable with those of \emph{case ${\cal B}$} pulsars having CO WD companions, assuming that the short, high-accretion rate recycling would efficiently screen the magnetic field during the mass transfer process. Afterwards, because of their relatively larger magnetic field, they would spin down more rapidly than \emph{case ${\cal B}$} pulsars. Thus, they would necessarily have longer spin periods and the true age is more likely to be in agreement with the measured characteristic age.

\subsection{AIC Channel}
Finally, a third scenario to explain the unusual properties of PSR~J1744$-$3922 is the accretion-induced collapse (AIC) of a massive ONeMg WD into a NS. Most likely AIC progenitors would be massive ONeMg WDs ($\gtrsim 1.15 \, \Msun$) accreting from MS donors in CV-like systems, or from red giants or He WDs in UCXB systems \citep[see][for more details]{taa04,it04}. \citet{nk91} have shown that for accretion rates $\gtrsim 0.001 \, \dot{M}_{Edd}$ and/or metal-rich accreted mass, the ONeMg WD would collapse to a NS rather than explode in a supernova. More recent calculations including Coulomb corrections to the equation of state by \citet{bg99} demonstrate that AIC is possible for critical densities of the accreting WD core that are $30\%$ lower than previously found by \citet{nk91}, thus facilitating the formation of neutron stars through this channel.

The properties predicted by the AIC scenario nicely agree with what we observe for the class we are proposing: the mass transfer prior to the NS formation would explain the low mass of the companion and the collapse is expected to be a quiet event during which almost no mass is lost in the system and only $\sim 0.2 \, \Msun$ is converted in binding energy into the NS. This would keep the final orbital period close to what it was prior to the collapse (which is small in most scenarios leading to AIC) and allow the eccentricity to be very small or, at least, circularize rapidly. The survival rate of such systems is probably higher than for standard systems which are more likely to be disrupted if a large amount of mass is lost during the supernova process. Although the initial properties of pulsars formed by AIC are not known, we may presume they resemble those of ``normal'' pulsars with magnetic fields in the $10^{11-12}$\,G range. Another interesting point to consider is that the AIC channel does not require a degenerate companion. As opposed to the other two proposed scenarios (\S\,\ref{s:j1744_magnetar} and \ref{s:ucxb}), the formation of a pulsar through AIC can interrupt the mass transfer, thus postponing further evolution. If the ``evolutionary quiescence'' is long enough, we would expect to find some non-degenerate companions around young AIC pulsars having circular orbits but with spin periods and magnetic fields that are more typical of isolated pulsars. Finally, as the companion continues to evolve to fill its now larger Roche lobe, a shortened accretion phase might occur, thus transfering a very small amount of mass to the pulsar and leaving it partly recycled despite its low-mass companion.

Some binary pulsars among the group we highlighted, like PSRs B1831$-$00 and B1718$-$19, have been proposed as AIC candidates in the past \citep{vdh95,erg93}. However, there is no firm evidence to support this. For instance, PSR~B1718$-$19 is presumably a member of the globular cluster NGC~6342. As \citet{jv05} show, although observations suggest AIC is possible, an encounter and tidal capture scenario cannot be ruled out and is very reasonable given the plausible globular cluster association. On the other hand, PSR~B1718$-$19 shares similarities with other binary pulsars in the Galactic field, especially with PSR~J1744$-$3922. PSR~B1718$-$19's younger age, larger magnetic field and spin period, as well as the fact that it has a non-degenerate companion, are all compatible with it being an AIC pulsar in the intermediate ``quiescent'' phase.

In this context, the other pulsars of our putative class would have reached the final evolutionary stage and, hence, display mildly recycled properties. If so, PSR~J1744$-$3922 likely has a very light He WD companion. This might explain the lack of traditional eclipses as in PSR~B1718$-$19 \citep{lbh+93}, large DM variations as in NGC~6397A \citep{dpm+01}, and orbital period derivatives as in other compact binary pulsars \citep{nat00} since tidal effects are important for non-degenerate companions. If the residual recycling phase left PSR~J1744$-$3922 with a long spin period, the cooling age of its hypothetical WD companion could be smaller than the characteristic age of the pulsar for the same reason described in the recycled high magnetic field pulsar scenario (see \S\,\ref{s:j1744_magnetar}). Also, pulsars forming through AIC can have lower masses than those made in standard supernovae and since they only accrete a small amount of mass from their companion afterwards, they might be less massive than conventionally formed millisecond pulsars.

\section{Conclusions} \label{s:conclusion}
This study highlights the unusual nature of the binary pulsar system PSR~J1744$-$3922. The puzzling radio flux modulation that it exhibits does not show typical nulling properties as displayed by some old isolated pulsars; specifically, we have found strong evidence that its variation is highly frequency dependent. Although our orbital modulation analysis does not show a significant correlation between orbital phase and flux, the modulation could still be caused by a process related to a wind from its companion, which results in short time scale variations grouped in extreme modulation ``events''. Additional monitoring of both the pulsar and of its companion may prove useful in this regard.

We pointed out that this pulsar has an unusual combination of characteristics: long spin period, very low-mass companion, high magnetic field and short orbital period, that are unexplained by standard binary pulsar evolution scenarios. We propose that PSR~J1744$-$3922, along with a several other binary pulsar systems, are part of a new class of low-mass binary pulsars which failed to be fully recyled. Specifically, we suggest three alternative scenarios for this class of binary pulsars. Distinguishing among them may be possible by improving our knowledge of the nature of their companions. We also reported the detection of a possible near-infrared counterpart to PSR~J1744$-$3922's companion, however, determining its nature will require detailed near-infrared/optical follow-up.


\chapter{The Eclipses of the Double Pulsar}\label{c:0737_eclipse}

\begin{flushright}
 \begin{singlespace}
 \emph{``A theory is something nobody believes, except the person who made it.\\
 An experiment is something everybody believes, except the person who made it.''}
 \end{singlespace}
 Albert Einstein
 \vspace{0.5in}
\end{flushright}

This chapter presents the results of an exhaustive study of the double pulsar PSR~J0737$-$3039A/B eclipses. Once per orbit, the pulsar `A' is eclipsed by its companion, pulsar `B', for about 30~s. Eclipse modeling allows us to reconstruct the geometrical orientation of pulsar B with respect to the orbit and measure the relativistic precession of its spin axis. This provides a test of general relativity and alternative theories of gravity in the strong-field regime as well as valuable information about the system's geometry.

Part of this work --- the 820\,MHz eclipse modeling and the derived measurement of relativistic spin precession --- was originally published as: R. P. Breton, V. M. Kaspi, M. Kramer, M. A. McLaughlin, M. Lyutikov, S. M. Ransom, I. H. Stairs, R. D. Ferdman, F. Camilo, and A. Possenti. \emph{Relativistic Spin Precession in the Double Pulsar}. Science, 321, 104, July 2008. The remaining part of this chapter will be published as a second paper complementing the Science paper and will focus on the phenomenological aspects of the eclipse, the multi-frequency observations and the geometrical consequences of the eclipse modeling.

\section{Introduction}
\subsection{Unique Eclipses}
As we described earlier in \S\,\ref{s:binary_population}, PSR~J0737$-$3039A/B consists of two neutron stars, both visible as radio pulsars, in a relativistic 2.45-hour orbit \citep{bdp+03,lbk+04}. High-precision timing of the pulsars, having spin periods of 23 ms and 2.8 s (hereafter called pulsars A and B, respectively), has already proven to be the most stringent test-bed for GR in the strong-field regime \citep{ksm+06} and enables four independent timing tests of gravity, more than any other binary system (see \S\,\ref{a:strong_field} for a discussion about the notion of strong-field regime and \S\,\ref{s:tests_gr} for a brief review of tests of general relativity involving binary pulsar timing).

The orbital inclination of the double pulsar is such that we observe the system almost perfectly edge-on. This coincidence causes pulsar A to be eclipsed by pulsar B at pulsar A's superior conjunction \citep{lbk+04}. The modestly frequency-dependent eclipse duration, about 30 s, corresponds to a region extending $\sim$1.5$\times 10^7$\,m \citep{krb+04}. The light curve of pulsar A during its eclipse shows flux intensity modulations that are spaced by half or integer numbers of pulsar B's rotational period \citep{mll+04}. This indicates that the material responsible for the eclipse corotates with pulsar B. The relative orbital motions of the two pulsars and the rotation of pulsar B thus allow a probe of different regions of pulsar B's magnetosphere in a plane containing the line of sight and the orbital motion.

Synchrotron resonance with relativistic electrons is the most likely mechanism for efficient absorption of radio emission over a wide range of frequencies. It has been proposed by \citet{lt05} that this absorbing plasma corotates with pulsar B and is confined within the closed field lines of a magnetic dipole truncated by the relativistic wind of pulsar A.

\subsection{Spin-Orbit Coupling}\label{s:so_coupling}
Spin is a fundamental property of most astrophysical bodies, making the study of its gravitational interaction an important challenge \citep{wil01}. Spin interaction manifests itself in different forms. For instance, we expect the spin of a compact rotating body in a binary system with another compact companion to couple gravitationally with the orbital angular momentum (relativistic spin-orbit coupling) and also with the spin of this companion (relativistic spin-spin coupling)\footnote{The contribution from a classical quadrupolar moment is negligible for compact bodies.} \citep{oco74}. Observing such phenomena provides important tests for theories of gravity, because every successful theory must be able to describe the couplings and to predict their observational consequences. In a binary system consisting of compact objects such as neutron stars, one can generally consider the spin-orbit contribution acting on each body to dominate greatly the spin-spin contribution --- in the double pulsar, the spin-spin interaction scales as only $\sim 0.0001\%$ of the spin-orbit interaction. This interaction results in a precession of the bodies' spin axis around the orbital angular momentum of the system, behavior we refer to as {\em relativistic spin precession}.

While relativistic spin precession is well studied theoretically in general relativity (GR), the same is not true of alternative theories of gravity and hence, quantitative predictions of deviations from GR spin precession do not yet exist \citep{dt92a}. For instance, it is expected that in alternative theories relativistic spin precession may also depend on strong self-gravitational effects, i.e.~the actual precession may depend on the structure of a gravitating body \citep{dt92a}. In the weak gravitational fields encountered in the solar system, these strong-field effects generally cannot be detected \citep{de92a,de92b,de96a}. Measurements in the strong-field regime near massive and compact bodies such as neutron stars and black holes are required. Relativistic spin precession has been observed in some binary pulsars \cite[e.g][]{wrt89,kra98,hbo05}, but it has usually only provided a qualitative confirmation of the effect. Recently, the binary pulsar PSR~B1534+12 has allowed the first quantitative measurement of this effect in a strong field, and although the spin precession rate was measured to low precision, it was consistent with the predictions of GR \citep{sta04b}.

We may suppose that if the modulation features in the eclipse profile of pulsar A are so tightly related to the spin phases of pulsar B, the relativistic spin precession of the latter may give rise to observable effects in the eclipse profile. In this respect, the \citet{lt05} model is making clear prediction about the time-evolution behavior of the eclipse profile that should allow to measure the precession rate if the model is successful. We shall test this hypothesis in this chapter.

\section{Observations and Data Reduction}\label{s:0737_observations}
We regularly observed the double pulsar from December 2003 to November 2007\footnote{The monitoring campaign is still on-going as of the time of writing this thesis.} as part of a multi-purpose monitoring campaign primarly aimed at the timing of the pulsars, but also at the investigation of the different phenomena displayed by this system such as the eclipses of pulsar A, the orbital modulation of pulsar B's flux and its subpulse drifting. Timing results were reported in \citet{ksm+06}. Data used for the analysis presented in this chapter were acquired at the Green Bank Telescope with the {\tt SPIGOT} and {\tt BCPM} instruments \citep{kel+05,bdz+97} (see \S\,\ref{s:radio_telescopes} for more details about data acquisition systems). Observations were conducted at central frequencies 325, 427, 820, 1400, 1950 and 2200\,MHz. The {\tt SPIGOT} back-end provides a 50\,MHz bandwidth segmented into 1024 channels for observations at 325, 427, 820\,MHz, and a 800\,MHz bandwidth with the same number of channels for observations at 1950\,MHz. The recording sample time is 81.92\,$\mu$s. The {\tt BCPM} back-end provides 96 channels over a bandwidth of 48\,MHz at 820\,MHz and lower frequencies, and 96\,MHz at 1400\,MHz and 2200\,MHz. The sampling time of {\tt BCPM} is 72\,$\mu$s. Typically, most data were recorded during biannual observing campaigns consisting of several (3-5) individual 4-7 hour long observations taking place over consecutive days. Occasionally, additional observations were made between these observing campaigns. Because the orbital period of the double pulsar is 2.45 hours, we usually recorded 1-3 eclipses per observation.

We performed the initial data reduction using the pulsar analysis packages {\tt PRESTO} \citep{rem02} and {\tt SIGPROC} \citep{sig} (see \S\,\ref{s:timing}). First, we dedispersed the data to correct for the frequency-dependent travel time in the ionized interstellar medium by adding time shifts to frequency channels\footnote{General details about pulsar timing can be found in \S\,\ref{s:timing} and specific information about the double pulsar timing were reported in \citet{ksm+06}.}. Then, we generated folded data products for both pulsars using the predicted spin periods of pulsar A and pulsar B. For these two steps, we used the timing solution presented in \citet{ksm+06}, which has slightly evolved over time since the earliest observations in 2003.

We generated a time series of pulsar A's pulsed flux for each segment of observation containing an eclipse of pulsar A. We first made a high signal-to-noise ratio pulse template of pulsar A from the folded data integrated over each observation. Then, we calculated the relative pulsed flux density of pulsar A by fitting the corresponding pulse profile template to the individual fold intervals. We have chosen the fold intervals to be the sum of four individual pulses of pulsar A ($\sim 91$\,ms) in order to have a good balance between signal-to-noise ratio and time resolution. Note that as opposed to {\tt SIGPROC}, which can fold at an integer number of pulses, {\tt PRESTO}'s folding algorithm is based on byte size and hence, despite the exact number of pulses being generally four, there is a handful of data points consisting of three pulses only. These data points were renormalized to correct for their shorter time integration. Finally, to facilitate the data analysis we normalized the time series such that the flux average in the out-of-eclipse region is unity.

Using the timing solution of \citet{ksm+06} we calculated the orbital phase corresponding to each data point in the flux time series. Throughout this work, we shall refer to the orbital phases using the superior conjunction of pulsar A (i.e. when pulsar A passes behind pulsar B) as the reference point. For eclipse modeling purposes, we also determined the spin phases of pulsar B associated to each data point of pulsar A's pulsed flux time series. We empirically determined the spin phases by calculating the phase shift to apply to the predicted spin phases from the integrated pulse profile over each observation.

Over the four years that our data span, we found significant changes in pulsar B's pulse profile (see Figure~\ref{f:profile_evolution}), likely due to the precession of its spin axis, which were also reported in \citet{bpm+05}. Around 2003, the average pulse profile was unimodal, resembling a Gaussian function. It evolved such that by 2007, it displayed two narrow peaks. Using the pulse peak maximum as a fiducial reference point is certainly not appropriate. We find, however, that the unimodal profile gradually became wider and then started to form a gap near the center of its peak. Since then, the outer edges of the pulse profile have not significantly changed but the gap evolved such that two peaks are now visible. This lets us presume that the underlying average profile is reminiscent of a Gaussian-like profile to which some ``absorption" feature has been superimposed near the center, leaving a narrow peak on each side. We therefore defined the fiducial reference point to lie at the center of the unimodal ``envelope" that we reconstructed from the first ten Fourier bins of the pulse profile, which contains 512 bins in total (see Figure~\ref{f:profile_evolution} for an illustration of the pulse profile evolution).

\afterpage{
\clearpage

\begin{figure}
 \centering
 \includegraphics[height=5.25in]{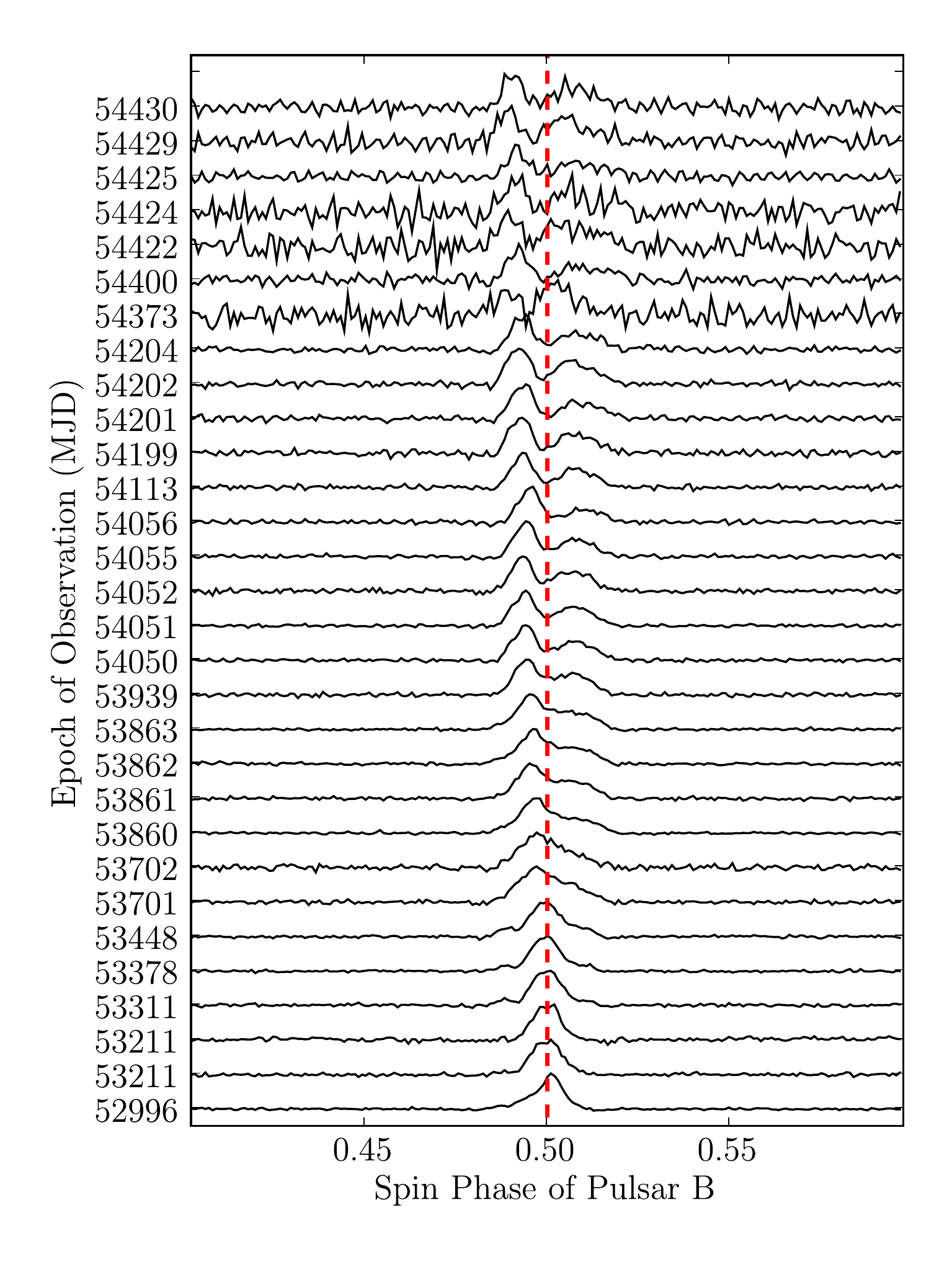}
 \caption[Pulsar B's pulse profile evolution]{Pulsar B's pulse profile consisting of the integrated flux at all orbital phases for each observation in our data set. Pulse profiles are normalized so the peak is unity and they are displayed with an incremental 1-unit vertical shift for clarity (the vertical axis does not show a linear time sequence). The vertical red dashed line marks the fiducial spin phase, which was determined by aligning the profiles using the first ten Fourier bins of the original 512-bin profile assuming that the two narrow peaks visible in the more recent data are the ``edges" of an underlying unimodal envelope reminiscent of the profile in the earlier observations. Note that the observation length and radio interference contamination slightly varies from epoch to epoch but the overall signal-to-noise ratio decreases in the latest observations due to pulsar B becoming weaker. MJD 52996 corresponds to Dec. 23, 2003, and MJD 54430 to Nov. 26, 2007.}
 \label{f:profile_evolution}
\end{figure}

\clearpage
}

\section{Eclipse Phenomenology}
The presence of eclipses in the double pulsar is not unique to this system. A handful of other binary pulsars are periodically occulted near superior conjunction when they pass behind their companion (see also \S\,\ref{ss:orbital}). There are numerous origins to these eclipses. Sometimes, the stellar wind of a \emph{Be} companion star screens the pulsar for part of the orbit. In other cases the orbital separation can be such that mass transfer from the companion to the pulsar may turn off the pulsar emission mechanism temporarily. In every case, binary pulsar eclipses do not appear to be caused by the surface of their companions, as in the case of solar and lunar eclipses for example.

The double pulsar eclipses are peculiar in that this is the first occurrence of such phenomenon in a double neutron star system. Mass transfer from degenerate objects like neutron stars is excluded unless the system is ultra-compact (i.e. an orbital separation of about 30\,km would be required for the neutron star to fill its Roche lobe)\footnote{In a double neutron star system, the mass ratio, $q$, is the order of unity. For mass transfer to happen, the Roche-lobe must be the size of the neutron star. Using the approximation of \citet{egg83}, the Roche-lobe radius of body 1, expressed in units of orbital separation $a_R$, is given by $R^1_L \approx \frac{0.49}{0.6 + q^{2/3} \ln (1 + q^{-1/3})}$, with $q = \frac{m_2}{m_1}$. This implies $R^1_L = 0.38 a_R$ for $q=1$, hence $a_R \simeq 30$\,km for a neutron star radius of about 10\,km.} and despite the fact that some pulsars can produce strong winds able to power pulsar wind nebulae, the energetics of this system is such that the spin-down energy released by pulsar A is about 3600 times larger than from pulsar B. Hence, if the interaction of pulsar A on pulsar B might be considerable, the converse is likely negligible. We therefore conclude that during the eclipses the radio emission from pulsar A has to be absorbed since pulsar B only plays a passive role. The short eclipse duration, $\sim 27$\,s, implies a projected cross-sectional eclipse region of about 18000\,km at the distance that separates pulsar B from pulsar A \citep{krb+04}. This physical scale of the eclipse fits well within the light-cylinder of pulsar B, $\sim 135000$\,km.

\subsection{Modulations}
The double pulsar eclipses present phenomenological aspects that clearly contrast with anything seen before in other astronomical eclipsing system. The eclipse light curve is strongly asymmetric \citep{krb+04} --- the egress occurs roughly 4 times faster than the ingress --- and very rapid flux variations were discovered in the high time resolution analysis of an 820\,MHz observation \citep{mll+04}. The most astounding property of this flux variability is that it shows a synchronicity with the rotational phase of pulsar B, which lies in front of pulsar A during the eclipse (see Figure~\ref{f:profile_with_phases} for a sample eclipse light curve). This kind of behavior has considerable implications for the physical mechanism creating the eclipses. It was initially proposed, before the modulations were found, that the eclipses could originate from the magnetosheath of pulsar B, which probably resembles that of the Earth \citep{abs+04,lyu04}. The relativistic shock of pulsar A's wind on pulsar B's magnetosphere share many similarities with the effect of the Solar wind on the Earth's magnetosphere. Such a model, however, was later excluded because it is unable to explain the rapid modulation of pulsar A's flux.

\afterpage{
\clearpage

\begin{figure}
 \centering
 \includegraphics[width=6in]{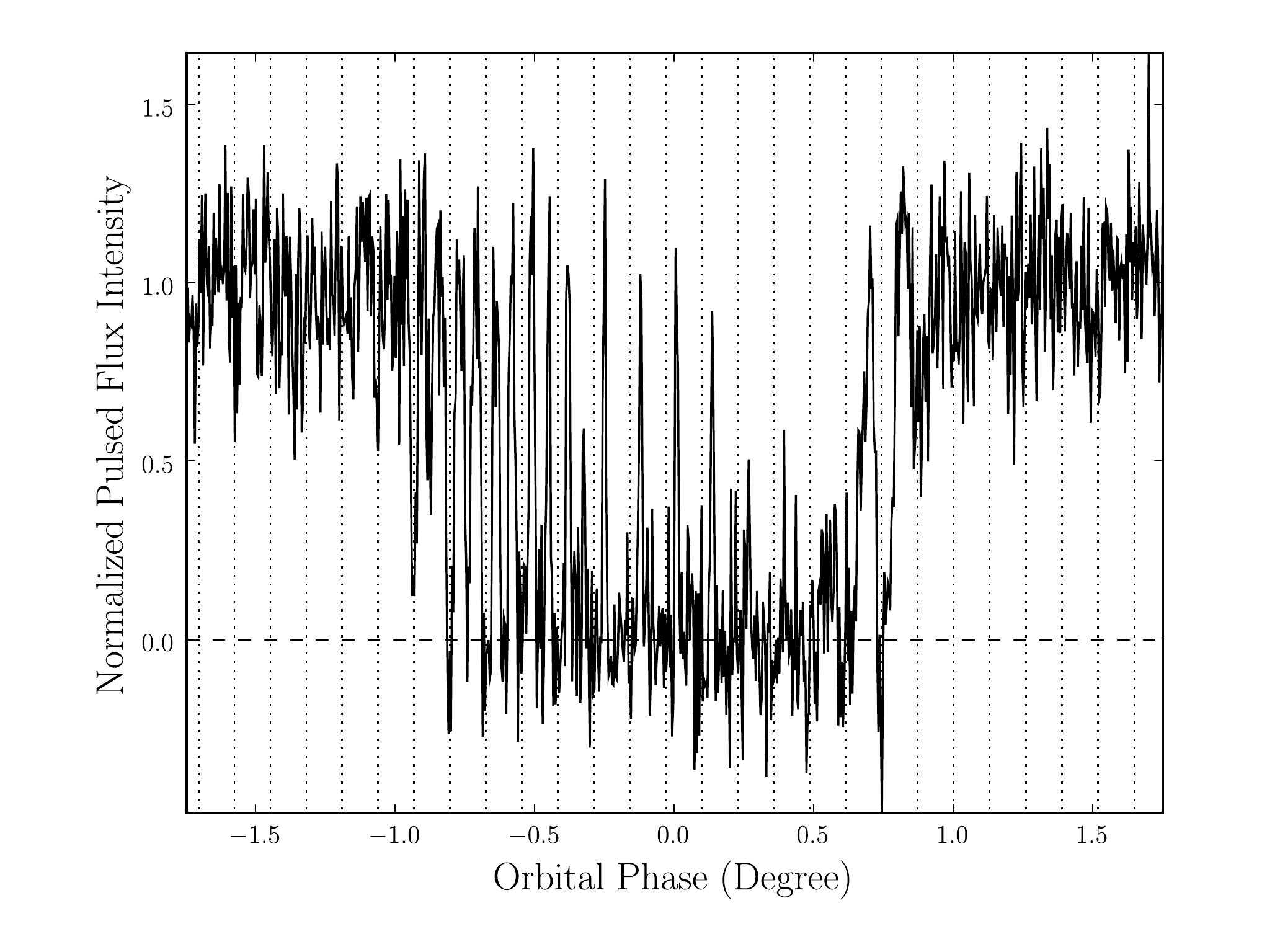}
 \caption[Eclipse profile of pulsar A with pulsar B's time of arrival]{Average eclipse profile of pulsar A consisting of eight eclipses observed at 820\,MHz over a five-day period around April 11, 2007. The relative pulsed flux density of pulsar A is normalized so the average level outside the eclipse region is unity. The resolution of each data point is $\sim$91\,ms while 1$^\circ$ in orbital phase corresponds to 24.5\,s. The times of arrival of pulsar B's radio pulsations are indicated with vertical dotted lines. As we can see the modulation features in the eclipse light curve are synchronized with pulsar B's rotational phase. In fact, before summing the individual time series to make the average light curve, we shifted them by up to $\pm P_B$ in order to make the nearest spin phases of pulsar B coincide. This technique developed by \citet{mll+04} avoids to wash out the modulation features.}
 \label{f:profile_with_phases}
\end{figure}

\clearpage
}

\subsubsection{Periodicity}
Even though the connection between the flux variability and the rotational phase of pulsar B is well established qualitatively \citep{mll+04}, no thorough quantitative analysis has been reported and the presence of such behavior at other radio frequencies, although logically expected, is yet to be confirmed. For various reasons, pulsar A is detected with a stronger signal-to-noise ratio at 820\,MHz than at other frequencies. Whereas the flux modulation is easily perceptible at this frequency, it is certainly more difficult to assess their presence at other frequencies.

To address these two questions, we performed a Fourier analysis of pulsar A's pulsed flux light curves. Their raw power spectra present a strong energy content at low frequencies because of the overall change in flux intensity during the eclipse (see Figure~\ref{f:eclipse_lightcurve}). For this reason, we subtracted a smoothed time series from the raw time series in order to obtain a high-pass filtered time series, which has an average flux level of zero. This enables us to conduct further analysis on the short term variability only (see Figure~\ref{f:eclipse_lightcurve}). From these filtered light curves, we found very significant excess power above the noise level at the exact spin frequency of pulsar B as well as at a number of harmonically related frequencies (see Figure~\ref{f:eclipse_powerspectrum}). This behavior is visible at all radio frequencies between 325 and 1950\,MHz, even when the modulations are not obviously distinguishable from the noise (see Figures~\ref{f:eclipse_325MHz}, \ref{f:eclipse_427MHz}, \ref{f:eclipse_1400MHz} and \ref{f:eclipse_1950MHz}). Furthermore, we observe that power in the first harmonic dominates the modulation content (i.e. twice the spin frequency of pulsar B). The only exception is the 2200\,MHz observation made with the {\tt BCPM} instrument, which does not show modulation (see Figure~\ref{f:eclipse_2200MHz}). The signal-to-noise ratio at this frequency is very poor because: 1) pulsar A becomes dimmer at increasing frequency, 2) RFI contamination and 3) smaller bandwidth of the {\tt BCPM} instrument compared to {\tt SPIGOT}. Since the 800\,MHz bandwidth of the {\tt SPIGOT} 1950\,MHz data overlaps with the {\tt BCPM} 2200\,MHz data, we conclude that the no-detection at this frequency is only an instrumental effect and that the flux modulation in the eclipse light curve is a frequency  independent phenomenon.

\afterpage{
\clearpage

\begin{figure}
 \centering
 \includegraphics[width=6in]{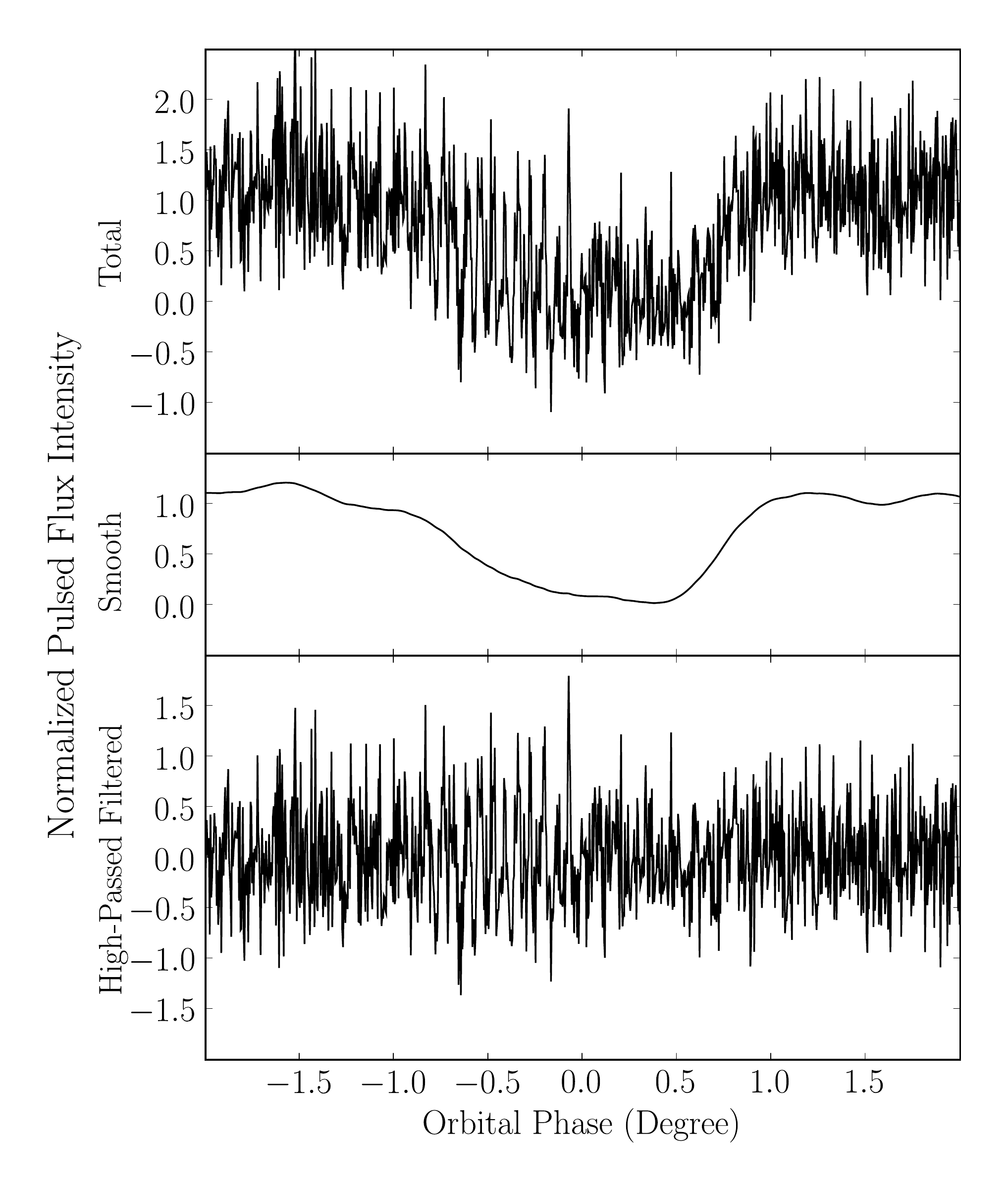}
 \caption[Eclipse profile of pulsar A observed on MJD\,52997]{Eclipse of pulsar A observed on MJD\,52996 at 820\,MHz with the {\tt SPIGOT} instrument at GBT. The upper, middle and lower panels display the total, the smoothed (i.e. total - high-pass filtered) and the high-passed filtered normalized pulsed flux intensities, respectively.}
 \label{f:eclipse_lightcurve}
\end{figure}

\clearpage

\begin{figure}
 \centering
 \includegraphics[width=6in]{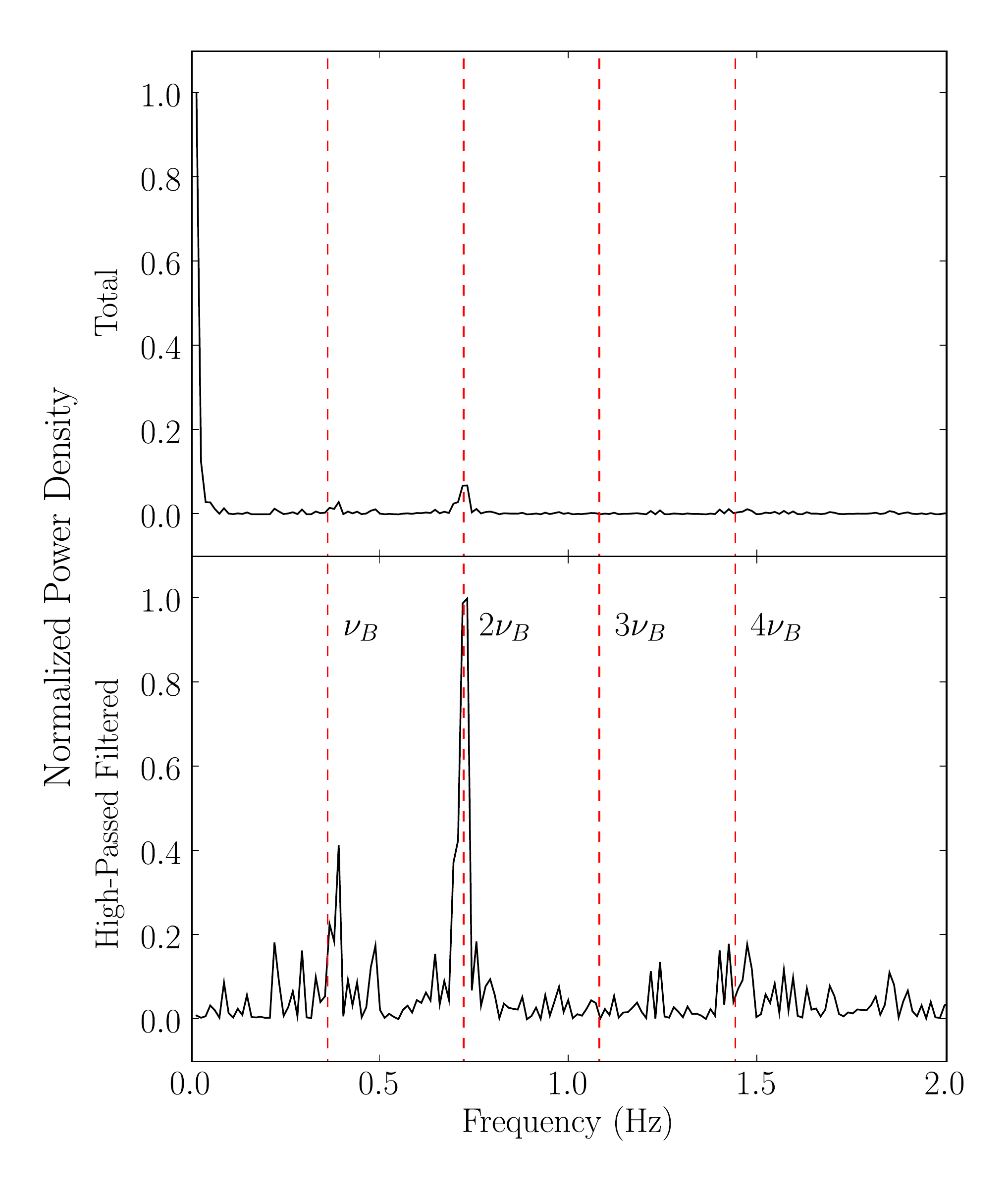}
 \caption[Power density spectrum of pulsar A's eclipse profile observed on MJD\,52997]{Power density spectrum of pulsar A's eclipse profile observed on MJD\,52997 at 820\,MHz with the {\tt SPIGOT} instrument at GBT (see Figure~\ref{f:eclipse_lightcurve} to see the light curve). The upper and the lower panels display the power density spectrum for the total and the high-pass filtered light curves, respectively.}
 \label{f:eclipse_powerspectrum}
\end{figure}

\clearpage

\begin{figure}
 \centering
 \includegraphics[width=6in]{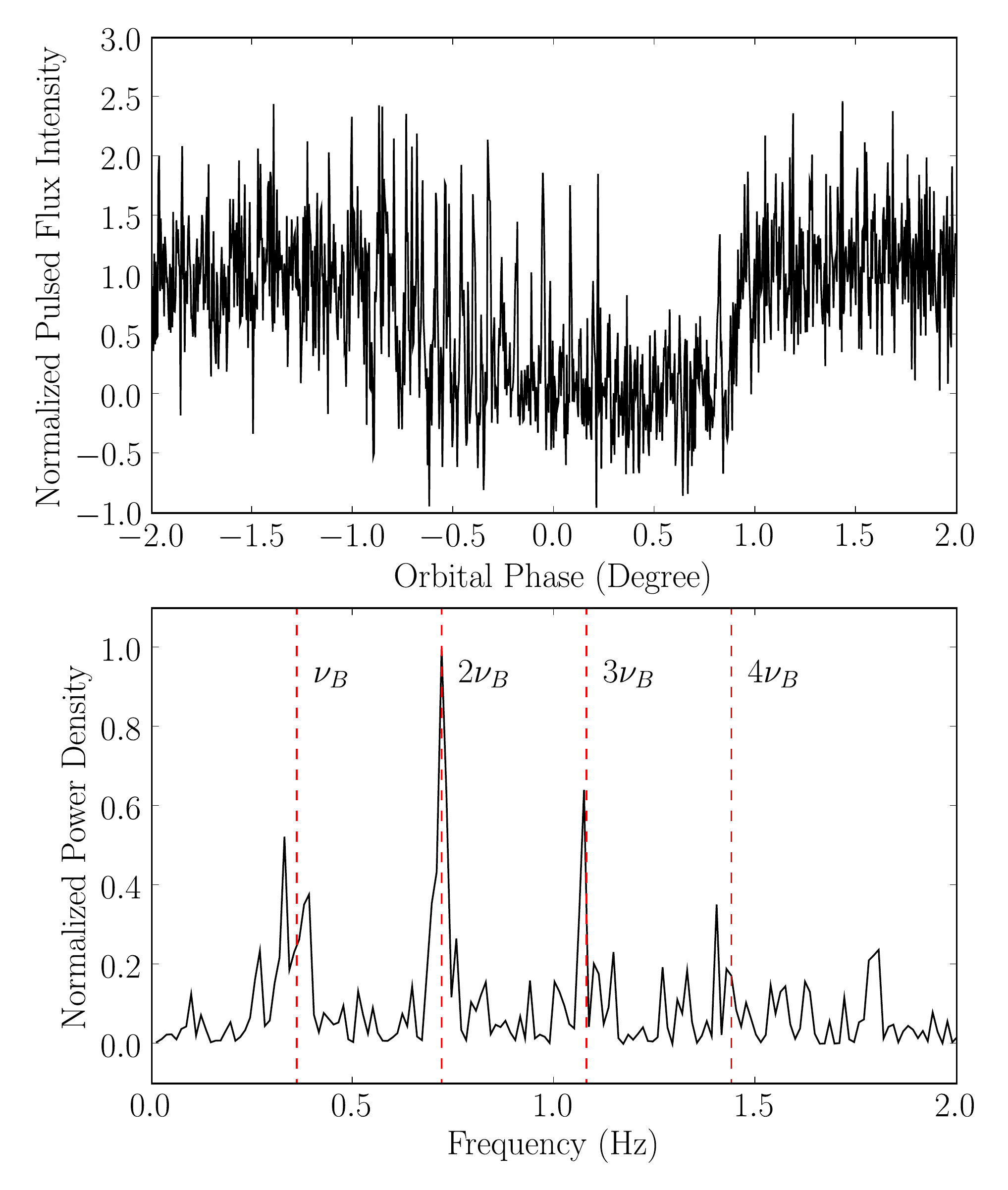}
 \caption[Eclipse light curve and power density spectrum of pulsar A at 325\,MHz]{Eclipse light curve of pulsar A observed on MJD\,53191 at 325\,MHz with the {\tt SPIGOT} instrument at GBT (upper panel) and corresponding power density spectrum for the high-pass filtered light curve (bottom panel).}
 \label{f:eclipse_325MHz}
\end{figure}

\clearpage

\begin{figure}
 \centering
 \includegraphics[width=6in]{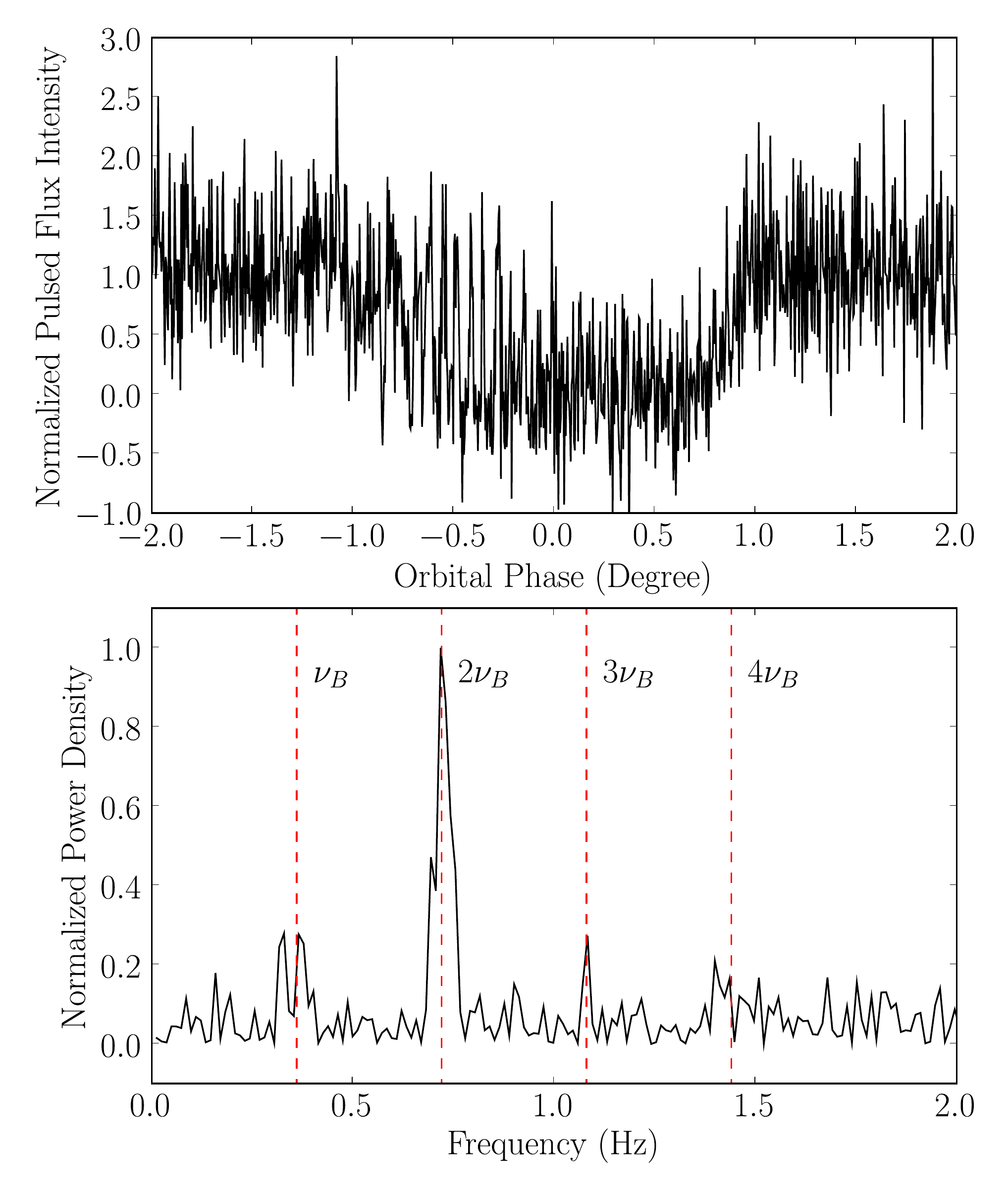}
 \caption[Eclipse light curve and power density spectrum of pulsar A at 427\,MHz]{Eclipse light curve of pulsar A observed on MJD\,53005 at 427\,MHz with the {\tt SPIGOT} instrument at GBT (upper panel) and corresponding power density spectrum for the high-pass filtered light curve (bottom panel).}
 \label{f:eclipse_427MHz}
\end{figure}

\clearpage

\begin{figure}
 \centering
 \includegraphics[width=6in]{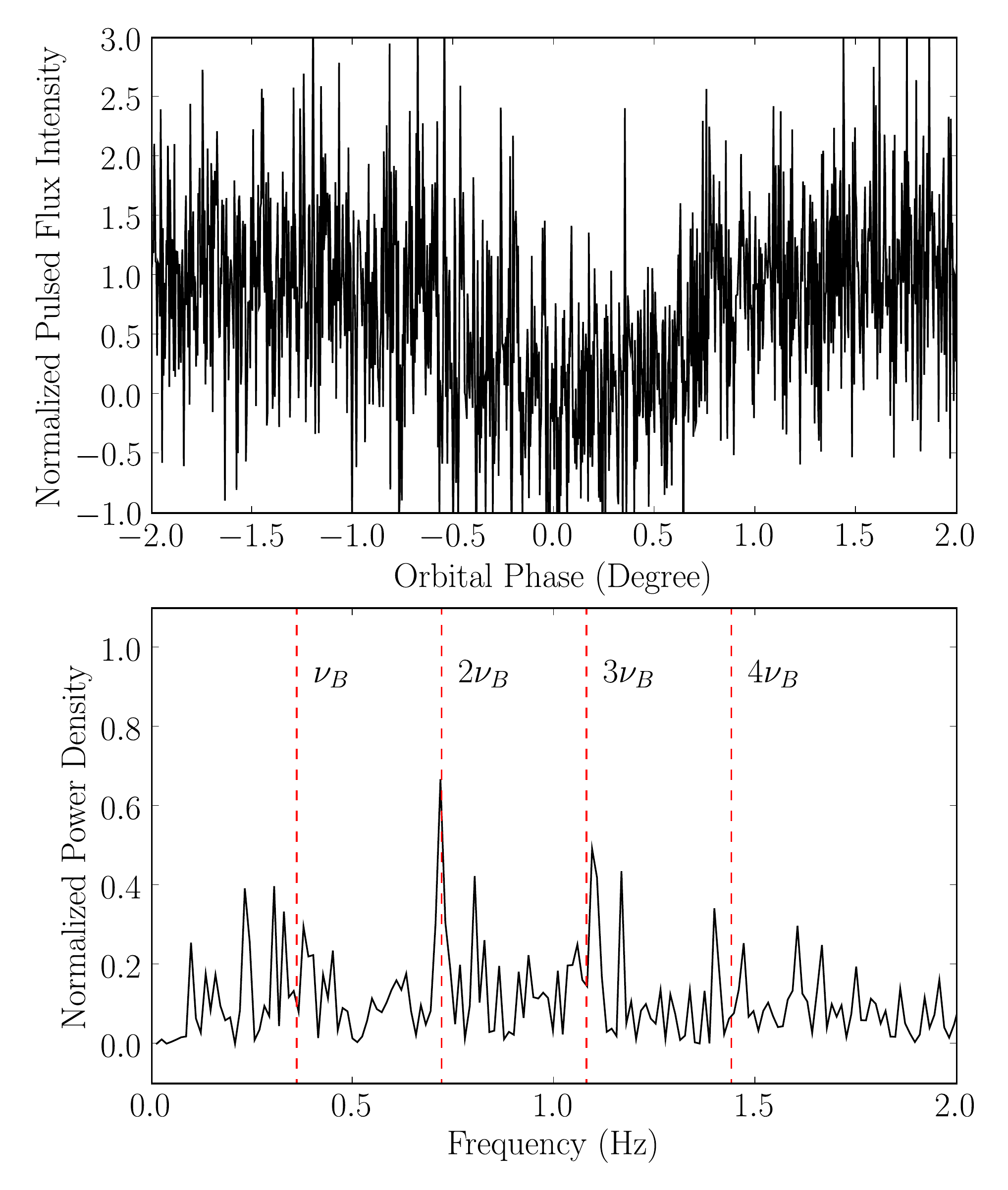}
 \caption[Eclipse light curve and power density spectrum of pulsar A at 1400\,MHz]{Eclipse light curve of pulsar A observed on MJD\,52984 at 1400\,MHz with the {\tt BCPM} instrument at GBT (upper panel) and corresponding power density spectrum for the high-pass filtered light curve (bottom panel).}
 \label{f:eclipse_1400MHz}
\end{figure}

\clearpage

\begin{figure}
 \centering
 \includegraphics[width=6in]{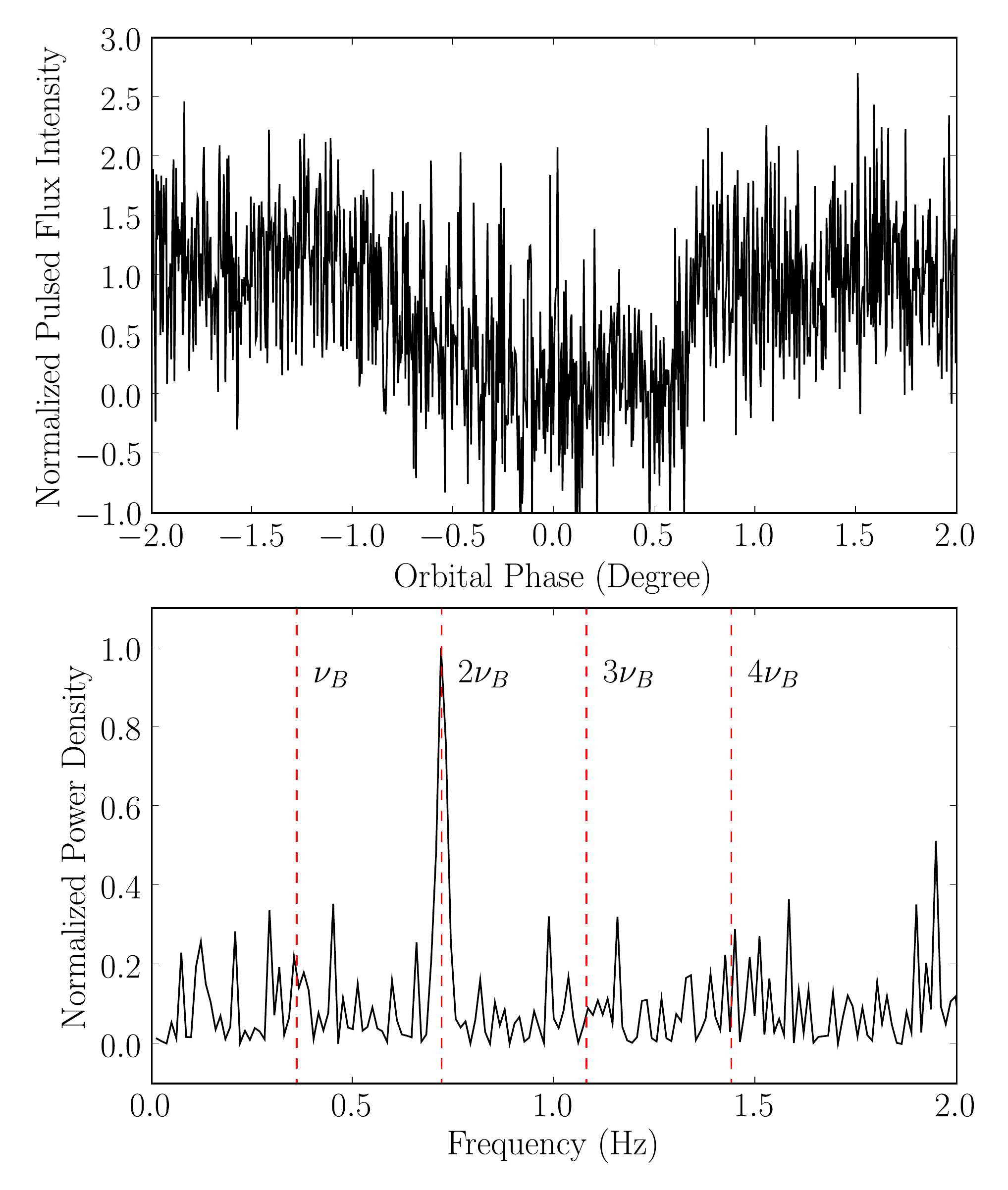}
 \caption[Eclipse light curve and power density spectrum of pulsar A at 1950\,MHz]{Eclipse light curve of pulsar A observed on MJD\,53378 at 1950\,MHz with the {\tt SPIGOT} instrument at GBT (upper panel) and corresponding power density spectrum for the high-pass filtered light curve (bottom panel).}
 \label{f:eclipse_1950MHz}
\end{figure}

\clearpage

\begin{figure}
 \centering
 \includegraphics[width=6in]{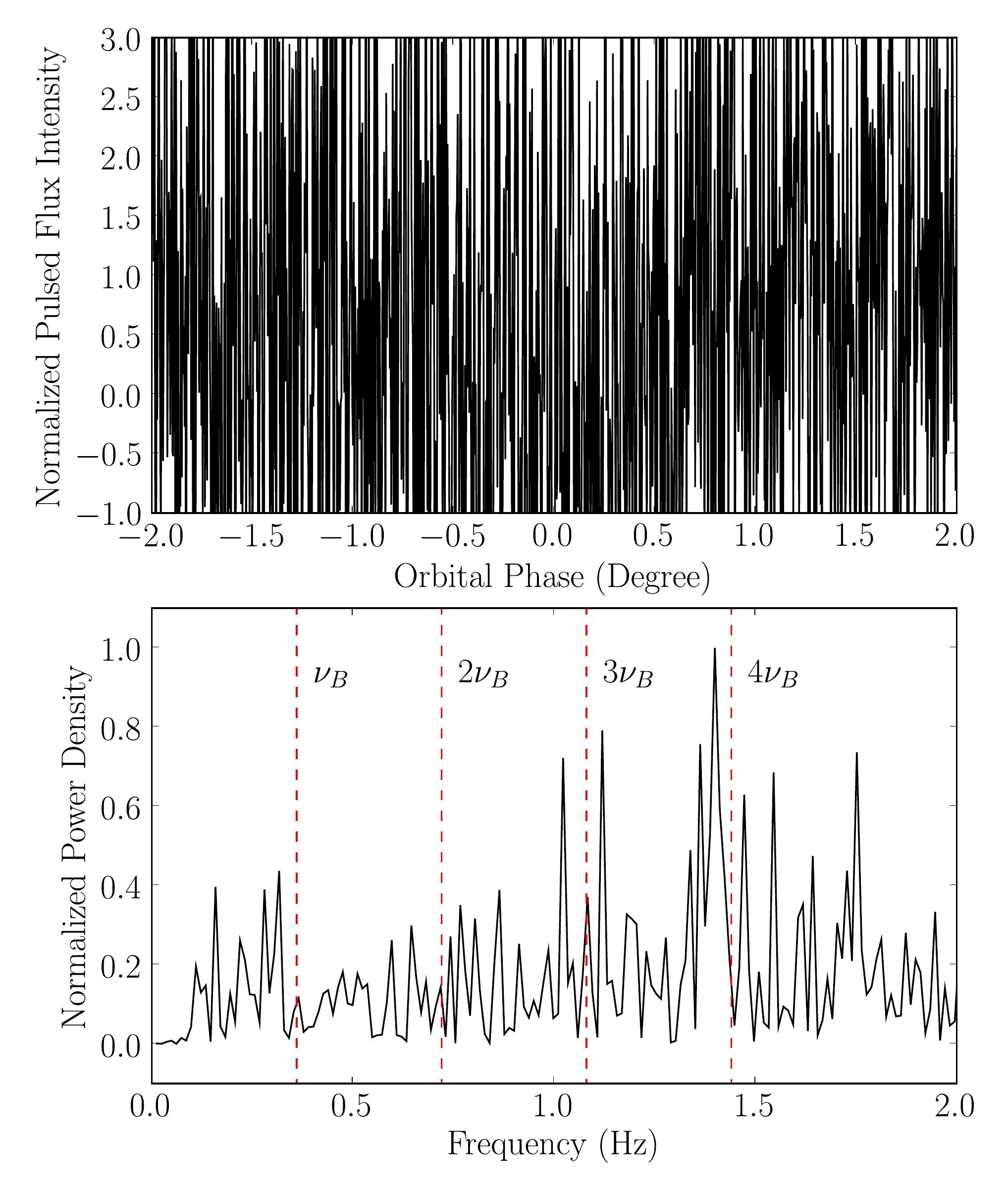}
 \caption[Eclipse light curve and power density spectrum of pulsar A at 2200\,MHz]{Eclipse light curve of pulsar A observed on MJD\,52996 at 2200\,MHz with the {\tt SPIGOT} instrument at GBT (upper panel) and corresponding power density spectrum for the high-pass filtered light curve (bottom panel). The signal-to-noise ratio is very poor because: 1) pulsar A becomes dimmer at increasing frequency, 2) RFI contamination and 3) smaller bandwidth of the {\tt BCPM} instrument compared to {\tt SPIGOT} and hence, despite that an eclipse trend is detectable, no modulations are seen.}
 \label{f:eclipse_2200MHz}
\end{figure}

\clearpage

}

More information about the modulation behavior can be obtained from a windowed  Fourier transform\footnote{The windowed Fourier transform is also referred to as dynamic Fourier transform.}. This technique consists in calculating the Fourier transform within a subsection of a time series. The Fourier transform is then recalculated after repeatedly translating the top-hat window kernel over the time series. This allows one to determine the dynamic dependence of the power content of a time series. A representative example of such a dynamic power spectrum obtained for an 820\,MHz observation is showed in Figure~\ref{f:dynamic}.

\afterpage{
\clearpage

\begin{figure}
 \centering
 \includegraphics[width=6in]{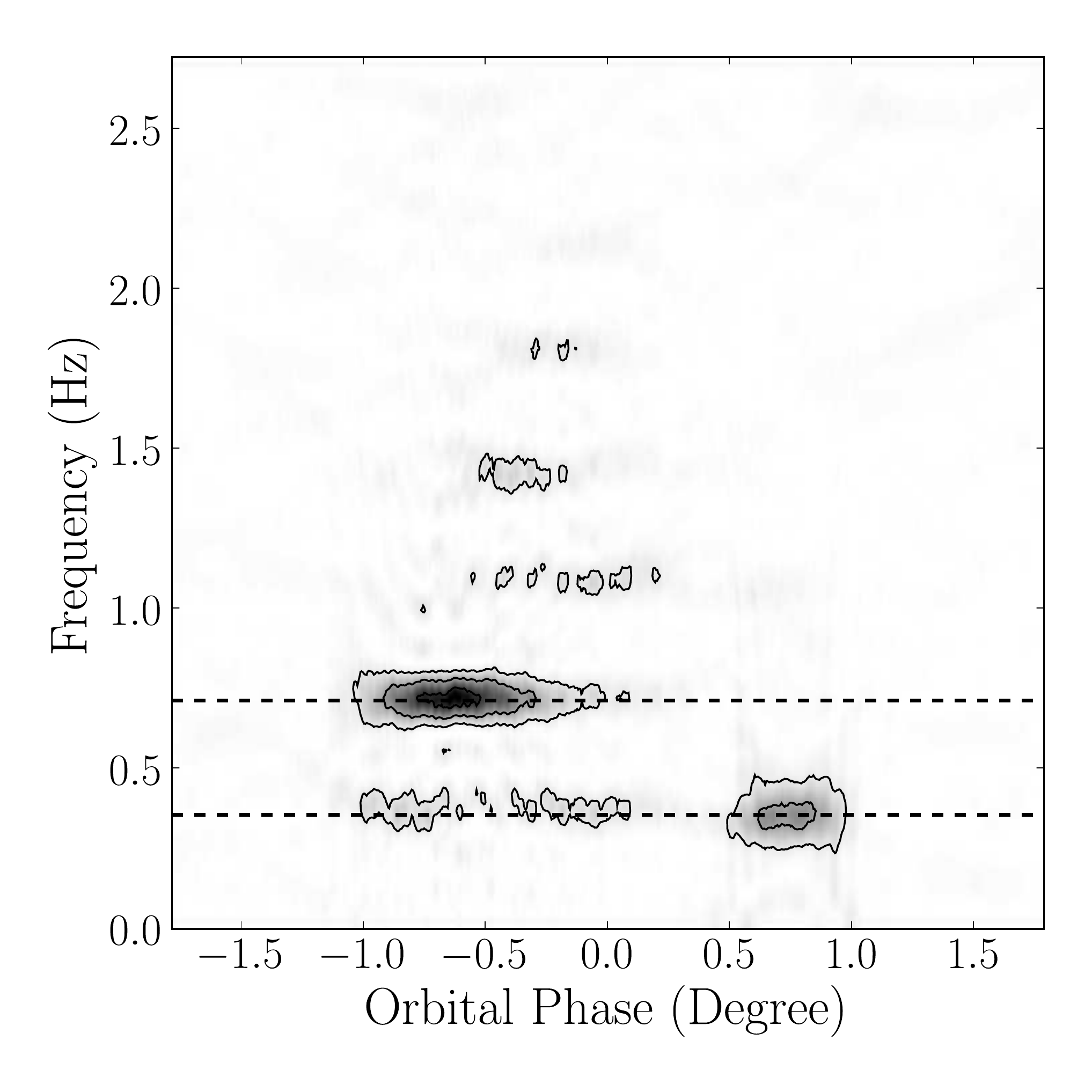}
 \caption[Dynamic power spectrum of the double pulsar eclipse]{Dynamic power spectrum of the combined eight eclipses observed at epoch MJD 54200. The figure shows the spectral energy density, vertically, as a function of orbital phase. Data were high-pass filtered in order to remove the low-frequency eclipse trend. The dashed lines indicate the fundamental and the first harmonic pulsar B's spin frequency at 0.36 and 0.72\,Hz, respectively. Contours identify constant power density levels.}
 \label{f:dynamic}
\end{figure}

\clearpage
}

Our dynamic power spectrum analysis yields two important results. First, outside the eclipse region the time series is well described by white noise and no periodic flux variations are found. Even though our pulsed flux time series are made from data folded at pulsar A's spin period and should consequently wash out the signature of pulsar B in the time series, the absence of modulation outside the eclipse region confirms that the modulation behavior is not an artifact of the data processing. The origin of the modulation is therefore directly connected to the physical process causing the eclipse. The second finding is that the power content significantly evolves over the course of the eclipse. From the December 2003 data presented in Figure~\ref{f:dynamic}, we observe that the eclipse ingress is characterized by modulation at twice the frequency of pulsar B while a `mode switching' appears to occur around orbital phase $-0.3$. The modulation then becomes a single peak per rotational period of pulsar B. Finally, at the eclipse egress, the modulation behavior briefly returns to the original double-peak mode. The two above conclusions are independent of the observed radio frequency.

\subsubsection{Folded Light Curve}
An interesting way of appreciating the connection between the pulsed flux modulation of pulsar A during the eclipse and the spin phase of pulsar B, as well as visualizing the mode switching behavior consists in folding the eclipse light curve at the spin period of pulsar B. With this method, we obtain a continuous eclipse profile of pulsar A for every spin phase of pulsar B. Figure~\ref{f:fold} illustrates such a folded eclipse profile made for the data presented in Figure~\ref{f:dynamic}. From this figure, we clearly see that the ingress starts very early and abruptly for regions corresponding to two specific spin phases of pulsar B. As the eclipse progresses, the opacity increases for all the spin phases of pulsar B except around phase 0.25 where the flux remains similar to the out-of-eclipse level for nearly the entire eclipse. This explains how the modulation switches from two bright regions per rotational phase of pulsar B to one region. Finally, the egress presents a fast rise at phase 0.25 as well as phase 0.75, thus causing the final mode switching.

\afterpage{
\clearpage

\begin{figure}
 \centering
 \includegraphics[width=6in]{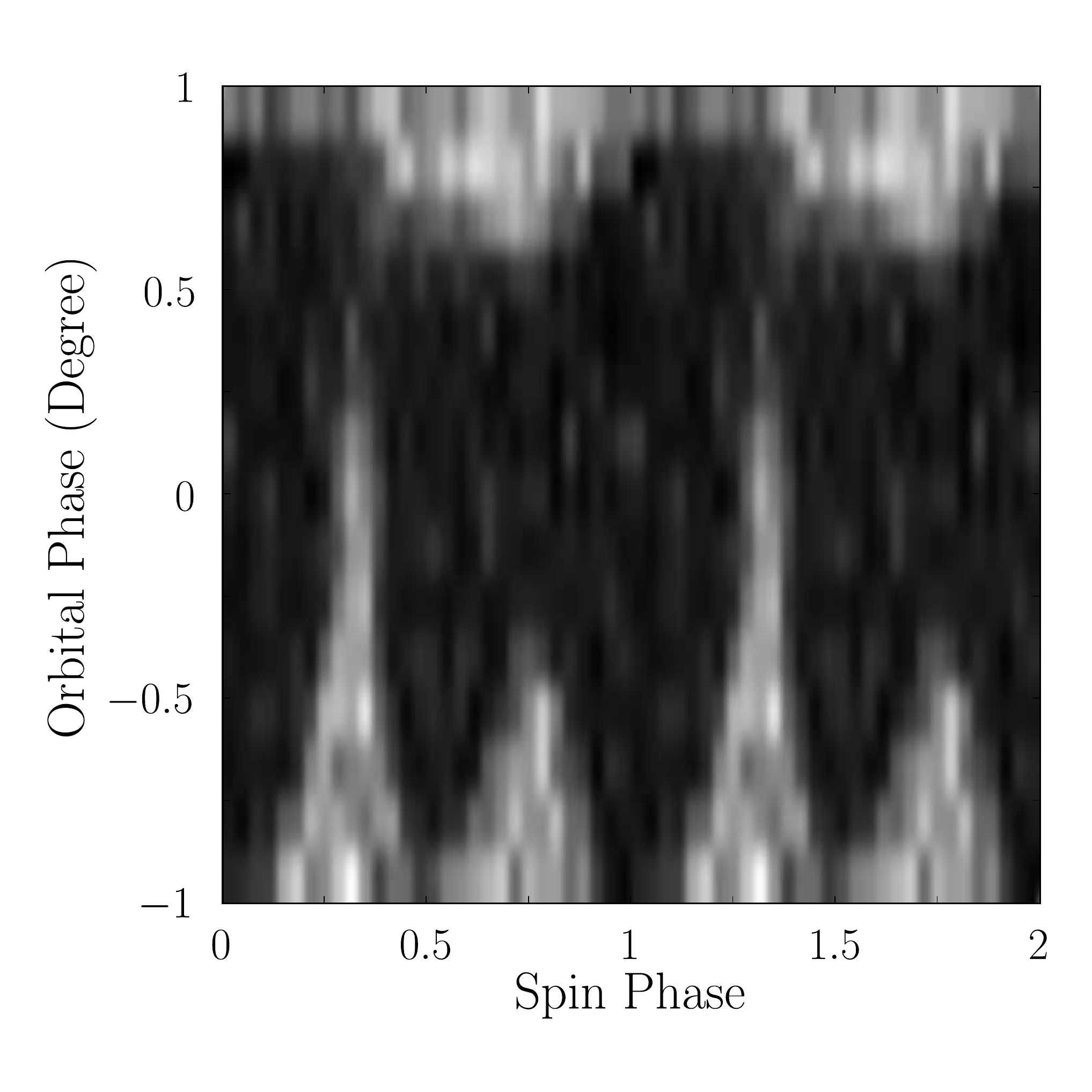}
 \caption[Folded eclipse profile]{Folded eclipse profile for the eight eclipses observed at epoch MJD 54200. Eclipses were combined before being folded at the spin period of pulsar B. Lighter regions have higher flux intensity levels.}
 \label{f:fold}
\end{figure}

\clearpage
}

We also observe that the enhanced pulsed flux features regularly return to the out-of-eclipse level. This question could not easily be addressed in the original eclipse modulation analysis because the time resolution was twice as long as the one we used here \citep{mll+04}. Several features are very narrow, and hence a coarser resolution has the effect of averaging the data points down to lower values. It appears, however, that some of the enhanced features do not reach the out-of-eclipse level.

\subsection{Duration}
\citet{krb+04} reported that the eclipse duration is nearly frequency independent in the range $427-1400$\,MHz. They separately fitted a Fermi function to each half of the eclipse and compared the inferred full width at half-max duration at different frequency. They found a small, non-zero linear decrease in the eclipse duration of $(-4.52 \pm 0.03) \times 10^{-7}$\,orbits\,MHz$^{-1}$. The low time resolution used in their analysis averages out the flux modulation, hence only providing information about the overall eclipse trend.

Because our data set includes observations made over a larger range of frequencies, we conducted a similar analysis in order to verify the behavior of the eclipse duration. We chose observations made around the same epoch in order to reduce effects of long-term evolution of the eclipse profile. Instead of fitting the ingress and the egress separately, we fitted the whole raw light curve to the following model:
\begin{equation}
 f(\phi) = \left( 1-f_{\rm min} \right) \left(\frac{1}{e^{(\phi-\phi_i)/w_i}+1} + \frac{1}{e^{(\phi_e - \phi)/w_e}+1}\right) + f_{\rm min} \,.
\end{equation}
This equation is simply a two-sided Fermi function having inflection points $\phi_i$ and $\phi_f$ at the ingress and egress, respectively, with corresponding sharpness factors $w_i$ and $w_e$. Because the flux may not drop completely to zero during the eclipse, we allow for an offset $f_{\rm min}$. We force the asymptotical upper value to be unity, however, since the eclipse light curves are normalized such that the out-of-eclipse level is unity. One can determine the eclipse duration, which is defined as the full width at half maximum (FWHM), by calculating $w_e-w_i$ directly. Results are presented in Table~\ref{t:duration}.

\afterpage{
\clearpage

\begin{landscape}
\begin{table}
\begin{tabular}{cccccccc}
\hline
Frequency & Number of Eclipses & $w_i$ & $w_e$ & $f_{\rm min}$ & $\phi_i$ & $\phi_e$ & FWHM \\
\small{(MHz)} & \,& \small{($10^{-4}$\,orbit)} & \small{($10^{-4}$\,orbit)} & \scriptsize{Normalized Pulsed Flux} & \small{($10^{-3}$\,orbit)} & \small{($10^{-3}$\,orbit)} & \small{($10^{-3}$\,orbit)} \\
\hline
325  & 2 & 6.7(9) & -1.5(4) & 0.09(3) & -2.1(1)  & 2.55(4) & 4.7(1)  \\
427  & 3 & 6.2(7) & -1.8(3) & 0.02(3) & -1.69(8) & 2.30(4) & 3.99(9) \\
820  & 3 & 5.6(6) & -1.7(3) & 0.02(3) & -1.65(7) & 2.06(3) & 3.70(7) \\
1400 & 3 & 4.6(8) & -2.8(6) & 0.02(5) & -1.57(8) & 1.96(7) & 3.5(1)  \\
1950 & 4 & 4.9(7) & -1.8(4) & 0.08(3) & -1.55(8) & 1.85(4) & 3.40(9) \\
2200 & 2 & 10(4)  & -2(2)   & -0.2(4) & -0.9(7)  & 1.5(2)  & 2.4(7)  \\
\hline
\end{tabular}
\caption[Fit results of the eclipse duration as a function of frequency.]{Fit results of the eclipse duration as a function of frequency. $w_i$, $w_e$, $f_{\rm min}$, $\phi_i$, $\phi_e$ and FWHM are the ingress and egress sharpness factors, the minimum pulsed flux level, the ingress and egress orbital phase at inflection point and the full width at half maximum, respectively. Numbers in brackets represent the 68\% confidence limit on the last significant digit and assumes fit reduced $\chi^2$ equal to unity. The reported values correspond to the joint constraint at each frequency, which is obtained as $\left< \phi_i \right> = \left( \sum_j \phi_{i_j}/\sigma_{\phi_{i_j}}^2 \right) / \left( \sum_j 1/\sigma_{\phi_{i_j}}^2 \right)$ for a parameter at a given frequency and $\left< \sigma_{\phi_i} \right> = \left( \sum_j 1/\sigma_{\phi_{i_j}}^2 \right)^{-1/2}$ for its uncertainty, with the index $j$ denoting the values measured for individual eclipses at each frequency.}
\label{t:duration}
\end{table}
\end{landscape}

\clearpage
}

From this analysis, we do not find any significant correlation between $f_{\rm min}$ and the observed frequency nor between the ratio $w_i/w_e$ and the observed frequency. We observe, however, that the eclipse FWHM does decrease with increasing observed frequency (see Figure~\ref{f:duration}). We fitted the duration for a linear relationship, FWHM $= m\nu + b$ and found: $m = -5.8(7) \times 10^{-7}$\,orbit\,\,MHz$^{-1}$ and $b = 0.00436(8)$\,orbit with a reduced $\chi^2 = 3.37$ (15 degrees of freedom). Our fitted slope is in agreement, within $2\sigma$ uncertainties, with the result from \citet{krb+04} who found $-4.52(3) \times 10^{-7}$\,orbit\,\,MHz$^{-1}$. The fitted intercepts, however, are different (\citet{krb+04} obtained 0.003412(2)\,orbit). The difference in our best-fit intercept may result from the fact that our time series have a much smaller time resolution --- a factor $\sim$22 --- that those of \citet{krb+04}. Since short timescale modulations are not visible in time series with coarser resolution, this also explain the smaller fit uncertainties reported by \citet{krb+04}. Another noticeable difference comes from the fact that we fit the eclipse as a whole instead of considering the ingress and the egress separately.

We also fitted the duration for a power-law relationship, FWHM $= k\nu^\alpha$ and obtained: $k = 1.0(1)$ and $\alpha = -0.17(2)$ with a reduced $\chi^2 = 2.09$ (15 degrees of freedom). It therefore appears that the eclipse duration more closely follows a power-law relationship with a small power-law index than a linear relationship. A possible reason why it did not appear in the original analysis by \citet{krb+04} is the shorter frequency baseline, which did not show the steep increase in duration at 325\,MHz and slower decrease above 1400\,MHz. It would be interesting to follow the eclipse behavior at much higher frequency in order to determine if there exists a cutoff in the eclipse duration. Unfortunately, pulsar A, like most radio pulsars, becomes very faint with increasing frequency and hence it would make such measurement impossible.

\afterpage{
\clearpage

\begin{figure}
 \centering
 \includegraphics[width=6in]{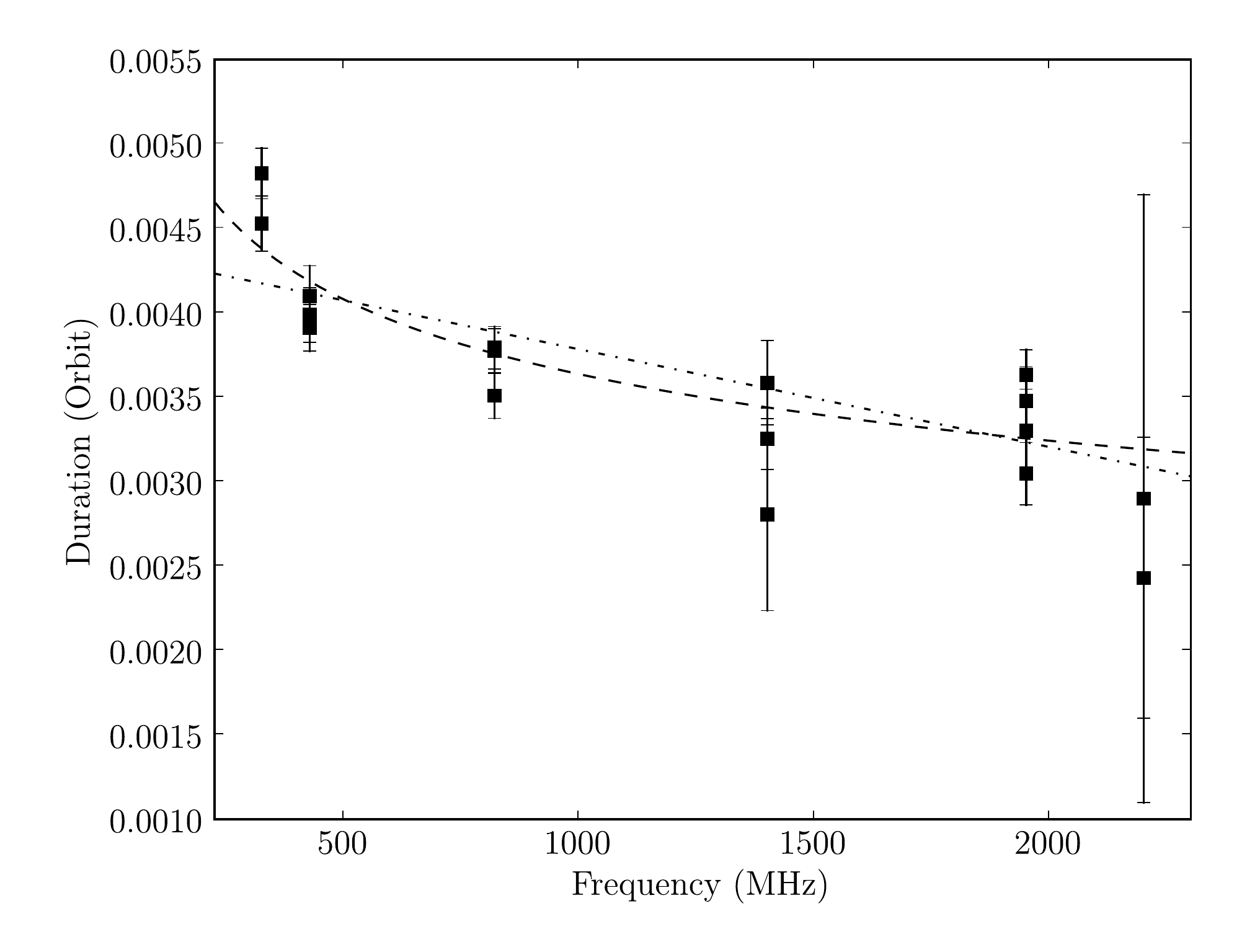}
 \caption[Eclipse duration as a function of observed radio frequency]{Eclipse duration (i.e. FWHM) as a function of observed radio frequency. A linear relationship (dot-dashed line) [FWHM $= -5.2(7) \times 10^{-7} \nu + 0.00435(8)$; reduced $\chi^2 = 4.49$ (12 d.o.f.)] and a power-law relationship (dashed line) are fitted to the duration [FWHM $= 0.98(12)\nu^{-0.16(2)}$; reduced $\chi^2 = 2.76$ (12 d.o.f.)].}
 \label{f:duration}
\end{figure}

\clearpage
}

\section{Eclipse Modeling}
\subsection{The Lyutikov and Thompson Model}
The presence of periodic flux modulation synchronized with the rotational phase of pulsar B is strong evidence that the eclipse material must co-rotate with the neutron star. Furthermore, an additional constraint on the eclipse mechanism comes from the fact that the physical process responsible for the radio absorption must be efficient over a wide range of frequencies. It was soon recognized that synchrotron absorption is the most likely physical process responsible for the radio absorption during the eclipses \citep{krb+04,mll+04,abs+04,lyu04}.

The model proposed by \citet{lt05} is particular in that it is tightly connected to geometrical aspects of the system. Hot relativistic electrons are confined within the closed field lines of pulsar B's magnetosphere, which is assumed to have a dipolar structure. Because magnetic bottling is efficient and the plasma cooling time is long, a large particle density can slowly build up to values that could reach $10^{5}$ times the Goldreich-Julian density according to \citet{lt05}. The total synchrotron optical depth along the line of sight to pulsar A varies as a function of the spin phase of pulsar B and the relative position of the two pulsars induced by the orbital motion. This naturally opens the possibility of generating flux modulation since the optical depth can vary rapidly on a time scale corresponding to the spin period of pulsar B.

Some extra ingredients must be added to the model in order to work. First, the plasma is expected to leak through the open field lines of pulsar B. Hence, we assume that the density profile of the plasma is constant within the magnetosphere up to some radius $R_{\rm mag}$ beyond which it drops abruptly, as a step-function, because not enough particles are retained to sustain a significant amount of synchrotron absorption. This `hard' boundary condition is supplemented by the extra condition that the local synchrotron frequency must be larger than the radio frequency of the emission from pulsar A, otherwise synchrotron absorption does not occur at this point of the space. On the other hand there is a radius, $R_{\rm min}$, from which particles can precipitate to the neutron star surface before cooling by cyclotron or synchrotron cooling. The size of this region is estimated to be a small fraction, $\sim 30$\% of $R_{\rm mag}$. Overall, the region of the magnetosphere opaque to the radio waves coming from pulsar A resembles a torus.

\subsection{Technical Definitions}\label{s:model_geometry}
\subsubsection{Geometry}
We conducted data modeling based on the work of \citet{lt05}. Here, we summarize critical technical aspects of this model and adapt it to our needs.

We define a cartesian coordinate system centered on pulsar B: the $x$-axis points in direction of Earth, along our line of sight, the $y$-axis is parallel to the projected orbital motion of pulsar A during the eclipse, and the $z$-axis is in the plane of the sky, coplanar with the orbital angular momentum vector (see Figure~\ref{f:schematic} for a schematic view of the geometry).

\afterpage{
\clearpage

\begin{figure}
 \centering
 \includegraphics[width=6in]{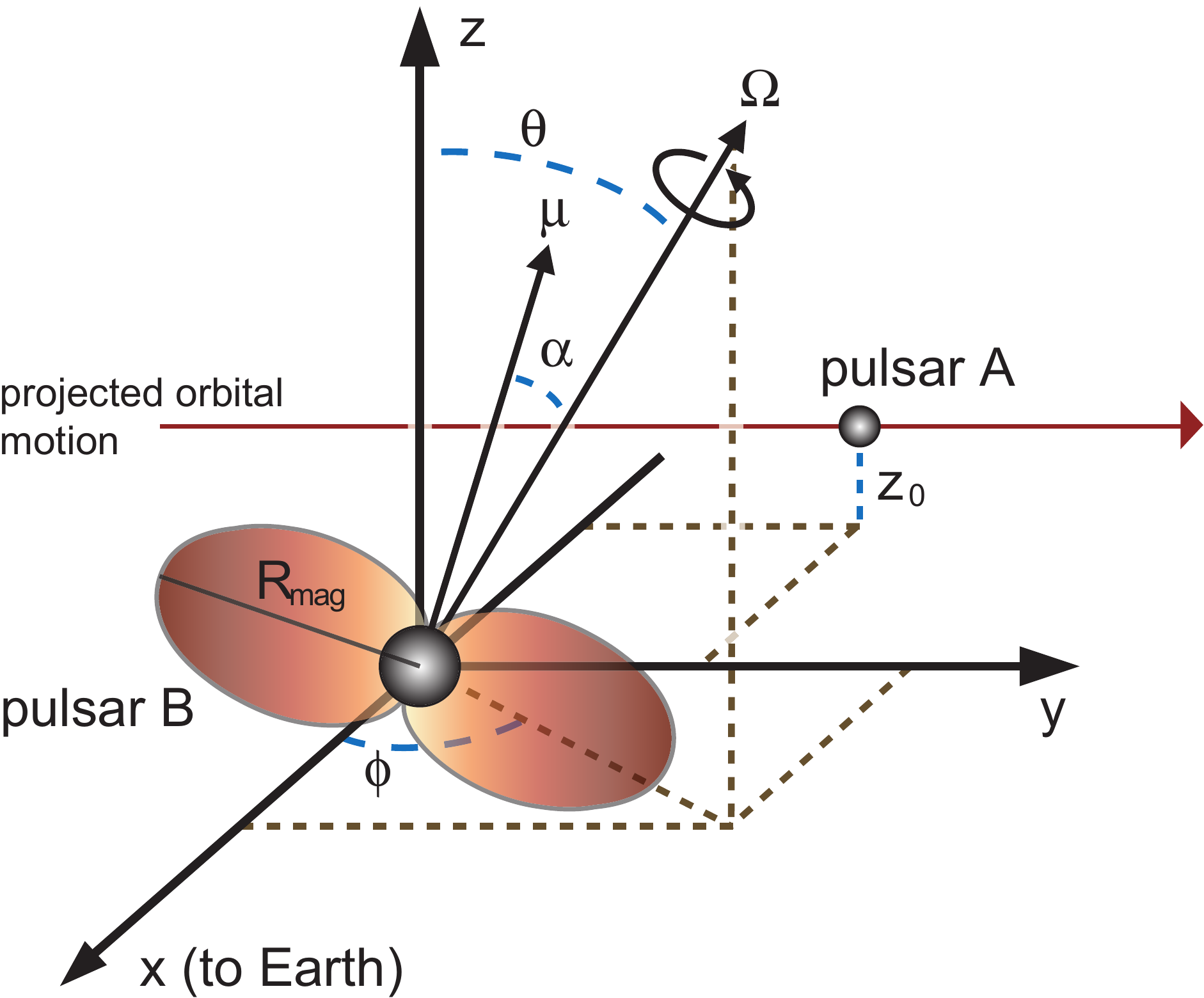}
 \caption[Schematic view of the double pulsar system at the eclipse]{Schematic view of the double pulsar system showing the important parameters for the modeling of pulsar A's eclipse (dimensions and angles are not to scale). Pulsar B is located at the origin of the cartesian coordinate system while the projected orbital motion of pulsar A during its eclipse is parallel to the $y$ axis at a constant $z_0$ as seen from Earth, which is located toward the positive $x$ axis. Note that since the orbital inclination is almost perfectly edge-on \citep{ksm+06}, we can approximate the $z$ axis to be coincident with the orbital angular momentum. The spin axis of pulsar B, whose spatial orientation is described by $\theta$ and $\phi$, is represented by the $\Omega$ vector. The magnetic axis of pulsar B corresponds to the $\mu$ vector and makes an angle $\alpha$ with respect to $\Omega$. Finally, the absorbing region of the dipolar magnetosphere of pulsar B, truncated at radius $R_{\rm mag}$, is shown as a shaded red region.}
 \label{f:schematic}
\end{figure}

\clearpage
}

We represent the orientation of pulsar B's spin axis in space using two angles: $\theta$ is the colatitude of the spin axis with respect to the $z$-axis and $\phi$ is the longitude of the spin axis with respect to the $x$-axis. We can relate this cartesian system with another one, also centered on pulsar B, but for which the $z^{\prime}$-axis is aligned with the orbital angular momentum of the system, while the $y^{\prime}$-axis is parallel to the $y$-axis and the $x^{\prime}$-axis is coplanar with the $x-z$ plane. In other words, the $x-y-z$ and the $x^{\prime}-y^{\prime}-z^{\prime}$ coordinate systems are related by a rotation about the $y$-axis or, equivalently, the $y^{\prime}$-axis. This rotation corresponds to $90^{\circ}-i$. If we denote $\delta$ and $\phi_{\rm so}$ the colatitude and the longitude of the spin axis in this coordinate system (following the notation introduced in \citet{dt92a}), then we can write:
\begin{eqnarray}
 \cos\theta &=& \cos(90^{\circ}-i) \cos\delta - \sin(90^{\circ}-i)
 \sin\delta \cos\phi_{\rm so} \, , \\
 \sin\phi &=& \frac{\sin\delta \sin\phi_{\rm so}}{\sin\theta} \,.
\end{eqnarray}

Because the orbital inclination of the system is very close to $90^\circ$ --- the timing yields $88\arcdeg 69 ^{+0\arcdeg 50}_{-0\arcdeg 76}$ \citep{ksm+06} --- we can make the following approximation:
\begin{eqnarray}
 \theta &\approx & \delta \, , \label{eqn:angle_a} \\
 \phi &\approx & \phi_{\rm so} \,. \label{eqn:angle_b}
\end{eqnarray}
This implies that the orientation of pulsar B's spin axis derived from the eclipse does not depend on the system's geometry derived from the timing.

We allow the dipole magnetic field of pulsar B to be misaligned with respect to its spin axis by an angle $\alpha$. Also, we define the truncation radius beyond which the synchrotron opacity becomes negligible as $R_{\rm mag}$.

Because the orbital plane is not exactly coincident with our line of sight vector, pulsar A describes an apparent motion during the eclipse along the $y$-axis at a fixed, non-zero $z$ position that we choose to be $z_0$.

\subsubsection{Dipolar Magnetosphere}\label{s:dipolar_magnetosphere}
For each point in the eclipse light curve, we calculate the synchrotron optical depth as follow:
\begin{equation}\label{eqn:tau}
 \tau = \frac{\mu}{\nu^{5/3}_{\rm GHz}} \int^{R_{\rm mag}}_{-R_{\rm mag}} \left( \frac{B \sin
 \kappa}{B_{\rm{mag}}} \right) d
 \left( \frac{x}{R_{\rm{mag}}} \right) \,.
\end{equation}
Here, $x$ is the position along the $x$-axis expressed in units of $R_{\rm mag}$ and $B$ is the local dipole magnetic field strength in unit of $B_{\rm mag}$, the magnetic field strength on the last field line that closes within $r=R_{\rm mag}$. The angle between our line of sight and the local dipole magnetic field direction is $\kappa$ and the observing frequency $\nu_{\rm GHz}$ in GHz. The scaling parameter $\mu$ accounts for the physical properties of the magnetosphere and is defined as \citep[see][Equation~52]{lt05}:
\begin{equation}
 \frac{4.5 \times 10^{-6} \lambda_{\rm mag}}{N^{1/4}_{\rm B}} \left(\frac{k_{\rm B} T_e}{10 m_e c^2}\right) \,.
\end{equation}

In the above equation, $\lambda_{\rm mag}$ is the electron multiplicity, $N_{\rm B}$ is a dimensionless parameter that rescales the magnetospheric radius size to account for the fact that the pulsar does not spin down in vacuum and is strongly perturbed by the relativistic wind of pulsar B, $k_{\rm B}$ is the Boltzmann constant, $T_e$ and $m_e$ are the electron temperature and mass, and $c$ the speed of light.

The eclipse intensity profile --- i.e. the transmitted flux from pulsar A --- corresponds to $e^{-\tau}$. We can readily conclude that $\mu$ modifies the overall eclipse ``depth'' but has no effect on the location of the modulation features, which are uniquely determined by the geometrical aspects of the model.

An additional free parameter, $\xi$, must be added in order to account for the mapping of the observed orbital phases to the coordinate system, which uses the magnetospheric size, $R_{\rm mag}$, as reference unit. This implies that $\xi$ scales the size of the magnetosphere relative to the orbit.

The dipole magnetic field strength is simply found using:
\begin{equation}
 B = \frac{\sqrt{1 + 3 \cos^2\theta_\mu}}{r^3} \mu_{\rm B} \,,
\end{equation}
where $r$ is the position in spherical coordinates, $\mu_{\rm B}$ is the dipole moment, and $\theta_\mu$ is the magnetic polar angle:
\begin{equation}
 \cos \theta_\mu = \frac{{\bf \hat{\mu} \cdot \vec{r}}}{r}.
\end{equation}

The components of the dipole unit vector are:
\begin{eqnarray} \label{eq:mu}
 \hat{\mu}_x &=& (\hat{\mu}_x^\Omega \cos \theta +
                 \hat{\mu}_z^\Omega \sin \theta) \cos \phi -
		 \hat{\mu}_y^\Omega \sin \phi \,, \\
 \hat{\mu}_y &=& (\hat{\mu}_x^\Omega \cos \theta + 
                 \hat{\mu}_z^\Omega \sin \theta) \sin \phi +
		 \hat{\mu}_y^\Omega \cos \phi \,, \\
 \hat{\mu}_z &=& \hat{\mu}_z^\Omega \cos \theta -
                 \hat{\mu}_x^\Omega \sin \theta \,,
\end{eqnarray}
with
\begin{eqnarray} \label{eq:mu_omega}
 \hat{\mu}_x^\Omega &=& \sin \alpha \cos (\frac{2\pi}{P_{\rm B}} t) \,, \\
 \hat{\mu}_y^\Omega &=& \sin \alpha \sin (\frac{2\pi}{P_{\rm B}} t) \,, \\
 \hat{\mu}_z^\Omega &=& \cos \alpha \,,
\end{eqnarray}
where $P_{\rm B}$ is the spin period of pulsar B.

\subsubsection{Rotational Phases of Pulsar B}
We should mention that this spin phase definition, originally introduced by \citet{lt05}, possesses the important caveat that it does not strictly correspond to the observed spin phase. Hence, the emission cone of pulsar B is not necessarily oriented toward Earth when $t = 0, P_{\rm B}, 2P_{\rm B}, ...$ for any value of $\theta$ and $\phi$. Let us consider the case of radio emission emitted parallel to the magnetic moment, which is orthogonal to the spin axis (e.g. $\alpha = 90^\circ$). For $\theta = 0^\circ$ and $\phi = 0^\circ$, the magnetic pole of pulsar B is indeed pointing toward Earth when $t = 0, P_{\rm B}, 2P_{\rm B}, ...$. If, however, $\theta = 0^\circ$ and $\phi = 90^\circ$, the magnetic pole is perpendicular to Earth when $t = P_{\rm B}/2, 3P_{\rm B}/2, ...$. Relating the observed spin phases to the model spin phases is essential not only to determine the correct model parameters from eclipse fitting, but also to ensure that a time variation of the geometric orientation of pulsar B would not be partly absorbed in a redefinition of the effective spin phase. While we do not make any assumption about the emission geometry, we suppose that radio pulses from pulsar B are observed when its magnetic pole is maximally oriented toward Earth. That is $t = t_0: {\it max} [{\bf \hat{\mu} \cdot \hat{x}}] \equiv \left(\frac{\partial \mu_x}{\partial t}\right)_{t_0} = 0$.

\subsection{Relativistic Spin Precession}\label{s:0737_precession}
Relativistic spin precession is expected to cause the spin angular momentum vector of pulsar B to precess around the total angular momentum of the system. Since this effect changes the geometry of the spin axis orientation, it should result in observable changes in the eclipse light curve that can be quantified using the modeling.

We argued in \S\,\ref{s:model_geometry} that the orbital angular momentum of the system and the $z$-axis are closely aligned and can be approximated to be the same (see Equations~\ref{eqn:angle_a} and \ref{eqn:angle_b}). In this case, the time evolution of the pulsar spin axis can be written as:
\begin{eqnarray}
 \label{eqn:model1}
 \theta &=& \theta_0 \, , \\
 \label{eqn:model2}
 \phi &=& \phi_0 - \Omega_{\rm B} t \, ,
\end{eqnarray}
where $\theta$ and $\phi$ are the angles defined in \S\,\ref{s:model_geometry}, $\Omega_{\rm B}$ is the precession rate, and $\phi_0$ is the longitude of the spin axis at the reference epoch.

\section{Eclipse Model Fitting}
The eclipse model comprises 7 free parameters: $\theta$, $\phi$, $\alpha$, $R_{\bf min}$, $z_0$, $\mu$ and $\xi$. Because of the large number of model parameters, it is technically challenging to search the full parameter space for a best-fit solution. Several least-squares and related maximization methods exist but most of them are not suited for the needs of our work. We require a method that can handle nonlinear models, is efficient with a large number of dimensions, can find a global maximum and can provide information about the topology of the \emph{a posteriori} probability distribution of the model parameters as well as allowing to derive confidence intervals for the best-fit solution.

\subsection{MCMC Analysis}
We identified Markov Chain Monte Carlo (MCMC) methods as very well suited to meet the needs that we enumerated above. MCMC methods are a class of algorithms designed to sample from a probability distribution that asymptotically converges toward a desired distribution by constructing a Markov Chain. As opposed to a `plain' Monte Carlo algorithm, which evenly samples all regions of the parameter space, an MCMC algorithm spends more time in the `interesting' regions of the parameter space --- to be more precise, the sampling rate is asymptotically proportional to the \emph{a posteriori} probability.

Many MCMC algorithms exist but one of the simplest to implement, and yet very efficient, is called the Metropolis-Hasting algorithm. The Metropolis-Hasting algorithm draws samples in the parameter space using a random walk and uses an adoption-rejection criteria to decide whether or not a new sample is added to the Markov Chain, hence ensuring convergence toward the targeted distribution.

The implementation of the Metropolis-Hasting algorithm is done as follows. Let $\vec{x}_i$ represents a vector of parameters in the parameter space at step $i$ of the Markov Chain and let $\vec{y}$ be a candidate parameter vector sampled from a proposal distribution $p(\vec{x}_i,\vec{y})$. Requirements for this proposal distribution are not very strigent \citep[see][for more details]{grs96} and a very common choice --- the one we used here --- is a random walk $\vec{y} = \vec{x}_i + \vec{\epsilon}$ where the random step $\vec{\epsilon}$ is chosen from a multivariate normal distribution ${\cal N}(0,\Sigma)$ having a 0 mean and a covariance matrix $\Sigma$. Let $\pi(\vec{x})$ be the target distribution. In our case, the target distribution is the \emph{a posteriori} density distribution of the model defined using the Bayes theorem as:
\begin{equation}
 \pi(\vec{x}) \equiv p(\vec{x} | D, I) = 
   \frac{p(I) p(D | \vec{x}, I)}{p(D)} \, ,
\end{equation}
where $I$ refers to the priors and $D$ to the data. In the present case, we simply assume flat priors and hence $p(I)$ can be ignored since it is absorbed in the normalizing constant $p(D)$, which ensures that the integrated probability of $\pi(\vec{x})$ is unity. The conditional probability of the data, $p(D | \vec{x}, I)$, is often referred to as the likelihood, ${\cal L}(\vec{x})$, which we define as
\begin{equation}
 {{\cal L}(\vec{x})} \equiv p(D | \vec{x}, I) = e^{-\frac{\chi^2}{2N}} \,.
\end{equation}
Here, $\chi^2$ is just the regular chi square, i.e. $\chi^2 = \sum_i^N \left(\frac{d_i - M_i(\vec{x})}{\sigma_i}\right)^2$.

The pseudo-code of the Metropolis-Hasting MCMC is:
\begin{enumerate}
 \item Initialize the Markov Chain with a random vector in the phase space,
$\vec{x}_0$.
 \item Choose a candidate vector, $\vec{y}$ from the proposal distribution
$p(\vec{x}_i,\vec{y}):$ \\ $\vec{y} = \vec{x}_i + {\cal N}(0,\Sigma)$.
 \item Choose a random number, $u$, in the interval $[0,1]$.
 \item Accept the candidate vector if it represents an improvement in the \emph{a posteriori} probability: $\pi(\vec{y}) > \pi(\vec{x}_i)$, i.e. ${\cal L}(\vec{y}) > {\cal L}(\vec{x}_i)$.

 Otherwise, accept it if:
  \begin{equation}
   u < \frac{p(\vec{x}_i,\vec{y}) \, \pi(\vec{y})}{p(\vec{y},\vec{x}_i)
       \, \pi(\vec{x}_i)} \,\,,\, {\rm i.e.} \quad u < \frac{{\cal L}(\vec{y})}{{\cal L}(\vec{x}_i)} \,.
  \end{equation}
  Set $\vec{x}_{i+1} = \vec{y}$ if the candidate vector is accepted and
$\vec{x}_{i+1} = \vec{x}_i$ otherwise. \emph{Our proposal distribution is symmetrical, $p(\vec{x}_i,\vec{y}) = p(\vec{y},\vec{x}_i)$, hence the simple form of the above equation.}
 \item Repeat steps 2 to 4 until convergence\footnote{One of the properties of Markov Chains is that they asymptotically converge toward a stationary distribution. In a more practical way, convergence means that the obtained distribution is `close' enough to the `real' underlying distribution. In this case, the calculated statistics such as the estimated mean, the moments and the standard distribution would yield comparable values to what one would obtain if these were calculated from the `real' distribution. Mathematically, there are a number of ways of assessing convergence \citep{grs96}. For example, one can compare the estimated mean in different segments of the Markov Chain or the variance of several independent chains having different starting points.} to the target distribution.
\end{enumerate}

We slightly modified the above algorithm in order to include a simulated annealing scheme that allows to boost the efficiency of our Metropolis-Hasting MCMC algorithm at exploring the parameter space. The concept of simulated annealing \citep{kgv83,cer85}, by analogy to the slow cooling of metals and crystal latices, employs a temperature parameter, $T$, that initially flattens the likelihood space when the temperature is high. This increases the acceptance rate of proposed moves, thus improving the mobility of the Markov Chain in the parameter space. Gradually, the temperature is lowered according to an annealing scheme and the algorithm converges toward the equilibrium temperature of 1. In a pure simulated annealing algorithm, the temperature is generally reduced below unity. One can show that, if $T = 0$ is reached in an infinite amount of time (or a time long enough compared to the ``thermodynamic'' time scale of the system), the algorithm will converge to the global maximum \citep{gg84}.

Step 4 of our above algorithm therefore becomes:
\begin{enumerate}
 \setcounter{enumi}{3}
 \item
 \begin{enumerate}
  \item The system temperature is adjusted using the following annealing scheme: $T_i = a^{i/b}T_0 + T_f$, where $a$ and $b$ are adjustment parameters\footnote{Typically, $a \lesssim 1$ ($a$ necessarily has to be less than 1 in order for the temperature to decrease) and $b > 1$. Tuning $a$ and $b$ allows one to determine how fast the temperature converges toward the final temperature.} for the rate for temperature change, and $T_0$ and $T_f$ are the initial and final temperature, respectively. 
  \item Accept the candidate vector if $\pi(\vec{y}) > \pi(\vec{x}_i)$. Otherwise, accept it if:
  \begin{equation}
   u < \left( \frac{p(\vec{x}_i,\vec{y}) \, \pi(\vec{y})}{p(\vec{y},\vec{x}_i)
       \, \pi(\vec{x}_i)} \right)^{T_i} \,.
  \end{equation}
  Set $\vec{x}_{i+1} = \vec{y}$ if the candidate vector is accepted and $\vec{x}_{i+1} = \vec{x}_i$ otherwise.
 \end{enumerate}
\end{enumerate}

\subsection{MCMC Results}\label{s:mcmc_results}
As we mentioned in \S\,\ref{s:dipolar_magnetosphere}, $\mu$ has no effect on the location of the modulation features and only modifies the deepness of the eclipse. Because the optical depth increases very sharply within the $R_{\rm mag}$ boundary, it is very difficult to probe the inner magnetosphere. Hence, it is impossible to obtain any reliable constraint on $R_{\rm min}$ --- any `reasonable' value between 0 and 0.5$R_{\rm mag}$ yields undistinguishable light curves considering the noise in the data. The large optical depth of the magnetosphere also implies that the constant electron distribution in the magnetosphere produces good results and, unfortunately, investigating the intricate details of the electron density profile will prove to be out of reach with the current sensitivity of the instruments. The parameters $z_0$ and $\xi$ can be constrained reasonably well. They mainly control the eclipse duration as well as the appearance of certain modulation features in the light curve. Finally, the parameters related to the geometry of pulsar B --- $\theta$, $\phi$ and $\alpha$ --- are those playing the most important role for the modeling since they govern the modulation behavior and the transition from the `double' modulation mode to the `single' modulation mode, and \emph{vice versa} (see Movie `Eclipses' at \url{http://www.physics.mcgill.ca/~bretonr/doublepulsar/} for an illustration of the eclipse modeling). They show covariance between themselves but little with the other parameters.

\afterpage{
\clearpage

\begin{figure}
 \centering
 \includegraphics[width=6in,angle=270]{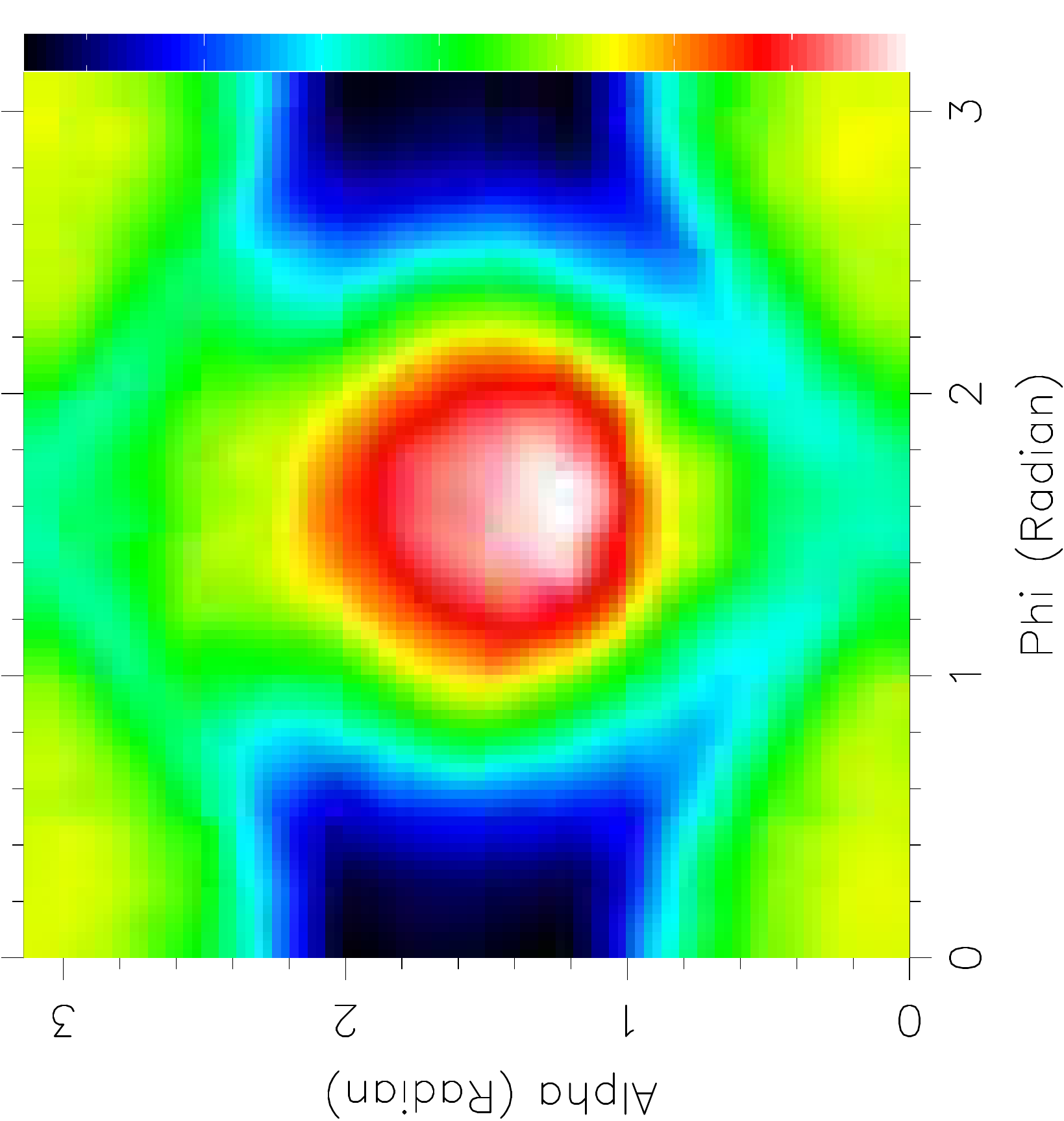}
 \caption[Joint \emph{a posteriori} probability of $\alpha$ and $\phi$]{Joint \emph{a posteriori} probability of $\alpha$ and $\phi$ for the three 820\,MHz observation made with the {\tt SPIGOT} instrument at GBT on MJD 52997 (December 24, 2003). Red regions are more likely than blue regions. The color scale is logarithmic in the probability. The 99.7\% confidence region ($3\sigma$) is very small and roughly corresponds to the delimitation between the light-red and red regions.}
 \label{f:chi_phi}
\end{figure}

\clearpage

\begin{figure}
 \centering
 \includegraphics[width=6in,angle=270]{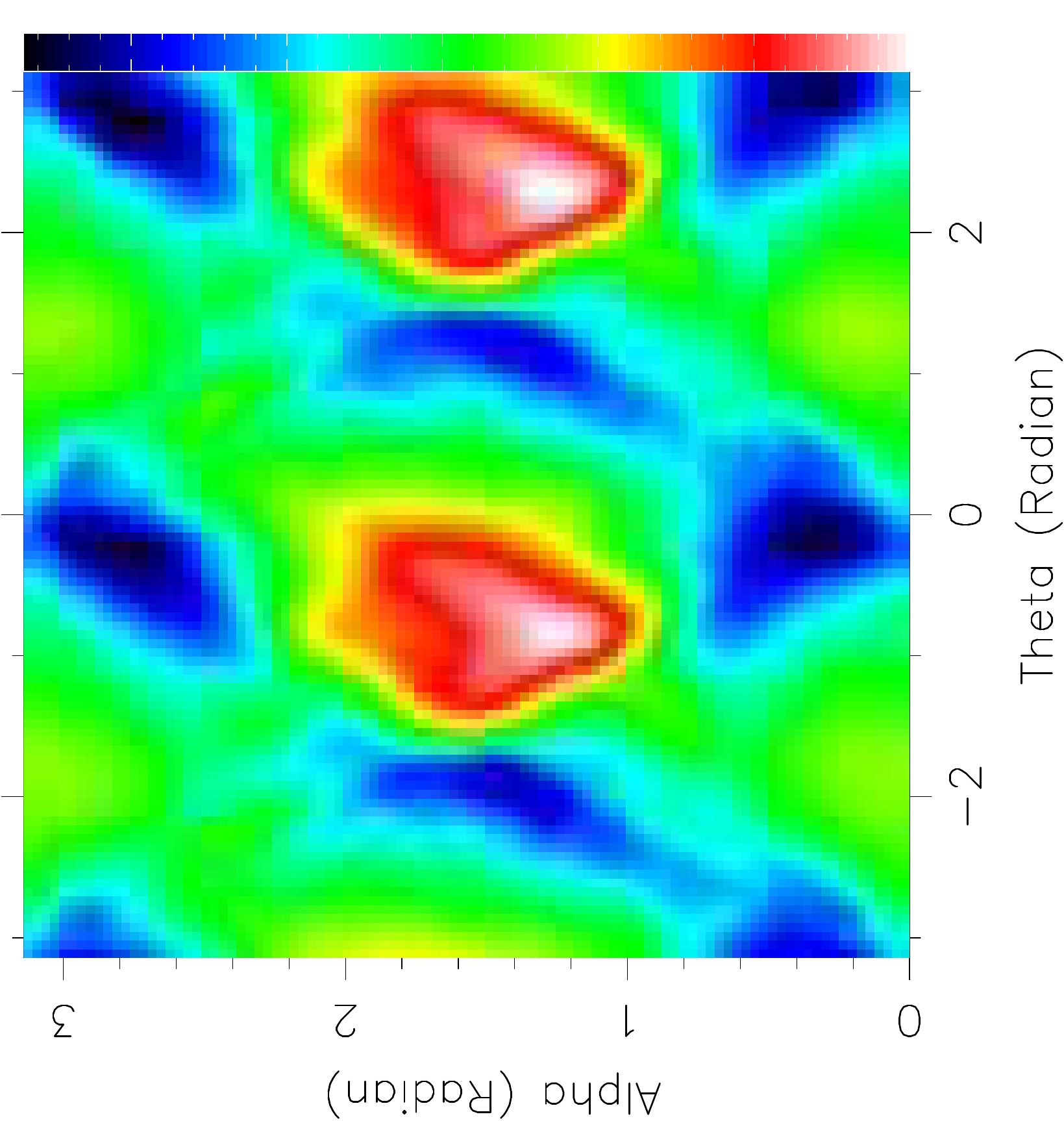}
 \caption[Joint \emph{a posteriori} probability of $\alpha$ and $\theta$]{Joint \emph{a posteriori} probability of $\alpha$ and $\theta$ for the three 820\,MHz observation made with the {\tt SPIGOT} instrument at GBT on MJD 52997 (December 24, 2003). Red regions are more likely than blue regions. The color scale is logarithmic in the probability. The 99.7\% confidence region ($3\sigma$) is very small and roughly corresponds to the delimitation between the light-red and red regions.}
 \label{f:chi_theta}
\end{figure}

\clearpage

\begin{figure}
 \centering
 \includegraphics[width=6in,angle=270]{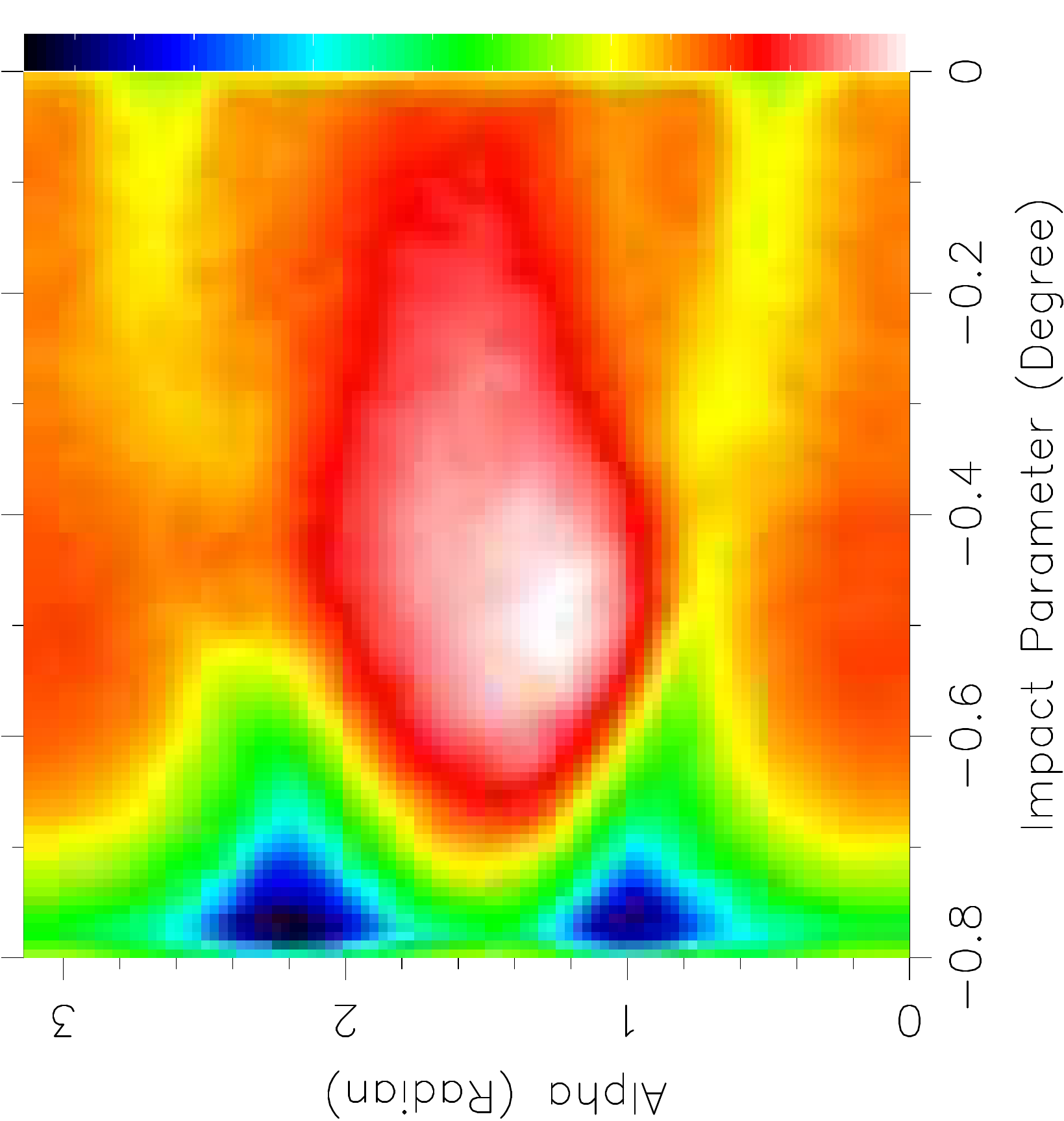}
 \caption[Joint \emph{a posteriori} probability of $\alpha$ and $z_0$]{Joint \emph{a posteriori} probability of $\alpha$ and $z_0$ for the three 820\,MHz observation made with the {\tt SPIGOT} instrument at GBT on MJD 52997 (December 24, 2003). Red regions are more likely than blue regions. The color scale is logarithmic in the probability. The 99.7\% confidence region ($3\sigma$) is very small and roughly corresponds to the delimitation between the light-red and red regions.}
 \label{f:chi_z0}
\end{figure}

\clearpage
}

\subsection{Grid Search}
We show in Figure~\ref{f:long_term_evolution} the 820\,MHz eclipse profile at different epochs over the last four years. Even though the signal-to-noise ratio varies depending on the number of eclipses used to make the plots, we observe that the modulation behavior changed over time. For instance, prominent ``absorption'' features gradually appeared in the egress where nothing was visible in the early days of our monitoring. The location of the transition of the modulation behavior --- from two to one flux enhancement per rotation of pulsar B --- has drifted toward the ingress. The \citet{lt05} model offers the possibility to quantify such kinds of long-term changes in eclipse profile and relate them to the geometry of pulsar B. This task requires the incorporation of the eclipse model in a framework accounting for the evolution of the parameters.

\afterpage{
\clearpage

\begin{figure}
 \centering
 \includegraphics[width=5in]{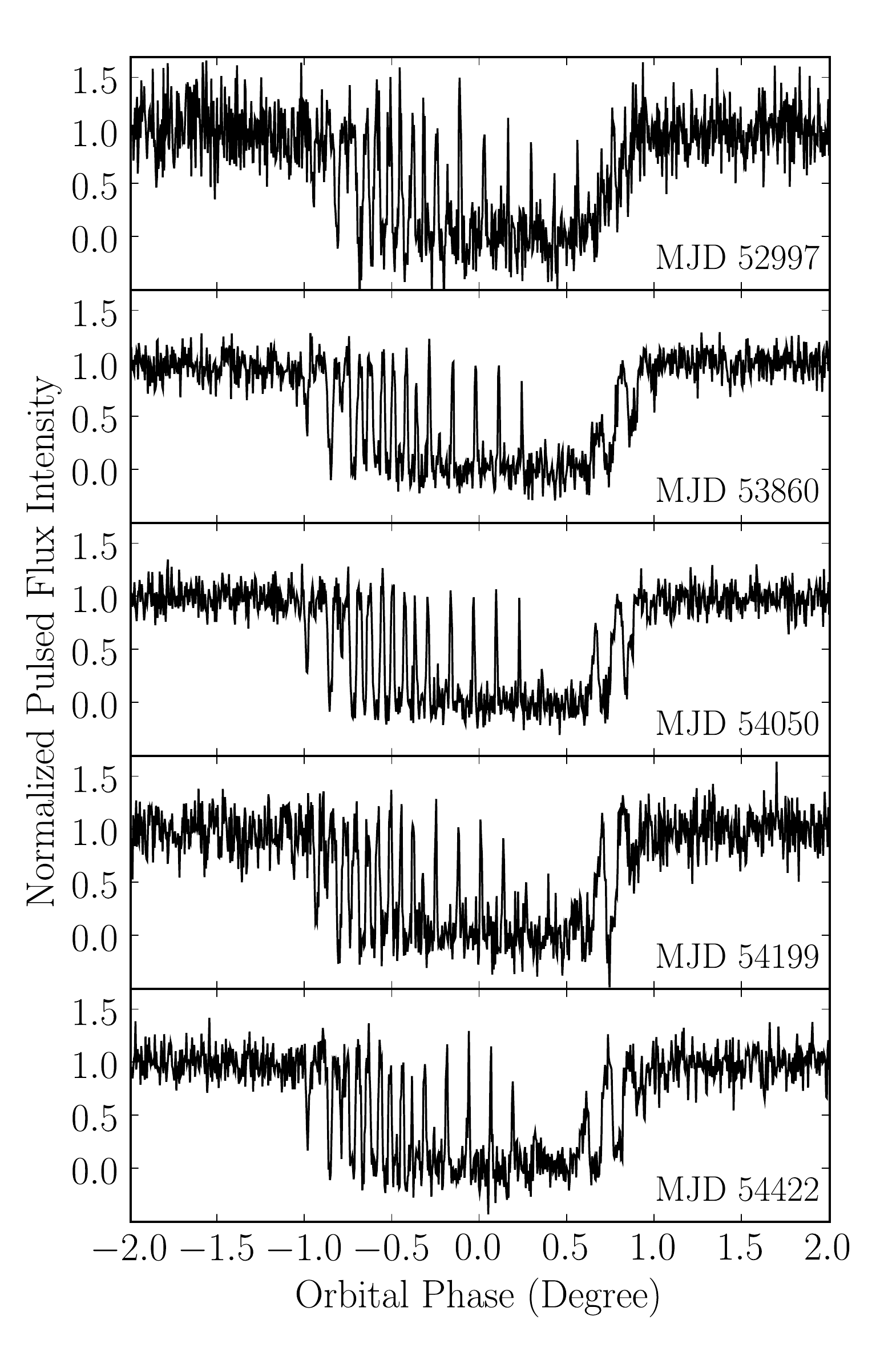}
 \caption[Long-term evolution of the eclipse profile at 820\,MHz]{Long-term evolution of the eclipse profile at 820\,MHz. The signal-to-noise varies from epoch to epoch because a different number of eclipse were combined --- 3, 8, 13, 8 and 10 for MJD 52997, 53860, 54050, 54199, 54422.}
 \label{f:long_term_evolution}
\end{figure}

\clearpage
}

Except for $\phi $, we do not find any significant secular evolution of the model parameters from their marginal \emph{a posteriori} probability\footnote{Marginalization is a common technique used in Bayesian analysis that consists of integrating the joint \emph{a posteriori} probability over the parameters that are not of interest so that it is reduced, in this case, to a one-dimensional probability for the parameter that we wish to consider \citep{gre05a}}. In principle, relativistic spin precession of pulsar B's spin axis around the total angular momentum should induce a secular change of the longitude of the spin axis, $\phi $ (see \S\,\ref{s:0737_precession}). Since the Markov Chain Monte Carlo technique poorly samples regions of low probability, we fixed $\mu = 2$, $\xi = 1.29^\circ \, R_{\rm mag}^{-1}$ (projected size of the magnetosphere in terms of orbital phase) and $z_0/R_{\rm mag} = -0.543$, their best-fit values, before making a deeper investigation of pulsar B's geometric evolution.

We performed a high-resolution mapping of the likelihood of this subspace in order to investigate subtle changes in the geometry and verify whether the pulsar could experience precession. Because of correlation between $\alpha$, $\theta$ and $\phi$, we jointly evaluated the best-fit geometry of pulsar B using a time-dependent model in which $\alpha=\alpha_0 $ and $\theta=\theta_0 $ are constants, and $\phi $ varies linearly with time, i.e. $\phi=\phi_0 - \Omega_{\rm B} t$, where $\Omega_{\rm B}$ is the rate of change of pulsar B's spin axis longitude and the epoch of $\phi = \phi_0$ is May 2, 2006 (MJD 53857).

\subsection{Grid Search Results}
Figure~\ref{f:eclipse_evolution} shows the time evolution of the parameters and the fit derived from this joint time-dependent model (Table~\ref{t:fit_result}, see also Figure~\ref{f:eclipse_pdf} for one and two-dimensional projections of the marginalized \emph{a posteriori} probability distributions). The precession rate $\Omega_{\rm B} $ of $4.77^{+0.66}_{-0.65}\,^{\circ}\rm{yr}^{-1}$ agrees\footnote{Unless otherwise stated, uncertainties are quoted at the 68\% confidence level.} with the precession rate predicted by GR \citep{bo75}, $5.0734 \pm 0.0007\,^{\circ}\rm{yr}^{-1}$, within an uncertainty of 13\% (68\% confidence level)\footnote{The uncertainty on the predicted GR spin precession rate arises because the value depends on the masses of the system, which are determined from two measured post-Keplerian parameters: the Shapiro delay $s$ parameter and the advance of periastron $\dot\omega$.}.

\afterpage{
\clearpage

\begin{figure}
 \centering
 \includegraphics[height=6in]{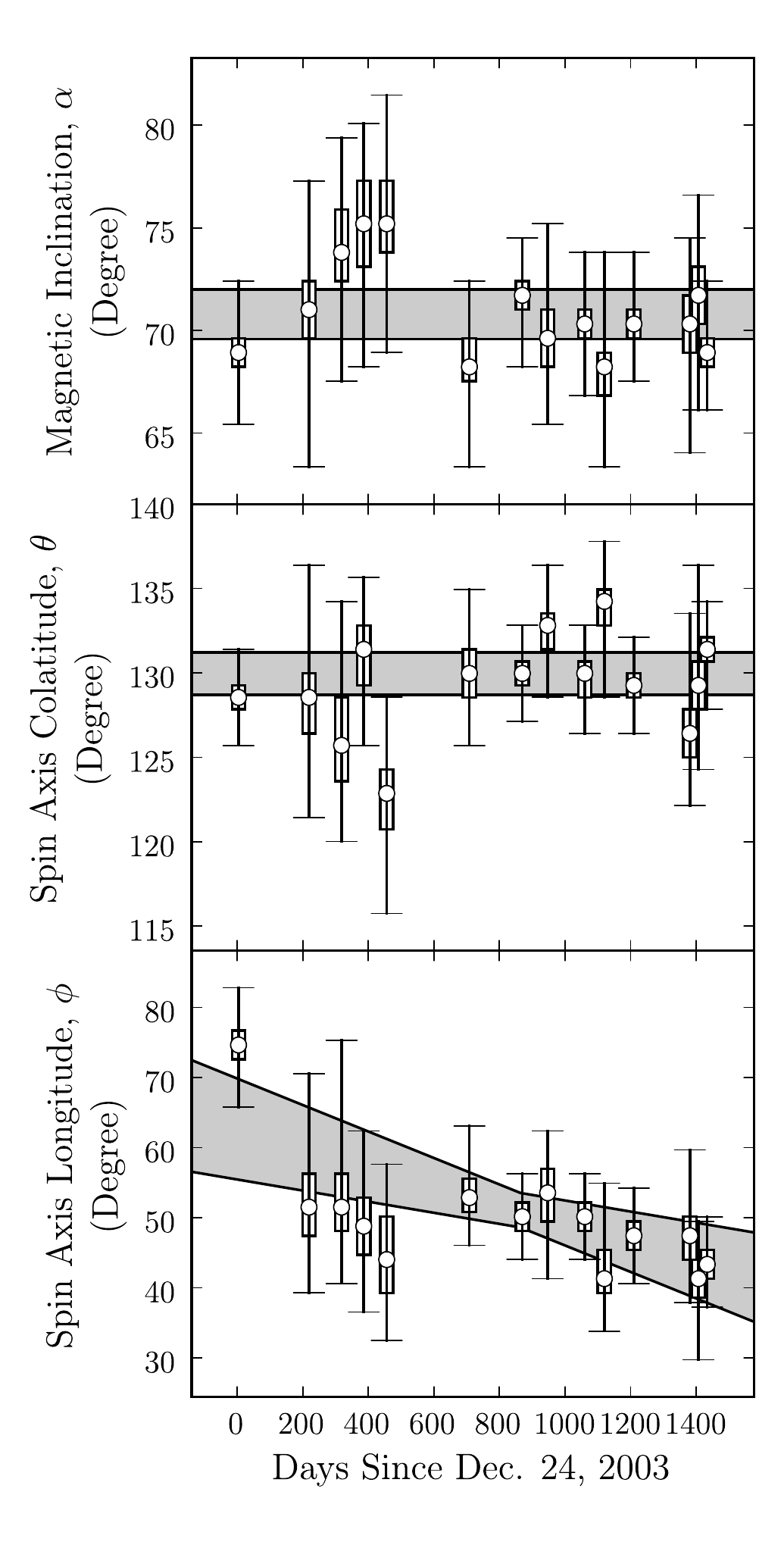}
 \caption[Evolution of pulsar B's geometry]{Evolution of pulsar B's geometry as a function of time. The marginalized \emph{a posteriori} probability distribution of the magnetic inclination ($\alpha$), the colatitude of the spin axis ($\theta$) and the longitude of the spin axis ($\phi$) of pulsar B are shown from top to bottom, respectively. For each data point, the circle represents the median value of the \emph{a posteriori} probability density while the box and the bar indicate the 1$\sigma$ and 3$\sigma$ confidence intervals, respectively. The gray regions are the 3$\sigma$ confidence regions derived from the joint time-dependent model fitting. Note that for clarity, multiple eclipses are displayed as single data points when observed over an interval of about a week.}
 \label{f:eclipse_evolution}
\end{figure}

\clearpage

\begin{table}
\begin{tabular}{lcccc}
\hline
Parameter & Mean & Median & $68.2\%$ Confidence & $99.7\%$ Confidence \\ 
\hline
$\alpha_0 $ & 70.92$^{\circ}$ & 70.94$^{\circ}$ & [70.49, 71.31]$^{\circ}$ & [69.68, 72.13]$^{\circ}$ \\
$\theta_0 $ & 130.02$^{\circ}$ & 130.02$^{\circ}$ & [129.58, 130.44]$^{\circ}$ & [128.79, 131.37]$^{\circ}$ \\
$\phi_0 $ & 51.21$^{\circ}$ & 51.20$^{\circ}$ & [50.39, 52.03]$^{\circ}$ & [48.80, 53.72]$^{\circ}$ \\
$\Omega_{\rm B} $ & 4.77$^{\circ}$yr$^{-1}$ & 4.76$^{\circ}$yr$^{-1}$ & [4.12, 5.43]$^{\circ}$yr$^{-1}$ & [2.89, 6.90]$^{\circ}$yr$^{-1}$ \\
\hline
\end{tabular}
\caption[Derived geometrical parameters of pulsar B]{Geometrical parameters of pulsar B derived from the eclipse model fitting. Note that the presented values include priors related to systematic uncertainties. The epoch of $\phi = \phi_0$ is May 2, 2006 (MJD 53857).}
\label{t:fit_result}
\end{table}

\clearpage
}

\subsection{Analysis of Systematics}
We investigated the importance of systematics in the eclipse modeling and concluded that two main effects should contribute to increasing the total uncertainty in our best-fit geometric parameters above the statistical value. First, we observe considerable changes in the pulse profile of pulsar B as shown in Figure~\ref{f:profile_evolution}. Since the spin phases of pulsar B are input data for the modeling, losing the fiducial reference to the neutron star surface will introduce additional error in the fitted eclipse parameters. While we do not require a measurement of the spin phases as accurate as for timing purposes, a few percent offset translates into slightly different geometrical parameters. The main effect of varying spin phases is to assign earlier or later rotational phases that mimic a slightly faster or slower precession rate. The pulse profile evolution of pulsar B is likely caused by the changing viewing geometry due to relativistic spin precession (see \S\,\ref{s:0737_precession}). Although it is not clear how pulsar B's pulse profile geometry is related to its surface, we are confident that the technique we used to determine the spin phases yields reliable results.

\afterpage{
\clearpage

\begin{figure}
 \centering
 \includegraphics[width=6in]{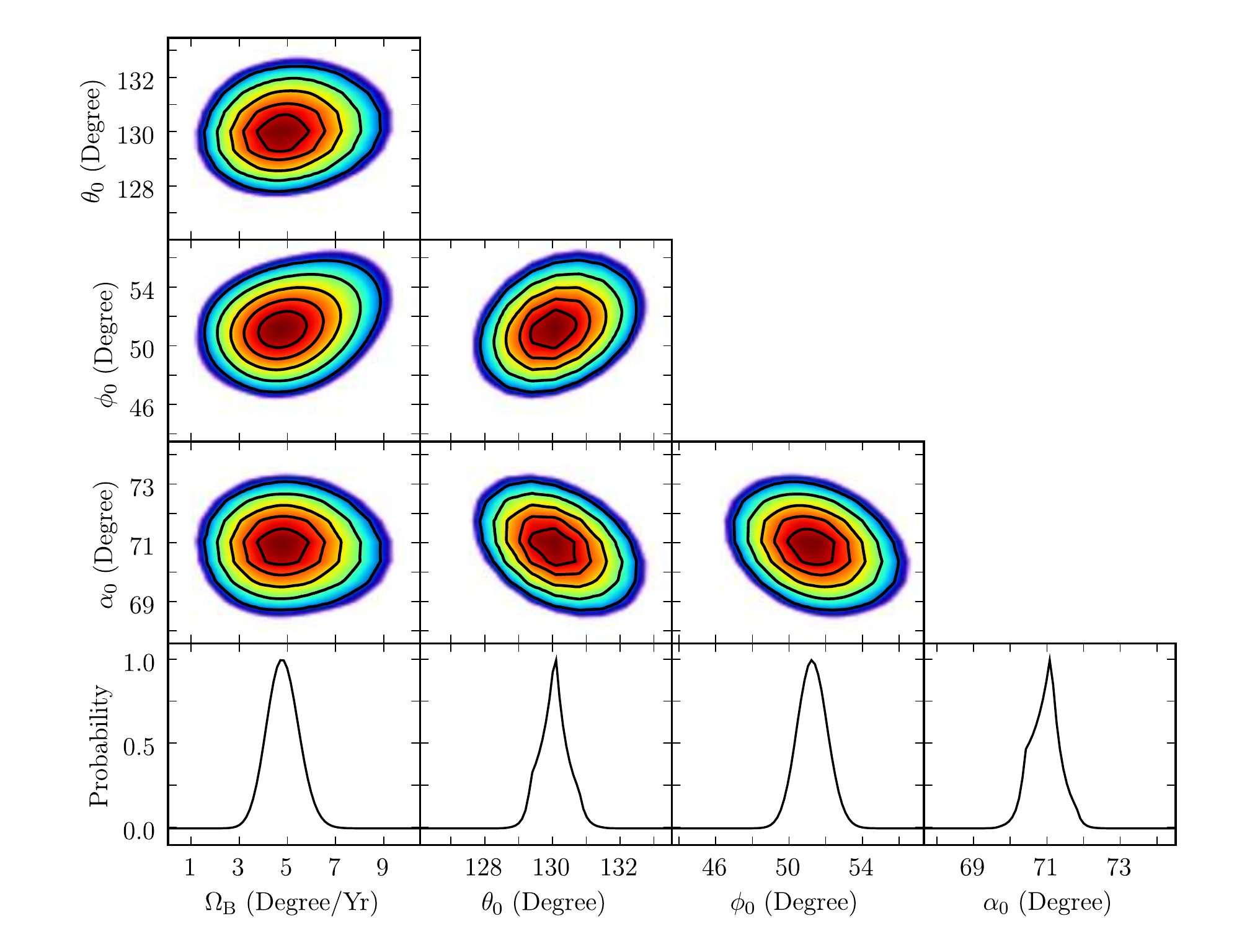}
 \caption[Marginalized \emph{a posteriori} probability distributions for the eclipse modeling]{One- and two-dimensional projections of the marginalized \emph{a posteriori} probability distributions for the joint fit of the parameters' evolution. Black contours in two-dimensional maps are joint $1,2,3,4$ and $5\sigma $ confidence regions, with the red color being associated with a higher likelihood value. The epoch of $\phi = \phi_0$ is May 2, 2006 (MJD 53857). Note that these probability distributions include priors related to systematic uncertainties.}
 \label{f:eclipse_pdf}
\end{figure}

\clearpage
}

A second source of systematics arises from the choice of the eclipse region to include in the fit. Changes in the eclipse light curve due to relativistic spin precession are not uniform and the eclipse model tends to perform better toward the eclipse center than at the ingress or the egress. As opposed to the eclipse center, where our sight line to pulsar A goes deep inside and outside the magnetosphere of pulsar B as it rotates, our sight line only briefly intersects the edge of the magnetosphere at the beginning and the end of the eclipse. Therefore, local distortions of pulsar B's magnetic field or variations of the plasma density may give rise to a slight departure from our model. Indeed, we observe that fitting the whole eclipse does not generally provide qualitatively good fits. The narrow and periodic modulations in the eclipse center are very important markers for the geometric orientation of pulsar B but they tend to be misfitted because broader features in the egress region lead to larger variations of the goodness-of-fit. We find that excluding the egress more accurately fits the overall light curves, without sacrificing critical information derived from narrow modulations, while still qualitatively reproducing the egress. Therefore, we chose to fit the eclipse in the range $[-1.0^{\circ},0.75^{\circ}]$ centered around conjunction (see Figure~\ref{f:eclipse_fit}).

\afterpage{
\clearpage

\begin{figure}
 \centering
 \includegraphics[width=6in]{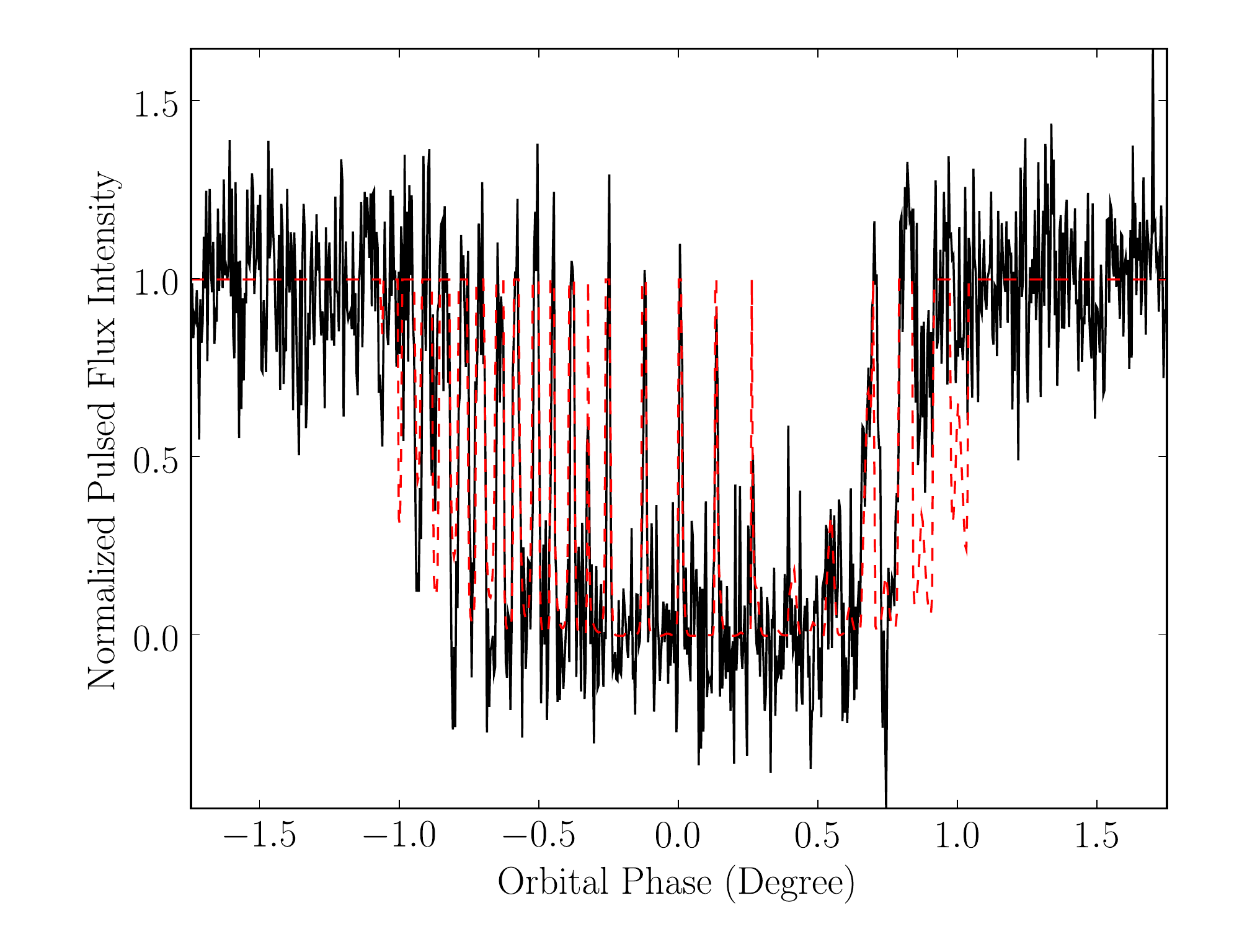}
 \caption[Average eclipse profile of pulsar A]{Average eclipse profile of pulsar A consisting of eight eclipses observed at 820\,MHz over a five-day period around April 11, 2007 (black line) along with a model eclipse profile (red dashed line). The relative pulsed flux density of pulsar A is normalized so the average level outside the eclipse region is unity. The resolution of each data point is $\sim$91\,ms while 1$^\circ$ in orbital phase corresponds to 24.5\,s. Note that near orbital phase 0.0 the spikes are separated by the spin period of pulsar B.}
 \label{f:eclipse_fit}
\end{figure}

\clearpage
}

Determining the boundaries of the region to fit is arbitrary and hence we estimated how much dispersion in the best-fit values is induced by other choices of limits. We compared our actual choice, $[-1.0^{\circ},0.75^{\circ}]$, with the full eclipse, $[-1.0^{\circ},1.0^{\circ}]$, the eclipse center, $[-0.6^{\circ},0.6^{\circ}]$, and the extended center, $[-0.7^{\circ},0.7^{\circ}]$, fits. In a Bayesian framework, we can easily incorporate the effect of systematics as priors on the model parameters. For simplicity and because the functional form of the systematics is poorly defined we assume Gaussian priors. Therefore, we can recast the \emph{a posteriori} probability distribution of our pre-systematics analysis work, for which we were assuming constant priors:
\begin{equation}
 p\left(\alpha,\theta,\phi | D\right) \propto {\cal L} \left(D | \alpha,\theta,\phi \right) \, ,
\end{equation}
as:
\begin{equation}
 p\left(\alpha,\theta,\phi | D\right) \propto
 	\int_{\alpha^{\prime}} {\cal N}(\alpha - \alpha^{\prime}, \sigma_{\alpha})
 	\int_{\theta^{\prime}} {\cal N}(\theta - \theta^{\prime}, \sigma_{\theta})
 	\int_{\phi^{\prime}} {\cal N}(\phi - \phi^{\prime}, \sigma_{\phi}) \,
 	{\cal L} \left( D | \alpha^{\prime},\theta^{\prime},\phi^{\prime} \right) \, d\alpha^{\prime} \, d\theta^{\prime} \, d\phi^{\prime}
 \, ,
\end{equation}
where ${\cal N}(\nu, \sigma_{\nu})$ is a Gaussian distribution of mean $\nu$ and standard deviation $\sigma_{\nu}$. The likelihood, ${\cal L} \left(D | \alpha,\theta,\phi \right)$, is defined as $\exp(-\chi^2_\nu / 2)$, with $\chi^2_\nu$ being the standard reduced chi-square. From the analysis of systematics due to the choice of the region to fit, along with the additional uncertainty in the spin phase of pulsar B due to the long-term pulse profile variations, we estimate that systematics contribute $\sigma_{\alpha} = 1^{\circ}$, $\sigma_{\theta} = 1^{\circ}$ and $\sigma_{\phi} = 2.0^{\circ}$. Note that incorporating Gaussian priors due to systematics has the effect of convolving the three-dimensional likelihood obtained from the eclipse fitting, ${\cal L} \left(D | \alpha,\theta,\phi \right)$, with a three-dimensional Gaussian. Note that the results reported above include these priors and so do the confidence interval reflect it.

\section{New Strong-Field Regime Test from Relativistic Spin Precession}
This relatively simple model \citep{lt05} is able to reproduce the complex phenomenology of the eclipses (see Figure~\ref{f:eclipse_fit}) except at the eclipse boundaries where slight magnetospheric distortions or variations in plasma density are likely to occur. Fits including the egress generally are poor in the central region where we observe narrow modulation features, which are critical for determining pulsar B's geometry. By analyzing the model quality in different portions of the eclipse, we have been able to asses credible confidence intervals that encompass excellent fits throughout the center region while still producing qualitatively good predictions near the eclipse ingress and egress. The overall success of the model implies that the geometry of pulsar B's magnetosphere is accurately described as predominantly dipolar; a pure quadrupole, for instance, does not reproduce the observed light curves. Although the model does not exclude the possibility that higher-order multipole components may exist close to the surface of pulsar B, our modeling supports the conclusions \citep{lt05} that these eclipses yield direct empirical evidence supporting the long-standing assumption that pulsars have mainly dipolar magnetic fields far from their surface.

The direct outcome from modeling the eclipse profile evolution is a measurement of the effect of relativistic spin precession (see Movie `Spin Precession' at \url{http://www.physics.mcgill.ca/~bretonr/doublepulsar/} for an illustration of the time evolution of the eclipse). We can use the inferred precession rate to test GR (see Figure~\ref{f:mass_mass}) and to further constrain alternative theories of gravity and the strong-field aspects of relativistic spin precession. We use the generic class of relativistic theories that are fully conservative (Lorentz-invariant) and based on a Lagrangian, as introduced by \citet{dt92a}. In this way we can study the constraints of our observations on theories of gravity by describing the spin-orbit interaction within a specific theory by coupling functions appearing in the corresponding part of the Lagrangian. In this framework, we can write the precession rate of pulsar B in a general form:
\begin{equation}\label{eqn:omegaB}
 \Omega_{\rm B} = \frac{\sigma_B L}{a^3_R (1-e^2)^{3/2}} \,,
\end{equation}
where $L$ is the orbital angular momentum of the system, $a_R$ is the semimajor axis of the relative orbit between the pulsars, $e$ the eccentricity of the orbit and $\sigma_B$ is a generic strong-field spin-orbit coupling constant. Since $L$ and $a_R$ are not directly measurable, it is more convenient to write the above expression using observable Keplerian and post-Keplerian parameters. While alternative forms generally involve a mixture of gravitational theory-dependent terms, the particular choice:
\begin{equation}
 \Omega_{\rm B} = \frac{x_A x_B}{s^2} \frac{n^3}{1 - e^2} \frac{c^2 \sigma_B}{\cal G} \,,
\end{equation}
is the only one that does not incorporate further theoretical terms other than the spin-orbit coupling constant, $\sigma_B$, the speed of light, $c$, and a generalized gravitational constant for the interaction between the two pulsars, ${\cal G}$. In this expression, the Keplerian parameters $e$ and $n=2\pi/P_b$, the angular orbital frequency, are easily measurable for any binary system. On the other hand, the post-Keplerian Shapiro delay shape parameter $s$, equivalent to the sine of the orbital inclination angle \citep{dt92a}, requires relatively edge-on orbits to be observed. Measurement of the projected semi-major axes of the two orbits\footnote{The projected semi-major axes are expressed in terms of light-travel time across the orbit.}, $x_A$ and $x_B$, found in the above equation, necessitates that each body must be timeable. Therefore, the double pulsar is the only relativistic binary system that allows a direct constraint on the spin-orbit coupling in general theories of gravity. Using the inferred precession rate of $\Omega_{\rm B}=4.77^{+0.66}_{-0.65}\,\,^{\circ}\,{\rm yr}^{-1}$, we derive $\left(\frac{c^2 \sigma_B}{\cal G}\right) = 3.38^{+0.49}_{-0.46}$. Every successful theory of gravity in the given generic framework must predict this value --- these observations provide a strong-field test of gravity that complements and goes beyond the weak-field tests of relativistic spin precession \citep{oco08}. In GR, we expect to measure $\left(\frac{c^2 \sigma_B}{\cal G}\right)_{\rm GR} = 2 + \frac{3}{2}\frac{m_A}{m_B} = 3.60677 \pm 0.00035$, where we have used the masses determined from the precisely observed orbital precession and the Shapiro delay shape parameter under the assumption that GR is correct \citep{ksm+06}. Comparing the observed value with GR's predictions, we find $\left( \frac{c^2 \sigma_B}{\cal G} \right)_{\rm obs} / \left(\frac{c^2 \sigma_B}{\cal G}\right)_{\rm GR} = 0.94 \pm 0.13$. Hence, GR passes this test of relativistic spin precession in a strong-field regime, confirming, within uncertainties, GR's effacement property of gravity even for spinning bodies, i.e.~the notion that strong internal gravitational fields do not prevent a compact rotating body from behaving just like a spinning test particle in an external weak field \citep{dam87}.

\afterpage{
\clearpage

\begin{figure}
 \centering
 \includegraphics[height=5.5in]{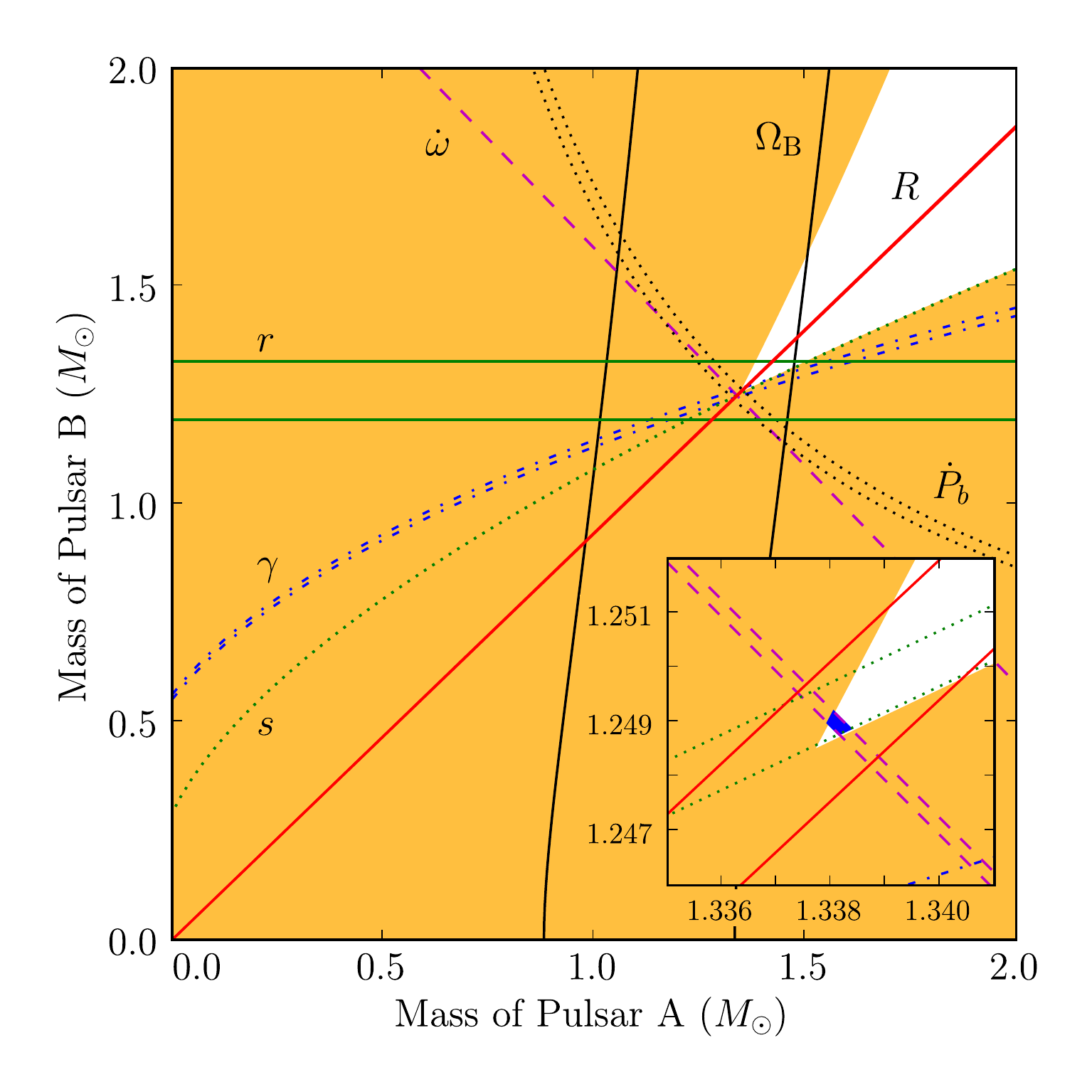}
 \caption[Mass-mass diagram for the double pulsar]{Mass-mass diagram illustrating the present tests constraining general relativity in the double pulsar system. The inset shows an expanded view of the region where the lines intersect. If general relativity is the correct theory of gravity, all lines should intersect at common values of masses. The mass ratio ($R = x_B/x_A$) and five post-Keplerian parameters ($s$ and $r$, Shapiro delay `shape' and `range'; $\dot \omega$, periastron advance; $\dot P_b$, orbital period decay due to the emission of gravitational waves; and $\gamma$, gravitational redshift and time dilation) were reported in \citet{ksm+06}. Shaded regions are unphysical solutions since $\sin i \le 1$, where $i$ is the orbital inclination. In addition to allowing a test of the strong-field parameter $\left(\frac{c^2 \sigma_B}{\cal G}\right)$, the spin precession rate of pulsar B, $\Omega_{\rm B}$, yields a new constraint on the mass-mass diagram.}
 \label{f:mass_mass}
\end{figure}

\clearpage
}

\subsection{Future Perspectives}
The spin precession rate, as well as the timing parameters entering in the calculation of $\left( \frac{c^2 \sigma_B}{\cal G}\right)$, are all independent of the assumed theory of gravity.  If the main contribution limiting the precision of this new strong-field test comes from the inferred spin precession rate, we expect that the statistical uncertainty should decrease significantly with time, roughly as the square of the monitoring baseline for similar quantity and quality of eclipse data. The contribution of systematics to the error budget should also decrease, but its functional time dependence is difficult to estimate. Although the orbital and spin phases of pulsar B are input variables to the eclipse model, our ability to determine the orientation of pulsar B in space does not require the degree of high-precision timing needed for measurement of post-Keplerian parameters; evaluating spin phases to the percent-level, for instance, is sufficient. Therefore, the intrinsic correctness of the model and its ability to reproduce future changes in the eclipse profile due to evolution of the geometry are the most likely limitations to improving the quality of this test of gravity, at least until the measured precession rate reaches a precision comparable with the timing parameters involved in the calculation of $\left(\frac{c^2 \sigma_B}{\cal G}\right)$. Better eclipse modeling could be achieved from more sensitive observations and thus new generation radio telescopes such as the proposed Square Kilometer Array could help make important progress. Pulsar A does not show evidence of precession \citep{mkp+05,fsk+08} likely because its spin axis is aligned with the orbital angular momentum; it should therefore always remain visible, thus allowing long-term monitoring of its eclipses. Pulsar B, however, could disappear if spin precession causes its radio beam to miss our line of sight \citep{bpm+05}. In this event, we would need to find a way to circumvent the lack of observable spin phases for pulsar B, which are necessary to the eclipse fitting.


\section{Lessons from the Geometry}
\subsection{Degenerate Solutions}\label{s:degeneracy}
\citet{lt05} demonstrated the potential of their eclipse model using a qualitative fit that reproduced the overall morphology of the eclipse. Although this result is by itself impressive given the complicated structure of the profile, it is important to assess whether only one or multiple solutions exist. We investigated the existance of degenerate solutions both analytically and using the results from MCMC presented in \S\,\ref{s:mcmc_results}.

From Equations~\ref{eq:mu} and \ref{eq:mu_omega}, we find four degenerate solutions that yield exactly the same eclipse light curve, two of which are not trivial transformations. Here is the list of parameter transformations that related these solutions:
\begin{enumerate}
 \item $(\theta,\phi,\alpha,z_0) \rightarrow (\theta,\phi,-\alpha,z_0)$:\\
 This degenerate solution is the same as the original one modulo half a rotation of pulsar B. This half spin period shift is however absorbed in a redefinition of the correspondence between the model spin phase, $\Omega_{\rm B}$, and the observed spin phase.
 \item $(\theta,\phi,\alpha,z_0) \rightarrow (\theta,\phi,\alpha + 180^\circ,z_0)$:\\
 This degenerate solution consists in flipping the north and south magnetic poles of the pulsar, which are undistinguishable here.
 \item $(\theta,\phi,\alpha,z_0) \rightarrow (\theta,\phi + 180^\circ,\alpha + 180^\circ,-z_0)$.
 This degenerate solution arises from the fact that the direction of the orbital inclination is not known because only $\sin i$ is measured. Therefore, $i = 90^\circ - \epsilon$ is undistinguishable from $i = 90^\circ + \epsilon$. In our geometrical framework, $z_0 > 0 \Rightarrow i < 90^\circ$ and conversely. Consequently, the above degeneracy controls the sign of $z_0$ and hence determines whether $i$ is greater or lower than $90^\circ$. Even though the true value of $i$ should theoretically alter $\delta$ and $\phi_{\rm so}$, the orbital inclination is too close to $90^\circ$ for such an effect to have an observable impact.
 \item $(\theta,\phi,\alpha,z_0) \rightarrow (\theta + 180^\circ,180^\circ - \phi,\alpha,z_0)$:\\
 This degenerate solution is a consequence of the fact the synchrotron absorption yields the same result upon a reversal of the light travel direction. A degeneracy therefore arise if the Earth and pulsar A are swapped. Practically, such an identical eclipse light curve is obtained by flipping the spin direction ($\theta \rightarrow \theta + 180^\circ$). It also requires to mirror the spin longitude with respect to the $x-z$ plane ($\phi \rightarrow 180^\circ - \phi$). It's important to note that reversing the spin direction itself can be distinguished by the model. However, there is no way to determine whether the spin direction is prograde (parallel with respect to the orbital angular momentum) or retrograde (anti-parallel with respect to the orbital angular momentum). If it was possible, this degeneracy would break down.
\end{enumerate}

From an inspection of the likelihood resulting from the MCMC analysis (\S\,\ref{s:mcmc_results}), we confirm the above degeneracies and assess that no other solution having significant probability exists. This is good since any credible theoretical model should be able to reproduce the observations using a minimal number of free parameters and have predicative behavior.

\subsection{Spin Direction and Direction of Precession}
In Chapter~\ref{c:0737_eclipse}, we report on the quantitative measurement of the relativistic spin precession of pulsar B. This work is more specifically targeting the spin precession rate. It is important to mention, however, that relativistic spin precession is a vector quantity defined as \citep{dt92a}\footnote{Here we present the quantities in brackets to denote that they are averaged over an orbit.}:
\begin{equation}\label{eqn:spin_precession_a}
 \left<\frac{d\vec{S}_{\rm B}}{dt}\right> = \left<\vec{\Omega}_{\rm B}\right> \times \vec{S}_{\rm B} \,.
\end{equation}
where $\vec{S}_{\rm B}$ is the spin angular momentum of pulsar B, and $\vec \Omega_{\rm B}$ has been defined in Equation~\ref{eqn:omegaB}.

As we can easily see, the spin precession vector is oriented along the orbital plane in a direction provided by $\vec{L} \times \vec{S}_{\rm B}$, that is, if both spin and orbital angular momentum are positive/parallel, the spin precession direction is positive, whereas if they have opposite signs, the spin precession direction is negative (i.e. opposite with respect to the spin direction). This interpretation is based on the assumption that other quantities than $\vec{L}$ and $\vec{S}_{\rm B}$ in Equation~\ref{eqn:spin_precession_a} are all positive quantities. Examination of Equation~\ref{eqn:omegaB} allows us to concluded that only $\sigma_{\rm B}$ could possibly be negative. In general relativity, this constant is strictly positive and has a value $\sigma_B \equiv \frac{3}{2} + \frac{m_A}{m_B}$, which depends on the masses in the system only. The definition of $\sigma_B$ in alternative theories of gravity could possibly take other forms and allow a negative value. Hence, although testing the spin precession rate is highly praised, it is also interesting to verify that the spin precession direction is consistent, or not, with $\sigma_B$ being positive as predicted by general relativity.

After examining the eclipse model, we conclude that it is possible to unambiguously assess the direction of the spin precession provided that there exists only one best-fit solution --- we argued in \S\,\ref{s:degeneracy} that this is the case and we shall demonstrate here that the degenerate solutions do not affect this conclusion. In the eclipse model, an arbitrary orbital motion direction is chosen according to the coordinate system. The spin direction is then fitted without any restriction to whether it is parallel or anti-parallel to the orbital angular momentum vector. Bear in mind that the colatitude of the spin axis, $\theta$, varies in the range $[0,\pi]$, while the longitude of the spin axis, $\phi$, does in range $[0,2\pi]$. As we pointed out in \S\,\ref{s:degeneracy}, a reversal of the spin direction has a noticeable effect on the light curve and hence the spin direction relative to the orbital motion direction can be uniquely determined.

As far as degenerate solutions are concerned, the first three solutions clearly do not change the spin direction. The fourth degeneracy, however, does invert the spin direction. It turns out that this shall not change the above statement because spin precession affects the spin axis geometry as follow: $\phi = \phi_0 + \Omega_{\rm B} t$, where $\Omega_{\rm B}$ is the precession rate. Since the two geometric configurations are related by a symmetry, $\phi \rightarrow 180^\circ - \phi$, it follows that time derivative of this angle will suffer a change in direction: $\frac{d\phi}{dt} \rightarrow - \frac{d\phi}{dt}$. This necessarily implies that regardless of what the right solution is, there is a way of testing the consistency of $\sigma_{\rm B}$ with general relativity. In our geometrical framework, the orbital motion direction is arbitrarily fixed but we should point out that reversing it obviously implies that the entire system geometry is inverted, thus providing the exact same solution as before.

Using the coordinate system defined in \S\,\ref{s:model_geometry}, we deduce that the chosen orbital angular momentum is negative, or clockwise as seen from above the $x-y$ plane. The solution we shall present in Chapter~\ref{c:0737_eclipse}, with $\theta = 130.02^\circ$, implies a negative spin angular momentum, i.e. counter-clockwise. Since the observed spin longitude goes toward decreasing values, we infer that the spin precession direction is also counter-clockwise. This imposes that $\Omega_{\rm B}$ and, by extension, $\sigma_B$ are both positive. This result is therefore consistent with the prediction of general relativity.

\subsection{Orbital Inclination}
In principle, the eclipse modeling permits to derive the orbital inclination of the system independently from the timing. The impact parameter $z_0$ measures the projected distance at conjunction between the two pulsars in units of $R_{\rm mag}$. Since $\xi$ measures the size of $R_{\rm mag}$ relative to the orbit, we can use the small angle approximation in order to find the following expression for the orbital inclination:
\begin{equation}\label{eqn:inclination}
 i = 90^\circ \pm | z_0 \xi | \,,
\end{equation}
where $\xi$ has units of degrees\,$R_{\rm mag}^{-1}$. From the eclipse modeling, we found $z_0 = -0.543 \, R_{\rm mag}$ and $\xi = 1.29^\circ \, R_{\rm mag}^{-1}$, which implies that $i = 89.3^\circ$ or $i = 90.7^\circ$. As we mentioned in the third degenerate solution of \S\,\ref{s:degeneracy}, we do not know whether the orbital angular momentum vector is inclined toward or away from us with respect to the sky plane, hence why the two possible inclination values. Although that we have not performed a thorough analysis of the \emph{a posteriori} probability distribution of $z_0$ and $\xi$ yet\footnote{This should be done as a follow up project to the work presented in this thesis. See \S\,\ref{c:conclusion} for more details.}, we estimate the relative uncertainties to be the order of 10\% in both cases. Under the assumption of Gaussian \emph{a posteriori} probability distributions, this implies that the relative uncertainty on the misalignment of our line of sight with the orbital plane of the system --- i.e. $|i-90^\circ|$ --- is about $\sqrt{2} \times 10\% \sim 14\%$, which means $\sim 0.1^\circ$.

According to our estimated values, our derived inclination angle agrees with the result reported by \citet{ksm+06}, $i = 88\fdg 69^{+0\fdg 50}_{-0\fdg 76}$, within the $1\sigma$ uncertainties. Note, however, that since $z_0/R_{\rm mag} \le 1$ it would be unlikely that we obtain $|i-90^\circ| > 1^\circ$ from the eclipse modeling unless $\xi$ was much larger than the above value. On the other hand, our measurement appears to exclude the inclination angle value derived from scintillation measurements obtained by \citet{cmr+05}, $|i-90^\circ| = 0.29^\circ \pm 0.14^\circ$, but is consistent with the one obtained by \citet{rkr+04}, $i = 88.7^\circ \pm 0.9^\circ$.


\subsection{Emission Geometry of Pulsar B}\label{s:geometry_psrB}
\subsubsection{Pulse Profile Changes}
In the traditional pulsar lighthouse model, the radio emission is emitted as a cone directed along the open field lines above the magnetic poles. The pulse profile owes its morphology to the emission structure in the plane intersected by the line of sight as the pulsar rotates. Phenomenological and semi-phenomenological models of pulsar emission have been proposed by many authors \citep[see, e.g.][]{rc69b,stu71,rs75,lm88,qlw+07,kj07}. The general picture is that the emission comes from the edge of one or several concentric hollow cones and/or from the central inner region of this cone. This model framework is qualitatively able to explain several observed features such as multipeak profiles showing symmetries and single `Gaussian'-looking peak \citep{lm88}.

One of the clear consequences of relativistic spin precession is the evolution of the pulse profile morphology caused by changes in the viewing geometry (see \S\,\ref{s:other_tests}). In the Hulse-Taylor pulsar, for example, the profile presents two peaks separated by about 40$^\circ$ in spin longitude having a bilateral symmetry and connected by a bridge of weaker emission. It is believed that the radio emission comes from a hollow cone \citep{tfm79,kra98}. \citet{wrt89} observed that the relative intensity of the two peaks was slowly varying with time and attributed it to relativistic spin precession. Although a change in the pulse component separation is also expected if our line of sight intercepts the emission cone at a different latitude, no significant change was found until \citet{kra98} revisited the problem about 10 years later. Using the predicted precession rate from general relativity, he determined the geometry of the pulsar and showed that no change in separation was initially seen because of the particular geometrical configuration at the time. Eventually the pulsar's radio beam should miss our line of sight and the pulsar will disappear until its spin axis gets oriented favorably for us to see it again. It is difficult to use these observations in order to obtain a quantitative measurement of relativistic spin precession since the beam geometry is not trivial in reality. Any attempt at measuring the precession rate is highly model dependent. Furthermore, it is not clear whether the emission does remain stable over an extended period of time. Nevertheless, one can simply assume that general relativity is correct and use the predicted precession rate to infer the emission geometry and the pulsar orientation. For PSR~B1913+16, \citet{cw08} recently showed that the beam shape appears to resemble an hourglass.

Pulse profile changes have also been observed in the double pulsar already. As reported in \citet{bkk+08} and earlier by \citet{bpm+05}, pulsar B's profile displayed dramatic morphologic changes over the last four years and its overall pulsed flux intensity has decreased at the same time (see also \S\,\ref{s:0737_observations}). This is clear evidence of the precession of its spin axis. On the other hand, while pulsar B's profile is evolving, the pulse profile of pulsar A has remained extremely stable over the same period \citep{mkp+05,fsk+08} and it seems to indicate that its spin axis is closely aligned with the total angular momentum of the system, hence explaining why no apparent precession is observed. We shall investigate this possibility in Chapter~\ref{c:0737_aberration}.

\subsubsection{Pulsar B Becomes `Invisible'}
As we showed in Figure~\ref{f:profile_evolution}, pulsar B appeared to become more difficult to detect toward the end of our observing campaign that it was initially in December 2003. Using the information about the geometry of pulsar B obtained from the eclipse modeling, we can calculate the minimum angle between our line of sight and the magnetic axis of pulsar B over the course of one rotation of the pulsar. This angle, also known as the impact parameter $\beta$, is widely used in the context of pulsar's linear polarization measurements (see \S\,\ref{s:rvm_prediction}) and corresponds to \citep[see, e.g., Equations~2.25b and 3.36a][]{dt92a}:
\begin{equation}
 \beta = \zeta - \alpha \,,
\end{equation}
where $\alpha$ is the magnetic inclination defined in \S\,\ref{s:model_geometry}, and $\zeta$ is the angle between our line of sight and the spin axis:
\begin{eqnarray}
 \cos (\pi-\zeta) & = & \cos \delta \cos i - \sin \delta \sin i \cos \phi_{\rm so} \\
  & \approx & \sin \theta \cos \phi \,.
\end{eqnarray}
Here the variables are the same as in \S\,\ref{s:model_geometry}. We make use of the fact that $i \sim 90^\circ$ and the identities defined in Equations~\ref{eqn:angle_a} and \ref{eqn:angle_b}.

Figure~\ref{f:los_separation} shows the value of $\beta$ as a function of time based on the values reported in Table~\ref{t:fit_result}. As we can see, our line of sight was almost intercepting the magnetic pole of pulsar B in 2003 --- $\beta \sim -5^\circ$ --- whereas it slowly drifted away since then such that the separation is now (as of December 2008) $\sim 20^\circ$. The behavior of pulsar B, which has gradually become dimmer, is therefore normal if the source of radio emission is located along or close to the magnetic axis. It appears, however, that the emission geometry of pulsar B strongly departs from a circular cone centered on the magnetic axis given that the width of the pulse is $\sim 3^\circ$ in spin longitude but was observed over a range of $\sim 15^\circ$ in spin latitude. It would therefore resemble a rather elongated oval shape. Such a non-circular emission beam geometry has also been observed in the relativistic binary pulsar PSR~B1913+16, which has been suggested to have an hourglass-shaped beam \citep{cw08}.

If the trend continues, it seems that the radio beam from pulsar B could miss our line of sight in a very near future. In this case, pulsar B would disappear and PSR~J0737$-$3039A/B would no longer be the `double pulsar', thus becoming a `normal' pulsar - neutron star binary like the Hulse-Taylor pulsar and the other known relativistic binary pulsars. Of course, this has considerable impact of the perspective of improving the current relativistic tests in the double pulsar and potentially measuring higher-order post-Keplerian parameters that could yield an estimate of the moment of inertia of pulsar A \citep{ls05,ior09}.

In the event that pulsar B would disappear, it would not be invisible forever. In fact, assuming that the emission geometry would remain stable over time, we could predict how long it would take until pulsar B reappears. This corresponds to the time interval between the node marking the time at which pulsar B disappears and the following node having the same impact parameter. For example, if pulsar B was to disappear during Fall 2008, it would reappear around Spring 2024 (see Figure~\ref{f:los_separation}), which implies that it would be invisible for slightly less than 16 years. If it disappears later, the invisibility period would be even shorter.

We can place a lower limit on the visibility duty cycle of pulsar B if we assume that it is visible if our line of sight lies within 20$^\circ$ (i.e. $|\beta| < 20^\circ$) of pulsar B's magnetic pole, which represents the approximate impact parameter at the time of taking our latest data in the Summer 2008 when it was still barely visible. Of course the caveat here is that we presume that the emission geometry is symmetrical for positive and negative impact parameters, which might not necessarily be the case. Under these assumptions, it appears that pulsar B would be visible for about 22 years during a full precession cycle of 71 years, which implies a duty cycle of $\sim 31$\%.

Conversely, our line of sight will describe a similar pattern with respect to the other magnetic pole of pulsar B (see Figure~\ref{f:los_separation}). We may therefore expect to see the other pole if it also produces radio pulsed emission. If we suppose that the pulsed emission from this pole is perfectly symmetric to the pole that we actually observe, its visibility duty cycle would also be $\sim 31$\%. The visibility periods for both poles are out of phase and slightly overlap. This has two implications. First, we might be able to see pulsed emission from both poles during the overlap periods. Second, this contributes to increase the detectability of pulsar B to about 41 years, which implies a $\sim 58$\% duty cycle. Our line of sight should enter within $\sim 20^\circ$ of this second pole's magnetic axis in mid-2033. One will have to wait until then to determine whether or not it also produces radio emission.

Because pulsar A's spin axis seems aligned with the orbital angular momentum of the system \citep{fsk+08} (see also Chapter~\ref{c:0737_aberration}), its does not display signs of precession and hence should always remain visible. The total probability of observing PSR~J0737$-$3039A/B as a double pulsar at a given random time is therefore $\sim 31$\% for emission from one pole and $\sim 58$\% if one allows both poles to be visible.

\afterpage{
\clearpage

\begin{figure}
 \centering
 \includegraphics[width=6.5in]{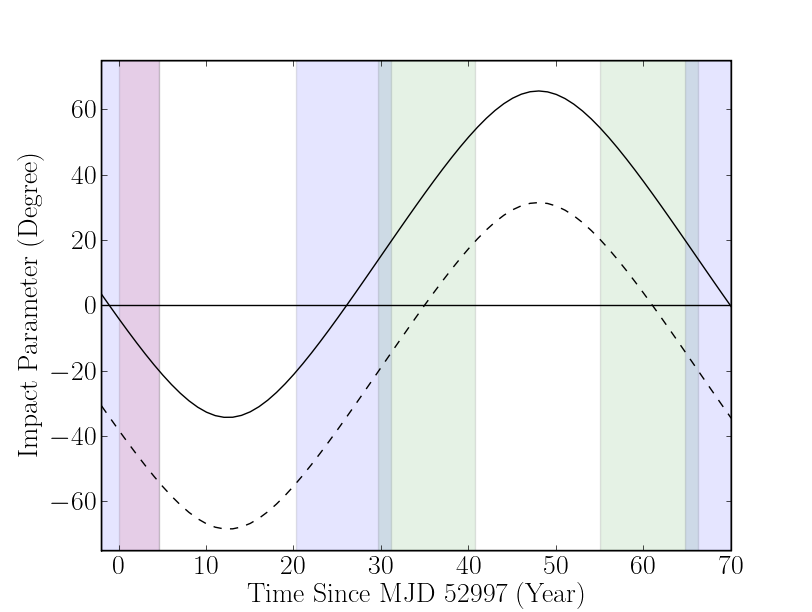}
 \caption[Angular separation between our line of sight and pulsar B's magnetic poles]{Angular separation between our line of sight and pulsar B's magnetic poles (also known as impact parameter). The plain curve shows the magnetic pole that we currently observe whereas the dashed curve shows the opposite pole. The time interval covered since the double pulsar's discovery indicated in purple. In light blue are the potential visibility periods for the observed pole assuming that the emission beam does not miss our line of sight if the separation is less than 20$^\circ$. In light green is the equivalent for opposite pole. Darker blue represents the overlapping regions.}
 \label{f:los_separation}
\end{figure}

\clearpage
}

\subsubsection{RVM Predictions}\label{s:rvm_prediction}
Since radio emission is produced by synchrotron emission of relativistic electrons gyrating along the open field lines, the linear polarization direction should be oriented radially with respect to the location of the magnetic pole. A simple pulse profile consisting of radio emission coming from one cone should, in principle, display an `S' swing in the position angle of its linear polarization as the line of sight crosses the leading, middle and trailing part of the cone \citep{rck+69}. This polarization model is referred to as the \emph{`rotating vector model'} (RVM) and we could make clear prediction on what the linear polarization of pulsar B should look like \citep{rc69}. Unfortunately, it appears that the radio emission from pulsar B only shows very weak linear polarization \citep{drb+04}, hence making the RVM's prediction hard, if not impossible, to verify.


\chapter{Latitudinal Aberration in the Double Pulsar}\label{c:0737_aberration}

\begin{flushright}
 \begin{singlespace}
 \emph{``All of physics is either impossible or trivial.\\
 It is impossible until you understand it, and then it becomes trivial.''}
 \end{singlespace}
 Ernest Rutherford
 \vspace{0.5in}
\end{flushright}

This chapter presents the results of a study of the stability of the radio pulse profile of pulsar A, PSR~J0737$-$3039A. We independently timed the two pulse components of this pulsar, which displays a bilateral symmetry and are separated by about half a spin period, in order to search for the presence of an orbital-dependent behavior in their separation. Such variations would result from the relativistic latitudinal aberration of the pulsed emission due to the orbital motion of the pulsar. The intricate details of this phenomenon are strongly related to the geometry of the pulsar in space and we have been able to constrain elements of its radio emission geometry and spatial orientation using it.

This work is part of a larger investigation of pulsar A's pulse profile, looking for signs of latitudinal aberration on the orbital time scale (presented here) but also for long term changes potentially related to relativistic spin precession. This additional work is independently made by Robert D. Ferdman and other collaborators and should eventually be published in a peer reviewed journal.

\section{Introduction}
In binary pulsars, effects of special and general relativity are easily observable in the timing. The TOAs of the pulses are normally altered as a consequence of the departure of the orbit from Newtonian motion and also from the travel of light in the gravitational field of its companion. In addition to changing the observed spin frequency, relativistic effects can also act on the structure of the pulse itself. Unfortunately, it is generally extremely difficult, if not impossible, to observe these phenomena because: 1. some effects, such as relativistic longitudinal aberration, are not measurable separately and are absorbed in a redefinition of other PK parameters \citep{dt92a}; 2. the implied changes are very small \citep{dt92a}; 3. disentangling relativistic pulse profile changes from those attributed to other intrinsic and extrinsic causes is challenging \citep{kra98}.

A typical example of profile changes is due by the relativistic spin precession of the pulsar's spin axis around the total angular momentum of the system. As we discussed in Chapter~\ref{c:0737_eclipse}, the change in the spin axis orientation implies that our line of sight intercepts the pulsar's radio emission beam at varying locations over time, thus causing the pulse profile to evolve. Pulse profile changes have been observed in several relativistic binary pulsars already --- PSR~B1913+16 \citep{wrt89}, PSR~B1534+12 \citep{arz95}, PSR~J1141$-$6545 \citep{hbo05}, PSR~J1906+0746 \citep{lsf+06} and PSR~J0737$-$3039B \citep{bpm+05,bkk+08} (see also Chapter~\ref{c:0737_eclipse}) --- and despite the fact that they provide strong evidence of spin precession, they do not yield constraining quantitative tests of relativity because the inferred precession rate depends on the presumed emission geometry and on the assumption that the profile itself remains stable over time \citep{kra98,cw08}. Nevertheless, \citet{sta04a} measured an orbital phase-dependent modulation of the total intensity profile of PSR~B1534+12 as well as its secular time evolution, which is a consequence of relativistic aberration. They combined this measurement to the secular change in the position angle of the linear polarization of the pulse profile in order to derive a quantitative, ``model-independent'' relativistic spin precession rate.

In this chapter, we present the results of a study of the pulse profile of PSR~J0737$-$3039A, hereafter `pulsar A'. We investigated the orbital phase-dependent stability of the pulse profile over three consecutive orbits. Although no significant changes were detected, our analysis allowed us to put meaningful upper limits on the relativistic latitudinal aberration phenomenon proposed by \citet{rl06b}. Furthermore, we discuss implications on the geometry of pulsar A.

\section{Theoretical Background}
\citet{rl06b} recently suggested that delays due to latitudinal aberration and gravitational lensing would modify the pulse profile of pulsar A in the double pulsar system PSR~J0737$-$3039A/B. These relativistic effects translate to a change of the colatitude of the emission vector and even though they are normally absorbed in the pulsar's timing, the profile of pulsar A, which presents a double-peak pulse, allows us to potentially observe it (see Figure~\ref{f:profile_pulsarA}). In fact, the latitudinal aberration and gravitational lensing do not affect every emission vector in the same way. We expect different components of the pulse profile to be distorted/shifted in a distinct way. In the particular case of pulsar A, we shall see that, as demonstrated by \citet{rl06b}, the separation between the two pulse peaks would in principle change as a function of the orbital phase.

\afterpage{
\clearpage

\begin{figure}
 \centering
 \includegraphics[width=6in]{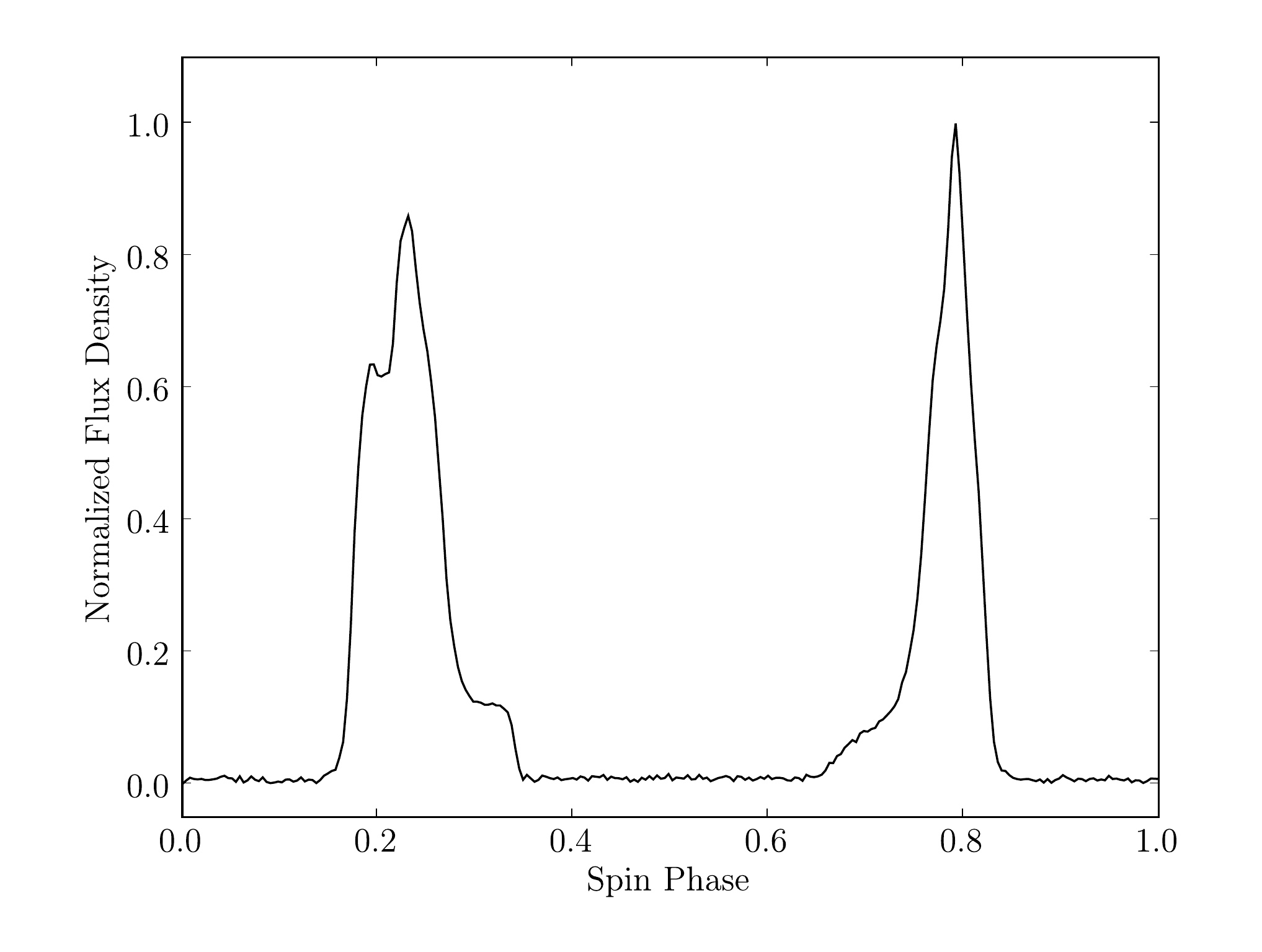}
 \caption[Pulse profile of pulsar A]{The pulse profile of pulsar A obtained at 820\,MHz during a $\sim 5$-hour observation with the {\tt SPIGOT} instrument at GBT. The profile presents two pulse components having a bilateral symmetry. The peak-to-peak distance between the components is 0.544 spin phase, i.e. $196^\circ$ in spin longitude units.
 \label{f:profile_pulsarA}}
\end{figure}

\clearpage
}

Let's consider an element of the pulse profile emitted at a colatitude $\zeta$ and spin longitude $\Phi_0$ from an emission cone having a half-opening angle $\rho$. The magnetic axis makes an angle $\alpha$ with respect to the spin axis\footnote{Please refer to Appendix~\ref{a:def_variables} for a comprehensive list of the variables used in this chapter.} (see Figure~\ref{f:geometry_aberration} for a schematic view of the geometry). Due to latitudinal aberration and gravitational lensing, the emission will suffer a shift $\Delta \Phi_0$ defined as \citep{rl06b}:
\begin{equation}\label{eqn:phi0}
   \Delta\Phi_0 = - \frac{\Delta \zeta}{\sin \zeta \tan \chi_0} \,
\end{equation}
where $\chi_0$ is the angle on the celestial sphere between the arc connecting our line of sight vector $\vec n_0$ and the spin axis vector $\vec s_p$ and the arc connecting $\vec n_0$ and the magnetic axis vector $\vec m$ the spin longitude $\Phi_0$. This also corresponds to the angle of linear polarization in the rotating vector model (RVM):
\begin{equation}
   \tan \chi_0 = \frac{\sin \alpha \sin \Phi_0}{\cos \alpha \sin \zeta -
\cos \Phi_0 \sin \alpha \cos \zeta} \,,
\end{equation}
and
\begin{equation}
   \cos \Phi_0 = \frac{\cos \rho - \cos \zeta \cos \alpha}{\sin \zeta \sin
   \alpha} \,.
\end{equation}

\afterpage{
\clearpage

\begin{figure}
 \centering
 \includegraphics[width=6in]{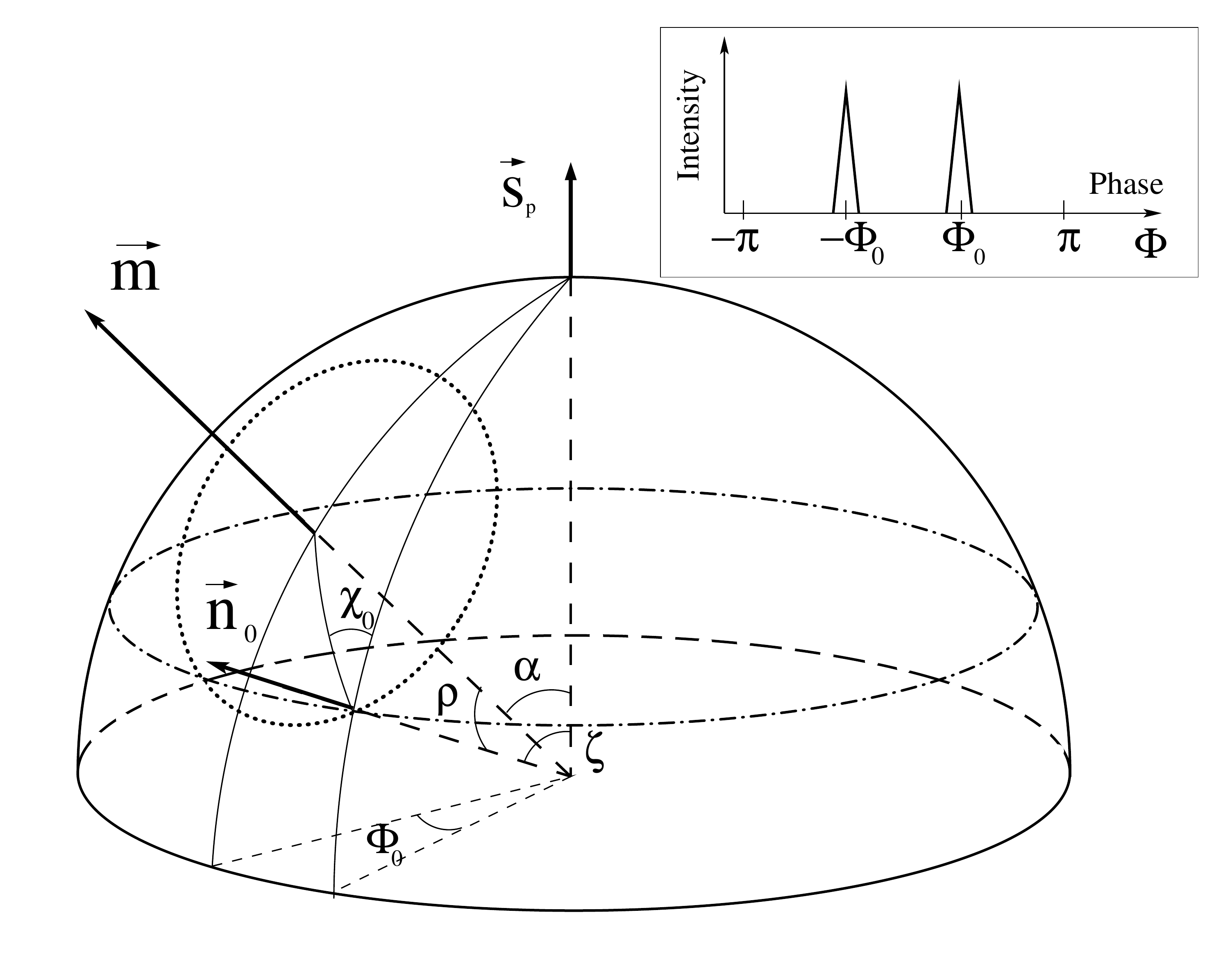}
 \caption[Schematic view of the Rafikov and Lai model geometry]{Schematic view of the Rafikov and Lai model geometry. The coordinate system is centered on and corotates with pulsar A. The direction to Earth is indicated by the vector $\vec n_0$ and describes a circle at colatitude $\zeta$ in pulsar A's sky. The pulsar's dipole moment, $\vec m$, is misaligned by an angle $\alpha$ with respect to the spin axis, $\vec s_p$ and defines the reference spin longitude. The emission cone has an opening angle $\rho$ and the position angle of the linear polarization of the emitted light, in the context of the rotating vector model, is given by $\chi_0$. Credit: \citet{rl06b}.
 \label{f:geometry_aberration}}
\end{figure}

\clearpage

\begin{figure}
 \centering
 \includegraphics[width=5.5in]{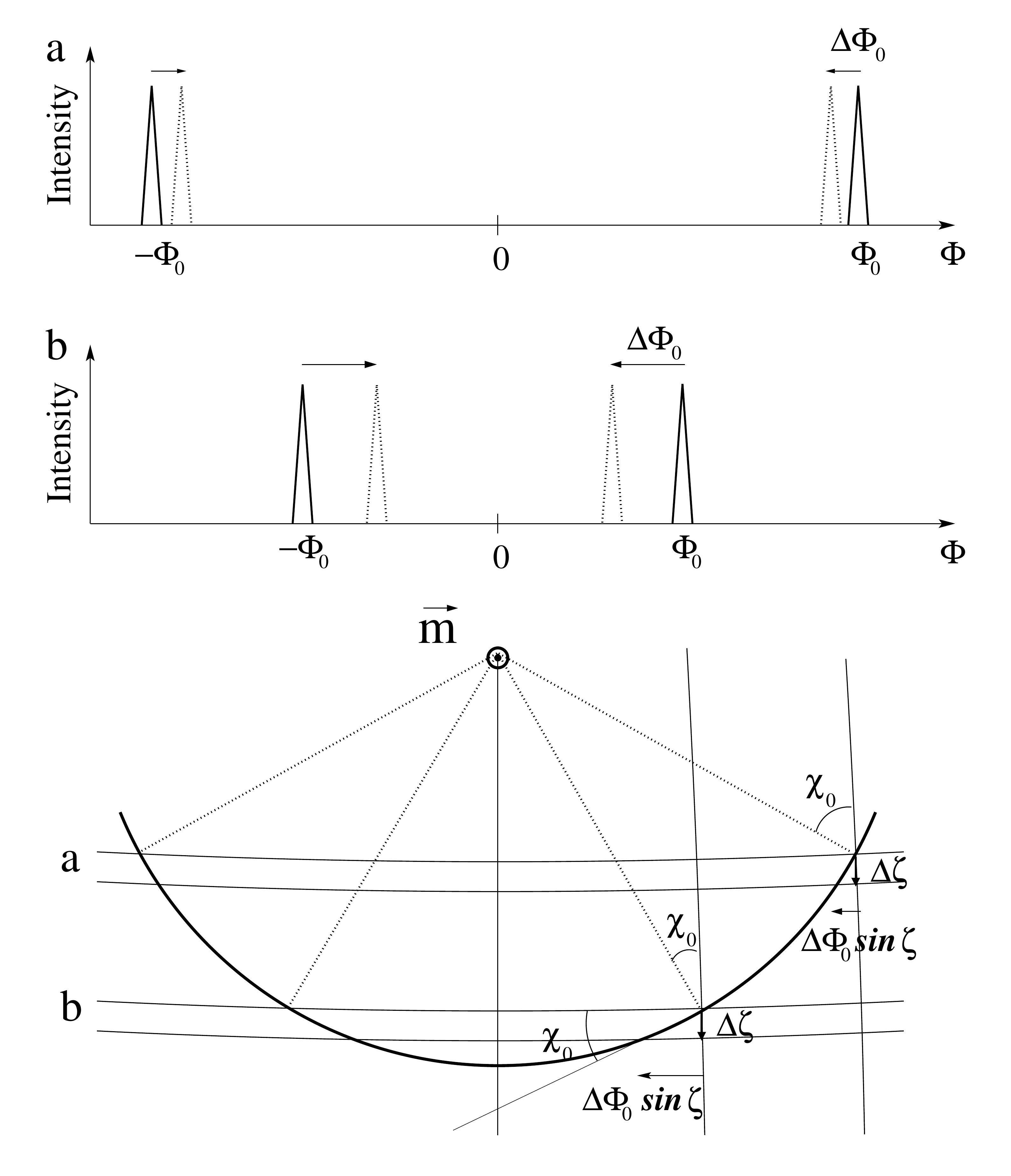}
 \caption[The effect of aberration and lensing on the pulsar A's profile]{Schematic view showing the effect of a change in the colatitude of the emission vector on the pulse profile. If the emission comes from the edge of a single hollow emission cone at spin longitude $\pm \Phi_0$, a change $\Delta \zeta$ in the emission colatitude will translate in a shift of the observed spin longitude $\pm \Delta \Phi_0$ for each component, therefore corresponding to a change of $2 \Phi_0$ in the pulse component separation. Credit: \citet{rl06b}.
 \label{f:aberration_profile}}
\end{figure}

}

The colatitudinal variation of the emission vector is \citep{rl06b}:
\begin{equation}\label{eqn:delta_zeta}
   \Delta \zeta = (\Delta \zeta)_A + (\Delta \zeta)_L \,
\end{equation}
where the aberration component is
\begin{equation}\label{eqn:deltaA}
   (\Delta \zeta)_A = \frac{\Omega_b a_p}{c \sqrt{1-e^2}}
                      [\cos i \sin \eta (\cos \psi + e \cos \omega)
		      - \cos \eta (\sin \psi + e \sin \omega)] \,,
\end{equation}
and the gravitational lensing component is
\begin{equation}\label{eqn:deltaL}
   (\Delta \zeta)_L = - \frac{\Delta R_\pm}{R} \frac{r}{a_{||}}
                      [\cos \eta \cos \psi + \cos i \sin \eta \sin \psi] \,,
\end{equation}
with $\Omega_b$ the orbital angular frequency; $a_p = a M_c/(M_c+M_p)$ where $a$ is the semi-major axis; $M_p$ and $M_c$ the pulsar of the pulsar and its companion, respectively; $c$ the speed of light; $e$ the eccentricity; $i$ the orbital inclination; $\eta$ is the angle between the ascending node and the projection of $\vec s_p$ on the sky plane; $\psi$ is the true anomaly measured from the ascending node; $\omega$ the longitude of the periastron; $r$ the distance between the two pulsars ($r = a (1-e^2)/(1+e \cos (\psi - \omega))$); and $a_{||}$ the distance at conjunction projected along our line of sight ($a_{||} = a \sin i (1-e^2)/(1+e \sin \omega)$).

The displacement of the pulsar image is defined as \citep{rl06b}:
\begin{equation}
   \Delta R_\pm = \frac{1}{2} \left( \pm \sqrt{R^2 + 4R_{E}^{2}} - R \right) \,,
\end{equation}
with
\begin{equation}
   R = r (1 - \sin^2 i \sin^2 \psi)^{1/2} \,,
\end{equation}
and the Einstein radius
\begin{equation}
   R_E = \sqrt{2 R_g a_{||}} = \sqrt{4 \frac{GM_c}{c^2} a_{||}}
         \simeq \sqrt{2 R_g a} \,.
\end{equation}

\emph{Note that the gravitational lensing component is important only in a short range near the conjunction.}

Equations~\ref{eqn:deltaA} and \ref{eqn:deltaL} show that latitudinal aberration and gravitational lensing delays are made of a pre-factor that acts like an amplitude term and depends on quantities known from the timing\footnote{Here we assume the orbital inclination value derived from the timing's Shapiro delay `s' parameter.}, as well as an orbital phase-dependent part ($\psi$). The only unknown variable remaining in these equations is $\eta$. This angle is expected to change with time as a result of relativistic spin precession \citep{dt92a}. At a given epoch, however, $\eta$ can be considered as a free, unknown constant parameter.

Even though the amplitudes of these delays are not likely to be very large (a few tens of $\mu$s, c.f. \citealt{rl06b}), they present a particular orbital phase-dependent signature that might be detectable if several orbits are analyzed coherently. Given the geometry derived from the timing (orbital inclination, longitude of periastron, etc.), we find that the amplitudes of the delays depend on pulsar A's orientation in space as well as on its the emission geometry.

\section{Observations and Data Reduction}\label{s:obs_reduc}
The data we used for this analysis consist of a $\sim$5-hour long 820\,MHz observation obtained with the {\tt SPIGOT} instrument \citep{kel+05} at GBT on December 23, 2003. More details about the observing setup can be found in Chapter~\ref{c:0737_eclipse}. We observed slightly more than two full orbits continuously, which include three eclipses of pulsar A. Data were folded with 256 bins per pulse period and 30\,s per time interval using the {\tt PRESTO} package \citep{rem02}. The length of the time intervals was chosen so that a relatively good signal-to-noise ratio is obtained, but no attempt to determine an optimal time integration has been performed (see below for more details).

Our data were incoherently dedispersed and thus provide limited sensitivity at resolving the pulse profile. For this reason, we focus our analysis on possible variations of the separation between the two main pulse peaks rather than on the individual properties of each component. Such an analysis will be performed separately \citep[see][for preliminary results]{fsk+08}. Also, we do not consider the contribution from the gravitational lensing term because it is likely impossible to detect. This is for two reasons. First, the orbital inclination angle derived from the timing \citep{ksm+06} is not close enough to $90^\circ$ to give rise to a measurable gravitational lensing effect --- it would likely be a few $\mu$s at the most. Second, because the effect only lasts a few seconds centered around conjunction, it would be difficult to observe since the eclipses of pulsar A by pulsar B strongly reduce the transmitted flux at the critical time. We therefore concentrate our efforts on the latitudinal aberration, which operates throughout the whole orbit and hence facilitate their detection.

\section{Simplified Model}\label{s:simple_model}
An important assumption that we are making in this analysis is that each pulse peak is shifted in a different way, hence giving rise to an observable orbital phase-dependent effect. Any shift of the overall pulse profile would otherwise be absorbed in the timing model. As we can see in Equations~\ref{eqn:deltaA} and \ref{eqn:deltaL}, at a given epoch, the only time-dependent terms are related to the orbital phase. This implies that, in a very general way, latitudinal aberration causes different components of the pulse profile to experience shifts having sinusoidal signatures. They all have the same phase and their period is equal the orbital period of the system. The only difference lies in the amplitude of the effect, which varies as a function of spin longitude and this is precisely what causes the aberration of the profile.

For the analysis, we will consider that distortions of the pulse profile structure are relatively minor but that the separation between the two peaks, which are presumably emitted on opposite edges of the emission cone, are easier to detect. Note, however, that we will not assume any particular emission geometry in the Bayesian analysis presented below in \S\,\ref{s:bayesian}.

Technically, we measure the TOA of a peak by cross-correlating the folded time interval with a high signal to noise template just like for regular timing (see \S\,\ref{s:timing}). The difference here is that the half-pulse window centered on each component is timed independently (see Figure~\ref{f:separation_fit} for a sample fit). This process produces two time series of observed TOAs that are converted to spin longitudes, $\Phi_{0_1}$ and $\Phi_{0_2}$. We then subtract one from the other in order to search for changes in the relative separation between the peaks as a function of orbital phase. The time series we obtained is presented in Figure~\ref{f:separation}.

\afterpage{
\clearpage

\begin{figure}
 \centering
 \includegraphics[width=6in]{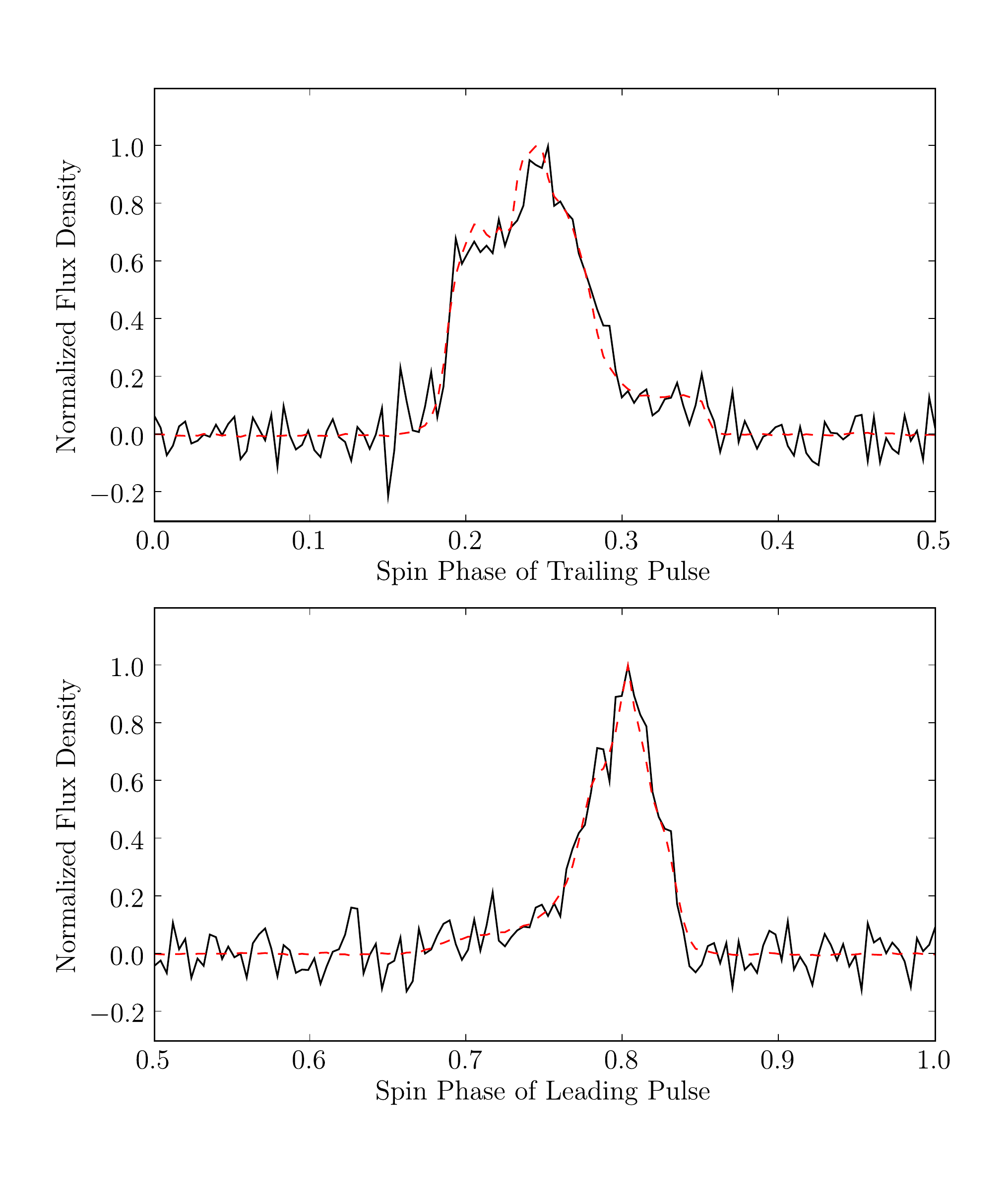}
 \caption[Sample fit of a fold interval to the pulse profile template of pulsar A]{Sample fit of a 30\,s fold interval (black) to the pulse profile template of pulsar A (dashed red). The relative spin phase of the leading and the trailing peaks of the pulse profile are independently found by cross-correlating the half-fold interval and the corresponding half-template.
 \label{f:separation_fit}}
\end{figure}

\clearpage
}

\afterpage{
\clearpage

\begin{figure}
 \centering
 \includegraphics[width=6in]{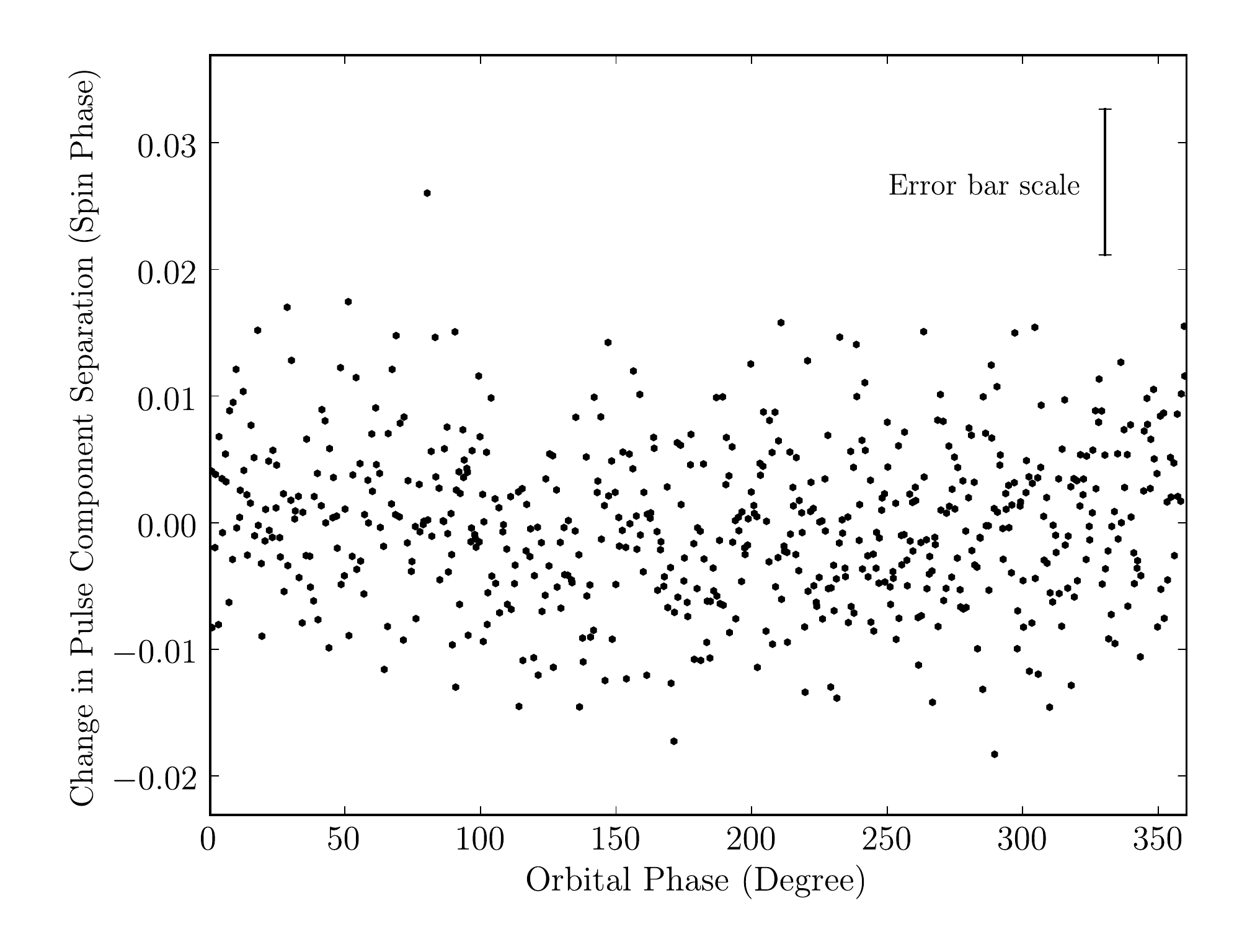}
 \caption[Measured relative separation between the two pulse components of pulsar A]{Change in relative separation between the two pulse components of pulsar A as a function of orbital phase. Each data point corresponds to an integration time of about 30 seconds.
 \label{f:separation}}
\end{figure}

\clearpage
}

We can rewrite the shift of a pulse component located at $\Phi_0$ (see Equation~\ref{eqn:phi0}) in a more practical form for fitting purposes. We ignore the lensing contribution (see \S\,\ref{s:obs_reduc}) and assume that the geometry of the system does not change --- this is the case for data obtained at a given epoch. In this case:
\begin{equation}\label{eqn:C}
   C \equiv \frac{1}{\sin \zeta \tan \chi_0} \,.
\end{equation}
This quantity is a constant that can be fitted and which acts like an amplitude parameter. More generally, if the emission geometry remains constant (i.e. $\alpha$ and $\rho$ are fixed), then we would expect $C$ to change only because of the evolution of the system geometry due to relativistic spin precession. This implies that $\zeta \to \zeta(t)$ and $\eta \to \eta(t)$. However, given the duration of our observation ($\sim 5$\,h) we can still assume that these two angles are constant since precession acts on a 75-year time scale. In this case, we can finally rewrite Equation~\ref{eqn:phi0} as:
\begin{equation}\label{eqn:phi0_simplified}
   \Delta \Phi_0 = \frac{CD}{\sin i} [\cos i \sin \eta (\cos \psi
                        + e \cos \omega) - \cos \eta (\sin \psi + e \sin
			\omega)] \,,
\end{equation}
where
\begin{equation}
   D = \frac{\Omega_b x_p}{\sqrt{1-e^2}}
\end{equation}
can be calculated directly since it depends on observed timing values only. Equation~\ref{eqn:phi0_simplified} leaves the model to be fitted with only two free parameters: $C and \eta$. Hence $\Delta \Phi_0 \to \Delta \Phi_0 (C, \eta)$. Both $C$ and $\eta$ are constant for a given observation and their predicted long time evolution due to relativistic spin precession can be derived quite easily.

\section{Bayesian Analysis}\label{s:bayesian}
In order to determine whether or not the change in separation between the two pulse components of pulsar A is detectable, and also to characterize the amplitude of this effect, we use a Bayesian analysis to perform model selection and parameter estimation.

From Baye's theorem \citep[see][for more details]{gre05a}:
\begin{equation}
   p(H_i|{\cal D},I) = \frac{p(H_i|I) p({\cal D}|H_i,I)}{p({\cal D}|I)} \,,
\end{equation}
where
\begin{eqnarray}
   p(H_i|{\cal D},I) & = & \textrm{posterior probability that hypothesis $H_i$ is
   true} \\
              & & \textrm{ given data ${\cal D}$ and prior information $I$.} \nonumber \\
   p(H_i|I)   & = & \textrm{prior probability of hypothesis $H_i$.} \\
   p({\cal D}|H_i,I) & = & \textrm{probability of obtaining data ${\cal D}$ given
   hypothesis $H_i$} \\
              & & \textrm{ and prior information $I$. $p({\cal D}|H_i,I)$ is also
   called} \nonumber \\
              & & \textrm{ likelihood.} \nonumber \\
   p({\cal D}|I)     & = & \sum_i p(H_i|I) p({\cal D}|H_i,I) \textrm{, probability of data
   ${\cal D}$ given} \\
              & & \textrm{ prior information $I$; normalization factor.} \nonumber
\end{eqnarray}

In the context of model comparison, the hypothesis $H_i$ becomes the model that is being considered.

\subsection{Constant Model, $M_0$}
The simplest model to consider is the constant separation model, which we refer as $M_0$. That is, for $M_0$, no change in pulse separation is present in the data and the measurement errors can account for the observed signal. Let $d_i$ be the measured half-separation between the pulse components from which the average has been subtracted and $\sigma_i$ the associated errors. Under the assumption of Gaussian errors the likelihood can be written as:
\begin{equation}
   p({\cal D}|M_0,I) = (2\pi)^{-\frac{N}{2}} \prod_i^N \sigma_i^{-1} \exp \left(-\sum_i^N
   \frac{d_i^2}{2 \sigma_i^2}\right) \,.
\end{equation}

The priors are simply $p(M_0|I) =  1$ because all models are assumed to be as probable as each other.

Therefore, the posterior probability\footnote{Note that for reasons that will become apparent later, we neglect the normalizing constant $p({\cal D}|I)$ in this equation and we write a pseudo posterior probability, $p^\prime(M_{0}|{\cal D},I)$, instead. The same thing will be done for the other considered models.} of this model is:
\begin{equation}
   \ln p^\prime(M_0|{\cal D},I) = \ln p(M_0|I) + \ln p({\cal D}|M_0,I) = 2179.46 \,.
\end{equation}

\subsection{Constant Model + Noise, $M_{0s}$}
We can imagine that our understanding of the signal is not perfect. One possibility is that our estimation of the errors does not reflect the real noise level. Another possibility is that a real signal, which is not necessarily understood, is present in the data and not accounted for by the constant separation model. For this reason, we can consider a constant separation model including an extra noise parameter having the following pseudo posterior probability:
\begin{equation}
   p^\prime(M_{0s}|{\cal D},I) = p(M_{0s}|I) \int_s ds\, p(s|M_{0s},I) p({\cal D}|M_{0s},s,I) \,.
\end{equation}

The pseudo posterior probability is calculated using the following quantities:
\begin{eqnarray}
   p(M_{0s}|I)     &=& 1 \,, \\
   p({\cal D}|M_{0s},s,I) &=& (2\pi)^{-\frac{N}{2}} \prod_i^N (\sigma_i^2 +
   s^2)^{-\frac{1}{2}} \exp \left(-\sum_i^N \frac{d_i^2}{2 (\sigma_i^2 +
   s^2)}\right) \,, \\
   p(s|M_{0s},I)   &=& \frac{1}{s + s_{low}} \frac{1}{\ln (\frac{s_{low} +
   s_{max}}{s_{low}})} \,.
\end{eqnarray}

We choose to use modified Jeffrey's priors for the noise parameter priors, $p(s|M_{0s},I)$. Jeffrey's priors imply an equal probability per decade (scale invariant), which is convenient for a noise parameter that can vary on different scales \citep{jef46,jay68}. Since we want to allow $s$ to be zero, the normal Jeffrey's priors ($1/s$) have to be modified in order to become linear at a sufficiently small $s_{low}$ value, otherwise the priors probability would blow up at $s=0$ \citep{gre05b}.

Because the typical uncertainty on the phase separation is $\left<\sigma_i\right>=0.0058$ in units of spin phase (see Figure~\ref{f:separation}), it is reasonable to let $s$ vary between $s_{min}=0$ and $s_{max}=0.05$, and set $s_{low}=0.005$ in units of spin phase\footnote{Typically, one wants to allow the unknown source of noise to scale between 0 and 10 times the average estimated uncertainty level in the data. A common choice for the linearization parameter, also known as the `knee', is to set it at 10\% the average estimated uncertainty level \citep[see, e.g.][]{gre05b}.}. We performed the numerical integration on $s$ using a simple 1000-point grid integration over the parameter space domain. We obtained:
\begin{equation}
   \ln p^\prime(M_{0s}|{\cal D},I) = 2185.77 \,.
\end{equation}

\subsection{Rafikov and Lai Model + Noise, $M_{2s}$}
Finally, the third model we investigated is the \citet{rl06b} model including a noise parameter. As we mentioned in \S\,\ref{s:simple_model}, the Rafikov and Lai model depends on two free parameters: $C$ and $\eta$. In this case, the pseudo posterior probability is written as:
\begin{equation}
   p^\prime(M_{2s}|{\cal D},I) = p(M_{2s}|I) \int_C dC \int_\eta d\eta \int_s ds\,
   p(C,\eta,s|M_{2s},I) p({\cal D}|M_{2s},C,\eta,s,I) \,,
\end{equation}
with
\begin{eqnarray}
   p(M_{2s}|I)            &=& 1 \,, \\
   p({\cal D}|M_{2s},C,\eta,s,I) &=& (2\pi)^{-\frac{N}{2}} \prod_i^N (\sigma_i^2 +
                              s^2)^{-\frac{1}{2}} \exp \left(-\sum_i^N
			      \frac{(d_i-\Delta \Phi_0)^2}{2 (\sigma_i^2 + 
			      s^2)}\right) \,, \\
   p(C,\eta,s|M_{2s},I)   &=& p(C|M_{2s},I)p(\eta|M_{2s},I)p(s|M_{2s},I) \,, \\
                          &=& \left (\frac{1}{C + C_{low}} \frac{1}{\ln
			      (\frac{C_{low} + C_{max}}{C_{low}})} \right) \times
			      \left (\frac{1}{\eta_{max}-\eta_{min}} \right) \times \nonumber \\
			              & & \left( \frac{1}{s + s_{low}} \frac{1}{\ln
			      (\frac{s_{low} + s_{max}}{s_{low}})} \right) \,.
\end{eqnarray}

Again, we choose modified Jeffrey's priors for the noise parameter priors, $p(s|M_{2s},I)$, as well as for the amplitude parameter priors, $p(C|M_{2s},I)$, since the amplitude is a scaling parameter. On the other hand, $\eta$ is an angle and thus flat priors are more adequate in this situation. $\Phi_0$ is calculated using Equation~\ref{eqn:phi0_simplified}.

For the noise parameter we again set $s_{min} = 0$, $s_{max} = 0.05$ and $s_{low} = 0.005$. For the amplitude parameter $C$ we set $C_{min} = 0$, $C_{max} = 5$ and $C_{low} = 0.01$. For the $\eta$ parameter we set $\eta_{min} = 0$ and $\eta_{max} = 2\pi$. The numerical integrations are done using a $100 \times 500 \times 100$-point grid for $\eta$, $C$ and $s$, respectively. We obtain:
\begin{equation}
   \ln p^\prime(M_{2s}|{\cal D},I) = 2185.48 \,.
\end{equation}

\subsection{Model Selection}
Model selection results from the calculation of the odds ratio of the different model. The results are:
\begin{eqnarray}
   &&O_{0s,0} = \frac{p(M_{0s}|{\cal D},I)}{p(M_{0}|{\cal D},I)} =
   \frac{p^\prime(M_{0s}|{\cal D},I)}{p^\prime(M_{0}|{\cal D},I)} = 549.59 \,, \\
   &&O_{2s,0} = \frac{p(M_{2s}|{\cal D},I)}{p(M_{0}|{\cal D},I)} =
      \frac{p^\prime(M_{2s}|{\cal D},I)}{p^\prime(M_{0}|{\cal D},I)} = 412.94 \,, \\
   &&O_{2s,0s} = \frac{p(M_{2s}|{\cal D},I)}{p(M_{0s}|{\cal D},I)} =
      \frac{p^\prime(M_{2s}|{\cal D},I)}{p^\prime(M_{0s}|{\cal D},I)} = 0.75 \,. \\
\end{eqnarray}

As we can see, the constant separation model $M_0$ is very unlikely to explain the data. However, the constant separation model having extra noise $M_{0s}$ and the Rafikov and Lai model having extra noise $M_{2s}$ are both relatively as likely, although the odds slightly favor $M_{0s}$. In this situation, the Occam's razor tells us that we should choose the simpler model and conclude that there is no evidence of a detectable change in the separation between the pulse peaks that can be explained by the Rafikov and Lai model.

\subsection{Upper Limit on the Amplitude Parameter $C$}\label{s:c_upper_limit}
Even though there is no evidence for a shift of the pulse components due to latitudinal aberration, we can put an upper limit on the amplitude of this effect. Here, we are interested in the marginal posterior probability function of the amplitude parameter $C$. The angle $\eta$ and the extra noise term $s$ can be considered as nuisance parameters and can thus be marginalized. Hence:
\begin{equation}
   p(C|M_{2s},{\cal D},I) = \frac{p(C|M_{2s},I) \int_\eta d\eta \int_s ds\,
   p(\eta,s|M_{2s},I) p({\cal D}|M_{2s},C,\eta,s,I)}{p({\cal D}|M_{2s},I)} \,.
\end{equation}

Figure~\ref{f:c_pdf} shows the resulting marginal posterior probability function of $C$. From this, we find the following upper limits: $C=0.13$, $1.35$ and $4.65$, which correspond to the $1\sigma$, $2\sigma$ and $3\sigma$ confidence levels, respectively.

\afterpage{
\clearpage

\begin{figure}
 \centering
 \includegraphics[width=6in]{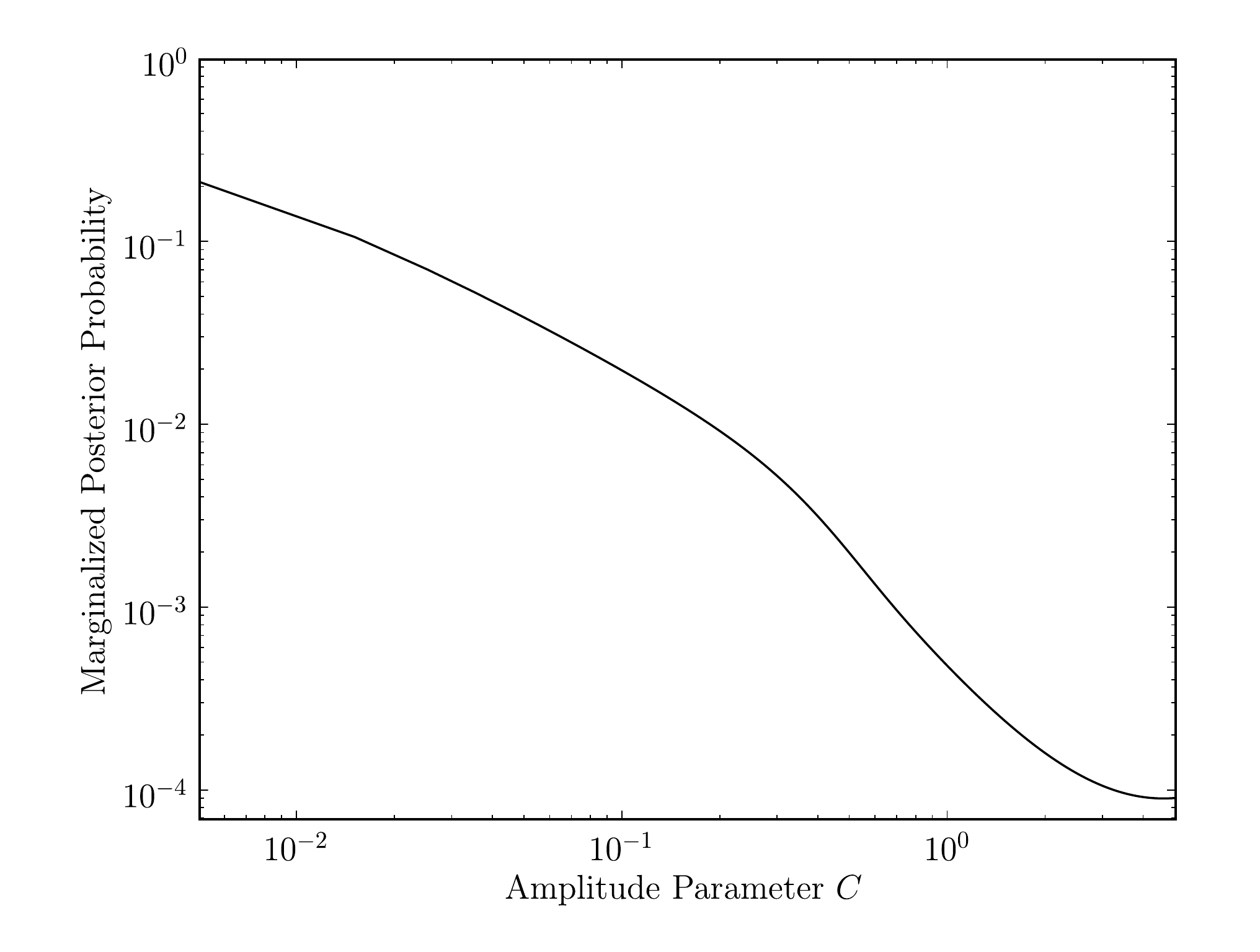}
 \caption[Marginal posterior probability function of the amplitude parameter $C$]{Marginal posterior probability function of the amplitude parameter $C$ for the Rafikov and Lai model including an extra noise parameter ($p(C|M_{2s},{\cal D},I)$).
 \label{f:c_pdf}}
\end{figure}

\clearpage
}

We observe that $p(C|M_{2s},{\cal D},I)$ drops rapidly and then reaches a plateau. This makes the $1\sigma$ upper limit relatively constraining but it leaves the $2\sigma$ and $3\sigma$ upper limits more unconstrained. This behavior is partly related to the high covariance between the amplitude parameter $C$ and the angle $\eta$ as we can see in Figure~\ref{f:cs_pdf}, which shows the joint posterior probability function $p(C,s|M_{2s},{\cal D},I)$ for these two parameters. When the angle $\eta$ approaches $\pi/2$ or $3\pi/2$ the colatitudinal variation of the emission vector $\Delta \zeta$ becomes close to zero and hence the phase shift $\Delta \Phi_0$ remains very small for a wide range of amplitudes $C$ (see Figure~\ref{f:eta} for an example).

It is interesting to note that the lack of detection of latitudinal aberration does not necessarily imply that, with the same observing setup, one could not detect it in the future. Indeed, $\eta$ is expected to change with time because of relativistic spin precession and hence one possibility is that no aberration was seen at the time that our data were obtained because of a (un)fortunate geometric configuration.

\afterpage{
\clearpage

\begin{figure}
 \centering
 \includegraphics[width=6in]{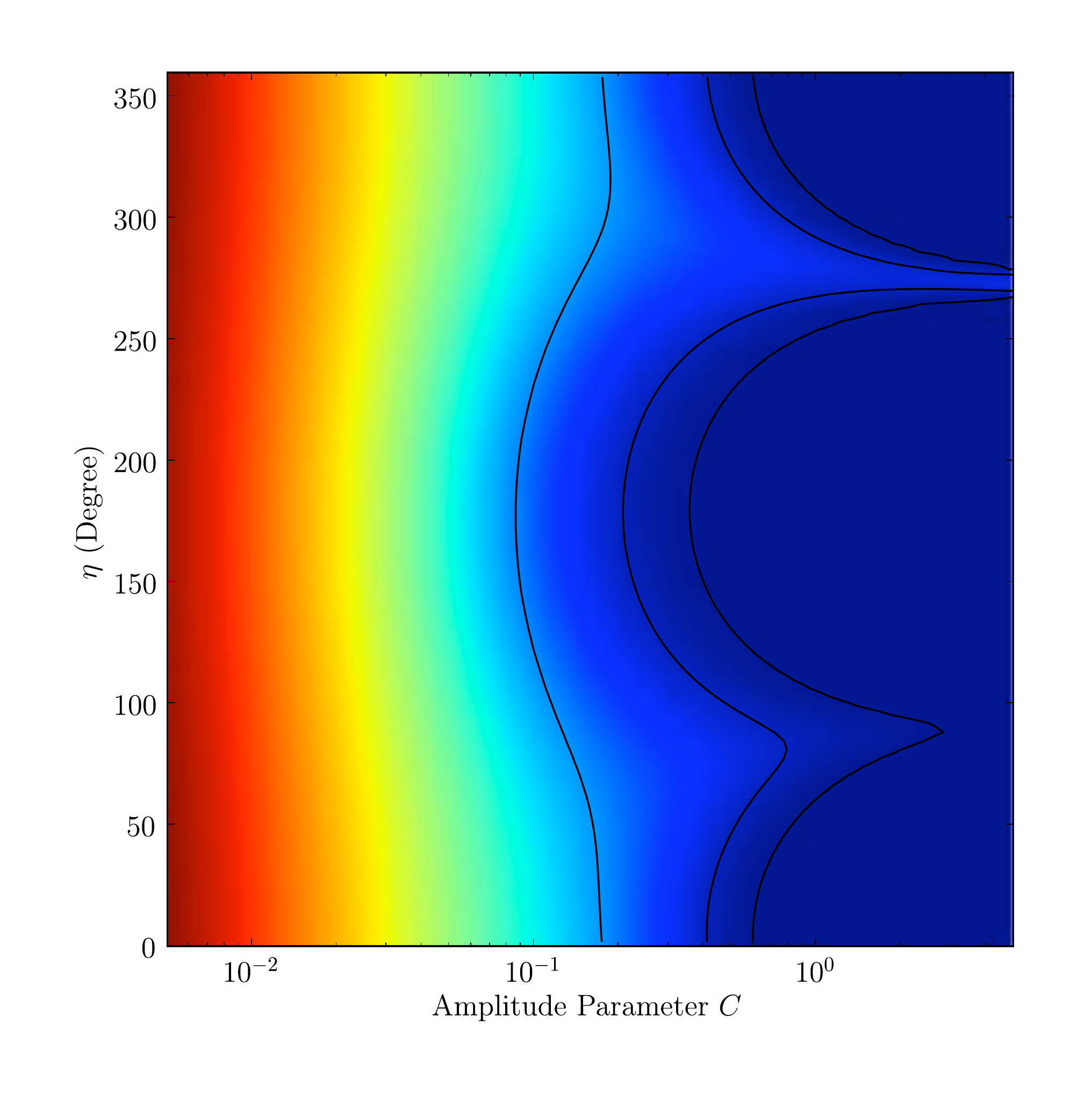}
 \caption[Joint posterior probability function of the amplitude parameter $C$ and the angle $\eta$]{Joint posterior probability function of the amplitude parameter $C$ and the angle $\eta$ for the Rafikov and Lai model including an extra noise parameter ($p(C,s|M_{2s},{\cal D},I)$). Dark red regions are the most probable and blue regions are the least probable. Contours are drawn for the $1\sigma$, $2\sigma$ and $3\sigma$ confidence regions.}
 \label{f:cs_pdf}
\end{figure}

\clearpage

\begin{figure}
 \centering
 \includegraphics[width=6in]{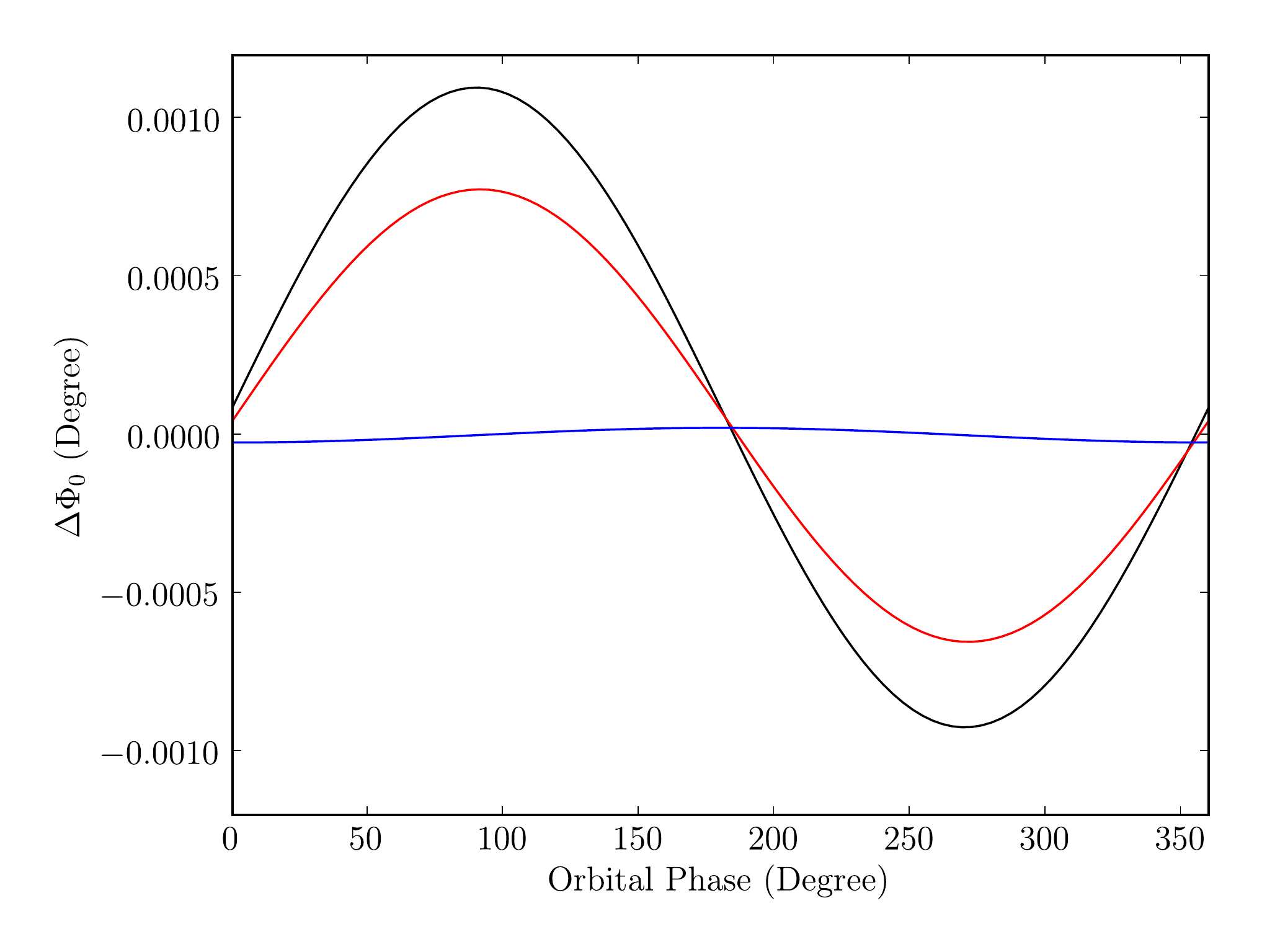}
 \caption[Effect of $\eta$ on the pulsed component separation]{$\Delta \Phi_0$ as a function of orbital phase for different values of $\eta$ and a fixed value of the amplitude parameter $C$. $\eta = 0$, $45$ and $90^\circ$ for the black, red and blue curves, respectively.}
 \label{f:eta}
\end{figure}

\clearpage
}

\section{Discussion}
In the previous Bayesian analysis, we refer to the $M_{2s}$ model as the Rafikov and Lai model. After working a quick algebraic rearrangement of Equation~\ref{eqn:phi0_simplified}, one can realize that it actually consists of a simple sinusoid having an unknown phase and amplitude. This leaves the interpretation of the results quite general though in the context of the Rafikov and Lai model, we can relate the parameters to physically meaningful quantities. The amplitude parameter $C$ depends on three quantities: the magnetic inclination, $\alpha$, the colatitude of the emission vector, $\zeta$, and the spin longitude $\Phi_0$. The other free parameter, $\zeta$, is related to the well-known impact parameter $\beta$, which is used in the rotating vector model (RVM), as following: $\zeta = \alpha + \beta$.

\subsection{Single-Cone Emission}
As pointed out by \citet{rl06b}, the simplest emission geometry to consider is that coming from a single cone. The remarkable symmetry between the two pulse peaks suggests that the emission is produced at the boundary of a hollow cone. As this cone sweeps across our line of sight it naturally produces two main features --- one at the leading edge and the other one at the trailing edge --- that are characterized by a steep rise on the outer side and slower decay in the inner side. The recent detection of X-ray pulsations from pulsar A by \citet{cgm+07} also appears to support this idea because the X-ray pulse profile, which is similar to its radio counterpart, presents hints of X-ray emission bridging the two peaks in what is believed to be the inner part of the cone.

In this context, interpreting the result of the Bayesian analysis is fairly trivial. Because of the symmetry in the pulse profile, we impose that the leading and trailing peaks are located at spin longitude $-\Phi_0$ and $\Phi_0$, respectively (see Figure~\ref{f:aberration_profile} for an schematic view of the model). Consequently, they experience shifts of $\Delta \Phi_0$ and $-\Delta \Phi_0$, respectively, and hence the separation between the two components is $2\Phi_0 + 2\Delta \Phi_0$. In the previous analysis we fitted the Rafikov and Lai model for variations around the mean of the half-separation, which implies that we measured $\Delta \Phi_0$. Therefore, the value of the amplitude parameter $C$ directly corresponds to that of Equation~\ref{eqn:C}. Furthermore, as we discussed above, the upper limit on the amplitude parameter $C$ given in \S\,\ref{s:c_upper_limit} yields a direct joint constraint on $\alpha$ and $\zeta$. Figure~\ref{f:alpha_zeta} shows the joint probability of $\alpha$ and $\zeta$ derived from the upper limit on $C$. This result is suggestive that the value of $\alpha$ is restricted to be close to $90^\circ$ whereas the value of $\zeta$ is left relatively unconstrained. Following from this joint constraint is the implied half-opening angle of the emission cone, which is obtained using the following relation:
\begin{equation}\label{eqn:rho}
 \cos \rho = \cos \Phi_0 \sin \zeta \sin \alpha + \cos \zeta \cos \alpha \,,
\end{equation}
where the observed half-separation between the two peaks is measured to be $\Phi_0 = 98^\circ$ (from peak to peak, see Figure~\ref{f:separation}). For comparison, the same joint constraint is also presented in Figure~\ref{f:alpha_zeta90} for $\Phi_0 = 90^\circ$.

\afterpage{
\clearpage

\begin{figure}
 \centering
 \includegraphics[width=6in]{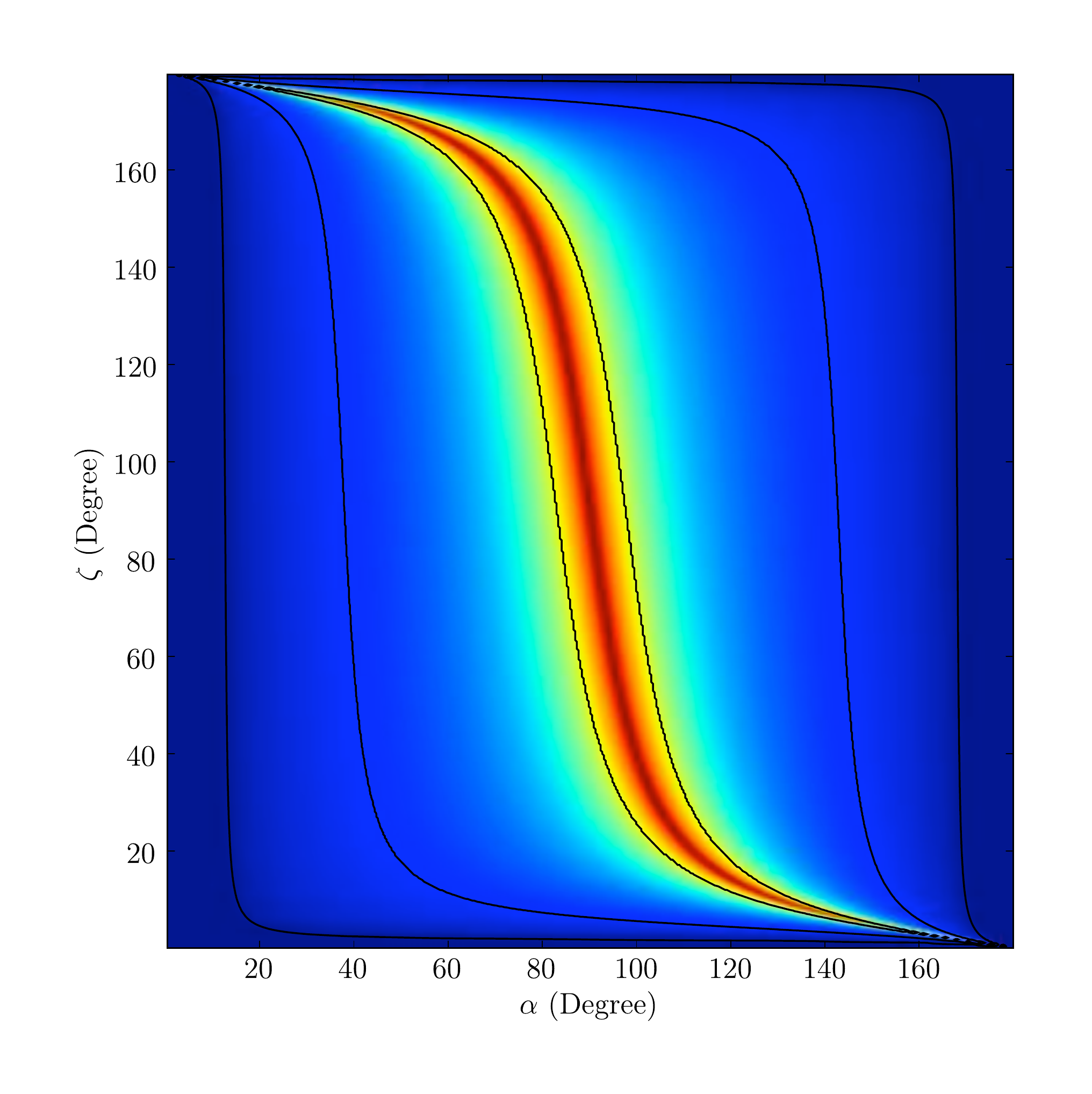}
 \caption[Joint probability of $\alpha$ and $\zeta$ (single cone, $\Phi_0 = 98^\circ$)]{Joint posterior probability of the magnetic inclination, $\alpha$, and the colatitude of the emission vector, $\zeta$, derived from the upper limit on the amplitude parameter $C$ for the single-cone emission geometry using $\Phi_0 = 98^\circ$. Dark red regions are the most probable and blue regions are the least probable. Contours are drawn for the $1\sigma$, $2\sigma$ and $3\sigma$ confidence regions.}
 \label{f:alpha_zeta}
\end{figure}

\clearpage

\begin{figure}
 \centering
 \includegraphics[width=6in]{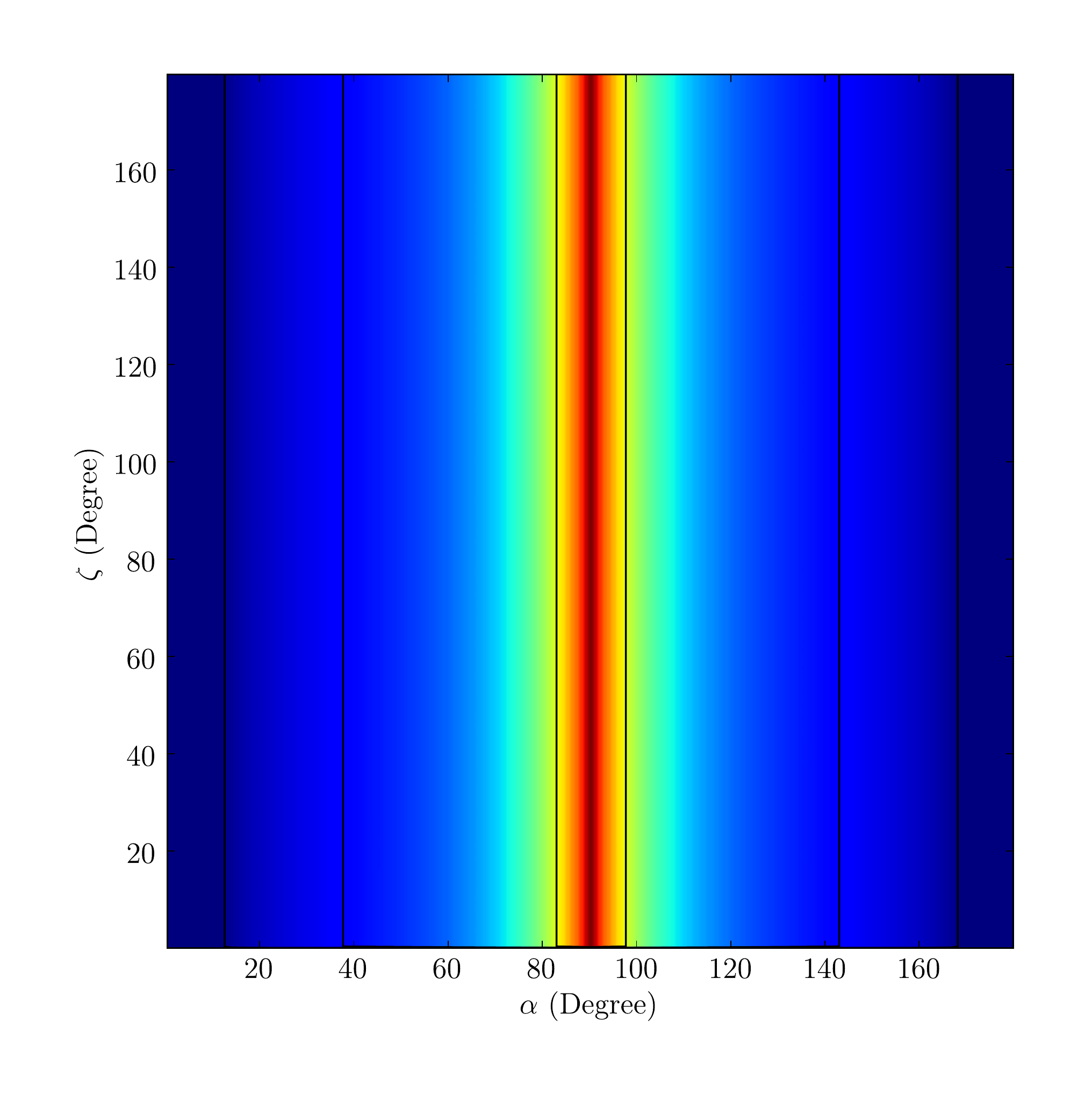}
 \caption[Joint probability of $\alpha$ and $\zeta$ (single cone, $\Phi_0 = 90^\circ$)]{Joint probability of the magnetic inclination, $\alpha$, and the colatitude of the emission vector, $\zeta$, derived from the upper limit on the amplitude parameter $C$ for the single-cone emission geometry using $\Phi_0 = 90^\circ$. Dark red regions are the most probable and blue regions are the least probable. Contours are drawn for the $1\sigma$, $2\sigma$ and $3\sigma$ confidence regions.}
 \label{f:alpha_zeta90}
\end{figure}

\clearpage
}

Although the joint constraint on $\alpha$ and $\zeta$ is relatively poor at a high confidence level, we can compare this result to the study by \citet{drb+04} of the linear polarization. They also report a joint constraint on $\alpha$ and $\zeta$ based on the application of the rotating vector model (RVM) to the position angle of the linear polarization (see Figure~\ref{f:demorest} for an illustration of the joint probability that they obtained).

\afterpage{
\clearpage

\begin{figure}
 \centering
 \includegraphics[width=6in]{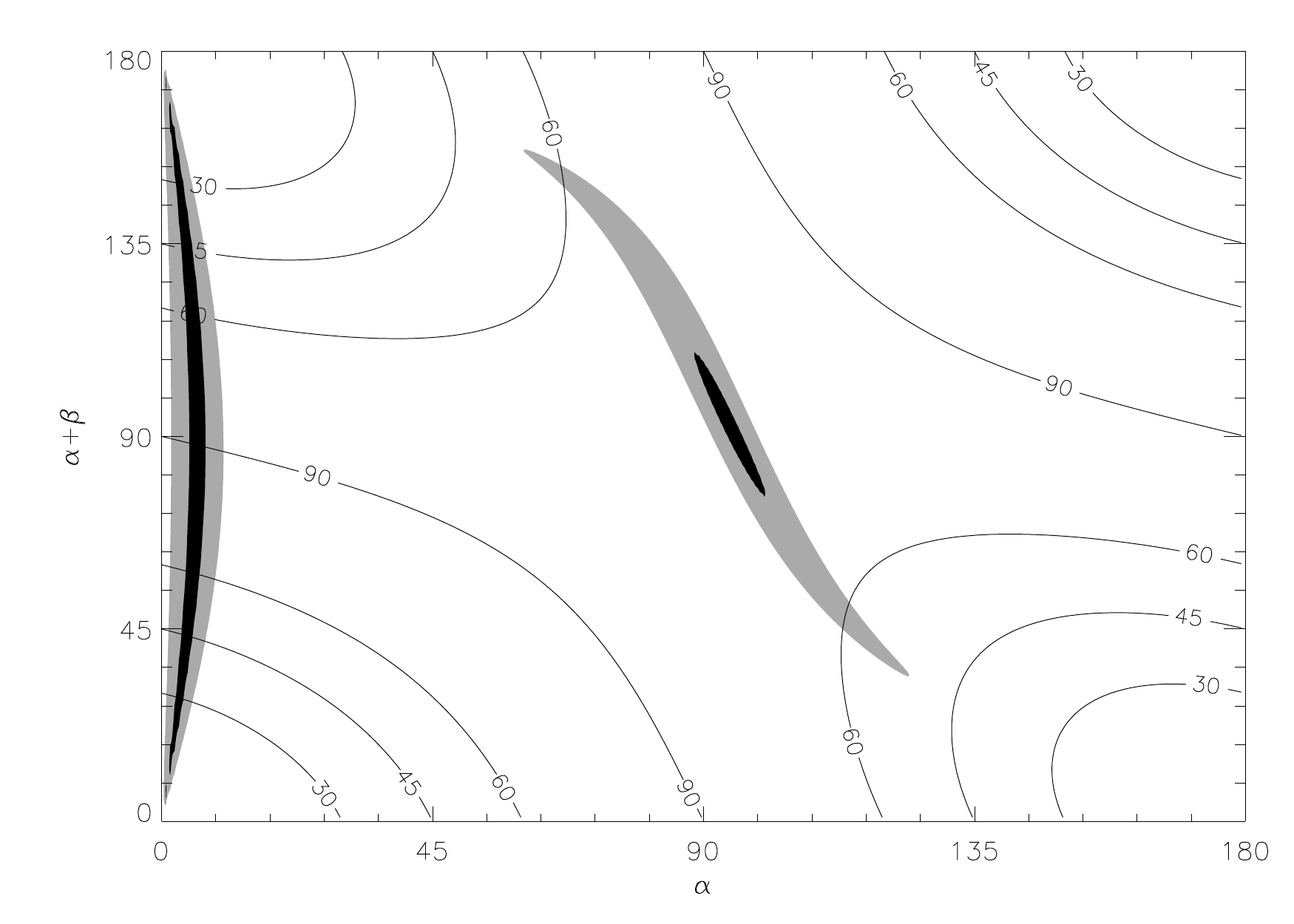}
 \caption[Geometrical constraints from the RVM]{The goodness of fit of the RVM applied to the position angle of the linear polarization of pulsar A's pulse profile. The filled contours represent the $1\sigma$ and $3\sigma$ confidence limits (dark and light gray, respectively). The line contours show the opening half-angle $\rho$ of the emission cone. Credit: \citet{drb+04}.}
 \label{f:demorest}
\end{figure}

\clearpage
}

The combined probability from our work and that of \citet{drb+04} makes the low $\alpha$ solution found by \citet{drb+04} very unlikely and strongly favors $\alpha \sim 90^\circ$. The colatitude of the emission vector $\zeta$ remains more or less unconstrained, although the probability reaches a maximum near $90^\circ$. This large $\alpha$ value, together with the observed spin longitude $\Phi_0 = 98^\circ$, yields $\rho \sim 90^\circ$. This value is relatively independent of $\zeta$ for $\alpha \sim 90^\circ$.

The implication of $\zeta \sim 90^{\circ}$ is that pulsar A's spin axis is parallel to the sky plane. Because no pulse profile changes were found \citep{mkp+05}, this naturally imposes that the spin axis is almost coincident with the orbital angular momentum since we know that we observe the system almost perfectly edge-on ($i \sim 90^{\circ}$). In the case of $\alpha, \zeta = 90^{\circ}$, pulsar A would be an orthogonal rotator. Our analysis also indicates that $\rho \sim 90^{\circ}$, which would imply an extremely wide emission beam that would in fact look like a ring around the pulsar's magnetic equator. Obviously, we question whether or not this kind of configuration is physically possible. In this case, it is unclear why we would not see the emission from both magnetic poles. Although our joint constraint on $\zeta$ and $\alpha$ is not stringent at a high confidence level, it appears unlikely that pulsar A has a narrow emission beam since narrow beam configurations favor larger aberration amplitudes, and hence are less probable.

\subsection{Two-Cone Emission}
Another possibility is that the two pulse components are emitted by different cones. In order to restrain the number of free parameters in this two-cone emission geometry, we can make the simple assumption that the north and the south poles are diametrically opposed to each other (i.e. perfect dipole) and that their associated emission cones are identical (i.e. they have the same half-opening angle). In this case, the pulse features are created by the intersection of our line of sight with the edge of each cone. It is interesting to note that unless the impact parameter, $\beta$, is precisely equal to the half-opening angle of the cones (see Figure 3 in \citet{rl06b}), our line of sight will intersect the edge of each cone twice. One would therefore expect four and not only two pulse components to arise from such a geometry. To keep the model as general as possible, we delay this discussion to later and simply assume that only one edge of each cone is ``bright'', i.e. is contributing to the observed pulse profile.

In this two-cone emission geometry framework, we consider the latitudinal aberration from each cone separately and deal with their associated pulse component, labeled 1 and 2, respectively. Hence, the equation for the variation in spin longitude of each pulse component can be written as:
\begin{equation}\label{eqn:delta_phi0_two}
 \Delta \Phi_{0_{12}} = - \Delta \zeta \left(
 \frac{1}{\sin \zeta_2 \tan \chi_{0_{2}}} -
 \frac{1}{\sin \zeta_1 \tan \chi_{0_{1}}}
 \right) \,.
\end{equation}

Since $\Delta \zeta$ is independent of the emission beam geometry (see Equation~\ref{eqn:delta_zeta} for a reference), we find that the two-cone emission geometry presents the same sinusoidal signature as in the single-cone emission geometry; i.e. it has the same period and phase. In this case, however, the physical interpretation of the ``amplitude'' parameter, $C$, is slightly different since it corresponds to the expression in brackets of Equation~\ref{eqn:delta_phi0_two}.

Because of the symmetries imposed to the model, we can relate the following quantities:
\begin{eqnarray}
 \rho_2 &=& \rho_1 \\
 \eta_2 &=& \eta_1 \\
 \zeta_2 &=& \zeta_1 \\
 \alpha_2 &=& \pi - \alpha_1 \\
 \Phi_{0_2} &=& \Phi_{0_1} + (s - \pi) \,,
\end{eqnarray}
where $s$ is the observed separation between the two pulse components.

This simplifies the equation for the amplitude parameter:
\begin{eqnarray}
 C &=& \frac{1}{\sin \zeta_2 \tan \chi_{0_{2}}} -
  \frac{1}{\sin \zeta_1 \tan \chi_{0_{1}}} \\
   &=& \frac{1}{\sin \zeta_1} \left(\frac{1}{\tan \chi_{0_{2}}} -
    \frac{1}{\tan \chi_{0_{1}}} \right) \,,
\end{eqnarray}
with
\begin{equation}\label{eqn:eq1}
 \tan \chi_{0_1} = \frac{\sin \alpha_1 \sin \Phi_{0_1}}{\cos \alpha_1 \sin
 \zeta_1 - \cos \Phi_{0_1} \sin \alpha_1 \cos \zeta_1} \,,
\end{equation}
and
\begin{equation}\label{eqn:eq2}
 \tan \chi_{0_2} = \frac{\sin \alpha_1 \sin (\Phi_{0_1} + (s - \pi))}
 {-\cos \alpha_1 \sin \zeta_1 -
 \cos \Phi_{0_1} \sin \alpha_1 \cos \zeta_1} \,.
\end{equation}

This leaves three unknown: $\zeta_1$, $\alpha_1$ and $\Phi_{0_1}$. However, we can make use of the $\rho_1 = \rho_2$ identity and its definition (Equation~\ref{eqn:rho}) to obtain one additional constraint:
\begin{equation}\label{eqn:eq3}
 \cos(\Phi_{0_1} + (s - \pi)) - \cos \Phi_{0_1} = \frac{2}{\tan \zeta_1
 \tan \alpha_1} \,,
\end{equation}
which reduces the number of free parameters to two. This is just as much as in the single-cone emission geometry. This means that for any combination of $\alpha_1$ and $\zeta_1$, we can solve Equations~\ref{eqn:eq1}, \ref{eqn:eq2} and \ref{eqn:eq3} in order to obtain an associated $C$ value. Provided an upper limit on $C$, we can derive credible joint $\alpha_1 - \zeta_1$ confidence regions (see the results in Figure~\ref{f:alpha_zeta_two_pole}). Note that Equation~\ref{eqn:eq3} imposes stringent restrictions on possible $\alpha_1$ and $\zeta_1$ values since any combination of these two parameters, given an $s$ value, do not necessarily lead to a valid solution. Also, for each possible $\alpha_1 - \zeta_1$ value, there are two associated $\Phi_{0_1}$, which implies two different $\rho_1$. The second solution, however, is trivially related to the first one --- $\alpha_2 = \pi - \alpha_1$ --- and corresponds to a simply interchange of the two poles.

\afterpage{
\clearpage

\begin{figure}
 \centering
 \includegraphics[width=6in]{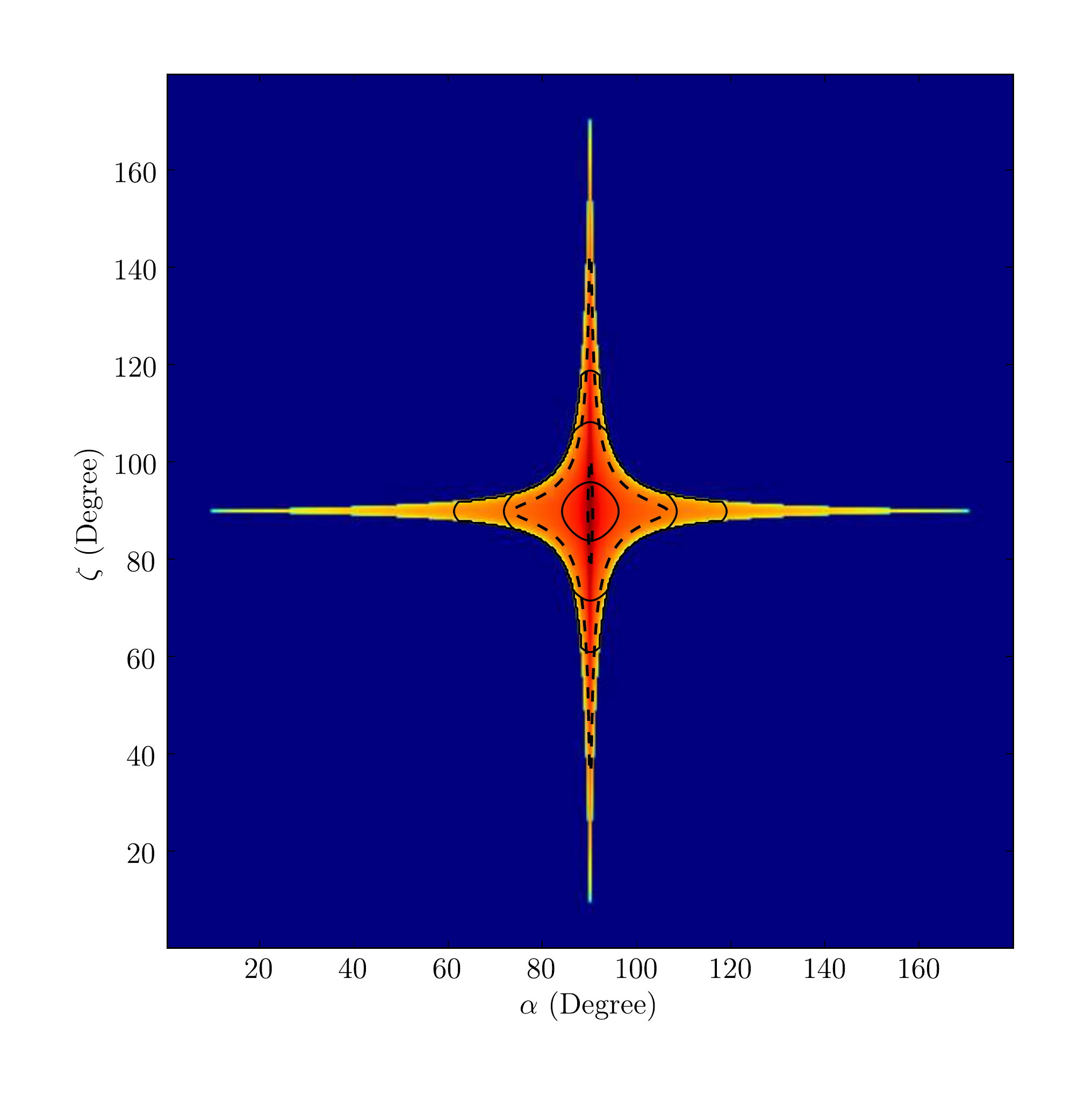}
 \caption[Joint probability of $\alpha$ and $\zeta$ (double cone, $\Phi_0 = 98^\circ$)]{Joint posterior probability of the magnetic inclination, $\alpha$, and the colatitude of the emission vector, $\zeta$, derived from the upper limit on the amplitude parameter $C$ for the two-cone emission geometry. The observed separation between the two pulse components is $s=196^{\circ}$ (i.e. $2 \times 98^{\circ}$, as assumed in the single-cone emission geometry). The color scale is such that red is more probable and green is less probable. The white area is excluded because it does not satisfy the geometrical constraint of the model. Dashed contours indicate the $1\sigma$ and $3\sigma$ confidence regions. Plain contours, going outward, indicate half-opening angles $\rho=10$,$20$ and $30^{\circ}$, respectively.}
 \label{f:alpha_zeta_two_pole}
\end{figure}

\clearpage
}

Surprisingly, the two-cone emission geometry also favors the orthogonal rotator scenario with a spin axis closely aligned with the orbital angular momentum. In this case, the implied emission cone is much smaller and is in better agreement with the commonly accepted pulsar emission picture. Given the span in spin longitude of each pulse component, it appears that the radio emission comes from a region that has a width comparable the emission cone opening. Hence, even though our line of sight crosses the edge of each emission cone twice, there is no need to fine tune the model since the emission can simply come for the whole slice of the cones rather than from their edges. The caveat of this scenario, however, is that it is more difficult to reconcile with the X-ray observations that show a bridge of emission between the two pulse components \citep{cgm+07}.

\section{Conclusion}
We performed an analysis of the pulse profile of pulsar A in the double pulsar PSR~J0737$-$3039A/B that was specifically targeted at identifying an orbital dependence in the separation between the two peaks of its pulse profile. A Bayesian analysis of the data showed that they are consistent with no sinusoidal variability of the pulse components separation with a period equal to the orbital period of the binary system. Such a variability would result from latitudinal aberration \citet{rl06b}. However, it appears that an additional noise term is required to account for the data variability. This is likely due to the fact that the uncertainties on the measured TOAs were underestimated.

From the constraint on the amplitude parameter $C$ of a putative sinusoidal variation in the separation between the two pulse components, we can infer interesting constraints on the emission geometry. First, in the case of a single-cone emission geometry, we find that pulsar A is likely an orthogonal rotator, whose spin axis is probably aligned with the orbital angular momentum of the system. This geometry argues against the possibility of a narrow emission beam, instead favoring an anomalously wide, ring-like cone. Second, we generalized the single-cone emission geometry proposed by \citet{rl06b} to a two-cone emission geometry. In this case, we also conclude that pulsar A is likely an orthogonal rotator and that its spin axis probably coincides with orbital angular momentum. This alternative emission geometry imposes more stringent constraints than the other one since both emission cones requires to be intercepted by our line of sight. It appears that much narrower emission cones would be possible in this two-cone model. With data having better sensitivity, such as what would be recorded by a coherent dedispersion back-ends, better geometrical constraints could be obtained. A similar analysis conducted on data acquired at a different epoch would also help improving the constraints since it would provide a coherent coverage of the $\eta$ parameter space, which is expected to vary because of relativistic spin precession.


\chapter{Conclusion}\label{c:conclusion}

\begin{flushright}
 \begin{singlespace}
 \emph{``I love to rove amidst the starry height,\\
 To leave the little scenes of Earth behind,\\
 And let Imagination wing her flight\\
 On eagle pinions swifter than the wind.\\
 I love the planets in their course to trace;\\
 To mark the comets speeding to the sun,\\
 Then launch into immeasurable space,\\
 Where, lost to human sight, remote they run.\\
 I love to view the moon, when high she rides\\
 Amidst the heav'ns, in borrowed lustre bright;\\
 To fathom how she rules the subject tides,\\
 And how she borrows from the sun her light.\\
 O! these are wonders of th' Almighty hand,\\
 Whose wisdom first the circling orbits planned.''}
 \end{singlespace}
 Love of Night, T. Rodd
 \vspace{0.5in}
\end{flushright}

Pulsars, and more particularly those in binary systems, are highly praised astronomical objects to study astrophysics but also to investigate fundamental physics. The reasons making pulsars so attractive can be summarized by a few key attributes: 1) small, massive and compact, 2) large magnetic field, 3) fast rotation and 4) beamed, coherent emission. The first three features allow one to study the physics of dense matter and a wide variety of astrophysical phenomena involving strong gravitational forces and electromagnetic processes. The last characteristic, about the nature of the pulsar emission itself, is central in the context that astrophysics is an observation-driven science. That pulsar emission is modulated at the rotational period of the pulsar enables a myriad of measurements that would not be possible if they were emitting persistent electromagnetic radiation only.

In this regard, pulsars in binary systems constitute, at some level, a perfect cocktail of physical and observational ingredients. The regular tick of pulsars permits one to deduce the presence of a companion and to infer its orbital motion with an astonishing precision. The large density and the small size of pulsars allow them to be found in compact, relativistic orbits. When these two elements are combined, one can expect to observe relativistic effects and test them against the prediction of general relativity and alternative theories of gravity.

So far, relativistic binary pulsars have been celebrated laboratories to probe general relativity via its effects on the orbital motion of pulsars. In this thesis, we presented in Chapter~\ref{c:0737_eclipse} the results of research attacking a very different ground: the effect of general relativity on the spin of a pulsar. In relativistic theories of gravity, the spin and the orbit of an object in a binary system are no longer independent and the coupling of spin to the orbit manifests itself, among other things, as a precession of the spin axis around the total angular momentum of the system.

Determining the spatial orientation of a 10-km pulsar located about 1700\,light-years away from Earth is not such a trivial task to perform. Thanks to an amazing cosmic coincidence, we have been to probe the precise geometry of a pulsar by modeling the eclipses visible in the double pulsar system PSR~J0737$-$3039A/B. Because two pulsars are visible in this system, we were able to use the modulation of the light from the pulsar behind, which is partly absorbed in the magnetosphere of the pulsar passing in front, in order to reconstruct the spatial orientation of the latter. Using data obtained over a four-year monitoring campaign, we were able to show that the pulsar precesses in a way consistent with general relativity. Also, with the help of other parameters measured from the timing of the pulsars in this system, we derived a ``theory-independent'' measurement of the strength of the spin-orbit coupling in a gravitating object. This work is complementary to parallel research conducted in the Solar System since pulsars in relativistic binary systems are so far the most extreme gravitational field environment in which these measurement were obtained. The manifestation of relativistic effects depends on the orbital periods, the masses and the orbital velocities involved in the problem (see \S\,\ref{s:relativistic_timing}) and in that respect, double neutron star systems lie one to several orders of magnitude ahead of our Solar System (see \S\,\ref{a:strong_field}). On the other hand, because of their large densities, pulsars also present a different regime of matter and the behavior of gravity could, in certain theories, depart from that of general relativity when tested in such extreme conditions. We also measured the direction of relativistic spin precession with respect to the spin and orbital angular momenta directions and showed that it also agrees with general relativity.

Obtaining such a knowledge about the geometry of the pulsar in space also has interesting astrophysical implications. We estimated the inclination of the orbit based of the eclipse geometry and found that it agrees, within uncertainties, with the value derived from the timing. As pulsars precess, their emission beam intercepts our line of sight at the same location. We can make use of this property to estimate the visibility duty cycle of such a pulsar. In the event that relativistic spin precession would cause the emission beam to miss our line of sight, we can make predictions as to when the pulsar will become visible again.

We plan to follow up this study of the eclipses. Among the goals that we pursue, we wish to reconstruct the emission beam of pulsar B using the changes in viewing geometry induced by relativistic spin precession. Although one can argue that the strong interaction between the two pulsars would make the results very specific to this system, it would certainly provide additional clues about the physical processes responsible for the radio pulsar emission. Our knowledge about the geometry of pulsar B could also provide essential information to understand the mechanism responsible for the X-ray pulsations of pulsar B that have recently been detected in specific portion of the orbit \citep{ptd+08}. We also wish to investigate in more details the consequences of our modeling on the magnetospheric physics of the pulsar. Potential work would involve tomographic reconstruction of the magnetosphere using the prior information about the orientation of the pulsar's spin axis. This could help to understand `second-order' deviation from the dipolar geometry, which we assume in the eclipse model.

Relativity can be tested using relativistic binary pulsars but one can also assume that relativity is correct and use its predictive power as a tool to study pulsars. We illustrated this in Chapter~\ref{c:0737_aberration} with a study of latitudinal aberration in the pulse profile of pulsar A in the double pulsar PSR~J0737$-$3039A/B. Despite the fact that no detection of orbital phase-dependent pulse profile variation was made, we showed that, even with data of restricted quality, obtaining an upper limit on the amplitude of this effect can yield interesting results. Given the large separation between the two pulse components, the non-detection of latitudinal aberration imposes restrictions on the radio emission geometry. It suggests that the canonical hollow cone model is not viable if both pulse components come from the same cone, unless the emission beam is surprisingly wide. Alternatively, the emission could be produced by two opposite poles. In this case the emission cones would have moderate openings, but the X-ray observations would be more difficult to reconcile. This paradox highlights that even though the pulsar toy model works surprisingly well in general, there are still many missing pieces to the puzzle.

Among the long-standing pulsar problems is the origin of the low magnetic field in binary pulsars and MSPs. Certainly, the concept of pulsar recycling in binary systems is well established, but several questions remain unanswered. What happened to the MSPs that are isolated? Is recycling the only cause for the low magnetic field pulsars or would a secondary mechanism be possible? Is the magnetic field temporarily buried/screened or it is permanently suppressed? How important are the initial conditions and the amount of accreted mass on the `recycling' process? Understanding the characteristics of the binary pulsar population is crucial to shed light on many of these questions and part of the solution likely lies in understanding the differences between `normal' binary systems and those presenting `odd' properties.

We presented a contribution to this quest in Chapter~\ref{c:1744} with the study of the binary pulsar PSR~J1744$-$3922. Our work demonstrated that, as opposed to the `standard' scenario, which predicts fast spin periods and low magnetic fields for binary pulsars with low-mass companions, some of these pulsars are found with long spin periods and large magnetic fields. We showed that there might be a small sub-population of such binary pulsars. Their exact formation mechanism is still uncertain though. We found that binary pulsars in circular orbits present a power-law relationship relating their magnetic fields and their spin periods. The possible `new' class of binary that we identified is no exception. Since this correlation does not exist in normal isolated pulsars and in eccentric binaries, this hints that all circular binaries must share a key evolution mechanism. Interesting clues might lie in the nature and properties of their companions. In this view, we are planning to focus future research on these binary pulsar companions. We have already obtained optical observations of PSR~J1744$-$3922's companion that should provide better constraints on its nature.

\subsubsection{Concluding Remark}
Pulsar research, like other branches of astrophysics, certainly contributes to pushing further the limits of our understanding of the Universe. Even though pulsar astrophysics is a very narrow area of research, it is amazing to realize how many breakthroughs have been made by studying these dead stars. In their afterlife, massive stars are far from being boring: they display a broad variety of extreme characteristics and behaviors that are challenging to understand. At the same time, pulsars also provide tools to address some fundamental physics questions such as the validity of general relativity. In this thesis, we attempted to make a modest step forward in both directions in order to decipher the nature of pulsars and use their properties to understand our Universe.

\appendix


\chapter{Remarks About the Notion of Strong-Field Regime}\label{a:strong_field}

The concept of a `strong-field' regime is arguably loosely defined. In one of its more `extreme' interpretations, `strong field' means that the body under consideration is located only a few Schwarzschild radii ($R_S$) away from the gravitational source. In the case of the double pulsar PSR~J0737$-$3039A/B, which is the most relativistic binary pulsar known to date, the mean orbital separation is of the order of $8 \times 10^{10}$\,cm. This is about $2 \times 10^{5} R_S$, far from a few Schwarzschild radii. However, to put this into perspective, Mercury is located $\sim 4 \times 10^{7} R_S$ from the Sun, which is 200 times larger than the double pulsar's orbital scale. This, and the fact that pulsars are more massive than planets, implies that the gravitational field of a binary pulsar is strong relative to that of the Solar System. For instance, relativistic effects yield large post-Keplerian orbital corrections, and second-order corrections are envisioned to be necessary for the double pulsar timing in the near future \citep{ksm+06}.

Another important factor to take into consideration is that pulsars are very compact objects. As we mentioned in \S\,\ref{s:ns_overview}, $\sim 10-20\%$ of the total mass of a neutron star consists of gravitational binding energy --- this is more than 8 orders of magnitude larger than that of the Earth! Studying gravity in the vicinity of strong self-gravitating bodies is particularly relevant for testing aspects of gravitational theories related to the Strong Equivalence Principle. A consequence of the Strong Equivalence Principle in general relativity is that the dynamics of gravitating bodies are uniquely determined by their external gravitational fields, provided that they are far enough apart from each other for classical tidal effects to be negligible \citep{wil01}. This concept is often referred to as the ``effacement" of the internal structure of gravitational bodies \citep{dam87}. In this context, pulsar tests are highly complementary to experiments made in the Solar System regime.

While we recognize that the observed binary pulsar systems do not possess the strong gravitational fields that theoreticians may encounter in simulations of black hole mergers and other kinds of high energy astrophysical phenomena, they nevertheless represent an important practical landmark for observational tests of gravitational theories. Even though black hole binaries and more compact neutron stars probably exist, it is not trivial that they offer accessible ways of studying their dynamical behavior in a fashion similar to binary pulsars. Although they lie somewhere between the weak-field regime and the `extremely' strong-field regime, binary pulsars can provide useful constraints on different aspects of gravitational theories. For example, they can limit the parameter space of tensor-biscalar theories and the non-linear parameters of gravitational theories (e.g. the ``1PN" $\alpha_0$ and $\beta_0$ parameters), as well as restrict the existence of preferred-frame effects and other violations of the Weak and Strong Equivalence Principles \citep{sta03a}.

For in-depth reviews of binary pulsar tests of gravitational theories, we recommend \citet{sta03a,wil01,wil93,de96b,de92a}.



\chapter{Definition of Variables Used in Chapter \ref{c:0737_aberration}}\label{a:def_variables}


\emph{By order of appearance.}

\begin{description}
\item[$\vec n_0$:] unit vector from the binary system to the observer.
\item[$\vec s_p$:] unit vector along pulsar A spin axis.
\item[$\vec m$:] unit vector along pulsar A magnetic axis.
\item[$\zeta$:] colatitude of the emission vector (angle between $\vec s_p$ and $\vec n_0$).
\item[$\Phi_0$:] spin longitude. $\Phi_0$ is defined as zero when $\vec s_p$, $\vec n_0$ and $\vec m$ are coplanar.
\item[$\rho$:] half opening angle of the emission cone.
\item[$\alpha$:] angle between $\vec s_p$ and $\vec m$.
\item[$\chi_0$:] angle between the arc connecting $\vec n_0$ and $\vec s_p$ and the arc connecting $\vec n_0$ and $\vec m$ at the edges of the pulse. This also corresponds to the angle of linear polarization in the rotating vector model (RVM).
\item[$\Omega_b$:] orbital angular frequency.
\item[$a_p$:] $a \frac{M_c}{M_c+M_p}$.
\item[$a$:] orbital semi-major axis.
\item[$M_p$:] pulsar mass.
\item[$M_c$:] companion mass.
\item[$c$:] speed of light.
\item[$e$:] orbital eccentricity.
\item[$i$:] orbital inclination.
\item[$\eta$:] angle between the ascending node of the orbit and the projection of $\vec s_p$ on the sky plane.
\item[$\psi$:] true anomaly measured from the ascending node.
\item[$\omega$:] angle of periastron measured from the ascending node.
\item[$r$:] distance between the two pulsars $\left(r = a \frac{1-e^2}{1+e \cos (\psi - \omega)}\right)$.
\item[$a_{||}$:] distance at the conjunction projected along our line of sight $\left(a_{||} = a \sin i \frac{1-e^2}{1+e \sin \omega}\right)$.
\item[$\zeta$:] angle between $\vec s_p$ and $\vec n_0$.
\item[$x_p$:] projected semi-major axis $\left(x_p =  \frac{a_p}{c} \sin i\right)$.
\end{description}

\setlength{\parskip}{0cm}
\addcontentsline{toc}{chapter}{\protect\hspace{3.5ex}Bibliography}
\setlength{\itemsep}{-20pt}
\renewcommand{\baselinestretch}{\smallstretch}\small\normalsize
\setlength{\parskip}{4pt}

\bibliographystyle{mystyles/plainnat}
\bibliography{breton_thesis}

\begin{thebibliography}{267}
\providecommand{\natexlab}[1]{#1}
\providecommand{\url}[1]{\texttt{#1}}
\expandafter\ifx\csname urlstyle\endcsname\relax
  \providecommand{\doi}[1]{doi: #1}\else
  \providecommand{\doi}{doi: \begingroup \urlstyle{rm}\Url}\fi

\bibitem[{Aharonian} et~al.(2006){Aharonian}, {Akhperjanian}, {Bazer-Bachi},
  {Beilicke}, {Benbow}, {Berge}, {Bernl{\"o}hr}, {Boisson}, {Bolz}, {Borrel},
  {Braun}, {Breitling}, {Brown}, {Chadwick}, {Chounet}, {Cornils},
  {Costamante}, {Degrange}, {Dickinson}, {Djannati-Ata{\"i}}, {Drury}, {Dubus},
  {Emmanoulopoulos}, {Espigat}, {Feinstein}, {Fontaine}, {Fuchs}, {Funk},
  {Gallant}, {Giebels}, {Gillessen}, {Glicenstein}, {Goret}, {Hadjichristidis},
  {Hauser}, {Heinzelmann}, {Henri}, {Hermann}, {Hinton}, {Hofmann}, {Holleran},
  {Horns}, {Jacholkowska}, {de Jager}, {Kh{\'e}lifi}, {Komin}, {Konopelko},
  {Latham}, {Le Gallou}, {Lemi{\`e}re}, {Lemoine-Goumard}, {Leroy}, {Lohse},
  {Martin}, {Martineau-Huynh}, {Marcowith}, {Masterson}, {McComb}, {de
  Naurois}, {Nolan}, {Noutsos}, {Orford}, {Osborne}, {Ouchrif}, {Panter},
  {Pelletier}, {Pita}, {P{\"u}hlhofer}, {Punch}, {Raubenheimer}, {Raue},
  {Raux}, {Rayner}, {Reimer}, {Reimer}, {Ripken}, {Rob}, {Rolland}, {Rowell},
  {Sahakian}, {Saug{\'e}}, {Schlenker}, {Schlickeiser}, {Schuster}, {Schwanke},
  {Siewert}, {Sol}, {Spangler}, {Steenkamp}, {Stegmann}, {Tavernet}, {Terrier},
  {Th{\'e}oret}, {Tluczykont}, {Vasileiadis}, {Venter}, {Vincent}, {V{\"o}lk},
  and {Wagner}]{aab+07}
F.~{Aharonian}, A.~G. {Akhperjanian}, A.~R. {Bazer-Bachi}, M.~{Beilicke},
  W.~{Benbow}, D.~{Berge}, K.~{Bernl{\"o}hr}, C.~{Boisson}, O.~{Bolz},
  V.~{Borrel}, et~al.
\newblock {The H.E.S.S. Survey of the Inner Galaxy in Very High Energy Gamma
  Rays}.
\newblock \emph{\apj}, 636:\penalty0 777--797, January 2006.

\bibitem[{Alcock} et~al.(1986){Alcock}, {Farhi}, and {Olinto}]{afo86}
C.~{Alcock}, E.~{Farhi}, and A.~{Olinto}.
\newblock {Strange stars}.
\newblock \emph{\apj}, 310:\penalty0 261--272, November 1986.

\bibitem[{Alpar} et~al.(1982){Alpar}, {Cheng}, {Ruderman}, and
  {Shaham}]{acr+82}
M.~A. {Alpar}, A.~F. {Cheng}, M.~A. {Ruderman}, and J.~{Shaham}.
\newblock {A new class of radio pulsars}.
\newblock \emph{\nat}, 300:\penalty0 728--730, December 1982.

\bibitem[{Arons} et~al.(2005){Arons}, {Backer}, {Spitkovsky}, and
  {Kaspi}]{abs+04}
J.~{Arons}, D.~C. {Backer}, A.~{Spitkovsky}, and V.~M. {Kaspi}.
\newblock {Probing Relativistic Winds: The Case of PSR J0737-3039 A and B}.
\newblock In F.~A. {Rasio} and I.~H. {Stairs}, editors, \emph{Binary Radio
  Pulsars}, volume 328 of \emph{Astronomical Society of the Pacific Conference
  Series}, pages 95--+, July 2005.

\bibitem[{Arzoumanian}(1995)]{arz95}
Z.~{Arzoumanian}.
\newblock \emph{{Radio Observations of Binary Pulsars: Clues to Binary
  Evolution and Tests of General Relativity.}}
\newblock PhD thesis, AA(PRINCETON UNIVERSITY.), 1995.

\bibitem[{Baade} and {Zwicky}(1934)]{bz34a}
W.~{Baade} and F.~{Zwicky}.
\newblock {On Super-novae}.
\newblock \emph{Proceedings of the National Academy of Science}, 20:\penalty0
  254--+, 1934.

\bibitem[Baade and Zwicky(1934)]{bz34b}
W.~Baade and F.~Zwicky.
\newblock Remarks on super-novae and cosmic rays.
\newblock \emph{Phys. Rev.}, 46\penalty0 (1):\penalty0 76--77, Jul 1934.

\bibitem[{Backer}(1970)]{bac70}
D.~C. {Backer}.
\newblock {Pulsar Nulling Phenomena}.
\newblock \emph{\nat}, 228:\penalty0 42--+, October 1970.

\bibitem[{Backer} et~al.(1993){Backer}, {Foster}, and {Sallmen}]{bfs93}
D.~C. {Backer}, R.~S. {Foster}, and S.~{Sallmen}.
\newblock {A Second Companion of the Millisecond Pulsar 1620-26}.
\newblock \emph{\nat}, 365:\penalty0 817--+, October 1993.

\bibitem[{Backer} et~al.(1997){Backer}, {Dexter}, {Zepka}, {Ng}, {Werthimer},
  {Ray}, and {Foster}]{bdz+97}
D.~C. {Backer}, M.~R. {Dexter}, A.~{Zepka}, D.~{Ng}, D.~J. {Werthimer}, P.~S.
  {Ray}, and R.~S. {Foster}.
\newblock {A Programmable 36-MHz Digital Filter Bank for Radio Science}.
\newblock \emph{\pasp}, 109:\penalty0 61--68, January 1997.

\bibitem[{Barker} and {O'Connell}(1975)]{bo75}
B.~M. {Barker} and R.~F. {O'Connell}.
\newblock {Gravitational two-body problem with arbitrary masses, spins, and
  quadrupole moments}.
\newblock \emph{\prd}, 12:\penalty0 329--335, July 1975.

\bibitem[{Bartel} et~al.(1981){Bartel}, {Sieber}, {Wielebinski}, {Kardashev},
  {Nikolaev}, {Popov}, {Soglasnov}, {Kuzmin}, and {Smirnova}]{bsw+81}
N.~{Bartel}, W.~{Sieber}, R.~{Wielebinski}, N.~S. {Kardashev}, N.~I.
  {Nikolaev}, M.~V. {Popov}, V.~A. {Soglasnov}, A.~D. {Kuzmin}, and T.~V.
  {Smirnova}.
\newblock {Simultaneous two-station single pulse observations of radio pulsars
  over a broad frequency range. I - With particular reference to PSR 0809+74}.
\newblock \emph{\aap}, 93:\penalty0 85--92, January 1981.

\bibitem[{Bassa} et~al.(2006){Bassa}, {van Kerkwijk}, {Koester}, and
  {Verbunt}]{bvk+06}
C.~G. {Bassa}, M.~H. {van Kerkwijk}, D.~{Koester}, and F.~{Verbunt}.
\newblock {The masses of PSR J1911-5958A and its white dwarf companion}.
\newblock \emph{\aap}, 456:\penalty0 295--304, September 2006.

\bibitem[{Belczynski} and {Taam}(2004)]{bt04}
K.~{Belczynski} and R.~E. {Taam}.
\newblock {Galactic Populations of Ultracompact Binaries}.
\newblock \emph{\apj}, 603:\penalty0 690--696, March 2004.

\bibitem[{Bell Burnell}(1977)]{bel77}
S.~J. {Bell Burnell}.
\newblock {Little Green Men, White Dwarfs or Pulsars}.
\newblock In \emph{Eighth Texas Symposium on Relativistic Astrophysics}, volume
  302 of \emph{Annals of the New York Academy of Sciences}, pages 685--689. New
  York Academy of Sciences, December 1977.

\bibitem[Benacquista(2006)]{ben06}
Matthew~J. Benacquista.
\newblock Relativistic binaries in globular clusters.
\newblock \emph{Living Reviews in Relativity}, 9\penalty0 (2), 2006.
\newblock URL \url{http://www.livingreviews.org/lrr-2006-2}.

\bibitem[{Benvenuto} and {De Vito}(2005)]{bd05}
O.~G. {Benvenuto} and M.~A. {De Vito}.
\newblock {The formation of helium white dwarfs in close binary systems - II}.
\newblock \emph{\mnras}, 362:\penalty0 891--905, September 2005.

\bibitem[{Bergeron}(2001)]{ber01}
P.~{Bergeron}.
\newblock {The Halo White Dwarf WD 0346+246 Revisited}.
\newblock \emph{\apj}, 558:\penalty0 369--376, September 2001.

\bibitem[{Bergeron} et~al.(1995){Bergeron}, {Saumon}, and {Wesemael}]{bsw95}
P.~{Bergeron}, D.~{Saumon}, and F.~{Wesemael}.
\newblock {New model atmospheres for very cool white dwarfs with mixed H/He and
  pure He compositions}.
\newblock \emph{\apj}, 443:\penalty0 764--779, April 1995.

\bibitem[{Bergeron} et~al.(2001){Bergeron}, {Leggett}, and {Ruiz}]{blr01}
P.~{Bergeron}, S.~K. {Leggett}, and M.~T. {Ruiz}.
\newblock {Photometric and Spectroscopic Analysis of Cool White Dwarfs with
  Trigonometric Parallax Measurements}.
\newblock \emph{\apjs}, 133:\penalty0 413--449, April 2001.

\bibitem[{Bhat} et~al.(2007){Bhat}, {Gupta}, {Kramer}, {Karastergiou}, {Lyne},
  and {Johnston}]{bgk+07}
N.~D.~R. {Bhat}, Y.~{Gupta}, M.~{Kramer}, A.~{Karastergiou}, A.~G. {Lyne}, and
  S.~{Johnston}.
\newblock {Simultaneous single-pulse observations of radio pulsars. V. On the
  broadband nature of the pulse nulling phenomenon in PSR B1133+16}.
\newblock \emph{\aap}, 462:\penalty0 257--268, January 2007.

\bibitem[{Bhattacharya} and {van den Heuvel}(1991)]{bv91}
D.~{Bhattacharya} and E.~P.~J. {van den Heuvel}.
\newblock {Formation and evolution of binary and millisecond radio pulsars.}
\newblock \emph{\physrep}, 203:\penalty0 1--124, 1991.

\bibitem[{Biggs}(1992)]{big92}
J.~D. {Biggs}.
\newblock {An analysis of radio pulsar nulling statistics}.
\newblock \emph{\apj}, 394:\penalty0 574--580, August 1992.

\bibitem[{Bisnovatyi-Kogan}(2006)]{bis06}
G.~{Bisnovatyi-Kogan}.
\newblock {Binary and recycled pulsars: 30 years after observational
  discovery}.
\newblock \emph{ArXiv Astrophysics e-prints}, November 2006.

\bibitem[{Branch}(1998)]{bra98}
D.~{Branch}.
\newblock {Type IA Supernovae and the Hubble Constant}.
\newblock \emph{\araa}, 36:\penalty0 17--56, 1998.

\bibitem[{Bravo} and {Garc{\'{\i}}a-Senz}(1999)]{bg99}
E.~{Bravo} and D.~{Garc{\'{\i}}a-Senz}.
\newblock {Coulomb corrections to the equation of state of nuclear statistical
  equilibrium matter: implications for SNIa nucleosynthesis and the
  accretion-induced collapse of white dwarfs}.
\newblock \emph{\mnras}, 307:\penalty0 984--992, August 1999.

\bibitem[{Breton} et~al.(2007){Breton}, {Roberts}, {Ransom}, {Kaspi}, {Durant},
  {Bergeron}, and {Faulkner}]{brr+07}
R.~P. {Breton}, M.~S.~E. {Roberts}, S.~M. {Ransom}, V.~M. {Kaspi}, M.~{Durant},
  P.~{Bergeron}, and A.~J. {Faulkner}.
\newblock {The Unusual Binary Pulsar PSR J1744-3922: Radio Flux Variability,
  Near-Infrared Observation, and Evolution}.
\newblock \emph{\apj}, 661:\penalty0 1073--1083, June 2007.

\bibitem[{Breton} et~al.(2008){Breton}, {Kaspi}, {Kramer}, {McLaughlin},
  {Lyutikov}, {Ransom}, {Stairs}, {Ferdman}, {Camilo}, and {Possenti}]{bkk+08}
R.~P. {Breton}, V.~M. {Kaspi}, M.~{Kramer}, M.~A. {McLaughlin}, M.~{Lyutikov},
  S.~M. {Ransom}, I.~H. {Stairs}, R.~D. {Ferdman}, F.~{Camilo}, and
  A.~{Possenti}.
\newblock {Relativistic Spin Precession in the Double Pulsar}.
\newblock \emph{Science}, 321:\penalty0 104--, July 2008.

\bibitem[{Brumfiel}(2007)]{bru07}
G.~{Brumfiel}.
\newblock {Air force had early warning of pulsars}.
\newblock \emph{\nat}, 448:\penalty0 2--+, 2007.

\bibitem[{Bucciantini}(2008)]{buc08}
N.~{Bucciantini}.
\newblock {Modeling Pulsar Wind Nebulae}.
\newblock \emph{Advances in Space Research}, 41:\penalty0 491--502, 2008.

\bibitem[{Burgay} et~al.(2003){Burgay}, {D'Amico}, {Possenti}, {Manchester},
  {Lyne}, {Joshi}, {McLaughlin}, {Kramer}, {Sarkissian}, {Camilo}, {Kalogera},
  {Kim}, and {Lorimer}]{bdp+03}
M.~{Burgay}, N.~{D'Amico}, A.~{Possenti}, R.~N. {Manchester}, A.~G. {Lyne},
  B.~C. {Joshi}, M.~A. {McLaughlin}, M.~{Kramer}, J.~M. {Sarkissian},
  F.~{Camilo}, et~al.
\newblock {An increased estimate of the merger rate of double neutron stars
  from observations of a highly relativistic system}.
\newblock \emph{\nat}, 426:\penalty0 531--533, December 2003.

\bibitem[{Burgay} et~al.(2005){Burgay}, {Possenti}, {Manchester}, {Kramer},
  {McLaughlin}, {Lorimer}, {Stairs}, {Joshi}, {Lyne}, {Camilo}, {D'Amico},
  {Freire}, {Sarkissian}, {Hotan}, and {Hobbs}]{bpm+05}
M.~{Burgay}, A.~{Possenti}, R.~N. {Manchester}, M.~{Kramer}, M.~A.
  {McLaughlin}, D.~R. {Lorimer}, I.~H. {Stairs}, B.~C. {Joshi}, A.~G. {Lyne},
  F.~{Camilo}, et~al.
\newblock {Long-Term Variations in the Pulse Emission from PSR J0737-3039B}.
\newblock \emph{\apjl}, 624:\penalty0 L113--L116, May 2005.

\bibitem[{Cameron} et~al.(2005){Cameron}, {Chandra}, {Ray}, {Kulkarni},
  {Frail}, {Wieringa}, {Nakar}, {Phinney}, {Miyazaki}, {Tsuboi}, {Okumura},
  {Kawai}, {Menten}, and {Bertoldi}]{ccr+05}
P.~B. {Cameron}, P.~{Chandra}, A.~{Ray}, S.~R. {Kulkarni}, D.~A. {Frail}, M.~H.
  {Wieringa}, E.~{Nakar}, E.~S. {Phinney}, A.~{Miyazaki}, M.~{Tsuboi}, et~al.
\newblock {Detection of a radio counterpart to the 27 December 2004 giant flare
  from SGR 1806 - 20}.
\newblock \emph{\nat}, 434:\penalty0 1112--1115, April 2005.

\bibitem[{Camilo} et~al.(2001){Camilo}, {Lyne}, {Manchester}, {Bell}, {Stairs},
  {D'Amico}, {Kaspi}, {Possenti}, {Crawford}, and {McKay}]{clm+01}
F.~{Camilo}, A.~G. {Lyne}, R.~N. {Manchester}, J.~F. {Bell}, I.~H. {Stairs},
  N.~{D'Amico}, V.~M. {Kaspi}, A.~{Possenti}, F.~{Crawford}, and N.~P.~F.
  {McKay}.
\newblock {Discovery of Five Binary Radio Pulsars}.
\newblock \emph{\apjl}, 548:\penalty0 L187--L191, February 2001.

\bibitem[{Campana} et~al.(2004){Campana}, {Possenti}, and {Burgay}]{cpb04}
S.~{Campana}, A.~{Possenti}, and M.~{Burgay}.
\newblock {XMM-Newton Observation of the Double Pulsar System J0737-3039}.
\newblock \emph{\apjl}, 613:\penalty0 L53--L56, September 2004.

\bibitem[{Carroll} and {Ostlie}(1996)]{co96}
B.~W. {Carroll} and D.~A. {Ostlie}.
\newblock \emph{{An Introduction to Modern Astrophysics}}.
\newblock Institute for Mathematics and Its Applications, 1996.

\bibitem[{Casali} and {Hawarden}(1992)]{ch92}
M.~M. {Casali} and T.~G. {Hawarden}.
\newblock {A set of faint JHK standards for UKIRT}.
\newblock \emph{JCMT–UKIRT Newsletter}, 4:\penalty0 33, 1992.

\bibitem[{Cerny}(1985)]{cer85}
V.~{Cerny}.
\newblock {Thermodynamical Approach to the Traveling Salesman Problem: An
  Efficient Simulation Algorithm}.
\newblock \emph{Journal of Optimization, Theory and Applications}, 45:\penalty0
  41--51, 1985.

\bibitem[{Champion} et~al.(2008){Champion}, {Ransom}, {Lazarus}, {Camilo},
  {Bassa}, {Kaspi}, {Nice}, {Freire}, {Stairs}, {van Leeuwen}, {Stappers},
  {Cordes}, {Hessels}, {Lorimer}, {Arzoumanian}, {Backer}, {Bhat},
  {Chatterjee}, {Cognard}, {Deneva}, {Faucher-Gigu{\`e}re}, {Gaensler}, {Han},
  {Jenet}, {Kasian}, {Kondratiev}, {Kramer}, {Lazio}, {McLaughlin},
  {Venkataraman}, and {Vlemmings}]{crl+08}
D.~J. {Champion}, S.~M. {Ransom}, P.~{Lazarus}, F.~{Camilo}, C.~{Bassa}, V.~M.
  {Kaspi}, D.~J. {Nice}, P.~C.~C. {Freire}, I.~H. {Stairs}, J.~{van Leeuwen},
  et~al.
\newblock {An Eccentric Binary Millisecond Pulsar in the Galactic Plane}.
\newblock \emph{Science}, 320:\penalty0 1309--, June 2008.

\bibitem[{Chandrasekhar}(1931{\natexlab{a}})]{cha31a}
S.~{Chandrasekhar}.
\newblock {The Density of White Dwarf Stars}.
\newblock \emph{Philosophical Magazine}, 11:\penalty0 592--596,
  1931{\natexlab{a}}.

\bibitem[{Chandrasekhar}(1931{\natexlab{b}})]{cha31b}
S.~{Chandrasekhar}.
\newblock {The Maximum Mass of Ideal White Dwarfs}.
\newblock \emph{\apj}, 74:\penalty0 81--+, July 1931{\natexlab{b}}.

\bibitem[{Chandrasekhar}(1931{\natexlab{c}})]{cha31c}
S.~{Chandrasekhar}.
\newblock {The highly collapsed configurations of a stellar mass}.
\newblock \emph{\mnras}, 91:\penalty0 456--466, March 1931{\natexlab{c}}.

\bibitem[{Chandrasekhar}(1935)]{cha35}
S.~{Chandrasekhar}.
\newblock {The highly collapsed configurations of a stellar mass (Second
  paper)}.
\newblock \emph{\mnras}, 95:\penalty0 207--225, January 1935.

\bibitem[{Chandrasekhar}(1967)]{cha67}
S.~{Chandrasekhar}.
\newblock \emph{{An introduction to the study of stellar structure}}.
\newblock New York: Dover, 1967, 1967.

\bibitem[{Chatterjee} et~al.(2007){Chatterjee}, {Gaensler}, {Melatos},
  {Brisken}, and {Stappers}]{cgm+07}
S.~{Chatterjee}, B.~M. {Gaensler}, A.~{Melatos}, W.~F. {Brisken}, and B.~W.
  {Stappers}.
\newblock {Pulsed X-Ray Emission from Pulsar A in the Double Pulsar System
  J0737-3039}.
\newblock \emph{\apj}, 670:\penalty0 1301--1306, December 2007.

\bibitem[{Clifton} and {Weisberg}(2008)]{cw08}
T.~{Clifton} and J.~M. {Weisberg}.
\newblock {A Simple Model for Pulse Profiles from Precessing Pulsars, with
  Special Application to Relativistic Binary PSR B1913+16}.
\newblock \emph{\apj}, 679:\penalty0 687--696, May 2008.

\bibitem[{Coles} et~al.(2005){Coles}, {McLaughlin}, {Rickett}, {Lyne}, and
  {Bhat}]{cmr+05}
W.~A. {Coles}, M.~A. {McLaughlin}, B.~J. {Rickett}, A.~G. {Lyne}, and N.~D.~R.
  {Bhat}.
\newblock {Probing the Eclipse of J0737-3039A with Scintillation}.
\newblock \emph{\apj}, 623:\penalty0 392--397, April 2005.

\bibitem[{Condon} and {Ransom}(2008)]{cr08}
J.~J. {Condon} and S.~M. {Ransom}.
\newblock {Essential Radio Astronomy}.
\newblock {Online Lecture Notes of the Course ASTR 534 at University of
  Virginia}, August 2008.

\bibitem[{Cordes} and {Lazio}(2002)]{cl02}
J.~M. {Cordes} and T.~J.~W. {Lazio}.
\newblock {NE2001.I. A New Model for the Galactic Distribution of Free
  Electrons and its Fluctuations}.
\newblock \emph{ArXiv Astrophysics e-prints}, July 2002.

\bibitem[{Cordes} and {Rickett}(1998)]{cr98}
J.~M. {Cordes} and B.~J. {Rickett}.
\newblock {Diffractive Interstellar Scintillation Timescales and Velocities}.
\newblock \emph{\apj}, 507:\penalty0 846--860, November 1998.

\bibitem[{Crawford} et~al.(2006){Crawford}, {Roberts}, {Hessels}, {Ransom},
  {Livingstone}, {Tam}, and {Kaspi}]{crh+06}
F.~{Crawford}, M.~S.~E. {Roberts}, J.~W.~T. {Hessels}, S.~M. {Ransom},
  M.~{Livingstone}, C.~R. {Tam}, and V.~M. {Kaspi}.
\newblock {A Survey of 56 Midlatitude EGRET Error Boxes for Radio Pulsars}.
\newblock \emph{\apj}, 652:\penalty0 1499--1507, December 2006.

\bibitem[{D'Amico} et~al.(2001){D'Amico}, {Possenti}, {Manchester},
  {Sarkissian}, {Lyne}, and {Camilo}]{dpm+01}
N.~{D'Amico}, A.~{Possenti}, R.~N. {Manchester}, J.~{Sarkissian}, A.~G. {Lyne},
  and F.~{Camilo}.
\newblock {An Eclipsing Millisecond Pulsar with a Possible Main-Sequence
  Companion in NGC 6397}.
\newblock \emph{\apjl}, 561:\penalty0 L89--L92, November 2001.

\bibitem[{Damour}(1987)]{dam87}
T.~{Damour}.
\newblock \emph{{The problem of motion in Newtonian and Einsteinian gravity.}},
  pages 128--198.
\newblock Three hundred years of gravitation, 1987.

\bibitem[{Damour} and {Esposito-Far{\`e}se}(1992{\natexlab{a}})]{de92a}
T.~{Damour} and G.~{Esposito-Far{\`e}se}.
\newblock {Testing local Lorentz invariance of gravity with binary-pulsar
  data}.
\newblock \emph{\prd}, 46:\penalty0 4128--4132, November 1992{\natexlab{a}}.

\bibitem[{Damour} and {Esposito-Far{\`e}se}(1992{\natexlab{b}})]{de92b}
T.~{Damour} and G.~{Esposito-Far{\`e}se}.
\newblock {Tensor-multi-scalar theories of gravitation.}
\newblock \emph{Classical and Quantum Gravity}, 9:\penalty0 2093--2176,
  September 1992{\natexlab{b}}.

\bibitem[{Damour} and {Esposito-Far{\`e}se}(1996{\natexlab{a}})]{de96a}
T.~{Damour} and G.~{Esposito-Far{\`e}se}.
\newblock {Testing gravity to second post-Newtonian order: A field-theory
  approach}.
\newblock \emph{\prd}, 53:\penalty0 5541--5578, May 1996{\natexlab{a}}.

\bibitem[{Damour} and {Esposito-Far{\`e}se}(1996{\natexlab{b}})]{de96b}
T.~{Damour} and G.~{Esposito-Far{\`e}se}.
\newblock {Tensor-scalar gravity and binary-pulsar experiments}.
\newblock \emph{\prd}, 54:\penalty0 1474--1491, July 1996{\natexlab{b}}.

\bibitem[{Damour} and {Taylor}(1992)]{dt92a}
T.~{Damour} and J.~H. {Taylor}.
\newblock {Strong-field tests of relativistic gravity and binary pulsars}.
\newblock \emph{\prd}, 45:\penalty0 1840--1868, March 1992.

\bibitem[{Darnley} et~al.(2006){Darnley}, {Bode}, {Kerins}, {Newsam}, {An},
  {Baillon}, {Belokurov}, {Calchi Novati}, {Carr}, {Cr{\'e}z{\'e}}, {Evans},
  {Giraud-H{\'e}raud}, {Gould}, {Hewett}, {Jetzer}, {Kaplan},
  {Paulin-Henriksson}, {Smartt}, {Tsapras}, and {Weston}]{dbk+06}
M.~J. {Darnley}, M.~F. {Bode}, E.~{Kerins}, A.~M. {Newsam}, J.~{An},
  P.~{Baillon}, V.~{Belokurov}, S.~{Calchi Novati}, B.~J. {Carr},
  M.~{Cr{\'e}z{\'e}}, et~al.
\newblock {Classical novae from the POINT-AGAPE microlensing survey of M31 -
  II. Rate and statistical characteristics of the nova population}.
\newblock \emph{\mnras}, 369:\penalty0 257--271, June 2006.

\bibitem[{de Loore} and {Doom}(1992)]{dd92}
C.~{de Loore} and C.~{Doom}, editors.
\newblock \emph{{Structure and evolution of single and binary stars}}, volume
  179 of \emph{Astrophysics and Space Science Library}, 1992.

\bibitem[{Demorest} et~al.(2004){Demorest}, {Ramachandran}, {Backer}, {Ransom},
  {Kaspi}, {Arons}, and {Spitkovsky}]{drb+04}
P.~{Demorest}, R.~{Ramachandran}, D.~C. {Backer}, S.~M. {Ransom}, V.~{Kaspi},
  J.~{Arons}, and A.~{Spitkovsky}.
\newblock {Orientations of Spin and Magnetic Dipole Axes of Pulsars in the
  J0737-3039 Binary Based on Polarimetry Observations at the Green Bank
  Telescope}.
\newblock \emph{\apjl}, 615:\penalty0 L137--L140, November 2004.

\bibitem[{Dewi} and {van den Heuvel}(2004)]{dv04}
J.~D.~M. {Dewi} and E.~P.~J. {van den Heuvel}.
\newblock {The formation of the double neutron star pulsar J0737 - 3039}.
\newblock \emph{\mnras}, 349:\penalty0 169--172, March 2004.

\bibitem[{Dodson} et~al.(2002){Dodson}, {McCulloch}, and {Lewis}]{dml02}
R.~G. {Dodson}, P.~M. {McCulloch}, and D.~R. {Lewis}.
\newblock {High Time Resolution Observations of the January 2000 Glitch in the
  Vela Pulsar}.
\newblock \emph{\apjl}, 564:\penalty0 L85--L88, January 2002.

\bibitem[{Driebe} et~al.(1998){Driebe}, {Schoenberner}, {Bloecker}, and
  {Herwig}]{dsb+98}
T.~{Driebe}, D.~{Schoenberner}, T.~{Bloecker}, and F.~{Herwig}.
\newblock {The evolution of helium white dwarfs. I. The companion of the
  millisecond pulsar PSR J1012+5307}.
\newblock \emph{\aap}, 339:\penalty0 123--133, November 1998.

\bibitem[{Drimmel} et~al.(2003){Drimmel}, {Cabrera-Lavers}, and
  {L{\'o}pez-Corredoira}]{dcl03}
R.~{Drimmel}, A.~{Cabrera-Lavers}, and M.~{L{\'o}pez-Corredoira}.
\newblock {A three-dimensional Galactic extinction model}.
\newblock \emph{\aap}, 409:\penalty0 205--215, October 2003.

\bibitem[{Duncan} and {Thompson}(1992)]{dt92b}
R.~C. {Duncan} and C.~{Thompson}.
\newblock {Formation of very strongly magnetized neutron stars - Implications
  for gamma-ray bursts}.
\newblock \emph{\apjl}, 392:\penalty0 L9--L13, June 1992.

\bibitem[{Edwards} and {Bailes}(2001)]{eb01}
R.~T. {Edwards} and M.~{Bailes}.
\newblock {Recycled Pulsars Discovered at High Radio Frequency}.
\newblock \emph{\apj}, 553:\penalty0 801--808, June 2001.

\bibitem[{Eggleton}(1983)]{egg83}
P.~P. {Eggleton}.
\newblock {Approximations to the radii of Roche lobes}.
\newblock \emph{\apj}, 268:\penalty0 368--+, May 1983.

\bibitem[{Ergma}(1993)]{erg93}
E.~{Ergma}.
\newblock {An Accretion Induced Collapse Model for the Eclipsing Binary Pulsar
  PSR:1718-19}.
\newblock \emph{\aap}, 273:\penalty0 L38+, June 1993.

\bibitem[{Ergma} and {Sarna}(2000)]{es00}
E.~{Ergma} and M.~J. {Sarna}.
\newblock {The eclipsing binary millisecond pulsar PSR B1744-24A - possible
  test for a magnetic braking mechanism}.
\newblock \emph{\aap}, 363:\penalty0 657--659, November 2000.

\bibitem[{Ergma} et~al.(1996){Ergma}, {Sarna}, and {Giersz}]{esg96}
E.~{Ergma}, M.~J. {Sarna}, and M.~{Giersz}.
\newblock {The possible evolutionary scenarios for the eclipsing binary system
  PSR 1718-19.}
\newblock \emph{\aap}, 307:\penalty0 768--774, March 1996.

\bibitem[{Faulkner} et~al.(2004){Faulkner}, {Stairs}, {Kramer}, {Lyne},
  {Hobbs}, {Possenti}, {Lorimer}, {Manchester}, {McLaughlin}, {D'Amico},
  {Camilo}, and {Burgay}]{fsk+04}
A.~J. {Faulkner}, I.~H. {Stairs}, M.~{Kramer}, A.~G. {Lyne}, G.~{Hobbs},
  A.~{Possenti}, D.~R. {Lorimer}, R.~N. {Manchester}, M.~A. {McLaughlin},
  N.~{D'Amico}, et~al.
\newblock {The Parkes Multibeam Pulsar Survey - V. Finding binary and
  millisecond pulsars}.
\newblock \emph{\mnras}, 355:\penalty0 147--158, November 2004.

\bibitem[{Ferdman} et~al.(2008){Ferdman}, {Stairs}, {Kramer}, {Manchester},
  {Lyne}, {Breton}, {McLaughlin}, {Possenti}, and {Burgay}]{fsk+08}
R.~D. {Ferdman}, I.~H. {Stairs}, M.~{Kramer}, R.~N. {Manchester}, A.~G. {Lyne},
  R.~P. {Breton}, M.~A. {McLaughlin}, A.~{Possenti}, and M.~{Burgay}.
\newblock {The double pulsar: evolutionary constraints from the system
  geometry}.
\newblock In \emph{40 Years of Pulsars: Millisecond Pulsars, Magnetars and
  More}, volume 983 of \emph{American Institute of Physics Conference Series},
  pages 474--478, February 2008.

\bibitem[{Ferrario} and {Wickramasinghe}(2005)]{fw95}
L.~{Ferrario} and D.~T. {Wickramasinghe}.
\newblock {Magnetic fields and rotation in white dwarfs and neutron stars}.
\newblock \emph{\mnras}, 356:\penalty0 615--620, January 2005.

\bibitem[{Freire} et~al.(2008){Freire}, {Wolszczan}, {van den Berg}, and
  {Hessels}]{fwv+08}
P.~C.~C. {Freire}, A.~{Wolszczan}, M.~{van den Berg}, and J.~W.~T. {Hessels}.
\newblock {A Massive Neutron Star in the Globular Cluster M5}.
\newblock \emph{\apj}, 679:\penalty0 1433--1442, June 2008.

\bibitem[{Fruchter} et~al.(1988{\natexlab{a}}){Fruchter}, {Gunn}, {Lauer}, and
  {Dressler}]{fgl+88}
A.~S. {Fruchter}, J.~E. {Gunn}, T.~R. {Lauer}, and A.~{Dressler}.
\newblock {Optical detection and characterization of the eclipsing pulsar's
  companion}.
\newblock \emph{\nat}, 334:\penalty0 686--689, August 1988{\natexlab{a}}.

\bibitem[{Fruchter} et~al.(1988{\natexlab{b}}){Fruchter}, {Stinebring}, and
  {Taylor}]{fst88}
A.~S. {Fruchter}, D.~R. {Stinebring}, and J.~H. {Taylor}.
\newblock {A millisecond pulsar in an eclipsing binary}.
\newblock \emph{\nat}, 333:\penalty0 237--239, May 1988{\natexlab{b}}.

\bibitem[{Funk} et~al.(2006){Funk}, {LAT Collab.~Pulsars}, {SNR}, and {Plerions
  group}]{fls06}
S.~{Funk}, G.~{LAT Collab.~Pulsars}, {SNR}, and {Plerions group}.
\newblock {Future GLAST Observations of Supernova Remnants and Pulsar Wind
  Nebulae}.
\newblock In \emph{Bulletin of the American Astronomical Society}, volume~38 of
  \emph{Bulletin of the American Astronomical Society}, pages 1107--+, December
  2006.

\bibitem[{Gaensler} et~al.(2005){Gaensler}, {Kouveliotou}, {Gelfand}, {Taylor},
  {Eichler}, {Wijers}, {Granot}, {Ramirez-Ruiz}, {Lyubarsky}, {Hunstead},
  {Campbell-Wilson}, {van der Horst}, {McLaughlin}, {Fender}, {Garrett},
  {Newton-McGee}, {Palmer}, {Gehrels}, and {Woods}]{gkg+05}
B.~M. {Gaensler}, C.~{Kouveliotou}, J.~D. {Gelfand}, G.~B. {Taylor},
  D.~{Eichler}, R.~A.~M.~J. {Wijers}, J.~{Granot}, E.~{Ramirez-Ruiz}, Y.~E.
  {Lyubarsky}, R.~W. {Hunstead}, et~al.
\newblock {An expanding radio nebula produced by a giant flare from the
  magnetar SGR 1806-20}.
\newblock \emph{\nat}, 434:\penalty0 1104--1106, April 2005.

\bibitem[{Gavriil} et~al.(2002){Gavriil}, {Kaspi}, and {Woods}]{gkw02}
F.~P. {Gavriil}, V.~M. {Kaspi}, and P.~M. {Woods}.
\newblock {Magnetar-like X-ray bursts from an anomalous X-ray pulsar}.
\newblock \emph{\nat}, 419:\penalty0 142--144, September 2002.

\bibitem[{Gavriil} et~al.(2008{\natexlab{a}}){Gavriil}, {Dib}, and
  {Kaspi}]{gdk08}
F.~P. {Gavriil}, R.~{Dib}, and V.~M. {Kaspi}.
\newblock {Activity From Magnetar Candidate 4U 0142+61: Bursts and Emission
  Lines}.
\newblock In C.~{Bassa}, Z.~{Wang}, A.~{Cumming}, and V.~M. {Kaspi}, editors,
  \emph{40 Years of Pulsars: Millisecond Pulsars, Magnetars and More}, volume
  983 of \emph{American Institute of Physics Conference Series}, pages
  234--238, February 2008{\natexlab{a}}.

\bibitem[{Gavriil} et~al.(2008{\natexlab{b}}){Gavriil}, {Gonzalez}, {Gotthelf},
  {Kaspi}, {Livingstone}, and {Woods}]{ggg+08}
F.~P. {Gavriil}, M.~E. {Gonzalez}, E.~V. {Gotthelf}, V.~M. {Kaspi}, M.~A.
  {Livingstone}, and P.~M. {Woods}.
\newblock {Magnetar-Like Emission from the Young Pulsar in Kes 75}.
\newblock \emph{Science}, 319:\penalty0 1802--, March 2008{\natexlab{b}}.

\bibitem[{Geman} and {Geman}(1984)]{gg84}
S.~{Geman} and D.~{Geman}.
\newblock {Stochastic relaxation, Gibbs distributions and the Bayesian
  restoration of images}.
\newblock \emph{IEEE Transactions on Pattern Analysis and Machine
  Intelligence}, 6\penalty0 (6):\penalty0 721--741, November 1984.
\newblock URL \url{http://dx.doi.org/10.1080/02664769300000058}.

\bibitem[{Geppert} et~al.(2006){Geppert}, {K{\"u}ker}, and {Page}]{gkp06}
U.~{Geppert}, M.~{K{\"u}ker}, and D.~{Page}.
\newblock {Temperature distribution in magnetized neutron star crusts. II. The
  effect of a strong toroidal component}.
\newblock \emph{\aap}, 457:\penalty0 937--947, October 2006.

\bibitem[{Ghosh}(2007)]{gho07}
P.~{Ghosh}.
\newblock \emph{{Rotation and Accretion Powered Pulsars}}.
\newblock Rotation and Accretion Powered Pulsars: World Scientific Series in
  Astronomy and Astrophysics -- Vol.~10.~Edited by Pranab Ghosh.~ISBN-10
  981-02-4744-3; ISBN-13 978-981-02-4744-7.~Published by World Scientific
  Publishing Co., Pte.~Ltd., Singapore, 2007., 2007.

\bibitem[{Gilks} and {Spiegelhalter}(1996)]{grs96}
R.~{Gilks}, W.~R.~{Richardson} and D.~J. {Spiegelhalter}.
\newblock Introducing markov chain monte carlo.
\newblock In R.~{Gilks}, W.~R.~{Richardson} and D.~J. {Spiegelhalter}, editors,
  \emph{Markov Chain Monte Carlo Methods in Practice}. Chapman and Hall,
  London, 1996.

\bibitem[{Gold}(1968)]{gol68}
T.~{Gold}.
\newblock {Rotating Neutron Stars as the Origin of the Pulsating Radio
  Sources}.
\newblock \emph{\nat}, 218:\penalty0 731--+, May 1968.

\bibitem[{Gonzalez} et~al.(2007){Gonzalez}, {Kaspi}, {Camilo}, {Gaensler}, and
  {Pivovaroff}]{gkc+07}
M.~E. {Gonzalez}, V.~M. {Kaspi}, F.~{Camilo}, B.~M. {Gaensler}, and M.~J.
  {Pivovaroff}.
\newblock {PSR J1119 6127 and the X-ray emission from high magnetic field radio
  pulsars}.
\newblock \emph{\apss}, 308:\penalty0 89--94, April 2007.

\bibitem[{Granot} and {M{\'e}sz{\'a}ros}(2004)]{gm04}
J.~{Granot} and P.~{M{\'e}sz{\'a}ros}.
\newblock {High-Energy Emission from the Double Pulsar System J0737-3039}.
\newblock \emph{\apjl}, 609:\penalty0 L17--L20, July 2004.

\bibitem[{Gregory}(2005{\natexlab{a}})]{gre05a}
P.~C. {Gregory}.
\newblock \emph{{Bayesian Logical Data Analysis for the Physical Sciences: A
  Comparative Approach with `Mathematica' Support}}.
\newblock Cambridge University Press, Cambridge, UK, 2005{\natexlab{a}}.

\bibitem[{Gregory}(2005{\natexlab{b}})]{gre05b}
P.~C. {Gregory}.
\newblock {A Bayesian Analysis of Extrasolar Planet Data for HD 73526}.
\newblock \emph{\apj}, 631:\penalty0 1198--1214, October 2005{\natexlab{b}}.

\bibitem[{Griffiths}(1995)]{gri95}
D.~J. {Griffiths}.
\newblock \emph{{Introduction to quantum mechanics}}.
\newblock Introduction to quantum mechanics/ David J.~Griffiths.~Englewood
  Cliffs, N.J.: Prentice Hall, c1995., 1995.

\bibitem[{Gunn} and {Ostriker}(1969)]{go69}
J.~E. {Gunn} and J.~P. {Ostriker}.
\newblock {Magnetic Dipole Radiation from Pulsars}.
\newblock \emph{\nat}, 221:\penalty0 454--+, 1969.

\bibitem[{G{\"u}ver} et~al.(2008){G{\"u}ver}, {{\"O}zel}, and {G{\"o}{\u
  g}{\"u}{\c s}}]{gog08}
T.~{G{\"u}ver}, F.~{{\"O}zel}, and E.~{G{\"o}{\u g}{\"u}{\c s}}.
\newblock {Physical Properties of the AXP 4U 0142+61 from X-Ray Spectral
  Analysis}.
\newblock \emph{\apj}, 675:\penalty0 1499--1504, March 2008.

\bibitem[{Haensel} et~al.(1986){Haensel}, {Zdunik}, and {Schaefer}]{hzs86}
P.~{Haensel}, J.~L. {Zdunik}, and R.~{Schaefer}.
\newblock {Strange quark stars}.
\newblock \emph{\aap}, 160:\penalty0 121--128, May 1986.

\bibitem[{Han} and {Manchester}(2001)]{hm01}
J.~L. {Han} and R.~N. {Manchester}.
\newblock {The shape of pulsar radio beams}.
\newblock \emph{\mnras}, 320:\penalty0 L35--L39, January 2001.

\bibitem[{Hansen} et~al.(2004){Hansen}, {Kawaler}, and {Trimble}]{hkt04}
C.~J. {Hansen}, S.~D. {Kawaler}, and V.~{Trimble}.
\newblock \emph{{Stellar interiors : physical principles, structure, and
  evolution}}.
\newblock 2004.

\bibitem[{Harding}(2007)]{har07}
A.~K. {Harding}.
\newblock {Pulsar High-Energy Emission From the Polar Cap and Slot Gap}.
\newblock \emph{ArXiv e-prints}, 710, October 2007.

\bibitem[{Harding} and {Lai}(2006)]{hl06}
A.~K. {Harding} and D.~{Lai}.
\newblock {Physics of strongly magnetized neutron stars.}
\newblock \emph{Reports of Progress in Physics}, 69:\penalty0 2631--2708, 2006.

\bibitem[{Haslam} et~al.(1982){Haslam}, {Salter}, {Stoffel}, and
  {Wilson}]{hss+82}
C.~G.~T. {Haslam}, C.~J. {Salter}, H.~{Stoffel}, and W.~E. {Wilson}.
\newblock {A 408 MHz all-sky continuum survey. II - The atlas of contour maps}.
\newblock \emph{\aaps}, 47:\penalty0 1--+, January 1982.

\bibitem[{Haslam} et~al.(1995){Haslam}, {Salter}, {Stoffel}, and
  {Wilson}]{hss+95}
C.~G.~T. {Haslam}, C.~J. {Salter}, H.~{Stoffel}, and W.~E. {Wilson}.
\newblock {408 MHz all-sky map}.
\newblock \emph{Astronomy Data Image Library}, pages 1--+, December 1995.

\bibitem[{Heggie}(1975)]{heg75}
D.~C. {Heggie}.
\newblock {Binary evolution in stellar dynamics}.
\newblock \emph{\mnras}, 173:\penalty0 729--787, December 1975.

\bibitem[{Heinke} et~al.(2006){Heinke}, {Rybicki}, {Narayan}, and
  {Grindlay}]{hrn+06}
C.~O. {Heinke}, G.~B. {Rybicki}, R.~{Narayan}, and J.~E. {Grindlay}.
\newblock {A Hydrogen Atmosphere Spectral Model Applied to the Neutron Star X7
  in the Globular Cluster 47 Tucanae}.
\newblock \emph{\apj}, 644:\penalty0 1090--1103, June 2006.

\bibitem[{Heiselberg}(2002)]{hei02}
H.~{Heiselberg}.
\newblock {Neutron Star Masses, Radii and Equation of State}.
\newblock \emph{ArXiv Astrophysics e-prints}, January 2002.

\bibitem[{Hessels} et~al.(2006){Hessels}, {Ransom}, {Stairs}, {Freire},
  {Kaspi}, and {Camilo}]{hrs+06}
J.~W.~T. {Hessels}, S.~M. {Ransom}, I.~H. {Stairs}, P.~C.~C. {Freire}, V.~M.
  {Kaspi}, and F.~{Camilo}.
\newblock {A Radio Pulsar Spinning at 716 Hz}.
\newblock \emph{Science}, 311:\penalty0 1901--1904, March 2006.

\bibitem[{Hewish}(1968)]{hew68}
A.~{Hewish}, 1968.
\newblock {Daily Telegraph 5 Mar 1968 21/3}.

\bibitem[{Hewish}(1972)]{hew72}
A.~{Hewish}.
\newblock {Three Years with Pulsars}.
\newblock \emph{Mitteilungen der Astronomischen Gesellschaft Hamburg},
  31:\penalty0 15--+, 1972.

\bibitem[{Hewish} et~al.(1964){Hewish}, {Scott}, and {Wills}]{hsw64}
A.~{Hewish}, P.~F. {Scott}, and D.~{Wills}.
\newblock {Interplanetary Scintillation of Small Diameter Radio Sources}.
\newblock \emph{\nat}, 203:\penalty0 1214--+, September 1964.

\bibitem[{Hewish} et~al.(1968){Hewish}, {Bell}, {Pilkington}, {Scott}, and
  {Collins}]{hbp+68}
A.~{Hewish}, S.~J. {Bell}, J.~D. {Pilkington}, P.~F. {Scott}, and R.~A.
  {Collins}.
\newblock {Observation of a Rapidly Pulsating Radio Source}.
\newblock \emph{\nat}, 217:\penalty0 709--+, February 1968.

\bibitem[{Hobbs} et~al.(2005){Hobbs}, {Lorimer}, {Lyne}, and {Kramer}]{hll+05}
G.~{Hobbs}, D.~R. {Lorimer}, A.~G. {Lyne}, and M.~{Kramer}.
\newblock {A statistical study of 233 pulsar proper motions}.
\newblock \emph{\mnras}, 360:\penalty0 974--992, July 2005.

\bibitem[{Hobbs} et~al.(2006{\natexlab{a}}){Hobbs}, {Lyne}, and
  {Kramer}]{hlk06}
G.~{Hobbs}, A.~{Lyne}, and M.~{Kramer}.
\newblock {Pulsar Timing Noise}.
\newblock \emph{Chinese Journal of Astronomy and Astrophysics Supplement},
  6\penalty0 (2):\penalty0 020000--175, December 2006{\natexlab{a}}.

\bibitem[{Hobbs} et~al.(2006{\natexlab{b}}){Hobbs}, {Edwards}, and
  {Manchester}]{hem06}
G.~B. {Hobbs}, R.~T. {Edwards}, and R.~N. {Manchester}.
\newblock {TEMPO2, a new pulsar-timing package - I. An overview}.
\newblock \emph{\mnras}, 369:\penalty0 655--672, June 2006{\natexlab{b}}.

\bibitem[{Hotan} et~al.(2005){Hotan}, {Bailes}, and {Ord}]{hbo05}
A.~W. {Hotan}, M.~{Bailes}, and S.~M. {Ord}.
\newblock {Geodetic Precession in PSR J1141-6545}.
\newblock \emph{\apj}, 624:\penalty0 906--913, May 2005.

\bibitem[{Hoyle} et~al.(1964){Hoyle}, {Narlikar}, and {Wheeler}]{hnw64}
F.~{Hoyle}, J.~V. {Narlikar}, and J.~A. {Wheeler}.
\newblock {Electromagnetic Waves from Very Dense Stars}.
\newblock \emph{\nat}, 203:\penalty0 914--+, August 1964.

\bibitem[{Hulse} and {Taylor}(1975)]{ht75}
R.~A. {Hulse} and J.~H. {Taylor}.
\newblock {Discovery of a pulsar in a binary system}.
\newblock \emph{\apjl}, 195:\penalty0 L51--L53, January 1975.

\bibitem[{Hurley} et~al.(2005){Hurley}, {Boggs}, {Smith}, {Duncan}, {Lin},
  {Zoglauer}, {Krucker}, {Hurford}, {Hudson}, {Wigger}, {Hajdas}, {Thompson},
  {Mitrofanov}, {Sanin}, {Boynton}, {Fellows}, {von Kienlin}, {Lichti}, {Rau},
  and {Cline}]{hbs+05}
K.~{Hurley}, S.~E. {Boggs}, D.~M. {Smith}, R.~C. {Duncan}, R.~{Lin},
  A.~{Zoglauer}, S.~{Krucker}, G.~{Hurford}, H.~{Hudson}, C.~{Wigger}, et~al.
\newblock {An exceptionally bright flare from SGR 1806-20 and the origins of
  short-duration {$\gamma$}-ray bursts}.
\newblock \emph{\nat}, 434:\penalty0 1098--1103, April 2005.

\bibitem[{Ibrahim} et~al.(2003){Ibrahim}, {Swank}, and {Parke}]{isp03}
A.~I. {Ibrahim}, J.~H. {Swank}, and W.~{Parke}.
\newblock {New Evidence of Proton-Cyclotron Resonance in a Magnetar Strength
  Field from SGR 1806-20}.
\newblock \emph{\apjl}, 584:\penalty0 L17--L21, February 2003.

\bibitem[{Ibrahim} et~al.(2004){Ibrahim}, {Markwardt}, {Swank}, {Ransom},
  {Roberts}, {Kaspi}, {Woods}, {Safi-Harb}, {Balman}, {Parke}, {Kouveliotou},
  {Hurley}, and {Cline}]{ims+04}
A.~I. {Ibrahim}, C.~B. {Markwardt}, J.~H. {Swank}, S.~{Ransom}, M.~{Roberts},
  V.~{Kaspi}, P.~M. {Woods}, S.~{Safi-Harb}, S.~{Balman}, W.~C. {Parke}, et~al.
\newblock {Discovery of a Transient Magnetar: XTE J1810-197}.
\newblock \emph{\apjl}, 609:\penalty0 L21--L24, July 2004.

\bibitem[{Iorio}(2009)]{ior09}
L.~{Iorio}.
\newblock {Prospects for measuring the moment of inertia of pulsar
  J0737-3039A}.
\newblock \emph{New Astronomy}, 14:\penalty0 40--43, January 2009.

\bibitem[Irion(2005)]{iri05}
Robert Irion.
\newblock {ASTROPHYSICS: Giant Neutron-Star Flare Blitzes the Galaxy With Gamma
  Rays}.
\newblock \emph{Science}, 307\penalty0 (5713):\penalty0 1178--1179, 2005.
\newblock URL \url{http://www.sciencemag.org}.

\bibitem[{Ivanova} and {Taam}(2004)]{it04}
N.~{Ivanova} and R.~E. {Taam}.
\newblock {Thermal Timescale Mass Transfer and the Evolution of White Dwarf
  Binaries}.
\newblock \emph{\apj}, 601:\penalty0 1058--1066, February 2004.

\bibitem[{Ivanova} et~al.(2005){Ivanova}, {Belczynski}, {Fregeau}, and
  {Rasio}]{ibf+05}
N.~{Ivanova}, K.~{Belczynski}, J.~M. {Fregeau}, and F.~A. {Rasio}.
\newblock {The evolution of binary fractions in globular clusters}.
\newblock \emph{\mnras}, 358:\penalty0 572--584, April 2005.

\bibitem[{Ivanova} et~al.(2008){Ivanova}, {Heinke}, {Rasio}, {Belczynski}, and
  {Fregeau}]{ihr+08}
N.~{Ivanova}, C.~O. {Heinke}, F.~A. {Rasio}, K.~{Belczynski}, and J.~M.
  {Fregeau}.
\newblock {Formation and evolution of compact binaries in globular clusters -
  II. Binaries with neutron stars}.
\newblock \emph{\mnras}, 386:\penalty0 553--576, May 2008.

\bibitem[{Jackson}(1975)]{jac75}
J.~D. {Jackson}.
\newblock \emph{{Classical electrodynamics}}.
\newblock 92/12/31, New York: Wiley, 1975, 2nd ed., 1975.

\bibitem[{Janssen} and {van Kerkwijk}(2005)]{jv05}
T.~{Janssen} and M.~H. {van Kerkwijk}.
\newblock {Observations of the companion to the pulsar PSR B1718-19. The role
  of tidal circularisation}.
\newblock \emph{\aap}, 439:\penalty0 433--441, August 2005.

\bibitem[Jaynes(1968)]{jay68}
E.T. Jaynes.
\newblock Prior probabilities.
\newblock \emph{Systems Science and Cybernetics, IEEE Transactions on},
  4\penalty0 (3):\penalty0 227--241, Sept. 1968.

\bibitem[{Jeffreys}(1946)]{jef46}
H.~{Jeffreys}.
\newblock An invariant form for the prior probability in estimation problems.
\newblock \emph{Proceedings of the Royal Society of London. Series A,
  Mathematical and Physical Sciences}, 186\penalty0 (1007):\penalty0 453--461,
  1946.
\newblock URL \url{http://www.jstor.org/stable/97883}.

\bibitem[{Jenet} et~al.(2006){Jenet}, {Hobbs}, {van Straten}, {Manchester},
  {Bailes}, {Verbiest}, {Edwards}, {Hotan}, {Sarkissian}, and {Ord}]{jhv+06}
F.~A. {Jenet}, G.~B. {Hobbs}, W.~{van Straten}, R.~N. {Manchester},
  M.~{Bailes}, J.~P.~W. {Verbiest}, R.~T. {Edwards}, A.~W. {Hotan}, J.~M.
  {Sarkissian}, and S.~M. {Ord}.
\newblock {Upper Bounds on the Low-Frequency Stochastic Gravitational Wave
  Background from Pulsar Timing Observations: Current Limits and Future
  Prospects}.
\newblock \emph{\apj}, 653:\penalty0 1571--1576, December 2006.

\bibitem[{Kaplan} et~al.(2005){Kaplan}, {Escoffier}, {Lacasse}, {O'Neil},
  {Ford}, {Ransom}, {Anderson}, {Cordes}, {Lazio}, and {Kulkarni}]{kel+05}
D.~L. {Kaplan}, R.~P. {Escoffier}, R.~J. {Lacasse}, K.~{O'Neil}, J.~M. {Ford},
  S.~M. {Ransom}, S.~B. {Anderson}, J.~M. {Cordes}, T.~J.~W. {Lazio}, and S.~R.
  {Kulkarni}.
\newblock {The Green Bank Telescope Pulsar Spigot}.
\newblock \emph{\pasp}, 117:\penalty0 643--653, June 2005.

\bibitem[{Karastergiou} and {Johnston}(2007)]{kj07}
A.~{Karastergiou} and S.~{Johnston}.
\newblock {An empirical model for the beams of radio pulsars}.
\newblock \emph{\mnras}, 380:\penalty0 1678--1684, October 2007.

\bibitem[{Kaspi}(2004)]{kas04}
V.~M. {Kaspi}.
\newblock {Soft {$\Gamma$} Repeaters and Anomalous X-ray Pulsars: Together
  Forever}.
\newblock In F.~{Camilo} and B.~M. {Gaensler}, editors, \emph{Young Neutron
  Stars and Their Environments}, volume 218 of \emph{IAU Symposium}, pages
  231--+, 2004.

\bibitem[{Kaspi}(2007)]{kas07}
V.~M. {Kaspi}.
\newblock {Recent progress on anomalous X-ray pulsars}.
\newblock \emph{\apss}, 308:\penalty0 1--4, April 2007.

\bibitem[{Kaspi} and {Helfand}(2002)]{kh02}
V.~M. {Kaspi} and D.~J. {Helfand}.
\newblock {Constraining the Birth Events of Neutron Stars}.
\newblock In P.~O. {Slane} and B.~M. {Gaensler}, editors, \emph{Neutron Stars
  in Supernova Remnants}, volume 271 of \emph{Astronomical Society of the
  Pacific Conference Series}, pages 3--+, 2002.

\bibitem[{Kaspi} and {McLaughlin}(2005)]{km05}
V.~M. {Kaspi} and M.~A. {McLaughlin}.
\newblock {Chandra X-Ray Detection of the High Magnetic Field Radio Pulsar PSR
  J1718-3718}.
\newblock \emph{\apjl}, 618:\penalty0 L41--L44, January 2005.

\bibitem[{Kaspi} et~al.(2004){Kaspi}, {Ransom}, {Backer}, {Ramachandran},
  {Demorest}, {Arons}, and {Spitkovsky}]{krb+04}
V.~M. {Kaspi}, S.~M. {Ransom}, D.~C. {Backer}, R.~{Ramachandran},
  P.~{Demorest}, J.~{Arons}, and A.~{Spitkovsky}.
\newblock {Green Bank Telescope Observations of the Eclipse of Pulsar ``A'' in
  the Double Pulsar Binary PSR J0737-3039}.
\newblock \emph{\apjl}, 613:\penalty0 L137--L140, October 2004.

\bibitem[{King} et~al.(2005){King}, {Beer}, {Rolfe}, {Schenker}, and
  {Skipp}]{kbr+05}
A.~R. {King}, M.~E. {Beer}, D.~J. {Rolfe}, K.~{Schenker}, and J.~M. {Skipp}.
\newblock {The population of black widow pulsars}.
\newblock \emph{\mnras}, 358:\penalty0 1501--1504, April 2005.

\bibitem[{Kippenhahn} and {Weigert}(1967)]{kw67}
R.~{Kippenhahn} and A.~{Weigert}.
\newblock {Entwicklung in engen Doppelsternsystemen I. Massenaustausch vor und
  nach Beendigung des zentralen Wasserstoff-Brennens}.
\newblock \emph{Zeitschrift fur Astrophysik}, 65:\penalty0 251--+, 1967.

\bibitem[{Kirkpatrick} et~al.(1983){Kirkpatrick}, {Gelatt}, and
  {Vecchi}]{kgv83}
S.~{Kirkpatrick}, C.~D. {Gelatt}, and M.~P. {Vecchi}.
\newblock {Optimization by Simulated Annealing}.
\newblock \emph{Science}, 220:\penalty0 671--680, 1983.

\bibitem[{Komissarov} and {Lyubarsky}(2004)]{kl04}
S.~{Komissarov} and Y.~{Lyubarsky}.
\newblock {MHD Simulations of Crab's Jet and Torus}.
\newblock \emph{\apss}, 293:\penalty0 107--113, September 2004.

\bibitem[{Konacki} et~al.(2003){Konacki}, {Wolszczan}, and {Stairs}]{kws03}
M.~{Konacki}, A.~{Wolszczan}, and I.~H. {Stairs}.
\newblock {Geodetic Precession and Timing of the Relativistic Binary Pulsars
  PSR B1534+12 and PSR B1913+16}.
\newblock \emph{\apj}, 589:\penalty0 495--502, May 2003.

\bibitem[{Konar} and {Choudhuri}(2004)]{ks04}
S.~{Konar} and A.~R. {Choudhuri}.
\newblock {Diamagnetic screening of the magnetic field in accreting neutron
  stars - II. The effect of polar cap widening}.
\newblock \emph{\mnras}, 348:\penalty0 661--668, February 2004.

\bibitem[{Kramer}(1998)]{kra98}
M.~{Kramer}.
\newblock {Determination of the Geometry of the PSR B1913+16 System by Geodetic
  Precession}.
\newblock \emph{\apj}, 509:\penalty0 856--860, December 1998.

\bibitem[{Kramer} and {Stairs}(2008)]{ks08}
M.~{Kramer} and I.~H. {Stairs}.
\newblock {The Double Pulsar}.
\newblock \emph{\araa}, 46:\penalty0 541--572, September 2008.

\bibitem[{Kramer} et~al.(2006){Kramer}, {Stairs}, {Manchester}, {McLaughlin},
  {Lyne}, {Ferdman}, {Burgay}, {Lorimer}, {Possenti}, {D'Amico}, {Sarkissian},
  {Hobbs}, {Reynolds}, {Freire}, and {Camilo}]{ksm+06}
M.~{Kramer}, I.~H. {Stairs}, R.~N. {Manchester}, M.~A. {McLaughlin}, A.~G.
  {Lyne}, R.~D. {Ferdman}, M.~{Burgay}, D.~R. {Lorimer}, A.~{Possenti},
  N.~{D'Amico}, et~al.
\newblock {Tests of General Relativity from Timing the Double Pulsar}.
\newblock \emph{Science}, 314:\penalty0 97--102, October 2006.

\bibitem[{Kulkarni} and {Hester}(1988)]{kh88}
S.~R. {Kulkarni} and J.~J. {Hester}.
\newblock {Discovery of a nebula around PSR1957+20}.
\newblock \emph{\nat}, 335:\penalty0 801--803, October 1988.

\bibitem[{Kulkarni} et~al.(1988){Kulkarni}, {Djorgovski}, and
  {Fruchter}]{kdf88}
S.~R. {Kulkarni}, S.~{Djorgovski}, and A.~S. {Fruchter}.
\newblock {Probable optical counterpart of the eclipsing millisecond pulsar
  system, 1957 + 20}.
\newblock \emph{\nat}, 334:\penalty0 504--506, August 1988.

\bibitem[{Lada}(2006)]{lad06}
C.~J. {Lada}.
\newblock {Stellar Multiplicity and the Initial Mass Function: Most Stars Are
  Single}.
\newblock \emph{\apjl}, 640:\penalty0 L63--L66, March 2006.

\bibitem[{Large} et~al.(1968){Large}, {Vaughan}, and {Mills}]{lvm68}
M.~I. {Large}, A.~E. {Vaughan}, and B.~Y. {Mills}.
\newblock {A Pulsar Supernova Association?}
\newblock \emph{\nat}, 220:\penalty0 340--+, October 1968.

\bibitem[{Lattimer}(2007)]{lat07}
J.~M. {Lattimer}.
\newblock {Equation of state constraints from neutron stars}.
\newblock \emph{\apss}, 308:\penalty0 371--379, April 2007.

\bibitem[{Lattimer} and {Prakash}(2001)]{lp01}
J.~M. {Lattimer} and M.~{Prakash}.
\newblock {Neutron Star Structure and the Equation of State}.
\newblock \emph{\apj}, 550:\penalty0 426--442, March 2001.

\bibitem[{Lattimer} and {Prakash}(2004)]{lp04}
J.~M. {Lattimer} and M.~{Prakash}.
\newblock {The Physics of Neutron Stars}.
\newblock \emph{Science}, 304:\penalty0 536--542, April 2004.

\bibitem[{Lattimer} and {Prakash}(2007)]{lp07}
J.~M. {Lattimer} and M.~{Prakash}.
\newblock {Neutron star observations: Prognosis for equation of state
  constraints}.
\newblock \emph{\physrep}, 442:\penalty0 109--165, April 2007.

\bibitem[{Lattimer} and {Schutz}(2005)]{ls05}
J.~M. {Lattimer} and B.~F. {Schutz}.
\newblock {Constraining the Equation of State with Moment of Inertia
  Measurements}.
\newblock \emph{\apj}, 629:\penalty0 979--984, August 2005.

\bibitem[{Laycock} et~al.(2005){Laycock}, {Corbet}, {Coe}, {Marshall},
  {Markwardt}, and {Lochner}]{lcc+05}
S.~{Laycock}, R.~H.~D. {Corbet}, M.~J. {Coe}, F.~E. {Marshall}, C.~{Markwardt},
  and J.~{Lochner}.
\newblock {Long-Term Behavior of X-Ray Pulsars in the Small Magellanic Cloud}.
\newblock \emph{\apjs}, 161:\penalty0 96--117, November 2005.

\bibitem[{Li} and {Wang}(1995)]{lw95}
X.-D. {Li} and Z.-R. {Wang}.
\newblock {Statistical analysis of pulse nulling in pulsars}.
\newblock \emph{Chinese Astronomy and Astrophysics}, 19:\penalty0 302--308,
  February 1995.

\bibitem[{Livingstone} et~al.(2007){Livingstone}, {Kaspi}, {Gavriil},
  {Manchester}, {Gotthelf}, and {Kuiper}]{lkg+07}
M.~A. {Livingstone}, V.~M. {Kaspi}, F.~P. {Gavriil}, R.~N. {Manchester},
  E.~V.~G. {Gotthelf}, and L.~{Kuiper}.
\newblock {New phase-coherent measurements of pulsar braking indices}.
\newblock \emph{\apss}, 308:\penalty0 317--323, April 2007.

\bibitem[{Lorimer}(2005)]{lor05}
D.~R. {Lorimer}.
\newblock {Binary and Millisecond Pulsars}.
\newblock \emph{Living Reviews in Relativity}, 8:\penalty0 7--+, November 2005.

\bibitem[{Lorimer}()]{sig}
D.~R. {Lorimer}.
\newblock {\texttt{SIGPROC} is freely available at
  \url{http://sigproc.sourceforge.net}}.

\bibitem[{Lorimer} and {Kramer}(2004)]{lk04}
D.~R. {Lorimer} and M.~{Kramer}.
\newblock \emph{{Handbook of Pulsar Astronomy}}.
\newblock vol.~4 of \emph{Cambridge observing handbooks for research
  astronomers}. Cambridge University Press, Cambridge, UK, December 2004.

\bibitem[{Lorimer} et~al.(2006){Lorimer}, {Stairs}, {Freire}, {Cordes},
  {Camilo}, {Faulkner}, {Lyne}, {Nice}, {Ransom}, {Arzoumanian}, {Manchester},
  {Champion}, {van Leeuwen}, {Mclaughlin}, {Ramachandran}, {Hessels},
  {Vlemmings}, {Deshpande}, {Bhat}, {Chatterjee}, {Han}, {Gaensler}, {Kasian},
  {Deneva}, {Reid}, {Lazio}, {Kaspi}, {Crawford}, {Lommen}, {Backer}, {Kramer},
  {Stappers}, {Hobbs}, {Possenti}, {D'Amico}, and {Burgay}]{lsf+06}
D.~R. {Lorimer}, I.~H. {Stairs}, P.~C. {Freire}, J.~M. {Cordes}, F.~{Camilo},
  A.~J. {Faulkner}, A.~G. {Lyne}, D.~J. {Nice}, S.~M. {Ransom},
  Z.~{Arzoumanian}, et~al.
\newblock {Arecibo Pulsar Survey Using ALFA. II. The Young, Highly Relativistic
  Binary Pulsar J1906+0746}.
\newblock \emph{\apj}, 640:\penalty0 428--434, March 2006.

\bibitem[{Lyne} and {Graham-Smith}(2006)]{lg06}
A.~G. {Lyne} and F.~{Graham-Smith}.
\newblock \emph{{Pulsar astronomy}}.
\newblock Pulsar astronomy, 3rd ed., by A.G.~Lyne and
  F.~Graham-Smith.~Cambridge astrophysics series.~Cambridge, UK: Cambridge
  University Press, 2006 ISBN 0521839548., 2006.

\bibitem[{Lyne} and {Manchester}(1988)]{lm88}
A.~G. {Lyne} and R.~N. {Manchester}.
\newblock {The shape of pulsar radio beams}.
\newblock \emph{\mnras}, 234:\penalty0 477--508, October 1988.

\bibitem[{Lyne} et~al.(1993){Lyne}, {Biggs}, {Harrison}, and {Bailes}]{lbh+93}
A.~G. {Lyne}, J.~D. {Biggs}, P.~A. {Harrison}, and M.~{Bailes}.
\newblock {A long-period globular-cluster pulsar in an eclipsing binary
  system}.
\newblock \emph{\nat}, 361:\penalty0 47--49, January 1993.

\bibitem[{Lyne} et~al.(2000){Lyne}, {Shemar}, and {Smith}]{lss00}
A.~G. {Lyne}, S.~L. {Shemar}, and F.~G. {Smith}.
\newblock {Statistical studies of pulsar glitches}.
\newblock \emph{\mnras}, 315:\penalty0 534--542, July 2000.

\bibitem[{Lyne} et~al.(2004){Lyne}, {Burgay}, {Kramer}, {Possenti},
  {Manchester}, {Camilo}, {McLaughlin}, {Lorimer}, {D'Amico}, {Joshi},
  {Reynolds}, and {Freire}]{lbk+04}
A.~G. {Lyne}, M.~{Burgay}, M.~{Kramer}, A.~{Possenti}, R.~N. {Manchester},
  F.~{Camilo}, M.~A. {McLaughlin}, D.~R. {Lorimer}, N.~{D'Amico}, B.~C.
  {Joshi}, et~al.
\newblock {A Double-Pulsar System: A Rare Laboratory for Relativistic Gravity
  and Plasma Physics}.
\newblock \emph{Science}, 303:\penalty0 1153--1157, February 2004.

\bibitem[{Lyubarsky}(2008)]{lyu08}
Y.~{Lyubarsky}.
\newblock {Pulsar emission mechanisms}.
\newblock In C.~{Bassa}, Z.~{Wang}, A.~{Cumming}, and V.~M. {Kaspi}, editors,
  \emph{40 Years of Pulsars: Millisecond Pulsars, Magnetars and More}, volume
  983 of \emph{American Institute of Physics Conference Series}, pages 29--37,
  February 2008.

\bibitem[{Lyutikov}(2004)]{lyu04}
M.~{Lyutikov}.
\newblock {On the nature of eclipses in binary pulsar J0737-3039}.
\newblock \emph{\mnras}, 353:\penalty0 1095--1106, October 2004.

\bibitem[{Lyutikov}(2005)]{lyu05}
M.~{Lyutikov}.
\newblock {Orbital modulation of emission of the binary pulsar J0737-3039B}.
\newblock \emph{\mnras}, 362:\penalty0 1078--1084, September 2005.

\bibitem[{Lyutikov} and {Thompson}(2005)]{lt05}
M.~{Lyutikov} and C.~{Thompson}.
\newblock {Magnetospheric Eclipses in the Double Pulsar System PSR J0737-3039}.
\newblock \emph{\apj}, 634:\penalty0 1223--1241, December 2005.

\bibitem[{Manchester} et~al.(2005{\natexlab{a}}){Manchester}, {Hobbs}, {Teoh},
  and {Hobbs}]{atnf}
R.~N. {Manchester}, G.~B. {Hobbs}, A.~{Teoh}, and M.~{Hobbs}.
\newblock {The Australia Telescope National Facility Pulsar Catalogue}.
\newblock \emph{\aj}, 129:\penalty0 1993--2006, April 2005{\natexlab{a}}.
\newblock {\url{http://www.atnf.csiro.au/research/pulsar/psrcat/}}.

\bibitem[{Manchester} et~al.(2005{\natexlab{b}}){Manchester}, {Kramer},
  {Possenti}, {Lyne}, {Burgay}, {Stairs}, {Hotan}, {McLaughlin}, {Lorimer},
  {Hobbs}, {Sarkissian}, {D'Amico}, {Camilo}, {Joshi}, and {Freire}]{mkp+05}
R.~N. {Manchester}, M.~{Kramer}, A.~{Possenti}, A.~G. {Lyne}, M.~{Burgay},
  I.~H. {Stairs}, A.~W. {Hotan}, M.~A. {McLaughlin}, D.~R. {Lorimer}, G.~B.
  {Hobbs}, et~al.
\newblock {The Mean Pulse Profile of PSR J0737-3039A}.
\newblock \emph{\apjl}, 621:\penalty0 L49--L52, March 2005{\natexlab{b}}.

\bibitem[{Maron} et~al.(2000){Maron}, {Kijak}, {Kramer}, and
  {Wielebinski}]{mkk+00}
O.~{Maron}, J.~{Kijak}, M.~{Kramer}, and R.~{Wielebinski}.
\newblock {Pulsar spectra of radio emission}.
\newblock \emph{\aaps}, 147:\penalty0 195--203, December 2000.

\bibitem[{McLaughlin}(2006)]{mcl06}
M.~A. {McLaughlin}.
\newblock {New Results on Rotating Radio Transients}.
\newblock \emph{On the Present and Future of Pulsar Astronomy, 26th meeting of
  the IAU, Joint Discussion 2, 16-17 August, 2006, Prague, Czech Republic,
  JD02, \#2}, 2, August 2006.

\bibitem[{McLaughlin} et~al.(2004{\natexlab{a}}){McLaughlin}, {Camilo},
  {Burgay}, {D'Amico}, {Joshi}, {Kramer}, {Lorimer}, {Lyne}, {Manchester}, and
  {Possenti}]{mcb+04}
M.~A. {McLaughlin}, F.~{Camilo}, M.~{Burgay}, N.~{D'Amico}, B.~C. {Joshi},
  M.~{Kramer}, D.~R. {Lorimer}, A.~G. {Lyne}, R.~N. {Manchester}, and
  A.~{Possenti}.
\newblock {X-Ray Emission from the Double Pulsar System J0737-3039}.
\newblock \emph{\apjl}, 605:\penalty0 L41--L44, April 2004{\natexlab{a}}.

\bibitem[{McLaughlin} et~al.(2004{\natexlab{b}}){McLaughlin}, {Kramer}, {Lyne},
  {Lorimer}, {Stairs}, {Possenti}, {Manchester}, {Freire}, {Joshi}, {Burgay},
  {Camilo}, and {D'Amico}]{mkl+04}
M.~A. {McLaughlin}, M.~{Kramer}, A.~G. {Lyne}, D.~R. {Lorimer}, I.~H. {Stairs},
  A.~{Possenti}, R.~N. {Manchester}, P.~C.~C. {Freire}, B.~C. {Joshi},
  M.~{Burgay}, et~al.
\newblock {The Double Pulsar System J0737-3039: Modulation of the Radio
  Emission from B by Radiation from A}.
\newblock \emph{\apjl}, 613:\penalty0 L57--L60, September 2004{\natexlab{b}}.

\bibitem[{McLaughlin} et~al.(2004{\natexlab{c}}){McLaughlin}, {Lyne},
  {Lorimer}, {Possenti}, {Manchester}, {Camilo}, {Stairs}, {Kramer}, {Burgay},
  {D'Amico}, {Freire}, {Joshi}, and {Bhat}]{mll+04}
M.~A. {McLaughlin}, A.~G. {Lyne}, D.~R. {Lorimer}, A.~{Possenti}, R.~N.
  {Manchester}, F.~{Camilo}, I.~H. {Stairs}, M.~{Kramer}, M.~{Burgay},
  N.~{D'Amico}, et~al.
\newblock {The Double Pulsar System J0737-3039: Modulation of A by B at
  Eclipse}.
\newblock \emph{\apjl}, 616:\penalty0 L131--L134, December 2004{\natexlab{c}}.

\bibitem[{McLaughlin} et~al.(2006){McLaughlin}, {Lyne}, {Lorimer}, {Kramer},
  {Faulkner}, {Manchester}, {Cordes}, {Camilo}, {Possenti}, {Stairs}, {Hobbs},
  {D'Amico}, {Burgay}, and {O'Brien}]{mll+06}
M.~A. {McLaughlin}, A.~G. {Lyne}, D.~R. {Lorimer}, M.~{Kramer}, A.~J.
  {Faulkner}, R.~N. {Manchester}, J.~M. {Cordes}, F.~{Camilo}, A.~{Possenti},
  I.~H. {Stairs}, et~al.
\newblock {Transient radio bursts from rotating neutron stars}.
\newblock \emph{\nat}, 439:\penalty0 817--820, February 2006.

\bibitem[{Miller} et~al.(1998){Miller}, {Lamb}, and {Psaltis}]{mlp98}
M.~C. {Miller}, F.~K. {Lamb}, and D.~{Psaltis}.
\newblock {Sonic-Point Model of Kilohertz Quasi-periodic Brightness
  Oscillations in Low-Mass X-Ray Binaries}.
\newblock \emph{\apj}, 508:\penalty0 791--830, December 1998.

\bibitem[{Nice}(2006)]{nic06}
D.~J. {Nice}.
\newblock {Neutron star masses derived from relativistic measurements of radio
  pulsars}.
\newblock \emph{Advances in Space Research}, 38:\penalty0 2721--2724, 2006.

\bibitem[{Nice} et~al.(2000){Nice}, {Arzoumanian}, and {Thorsett}]{nat00}
D.~J. {Nice}, Z.~{Arzoumanian}, and S.~E. {Thorsett}.
\newblock {Binary Eclipsing Millisecond Pulsars: A Decade of Timing}.
\newblock In M.~{Kramer}, N.~{Wex}, and R.~{Wielebinski}, editors, \emph{IAU
  Colloq. 177: Pulsar Astronomy - 2000 and Beyond}, volume 202 of
  \emph{Astronomical Society of the Pacific Conference Series}, pages 67--+,
  2000.

\bibitem[{Nice} et~al.(2008){Nice}, {Stairs}, and {Kasian}]{nsk08}
D.~J. {Nice}, I.~H. {Stairs}, and L.~E. {Kasian}.
\newblock {Masses of Neutron Stars in Binary Pulsar Systems}.
\newblock In C.~{Bassa}, Z.~{Wang}, A.~{Cumming}, and V.~M. {Kaspi}, editors,
  \emph{40 Years of Pulsars: Millisecond Pulsars, Magnetars and More}, volume
  983 of \emph{American Institute of Physics Conference Series}, pages
  453--458, February 2008.

\bibitem[{Nomoto} and {Kondo}(1991)]{nk91}
K.~{Nomoto} and Y.~{Kondo}.
\newblock {Conditions for accretion-induced collapse of white dwarfs}.
\newblock \emph{\apjl}, 367:\penalty0 L19--L22, January 1991.

\bibitem[{O'Connell}(2008)]{oco08}
R.~F. {O'Connell}.
\newblock {Gravito-Magnetism in one-body and two-body systems: Theory and
  Experiments}.
\newblock \emph{ArXiv e-prints: 0804.3806}, April 2008.

\bibitem[{O'Connell}(1974)]{oco74}
R.~F. {O'Connell}.
\newblock {Spin, Rotation and C, P, and T Effects in the Gravitational
  Interaction and Related Experiments}.
\newblock In B.~{Bertotti}, editor, \emph{Experimental Gravitation: Proceedings
  of Course 56 of the International School of Physics "Enrico Fermi"}, page
  496. Academic Press, 1974.

\bibitem[{Oosterbroek} et~al.(2008){Oosterbroek}, {Cognard}, {Golden},
  {Verhoeve}, {Martin}, {Erd}, {Schulz}, {St{\"u}we}, {Stankov}, and
  {Ho}]{ocg+08}
T.~{Oosterbroek}, I.~{Cognard}, A.~{Golden}, P.~{Verhoeve}, D.~D.~E. {Martin},
  C.~{Erd}, R.~{Schulz}, J.~A. {St{\"u}we}, A.~{Stankov}, and T.~{Ho}.
\newblock {Simultaneous absolute timing of the Crab pulsar at radio and optical
  wavelengths}.
\newblock \emph{\aap}, 488:\penalty0 271--277, September 2008.

\bibitem[Oppenheimer and Volkoff(1939)]{ov39}
J.~R. Oppenheimer and G.~M. Volkoff.
\newblock On massive neutron cores.
\newblock \emph{Phys. Rev.}, 55\penalty0 (4):\penalty0 374--381, Feb 1939.

\bibitem[{Pacini}(1967)]{pac67}
F.~{Pacini}.
\newblock {Energy Emission from a Neutron Star}.
\newblock \emph{\nat}, 216:\penalty0 567--+, November 1967.

\bibitem[{Pacini}(1968)]{pac68}
F.~{Pacini}.
\newblock {Rotating Neutron Stars, Pulsars and Supernova Remnants}.
\newblock \emph{\nat}, 219:\penalty0 145--+, July 1968.

\bibitem[{Paczy{\'n}ski}(1971)]{pac71}
B.~{Paczy{\'n}ski}.
\newblock {Evolutionary Processes in Close Binary Systems}.
\newblock \emph{\araa}, 9:\penalty0 183--+, 1971.

\bibitem[{Palmer} et~al.(2005){Palmer}, {Barthelmy}, {Gehrels}, {Kippen},
  {Cayton}, {Kouveliotou}, {Eichler}, {Wijers}, {Woods}, {Granot}, {Lyubarsky},
  {Ramirez-Ruiz}, {Barbier}, {Chester}, {Cummings}, {Fenimore}, {Finger},
  {Gaensler}, {Hullinger}, {Krimm}, {Markwardt}, {Nousek}, {Parsons}, {Patel},
  {Sakamoto}, {Sato}, {Suzuki}, and {Tueller}]{pbg+05}
D.~M. {Palmer}, S.~{Barthelmy}, N.~{Gehrels}, R.~M. {Kippen}, T.~{Cayton},
  C.~{Kouveliotou}, D.~{Eichler}, R.~A.~M.~J. {Wijers}, P.~M. {Woods},
  J.~{Granot}, et~al.
\newblock {A giant {$\gamma$}-ray flare from the magnetar SGR 1806 - 20}.
\newblock \emph{\nat}, 434:\penalty0 1107--1109, April 2005.

\bibitem[{Payne}(2005)]{pay05}
D.~J.~B. {Payne}.
\newblock \emph{{Magnetic field evolution in accreting neutron stars}}.
\newblock PhD thesis, {The University of Melbourne}, 2005.

\bibitem[{Payne} and {Melatos}(2004)]{pm04}
D.~J.~B. {Payne} and A.~{Melatos}.
\newblock {Burial of the Polar Magnetic Field of an Accreting Neutron Star - I.
  Self-consistent Analytic and Numerical Equilibria}.
\newblock \emph{\mnras}, 351:\penalty0 569--584, June 2004.

\bibitem[{Pellizzoni} et~al.(2004){Pellizzoni}, {De Luca}, {Mereghetti},
  {Tiengo}, {Mattana}, {Caraveo}, {Tavani}, and {Bignami}]{pdm+04}
A.~{Pellizzoni}, A.~{De Luca}, S.~{Mereghetti}, A.~{Tiengo}, F.~{Mattana},
  P.~{Caraveo}, M.~{Tavani}, and G.~F. {Bignami}.
\newblock {A First XMM-Newton Look at the Relativistic Double Pulsar PSR
  J0737-3039}.
\newblock \emph{\apjl}, 612:\penalty0 L49--L52, September 2004.

\bibitem[{Pellizzoni} et~al.(2008){Pellizzoni}, {Tiengo}, {De Luca},
  {Esposito}, and {Mereghetti}]{ptd+08}
A.~{Pellizzoni}, A.~{Tiengo}, A.~{De Luca}, P.~{Esposito}, and S.~{Mereghetti}.
\newblock {PSR J0737-3039: Interacting Pulsars in X-Rays}.
\newblock \emph{\apj}, 679:\penalty0 664--674, May 2008.

\bibitem[{Phinney}(1992)]{phi92}
E.~S. {Phinney}.
\newblock {Pulsars as Probes of Newtonian Dynamical Systems}.
\newblock \emph{Phil. Trans. Roy. Soc. A}, 341:\penalty0 39, 1992.

\bibitem[{Phinney} et~al.(1988){Phinney}, {Evans}, {Blandford}, and
  {Kulkarni}]{peb+88}
E.~S. {Phinney}, C.~R. {Evans}, R.~D. {Blandford}, and S.~R. {Kulkarni}.
\newblock {Ablating dwarf model for eclipsing millisecond pulsar 1957 + 20}.
\newblock \emph{\nat}, 333:\penalty0 832--834, June 1988.

\bibitem[{Pilkington} et~al.(1968){Pilkington}, {Hewish}, {Bell}, and
  {Cole}]{phb+68}
J.~D.~H. {Pilkington}, A.~{Hewish}, S.~J. {Bell}, and T.~W. {Cole}.
\newblock {Observations of some further Pulsed Radio Sources}.
\newblock \emph{\nat}, 218:\penalty0 126--129, April 1968.

\bibitem[{Possenti} et~al.(2008){Possenti}, {Rea}, {McLaughlin}, {Camilo},
  {Kramer}, {Burgay}, {Joshi}, and {Lyne}]{prm+08}
A.~{Possenti}, N.~{Rea}, M.~A. {McLaughlin}, F.~{Camilo}, M.~{Kramer},
  M.~{Burgay}, B.~C. {Joshi}, and A.~G. {Lyne}.
\newblock {The Very Soft X-Ray Spectrum of the Double Pulsar System
  J0737-3039}.
\newblock \emph{\apj}, 680:\penalty0 654--663, June 2008.

\bibitem[{Qiao} et~al.(2007){Qiao}, {Lee}, {Wang}, and {Xu}]{qlw+07}
G.~J. {Qiao}, K.~J. {Lee}, H.~G. {Wang}, and R.~X. {Xu}.
\newblock {A Model of Pulsar Radio and Gamma-ray Emissions}.
\newblock In Y.~W. {Kang}, H.-W. {Lee}, K.-C. {Leung}, and K.-S. {Cheng},
  editors, \emph{The Seventh Pacific Rim Conference on Stellar Astrophysics},
  volume 362 of \emph{Astronomical Society of the Pacific Conference Series},
  pages 126--+, June 2007.

\bibitem[{Radhakrishnan} and {Cooke}(1969)]{rc69b}
V.~{Radhakrishnan} and D.~J. {Cooke}.
\newblock {Magnetic Poles and the Polarization Structure of Pulsar Radiation}.
\newblock \emph{\aplett}, 3:\penalty0 225--+, 1969.

\bibitem[{Radhakrishnan} and {Manchester}(1969)]{rm69}
V.~{Radhakrishnan} and R.~N. {Manchester}.
\newblock {Detection of a change of state in the Pulsar PSR 0833-45}.
\newblock \emph{\nat}, 222:\penalty0 228--+, April 1969.

\bibitem[{Radhakrishnan} and {Srinivasan}(1982)]{rs82}
V.~{Radhakrishnan} and G.~{Srinivasan}.
\newblock {On the origin of the recently discovered ultra-rapid pulsar}.
\newblock \emph{Current Science}, 51:\penalty0 1096--1099, December 1982.

\bibitem[{Radhakrishnan} et~al.(1969){Radhakrishnan}, {Cooke}, {Komesaroff},
  and {Morris}]{rck+69}
V.~{Radhakrishnan}, D.~J. {Cooke}, M.~M. {Komesaroff}, and D.~{Morris}.
\newblock {Evidence in Support of a Rotational Model for the Pulsar PSR
  0833-45}.
\newblock \emph{\nat}, 221:\penalty0 443--+, 1969.

\bibitem[{Rafikov} and {Lai}(2006)]{rl06b}
R.~R. {Rafikov} and D.~{Lai}.
\newblock {Effects of Pulsar Rotation on Timing Measurements of the Double
  Pulsar System J0737-3039}.
\newblock \emph{\apj}, 641:\penalty0 438--446, April 2006.

\bibitem[{Ransom} et~al.(2002){Ransom}, {Eikenberry}, and {Middleditch}]{rem02}
S.~M. {Ransom}, S.~S. {Eikenberry}, and J.~{Middleditch}.
\newblock {Fourier Techniques for Very Long Astrophysical Time-Series
  Analysis}.
\newblock \emph{\aj}, 124:\penalty0 1788--1809, September 2002.
\newblock {\texttt{PRESTO} is freely available at
  \url{http://www.cv.nrao.edu/~sransom/presto/}}.

\bibitem[{Ransom} et~al.(2004){Ransom}, {Kaspi}, {Ramachandran}, {Demorest},
  {Backer}, {Pfahl}, {Ghigo}, and {Kaplan}]{rkr+04}
S.~M. {Ransom}, V.~M. {Kaspi}, R.~{Ramachandran}, P.~{Demorest}, D.~C.
  {Backer}, E.~D. {Pfahl}, F.~D. {Ghigo}, and D.~L. {Kaplan}.
\newblock {Green Bank Telescope Measurement of the Systemic Velocity of the
  Double Pulsar Binary J0737-3039 and Implications for Its Formation}.
\newblock \emph{\apjl}, 609:\penalty0 L71--L74, July 2004.

\bibitem[{Ransom} et~al.(2005){Ransom}, {Hessels}, {Stairs}, {Freire},
  {Camilo}, {Kaspi}, and {Kaplan}]{rhs+05}
S.~M. {Ransom}, J.~W.~T. {Hessels}, I.~H. {Stairs}, P.~C.~C. {Freire},
  F.~{Camilo}, V.~M. {Kaspi}, and D.~L. {Kaplan}.
\newblock {Twenty-One Millisecond Pulsars in Terzan 5 Using the Green Bank
  Telescope}.
\newblock \emph{Science}, 307:\penalty0 892--896, February 2005.

\bibitem[{Ransom} et~al.(2008){Ransom}, {Roberts}, {Hessels}, {Livingstone},
  {Crawford}, {Tam}, and {Kaspi}]{rrh+08}
S.~M. {Ransom}, M.~S.~E {Roberts}, J.~W.~T. {Hessels}, M.~{Livingstone},
  F.~{Crawford}, C.~{Tam}, and V.~M. {Kaspi}, 2008.

\bibitem[{Rappaport} et~al.(1995){Rappaport}, {Podsiadlowski}, {Joss}, {Di
  Stefano}, and {Han}]{rpj+95}
S.~{Rappaport}, P.~{Podsiadlowski}, P.~C. {Joss}, R.~{Di Stefano}, and
  Z.~{Han}.
\newblock {The relation between white dwarf mass and orbital period in wide
  binary radio pulsars}.
\newblock \emph{\mnras}, 273:\penalty0 731--741, April 1995.

\bibitem[{Rea} et~al.(2003){Rea}, {Israel}, {Stella}, {Oosterbroek},
  {Mereghetti}, {Angelini}, {Campana}, and {Covino}]{ris+03}
N.~{Rea}, G.~L. {Israel}, L.~{Stella}, T.~{Oosterbroek}, S.~{Mereghetti},
  L.~{Angelini}, S.~{Campana}, and S.~{Covino}.
\newblock {Evidence of a Cyclotron Feature in the Spectrum of the Anomalous
  X-Ray Pulsar 1RXS J170849-400910}.
\newblock \emph{\apjl}, 586:\penalty0 L65--L69, March 2003.

\bibitem[{Richards}(1968)]{ric68}
D.~W. {Richards}.
\newblock {Np 0532.}
\newblock \emph{\iaucirc}, 2114:\penalty0 1--+, 1968.

\bibitem[{Richards} and {Comella}(1969)]{rc69}
D.~W. {Richards} and J.~M. {Comella}.
\newblock {The Period of Pulsar NP0532}.
\newblock \emph{\nat}, 222:\penalty0 551--+, May 1969.

\bibitem[{Rickett}(1970)]{ric70}
B.~J. {Rickett}.
\newblock {Interstellar scintillation and pulsar intensity variations}.
\newblock \emph{\mnras}, 150:\penalty0 67--+, 1970.

\bibitem[{Rieke} and {Lebofsky}(1985)]{rl85}
G.~H. {Rieke} and M.~J. {Lebofsky}.
\newblock {The interstellar extinction law from 1 to 13 microns}.
\newblock \emph{\apj}, 288:\penalty0 618--621, January 1985.

\bibitem[{Rigaut} et~al.(1998){Rigaut}, {Salmon}, {Arsenault}, {Thomas}, {Lai},
  {Rouan}, {V{\'e}ran}, {Gigan}, {Crampton}, {Fletcher}, {Stilburn}, {Boyer},
  and {Jagourel}]{rsa+98}
F.~{Rigaut}, D.~{Salmon}, R.~{Arsenault}, J.~{Thomas}, O.~{Lai}, D.~{Rouan},
  J.~P. {V{\'e}ran}, P.~{Gigan}, D.~{Crampton}, J.~M. {Fletcher}, et~al.
\newblock {Performance of the Canada-France-Hawaii Telescope Adaptive Optics
  Bonnette}.
\newblock \emph{\pasp}, 110:\penalty0 152--164, February 1998.

\bibitem[{Ruderman} and {Sutherland}(1975)]{rs75}
M.~A. {Ruderman} and P.~G. {Sutherland}.
\newblock {Theory of pulsars - Polar caps, sparks, and coherent microwave
  radiation}.
\newblock \emph{\apj}, 196:\penalty0 51--72, February 1975.

\bibitem[{Safi-Harb}(2008)]{saf08}
S.~{Safi-Harb}.
\newblock {An X-ray View of the High Magnetic Field Radio Pulsar J1119-6127:
  Any Link to Magnetars?}
\newblock In C.~{Bassa}, Z.~{Wang}, A.~{Cumming}, and V.~M. {Kaspi}, editors,
  \emph{40 Years of Pulsars: Millisecond Pulsars, Magnetars and More}, volume
  983 of \emph{American Institute of Physics Conference Series}, pages
  213--215, February 2008.

\bibitem[{Schisler}(2008)]{sch08}
C.~{Schisler}.
\newblock {An Independent 1967 Discovery of Pulsars}.
\newblock In C.~{Bassa}, Z.~{Wang}, A.~{Cumming}, and V.~M. {Kaspi}, editors,
  \emph{40 Years of Pulsars: Millisecond Pulsars, Magnetars and More}, volume
  983 of \emph{American Institute of Physics Conference Series}, pages
  642--645, February 2008.

\bibitem[Schwarzschild(1916{\natexlab{a}})]{sch16a}
K.~Schwarzschild.
\newblock \"uber das gravitationsfeld eines massenpunktes nach der
  einsteinschen theorie.
\newblock \emph{Sitzungsber. K. Preuss. Akad. Wiss., Phys.-Math. Kl.},
  1916\penalty0 (XII):\penalty0 189--196, 1916{\natexlab{a}}.

\bibitem[Schwarzschild(1916{\natexlab{b}})]{sch16b}
K.~Schwarzschild.
\newblock \"uber das gravitationsfeld einer kugel aus inkompressibler
  fl\"ussigkeit nach der einsteinschen theorie.
\newblock \emph{Sitzungsber. K. Preuss. Akad. Wiss., Phys.-Math. Kl.},
  1916\penalty0 (III):\penalty0 424--434, 1916{\natexlab{b}}.

\bibitem[{Shklovsky}(1967)]{shk07}
I.~S. {Shklovsky}.
\newblock {On the Nature of the Source of X-Ray Emission of SCO XR-1.}
\newblock \emph{\apjl}, 148:\penalty0 L1+, April 1967.

\bibitem[{Skrutskie} et~al.(2006){Skrutskie}, {Cutri}, {Stiening}, {Weinberg},
  {Schneider}, {Carpenter}, {Beichman}, {Capps}, {Chester}, {Elias}, {Huchra},
  {Liebert}, {Lonsdale}, {Monet}, {Price}, {Seitzer}, {Jarrett}, {Kirkpatrick},
  {Gizis}, {Howard}, {Evans}, {Fowler}, {Fullmer}, {Hurt}, {Light}, {Kopan},
  {Marsh}, {McCallon}, {Tam}, {Van Dyk}, and {Wheelock}]{scs+06}
M.~F. {Skrutskie}, R.~M. {Cutri}, R.~{Stiening}, M.~D. {Weinberg},
  S.~{Schneider}, J.~M. {Carpenter}, C.~{Beichman}, R.~{Capps}, T.~{Chester},
  J.~{Elias}, et~al.
\newblock {The Two Micron All Sky Survey (2MASS)}.
\newblock \emph{\aj}, 131:\penalty0 1163--1183, February 2006.

\bibitem[{Slane}(2008)]{sla08}
P.~{Slane}.
\newblock {Recent Progress in Studies of Pulsar Wind Nebulae}.
\newblock In Y.-F. {Yuan}, X.-D. {Li}, and D.~{Lai}, editors,
  \emph{Astrophysics of Compact Objects}, volume 968 of \emph{American
  Institute of Physics Conference Series}, pages 143--150, January 2008.

\bibitem[{Stairs}(2003)]{sta03a}
I.~H. {Stairs}.
\newblock {Testing General Relativity with Pulsar Timing}.
\newblock \emph{Living Reviews in Relativity}, 6:\penalty0 5--+, September
  2003.

\bibitem[{Stairs}(2004)]{sta04b}
I.~H. {Stairs}.
\newblock {Pulsars in Binary Systems: Probing Binary Stellar Evolution and
  General Relativity}.
\newblock \emph{Science}, 304:\penalty0 547--552, April 2004.

\bibitem[{Stairs} et~al.(2004){Stairs}, {Thorsett}, and {Arzoumanian}]{sta04a}
I.~H. {Stairs}, S.~E. {Thorsett}, and Z.~{Arzoumanian}.
\newblock {Measurement of Gravitational Spin-Orbit Coupling in a Binary-Pulsar
  System}.
\newblock \emph{Physical Review Letters}, 93\penalty0 (14):\penalty0 141101--+,
  September 2004.

\bibitem[{Stairs} et~al.(2006){Stairs}, {Thorsett}, {Dewey}, {Kramer}, and
  {McPhee}]{std+06}
I.~H. {Stairs}, S.~E. {Thorsett}, R.~J. {Dewey}, M.~{Kramer}, and C.~A.
  {McPhee}.
\newblock {The formation of the double pulsar PSR J0737-3039A/B}.
\newblock \emph{\mnras}, 373:\penalty0 L50--L54, November 2006.

\bibitem[{Stappers} et~al.(1996){Stappers}, {Bessell}, and {Bailes}]{sbb96}
B.~W. {Stappers}, M.~S. {Bessell}, and M.~{Bailes}.
\newblock {Detection of an Irradiated Pulsar Companion}.
\newblock \emph{\apjl}, 473:\penalty0 L119+, December 1996.

\bibitem[{Staubert}(2003)]{sta03b}
R.~{Staubert}.
\newblock {Magnetic fields of accreting X-ray pulsars}.
\newblock \emph{Chinese Journal of Astronomy and Astrophysics Supplement},
  3:\penalty0 270--280, December 2003.

\bibitem[{Stetson}(1987)]{ste87}
P.~B. {Stetson}.
\newblock {DAOPHOT - A computer program for crowded-field stellar photometry}.
\newblock \emph{\pasp}, 99:\penalty0 191--222, March 1987.

\bibitem[{Sturrock}(1971)]{stu71}
P.~A. {Sturrock}.
\newblock {A Model of Pulsars}.
\newblock \emph{\apj}, 164:\penalty0 529--+, March 1971.

\bibitem[{Sutantyo} and {Li}(2000)]{sl00}
W.~{Sutantyo} and X.-D. {Li}.
\newblock {Formation of binary millisecond pulsars with relatively high surface
  dipole magnetic fields}.
\newblock \emph{\aap}, 360:\penalty0 633--636, August 2000.

\bibitem[{Taam}(2004)]{taa04}
R.~E. {Taam}.
\newblock {The Formation and Evolution of Compact Stars in Binaries}.
\newblock In G.~{Tovmassian} and E.~{Sion}, editors, \emph{Revista Mexicana de
  Astronomia y Astrofisica Conference Series}, volume~20 of \emph{Revista
  Mexicana de Astronomia y Astrofisica Conference Series}, pages 81--84, July
  2004.

\bibitem[{Tauris} and {Savonije}(1999)]{ts99}
T.~M. {Tauris} and G.~J. {Savonije}.
\newblock {Formation of millisecond pulsars. I. Evolution of low-mass X-ray
  binaries with $P_{orb} > 2$ days}.
\newblock \emph{\aap}, 350:\penalty0 928--944, October 1999.

\bibitem[{Taylor} and {Weisberg}(1982)]{tw82}
J.~H. {Taylor} and J.~M. {Weisberg}.
\newblock {A new test of general relativity - Gravitational radiation and the
  binary pulsar PSR 1913+16}.
\newblock \emph{\apj}, 253:\penalty0 908--920, February 1982.

\bibitem[{Taylor} et~al.(1979){Taylor}, {Fowler}, and {McCulloch}]{tfm79}
J.~H. {Taylor}, L.~A. {Fowler}, and P.~M. {McCulloch}.
\newblock {Measurements of general relativistic effects in the binary pulsar
  PSR 1913+16}.
\newblock \emph{\nat}, 277:\penalty0 437--440, February 1979.

\bibitem[{Thompson} and {Duncan}(1993)]{td93}
C.~{Thompson} and R.~C. {Duncan}.
\newblock {Neutron star dynamos and the origins of pulsar magnetism}.
\newblock \emph{\apj}, 408:\penalty0 194--217, May 1993.

\bibitem[{Thompson} and {Duncan}(1995)]{td95}
C.~{Thompson} and R.~C. {Duncan}.
\newblock {The soft gamma repeaters as very strongly magnetized neutron stars -
  I. Radiative mechanism for outbursts}.
\newblock \emph{\mnras}, 275:\penalty0 255--300, July 1995.

\bibitem[{Thompson} et~al.(2002){Thompson}, {Lyutikov}, and {Kulkarni}]{tlk02}
C.~{Thompson}, M.~{Lyutikov}, and S.~R. {Kulkarni}.
\newblock {Electrodynamics of Magnetars: Implications for the Persistent X-Ray
  Emission and Spin-down of the Soft Gamma Repeaters and Anomalous X-Ray
  Pulsars}.
\newblock \emph{\apj}, 574:\penalty0 332--355, July 2002.

\bibitem[{Thorsett} and {Chakrabarty}(1999)]{tc99}
S.~E. {Thorsett} and D.~{Chakrabarty}.
\newblock {Neutron Star Mass Measurements. I. Radio Pulsars}.
\newblock \emph{\apj}, 512:\penalty0 288--299, February 1999.

\bibitem[Tolman(1939)]{tol39}
Richard~C. Tolman.
\newblock Static solutions of einstein's field equations for spheres of fluid.
\newblock \emph{Phys. Rev.}, 55\penalty0 (4):\penalty0 364--373, Feb 1939.

\bibitem[{van den Heuvel}(2007)]{vdh07}
E.~P.~J. {van den Heuvel}.
\newblock {Double Neutron Stars: Evidence For Two Different Neutron-Star
  Formation Mechanisms}.
\newblock In T.~{di Salvo}, G.~L. {Israel}, L.~{Piersant}, L.~{Burderi},
  G.~{Matt}, A.~{Tornambe}, and M.~T. {Menna}, editors, \emph{The Multicolored
  Landscape of Compact Objects and Their Explosive Origins}, volume 924 of
  \emph{American Institute of Physics Conference Series}, pages 598--606,
  August 2007.

\bibitem[{van den Heuvel}(1995)]{vdh95}
E.~P.~J. {van den Heuvel}.
\newblock {Scenarios for the Formation of Binary and Millisecond Pulsars - a
  Critical Assessment}.
\newblock \emph{Journal of Astrophysics and Astronomy}, 16:\penalty0 255--+,
  June 1995.

\bibitem[{van den Heuvel} and {van Paradijs}(1988)]{vv88}
E.~P.~J. {van den Heuvel} and J.~{van Paradijs}.
\newblock {Fate of the companion stars of ultra-rapid pulsars}.
\newblock \emph{\nat}, 334:\penalty0 227--+, July 1988.

\bibitem[{van der Klis}(2006)]{vdk06}
M.~{van der Klis}.
\newblock {Overview of QPOs in neutron-star low-mass X-ray binaries}.
\newblock \emph{Advances in Space Research}, 38:\penalty0 2675--2679, 2006.

\bibitem[{van der Sluys}(2006)]{vds06}
M.~{van der Sluys}.
\newblock \emph{{Formation and evolution of compact binaries}}.
\newblock PhD thesis, Utrecht University, 2006.

\bibitem[{van Kerkwijk} et~al.(2000){van Kerkwijk}, {Kaspi}, {Klemola},
  {Kulkarni}, {Lyne}, and {Van Buren}]{vkk+00}
M.~H. {van Kerkwijk}, V.~M. {Kaspi}, A.~R. {Klemola}, S.~R. {Kulkarni}, A.~G.
  {Lyne}, and D.~{Van Buren}.
\newblock {Optical Observations of the Binary Pulsar System PSR B1718-19:
  Implications for Tidal Circularization}.
\newblock \emph{\apj}, 529:\penalty0 428--434, January 2000.

\bibitem[{van Kerkwijk} et~al.(2005){van Kerkwijk}, {Bassa}, {Jacoby}, and
  {Jonker}]{vbj+05}
M.~H. {van Kerkwijk}, C.~G. {Bassa}, B.~A. {Jacoby}, and P.~G. {Jonker}.
\newblock {Optical Studies of Companions to Millisecond Pulsars}.
\newblock In F.~A. {Rasio} and I.~H. {Stairs}, editors, \emph{Binary Radio
  Pulsars}, volume 328 of \emph{Astronomical Society of the Pacific Conference
  Series}, pages 357--+, July 2005.

\bibitem[{van Paradijs} et~al.(1988){van Paradijs}, {Allington-Smith},
  {Callanan}, {Hassall}, and {Charles}]{vac+88}
J.~{van Paradijs}, J.~{Allington-Smith}, P.~{Callanan}, B.~J.~M. {Hassall}, and
  P.~A. {Charles}.
\newblock {Optical observations of the eclipsing binary radio pulsar PSR1957 +
  20}.
\newblock \emph{\nat}, 334:\penalty0 684--686, August 1988.

\bibitem[{Verbunt}(1993)]{ver93}
F.~{Verbunt}.
\newblock {Origin and evolution of X-ray binaries and binary radio pulsars}.
\newblock \emph{\araa}, 31:\penalty0 93--127, 1993.

\bibitem[{Volpi} et~al.(2007){Volpi}, {Del Zanna}, {Amato}, and
  {Bucciantini}]{vda+07}
D.~{Volpi}, L.~{Del Zanna}, E.~{Amato}, and N.~{Bucciantini}.
\newblock {Synchrotron emission simulated maps: Pulsar Wind Nebulae.}
\newblock \emph{Memorie della Societa Astronomica Italiana}, 78:\penalty0
  662--+, 2007.

\bibitem[{Webb} and {Barret}(2007)]{wb07}
N.~A. {Webb} and D.~{Barret}.
\newblock {Constraining the Equation of State of Supranuclear Dense Matter from
  XMM-Newton Observations of Neutron Stars in Globular Clusters}.
\newblock \emph{\apj}, 671:\penalty0 727--733, December 2007.

\bibitem[{Weber} et~al.(2007){Weber}, {Negreiros}, {Rosenfield}, and
  {Stejner}]{wnr07}
F.~{Weber}, R.~{Negreiros}, P.~{Rosenfield}, and M.~{Stejner}.
\newblock {Pulsars as astrophysical laboratories for nuclear and particle
  physics}.
\newblock \emph{Progress in Particle and Nuclear Physics}, 59:\penalty0
  94--113, July 2007.

\bibitem[{Weisberg} and {Taylor}(2005)]{wt05}
J.~M. {Weisberg} and J.~H. {Taylor}.
\newblock {The Relativistic Binary Pulsar B1913+16: Thirty Years of
  Observations and Analysis}.
\newblock In F.~A. {Rasio} and I.~H. {Stairs}, editors, \emph{Binary Radio
  Pulsars}, volume 328 of \emph{Astronomical Society of the Pacific Conference
  Series}, pages 25--+, July 2005.

\bibitem[{Weisberg} et~al.(1989){Weisberg}, {Romani}, and {Taylor}]{wrt89}
J.~M. {Weisberg}, R.~W. {Romani}, and J.~H. {Taylor}.
\newblock {Evidence for geodetic spin precession in the binary pulsar 1913 +
  16}.
\newblock \emph{\apj}, 347:\penalty0 1030--1033, December 1989.

\bibitem[{Weltevrede} et~al.(2006){Weltevrede}, {Stappers}, {Rankin}, and
  {Wright}]{wsr+06}
P.~{Weltevrede}, B.~W. {Stappers}, J.~M. {Rankin}, and G.~A.~E. {Wright}.
\newblock {Is Pulsar B0656+14 a Very Nearby Rotating Radio Transient?}
\newblock \emph{\apjl}, 645:\penalty0 L149--L152, July 2006.

\bibitem[Whitaker(1996)]{whi96}
Jerry~C. Whitaker, editor.
\newblock \emph{The Electronics Handbook}.
\newblock CRC Press, Boca Raton, 1996.
\newblock ISBN 0-8493-8345-5.

\bibitem[{Wijers}(1997)]{wij97}
R.~A.~M.~J. {Wijers}.
\newblock {Evidence against field decay proportional to accreted mass in
  neutron stars}.
\newblock \emph{\mnras}, 287:\penalty0 607--614, May 1997.

\bibitem[{Will}(2001)]{wil01}
C.~{Will}.
\newblock {The Confrontation between General Relativity and Experiment}.
\newblock \emph{Living Reviews in Relativity}, 4:\penalty0 4--+, May 2001.

\bibitem[{Will}(1993)]{wil93}
C.~M. {Will}.
\newblock \emph{{Theory and Experiment in Gravitational Physics}}.
\newblock Cambridge University Press, Cambridge, UK, March 1993.

\bibitem[{Wilson} and {Fishman}(1983)]{wf83}
R.~B. {Wilson} and G.~J. {Fishman}.
\newblock {The pulse profile of the Crab pulsar in the energy range 45 keV-1.2
  MeV}.
\newblock \emph{\apj}, 269:\penalty0 273--280, June 1983.

\bibitem[{Wolszczan} and {Frail}(1992)]{wf92}
A.~{Wolszczan} and D.~A. {Frail}.
\newblock {A planetary system around the millisecond pulsar PSR1257 + 12}.
\newblock \emph{\nat}, 355:\penalty0 145--147, January 1992.

\bibitem[{Woods} and {Thompson}(2006)]{wt06}
P.~M. {Woods} and C.~{Thompson}.
\newblock \emph{{Soft gamma repeaters and anomalous X-ray pulsars: magnetar
  candidates}}, pages 547--586.
\newblock Compact stellar X-ray sources, April 2006.

\bibitem[{Yakovlev} and {Pethick}(2004)]{yp04}
D.~G. {Yakovlev} and C.~J. {Pethick}.
\newblock {Neutron Star Cooling}.
\newblock \emph{\araa}, 42:\penalty0 169--210, September 2004.

\bibitem[{Young} et~al.(1999){Young}, {Manchester}, and {Johnston}]{ymj99}
M.~D. {Young}, R.~N. {Manchester}, and S.~{Johnston}.
\newblock {A radio pulsar with an 8.5-second period that challenges emission
  models}.
\newblock \emph{\nat}, 400:\penalty0 848--849, August 1999.

\bibitem[{Zavlin} et~al.(1996){Zavlin}, {Pavlov}, and {Shibanov}]{zps96}
V.~E. {Zavlin}, G.~G. {Pavlov}, and Y.~A. {Shibanov}.
\newblock {Model neutron star atmospheres with low magnetic fields. I.
  Atmospheres in radiative equilibrium.}
\newblock \emph{\aap}, 315:\penalty0 141--152, November 1996.

\bibitem[{Zhang} et~al.(2007){Zhang}, {Yin}, {Kojima}, {Chang}, {Xu}, {Li},
  {Zhang}, and {Kiziltan}]{zyk+07}
C.~M. {Zhang}, H.~X. {Yin}, Y.~{Kojima}, H.~K. {Chang}, R.~X. {Xu}, X.~D. {Li},
  B.~{Zhang}, and B.~{Kiziltan}.
\newblock {Measuring neutron star mass and radius with three mass-radius
  relations}.
\newblock \emph{\mnras}, 374:\penalty0 232--236, January 2007.

\end{thebibliography}

\end{document}